\documentclass[aps,prd,onecolumn,groupedaddress,showpacs,nofootinbib,amssymb
]{revtex4}
\usepackage[dvips]{graphicx}
\usepackage{amssymb}
\usepackage{amsmath}
\usepackage{graphicx,,color}
\usepackage{amsfonts}
\usepackage{bm}

\allowdisplaybreaks[4]

\begin{document}

\title{Modified Gravity Theories on a Nutshell: Inflation, Bounce
and Late-time Evolution}
\author{S.~Nojiri,$^{1,2,3}$\,\thanks{nojiri@gravity.phys.nagoya-u.ac.jp}
S.~D.~Odintsov,$^{4,5}$\,\thanks{odintsov@ieec.uab.es}
V.K.~Oikonomou,$^{6,7}$\,\thanks{v.k.oikonomou1979@gmail.com}}

\affiliation{$^{1)}$ Department of Physics, Nagoya University,
Nagoya 464-8602, Japan \\
$^{2)}$ Kobayashi-Maskawa Institute for the Origin of Particles and
the Universe, Nagoya University, Nagoya 464-8602, Japan \\
$^{3)}$KEK Theory Center, High Energy Accelerator Research
Organization (KEK), Oho 1-1, Tsukuba, Ibaraki 305-0801, Japan \\
$^{4)}$Institut de Ciencies de lEspai (IEEC-CSIC),
Carrer de Can Magrans s/n, 08193 Barcelona, SPAIN \\
$^{5)}$ ICREA, Passeig Companys, 23,
08010 Barcelona, Spain \\
$^{6)}$ Laboratory for Theoretical Cosmology, Tomsk State University
of Control Systems and Radioelectronics (TUSUR), 634050 Tomsk, Russia \\
$^{7)}$ Tomsk State Pedagogical University, 634061 Tomsk, Russia \\
}

\tolerance=5000

\begin{abstract}
We systematically review some standard issues and also the latest
developments of modified gravity in cosmology, emphasizing on
inflation, bouncing cosmology and late-time acceleration era.
Particularly, we present the formalism of standard modified gravity
theory representatives, like $F(R)$, $F(\mathcal{G})$ and $F(T)$
gravity theories, but also several alternative theoretical proposals
which appeared in the literature during the last decade. We
emphasize on the formalism developed for these theories and we
explain how these theories can be considered as viable descriptions
for our Universe. Using these theories, we present how a viable
inflationary era can be produced in the context of these theories,
with the viability being justified if compatibility with the latest
observational data is achieved. Also we demonstrate how bouncing
cosmologies can actually be described by these theories. Moreover,
we systematically discuss several qualitative features of the dark
energy era by using the modified gravity formalism, and also we
critically discuss how a unified description of inflation with dark
energy era can be described by solely using the modified gravity
framework. Finally, we also discuss some astrophysical solutions in
the context of modified gravity, and several qualitative features of
these solutions. The aim of this review is to gather the different
modified gravity techniques and form a virtual modified gravity
``toolbox'', which will contain all the necessary information on
inflation, dark energy and bouncing cosmologies in the context of
the various forms of modified gravity.
\end{abstract}

\pacs{04.50.Kd, 95.36.+x, 98.80.-k, 98.80.Cq,11.25.-w}

\maketitle

\def\pp{{\, \mid \hskip -1.5mm =}}
\def\cL{\mathcal{L}}
\def\be{\begin{equation}}
\def\ee{\end{equation}}
\def\bea{\begin{eqnarray}}
\def\eea{\end{eqnarray}}
\def\tr{\mathrm{tr}\, }
\def\nn{\nonumber \\}
\def\e{\mathrm{e}}

\newpage

\tableofcontents

\section{Introduction}

With this work we shall try to provide a concise pedagogical review
on the latest developments in modified gravity aspects of inflation,
dark energy and bouncing cosmology. The attempt is challenging and
it is conceivable that it is impossible to cover all the different
approaches in the field, since the number of these is vast. Hence we
focus on modified gravity aspects of early, late-time acceleration
and bounce dynamics and we shall try to provide a pedagogical text
accessible by non-experts but also useful to experts.

One of the most fundamental questions in modern theoretical
cosmology is, whether the genesis of the Universe was singular or
non-singular. This question is equivalent in asking if the Big Bang
theory or the Big Bounce theory actually describes the evolution of
our Universe. Naturally thinking, the initial singularity described
by the Big Bang theory, is a mentally more convenient description,
since we can easily imagine a zero sized Universe, with infinite
temperature and energy density, and also in which all fundamental
interactions are unified under the yet unknown same theoretical
framework. However, no one can actually exclude a cyclic
cosmological evolution, in which the Universe never shrinks to zero.
Actually the latter perspective seems to be supported by quantum
cosmologies, as we discuss later on. One of the main purposes of
this review article is to present the tools to study these two types
of cosmological scenarios, in the context of modified gravity.

During the last 25 years, the cosmologists community has experienced
great surprises, since the observations indicated in the late 90's
\cite{Riess:1998cb} that the Universe is expanding, but in an
accelerating way. This observation utterly changed the cosmologists
way of thinking and also it revived theories containing a
cosmological constant. It was Einstein that had firstly proposed a
theory of cosmological evolution with a cosmological constant, and
later on he admitted that this was the greatest scientific mistake
of his life. However, it seems that nature can actually incorporate
such a cosmological constant, so it seems eventually that Einstein
was not wrong to some extent. We need to note that finding the
correct model for the late-time acceleration is not an easy task at
all. The late-time acceleration era is known as the dark energy era
\cite{Carroll:2003wy,Peebles:2002gy,Durrer:2008in,Caldwell:2009ix,Caldwell:2009zzb,
Li:2011sd,Bamba:2012cp,Lobo:2008sg,Gubitosi:2012hu,Bloomfield:2012ff,
Gleyzes:2013ooa,Li:2012dt,Sami:2009jx,Balakin:2016cbe} (see also
\cite{Weinberg:2002rd,Perivolaropoulos:2006ce,Dobado:2008xn,Dunsby:2015ers}),
and up to date there are many proposals that try to model this era,
with other using a scalar field, known as quintessence models
\cite{Zlatev:1998tr,Carroll:1998zi,Wang:1998gt,Chiba:1999wt,Barreiro:1999zs,
Chiba:1999ka,Haiman:2000bw,Capozziello:2002rd,Sahni:2002kh,
Capozziello:2003gx,Hao:2004ky,Bassett:2004wz,Perivolaropoulos:2004yr,
Zhang:2005rg,Olivares:2005tb,Guo:2005ata,
Caldwell:2005tm,Scherrer:2007pu,Creminelli:2008wc}, while other
models use modified gravity in its various forms. With regards to
quintessence, there exist various alternative proposals on this
topic, see for example
\cite{Capozziello:2003tk,Kamenshchik:2001cp,Banerjee:2000mj,
Sahni:1999qe,Perrotta:1999am}, and also approaches with non-minimal
coupling, see
\cite{Faraoni:2000wk,Torres:2002pe,Pettorino:2008ez,Bertolami:1999dp}.
 This field of research is still developing rapidly up to date.

Apart from the late-time acceleration era, the Universe experienced
another acceleration era during its first stages of evolution, and
the latest observational data have actually constrained this era
\cite{Ade:2015lrj,Array:2015xqh}, which is known as inflation
\cite{Linde:2007fr,Gorbunov:2011zzc,Lyth:1998xn,
Linde:1983gd,Linde:1985ub,Albrecht:1982wi,Linde:1993cn,Sasaki:1995aw,Turok:2002yq,
Linde:2005dd,Kachru:2003sx,
Brandenberger:2016uzh,Bamba:2015uma,Martin:2013tda,
Martin:2013nzq,Baumann:2014nda,Baumann:2009ds,Linde:2014nna,
Pajer:2013fsa,Yamaguchi:2011kg,Byrnes:2010em}. During the
inflationary era, the Universe increased its size at an exponential
rate, so it expanded quite quickly and gained a large size in a
relatively small amount of time. From a theoretical point of view,
the inflationary description of the early time Universe, solved
quite many theoretical problems that the original Big-Bang theory
had, for example the horizon problem etc. The inflationary evolution
had originally two different forms, which now are known as the old
inflationary scenario \cite{Guth:1980zm}, and the new inflationary
scenario \cite{Linde:1983gd,Albrecht:1982wi}. From the time that the
new inflationary scenario was introduced, many models where proposed
that could actually describe in a successful way the inflationary
era, with many of the models using a scalar field or multiple scalar
fields, see \cite{Byrnes:2010em} for details. Also for supergravity
models of inflation, see the review \cite{Yamaguchi:2011kg}, and for
string cosmology consult Refs.~\cite{Linde:2005dd,Kachru:2003sx}.
There also exist alternative scenarios for inflation, involving the
axion field, see for example \cite{Pajer:2013fsa} for a review. In
addition for inflationary theories in the context of modified
gravity in its various forms, see \cite{Bamba:2015uma}.


However, a successful model at present time (2017), has to be
confronted with the observational data coming from the Planck
satellite, which severely constraints the inflationary era
\cite{Ade:2015lrj}. The Planck data have excluded many inflationary
models from the viable inflationary models list, and nowadays these
data are a benchmark with regard to the viability of an inflationary
model. Thus, it is compelling that a model has to be compatible with
the Planck data, before it can be considered as a candidate for
inflation. Apart from the scalar-tensor models of inflation, there
are alternative proposals in the literature that use modified
gravity in its various forms
\cite{Bamba:2015uma,Nojiri:2013zza,Nojiri:2006ri,Capozziello:2011et,
Capozziello:2010zz,Capozziello:2009nq,Nojiri:2010wj,Clifton:2011jh},
in order to describe this early-time era. We need to note that in
this review we shall adopt the metric modified gravity formalism,
but there also exists the Palatini formalism, for which we refer the
reader to
Refs.~\cite{Ferraris:1992dx,Meng:2003en,Amarzguioui:2005zq,Flanagan:2003rb,
Koivisto:2005yc,Capozziello:2012ny,Capozziello:2013uya,Olmo:2011uz,
Olmo:2009xy,Harko:2011nh,Makarenko:2014lxa}.

Most of the metric modified gravity descriptions provide a
consistent framework for the description of the early-time
acceleration, and also the compatibility of the model with the data
can be achieved in many of these models. Moreover, each modification
of Einstein's theory of general relativity eventually is confronted
with the successes of general relativity. Hence all the constraints
on modified gravity imposed by local astrophysical data but also
from global constraints, have to be satisfied to an adequate level.
Therefore it is conceivable that a modified gravity has many
challenges and obstacles to overcome in order to be considered as a
viable cosmological theory. Apart from the constraints, the ultimate
goal in modified gravity is to offer a self-consistent theoretical
framework in the context of which the early-time and late-time
acceleration will be described by the same theory. In this direction
there are many works using various theoretical contexts, that
provide such a framework
\cite{Capozziello:2005tf,Nojiri:2005pu,Carter:2005fu,Liddle:2006qz,
Chen:2006qy,Nojiri:2007as,Appleby:2007vb,Nojiri:2007cq,Cognola:2007zu,Cognola:2008zp,
Koivisto:2009fb,Koivisto:2008xf,Xia:2007me,Liddle:2008bm,Elizalde:2009gx,Elizalde:2010ep,Makarenko:2014nca,
deHaro:2016hsh,Beltran:2015hja}. Also dark energy
\cite{Elizalde:2007kb,Nojiri:2007te,Huterer:2006mva,Nojiri:2006jy,
Capozziello:2005ku,Borowiec:2006qr,Nojiri:2005am,Szydlowski:2006ay,
Borowiec:2006hk,Allemandi:2004wn} and other aspects of cosmological
evolution and cosmological implications
\cite{Easson:2004fq,Carloni:2004kp,Clifton:2005aj,Sanyal:2006wi,Appleby:2008tv,Capozziello:2008qc,Evans:2007ch,Capozziello:2007ec,
Li:2007jm,Bertolami:2007gv,Li:2007xn,Song:2006ej,Arbuzova:2011fu}
are also explained in an appealing way by modified gravity. In
Refs.~\cite{Nojiri:2007as,Appleby:2007vb,Nojiri:2007cq,
Cognola:2007zu,Artymowski:2014gea,Fay:2007uy} the unification of
$\Lambda$-Cold Dark Matter model ($\Lambda$CDM hereafter) with
inflation was developed, in the context of $F(R)$ gravity, while in
Refs.~\cite{Capozziello:2005tf,Nojiri:2005pu}, the same issue was
addressed in the context of phantom cosmology. In principle the
possibility of phantom inflation always exists in various
cosmological contexts
\cite{Yurov:2003zt,Piao:2004tq,GonzalezDiaz:2004df,Izumi:2010wm,
Feng:2010ya,Liu:2012iba}.

The first unified description of the inflation with dark energy in
modified gravity was proposed in Ref.~\cite{Nojiri:2003ft}, while in
Ref.~\cite{Liddle:2006qz} the unified description of dark energy and
inflation was developed in the context of string theory, and in
\cite{Nojiri:2005pu,Chen:2006qy} a holographic approach to the same
issue was performed. Also in Refs.~\cite{Xia:2007me} the problem
under discussion was addressed in the context of a dynamical
involving dark energy component, and in \cite{Elizalde:2010ep} the
problem was addressed in the context of modified $F(R)$
Horava-Lifshitz gravity. The most successful theory will be
consistent with observations, with the local and global constraints
and will describe simultaneously all or at least most of the
evolution eras of our Universe. Finally for some different scenarios
see \cite{Koivisto:2009fb,Koivisto:2008xf,Xia:2007me,Liddle:2008bm,
Rinaldi:2014yta,Nojiri:2016vhu,Borowiec:2008js,
Elizalde:2013paa,Cai:2008gk,Geng:2011aj,Vacaru:2015iga}. Also
modified gravity usually brings along various exotic features, for
example the equivalence principle can be different in comparison to
the ordinary Einstein gravity \cite{Olmo:2006zu}, or the
energy conditions may differ \cite{Santos:2007bs} and in addition
the thermodynamic interpretation is different \cite{Bamba:2016aoo},
see also \cite{Zubair:2016bpi}. But inevitably modified gravity has
eventually to confront the successes of the Einstein-Hilbert
gravity, so stringent rules are imposed from solar system tests.
Some relevant studies with regards to solar system implications and
constraints of modified gravity can be found in
Refs.~\cite{Nojiri:2007as,Olmo:2005zr,Faraoni:2006hx,Zhang:2007ne,Olmo:2006eh,
Allemandi:2006bm,Erickcek:2006vf,Lin:2010hk,
Iorio:2010tp,Olmo:2005hc}, while the Newtonian limit of $F(R)$
gravity was studied in Ref.~\cite{Capozziello:2007ms}. Also some
studies on the stability of $F(R)$ can be found for example in
\cite{Faraoni:2005vk,Faraoni:2006sy,Cognola:2007vq,Sawicki:2007tf}
and a recent study on the observational constraints on $F(R)$
gravity from cosmic chronometers can be found in
\cite{Nunes:2016drj}. Finally, we need to note that the phase space
of modified gravity theories is quite richer in geometric
structures, in comparison to the Einstein gravity phase
space, see for example \cite{Faraoni:2005vc,Carloni:2017ucm}.

An alternative description to the inflationary paradigm and to the
Big Bang cosmology, is offered by bouncing cosmologies, see Ref.
\cite{Tolman:1931zz} for the original idea of a bounce cosmology,
and for an important stream of reviews see
Refs.~\cite{Brandenberger:2012zb,Brandenberger:2016vhg,Battefeld:2014uga,Novello:2008ra,Cai:2014bea,deHaro:2015wda,Lehners:2011kr,Lehners:2008vx,
Cheung:2016wik,Cai:2016hea}. Particularly, in Ref.
\cite{Brandenberger:2012zb}, the study was focused on the matter
bounce scenario and various ways that may realize this scenario were
discussed. Also several observational signatures that may
distinguish the matter bounce from the standard inflationary
paradigm were also discussed. A more focused review on bouncing
cosmologies is Ref.~\cite{Brandenberger:2016vhg}, were the origin of
primordial perturbations was discussed in the context of bouncing
cosmologies, and also several examples that realize bounce
cosmologies, including pre-big-bang scenarios, ekpyrotic scenarios,
string gas cosmology, bouncing cosmologies from modified gravity and
string theory, were also discussed. Moreover, the observational
signatures that may distinguish a bounce from the inflationary
paradigm were examined too, like for example the existence of
non-Gaussianities. In Ref.~\cite{Battefeld:2014uga}, the bouncing
cosmologies were discussed and in addition certain potentially fatal
effects that undermine non-singular bouncing models were pointed
out. Also, the unstable growth of curvature fluctuations and the
growth of the quantum induced anisotropy, in conjunction with the
study of various gravitational instabilities, were discussed too. In
Ref.~\cite{Novello:2008ra}, the study performed covered the topics
of higher-order gravitational theories, theories with a scalar
field, bounces in the braneworld scenarios and several quantum
cosmology scenarios. Also the cyclic Universes were discussed and
the issue of perturbations in the context of bouncing Universes were
addressed too. In Ref.~\cite{Cai:2014bea}, the matter-ekpyrotic
bounce scenario was extensively studied, and also various
realizations of this scenario were presented, like for example the
two scalar field realization. Also the observational constraints
were thoroughly discussed and also several mechanisms for generating
a red tilt for primordial perturbations were presented too. In Ref.
\cite{deHaro:2015wda}, the matter bounce scenario was also
presented, and its realization was achieved by using a single scalar
field with a nearly exponential potential. The main result was that
the rolling of the scalar field leads to a running of the spectral
index, and specifically a negative running is obtained. Also
possible theories that realize such a scenario are discussed, such
as holonomy corrected loop quantum cosmology theories and also
teleparallel $F(T)$ gravity. In addition, and insightful study on
the reheating process is discussed too. In Ref.
\cite{Lehners:2008vx}, several issues concerning cyclic cosmologies
were discussed, including, the ekpyrotic phase of a bounce, how to
avoid chaos in such models, the Milne Universe and finally several
ekpyrotic models were presented. Also, the scalar and tensor
perturbations were addressed, and in addition the link of these
theories to a more fundamental theory, like heterotic M-theory was
discussed too. Also in Ref.~\cite{Cai:2016hea}, matter bounce
scenarios in which the matter content consists of dark energy and
dark matter were reviewed. Specifically, the $\Lambda$CDM bounce
scenario was discussed, and also theories with interacting dark
matter and dark energy were addressed too. Moreover, the
observational signatures that may distinguish bounces from the
inflationary paradigm were discussed, and also several theories that
may realize a bounce were also presented, including, loop quantum
cosmology, string Cosmology, $F(R)$ gravity, kinetic gravity
braiding theories and finally the Fermi bounce mechanism.
Furthermore, some interesting information for the occurrence of
bounces can be found in \cite{Cattoen:2005dx} and for a pioneer
version of the non-singular bounce in the context of modified
gravity see \cite{Mukhanov:1991zn}. The Big Bounce cosmology
\cite{Li:2014era,Brizuela:2009nk,Cai:2013kja,Quintin:2014oea} is an
appealing alternative to inflation, since the initial singularity
which haunts the Big Bang cosmology is absent, hence these
cosmologies are essentially non-singular
\cite{Cai:2013vm,Poplawski:2011jz,Koehn:2015vvy}. However, other
types of singular bounces appeared in the literature, in which case
the singularity which occurs at the bouncing point is a soft type
singularity
\cite{Odintsov:2015zza,Nojiri:2016ygo,Oikonomou:2015qha,Odintsov:2015ynk}.
In the context of bouncing cosmologies there are various scenarios
in the literature, and bouncing cosmologies are often studied in
ekpyrotic scenarios of some sort
\cite{Koehn:2013upa,Battarra:2014kga,Martin:2001ue} (see
Refs.~\cite{Khoury:2001wf,Buchbinder:2007ad} for the ekpyrotic
scenario per se). In the case of a bounce, the Universe is described
by a repeating cycle of evolution, in which initially the Universe
contracts, until a minimal radius is reached, which is called the
bouncing point. After that point, the Universe starts to expand
again and this cycle is continuously repeated. Hence bouncing
cosmologies are essentially cyclic cosmologies or equivalently
oscillating cosmologies
\cite{Brown:2004cs,Hackworth:2004xb,Nojiri:2006ww,Johnson:2011aa}.
Cosmological perturbations in bouncing cosmologies are generated
usually during the contracting phase \cite{Brandenberger:2016vhg},
however this is not always true, see for example the singular bounce
of Ref.~\cite{Odintsov:2015ynk}, where the primordial perturbations
are actually generated during the expanding phase after the Type IV
singular bouncing point. For some very relevant studies of
perturbations in bouncing cosmologies, see
Refs.~\cite{Peter:2002cn,Gasperini:2003pb,Creminelli:2004jg,Lehners:2015mra}.
In principle, a scale invariant or nearly scale invariant power
spectrum can be generated by a bounce cosmology
\cite{Brandenberger:2016vhg,Odintsov:2015ynk}, and also the recent
observational data can be consistent with cyclic cosmologies
\cite{Mielczarek:2010ga,Lehners:2013cka,Cai:2014xxa}. There are many
bouncing cosmologies in the literature and some of these scenarios
naturally occur in Loop Quantum Cosmology
\cite{Laguna:2006wr,Corichi:2007am,Bojowald:2008pu,Singh:2006im,
Date:2004fj,deHaro:2012xj,Cianfrani:2010ji,Cai:2014zga,
Mielczarek:2008zz,Mielczarek:2008zv,Diener:2014mia,Haro:2015oqa,
Zhang:2011qq,Zhang:2011vi,Cai:2014jla,WilsonEwing:2012pu}. One quite
well known bounce cosmology is the matter bounce scenario
\cite{deHaro:2015wda,Finelli:2001sr,Quintin:2014oea,Cai:2011ci,
Haro:2015zta,Cai:2011zx,Cai:2013kja,
Haro:2014wha,Brandenberger:2009yt,deHaro:2014kxa,Odintsov:2014gea,
Qiu:2010ch,Oikonomou:2014jua,Bamba:2012ka,deHaro:2012xj,
WilsonEwing:2012pu}, which is known to provide a scale invariant
spectrum during the contracting phase, see for example
\cite{Brandenberger:2016vhg}. Also for some alternative scenarios in
the context of cosmological bounces, see the informative
Refs.~\cite{Cai:2007qw,Cai:2010zma,Avelino:2012ue,Barrow:2004ad,
Haro:2015zda,Elizalde:2014uba}. An interesting bouncing cosmology
scenario appeared in \cite{Cai:2007qw}, called quintom scenario, see
Ref.~\cite{Cai:2009zp} for a comprehensive review on quintom
cosmology. The quintom scenario is highly motivated by the current
observations which indicate that the dark energy equation of state
crosses the phantom divide line. In order that the quintom scenario
is realized two scalar fields are required, since a no-go theorem
forbids the single scalar field realization of the quintom scenario
\cite{Cai:2009zp}. In the two scalar field realization of the
quintom scenario, one scalar field is quintessential and the other
scalar field is a phantom one, and the drawback of these theories is
the existence of ghost degrees of freedom. Also the quintom scenario
may be obtained from string theory motivated higher derivative
scalar field theories and from braneworld scenarios
\cite{Cai:2009zp}. The quintom bounce can be realized if the null
energy condition is violated, and as we already mentioned, the no-go
theorem in quintom cosmology makes compelling to use two scalar
fields. One important feature of the two scalar field realization of
the quintom bounce is the fact that for each scalar component, the
effective equation of state needs not to cross the phantom divide
line, and thus the classical perturbations remain stable.

Some bouncing cosmologies scenarios have been proposed to describe
the pre-inflationary era \cite{Cai:2015nya,Piao:2003zm,Piao:2005ag},
and thus in these scenarios the inflationary paradigm is combined
with cosmological bounces. So bouncing cosmologies can produce an
exact scale invariant power spectrum of primordial curvature
perturbations, for example the matter bounce scenario
\cite{Brandenberger:2016vhg} during the contraction era. However, in
such cases, during the expansion era, entropy is produced and the
perturbation modes grow with the cosmic time \cite{Finelli:2001sr}.
Such a continuous cycle of cosmological bounces can be stopped if a
crushing singularity occurs at the end of the expanding era, as for
example in the deformed matter bounce scenario studied in
Ref.~\cite{Odintsov:2016tar}, where it was shown that the infinite
repeating evolution of the Universe stops at the final attractor of
the theory, which is a Big Rip singularity
\cite{Caldwell:2003vq,McInnes:2001zw,Nojiri:2003vn,Nojiri:2005sr,
Gorini:2002kf,Elizalde:2004mq,Faraoni:2001tq,Singh:2003vx,Csaki:2004ha,
Wu:2004ex,Nesseris:2004uj,Sami:2003xv,Stefancic:2003rc,
Chimento:2003qy,Chimento:2004ps,Hao:2004ky,Babichev:2004qp,
Zhang:2005eg,Dabrowski:2004hx,Lobo:2005us,Cai:2005ie,Aref'eva:2005fu,Lu:2005qy,Godlowski:2005tw,Guberina:2005mp,Dabrowski:2006dd,Chimento:2015gga,
Barrow:2009df,Yurov:2007tw,BouhmadiLopez:2007qb,
BouhmadiLopez:2006fu,Briscese:2006xu}. Modified gravity in general
offers a consistent theoretical framework in the context of which
bouncing cosmologies can be realized, without the need to satisfy
specific constraints which are compelling in the case of the
standard general relativity approach. Hence the study of bouncing
cosmologies in the context of modified gravity is important, and
these cosmologies need to be critically examined with regards to
their observational consequences. For some recent studies on
bouncing cosmologies in the context of modified gravity, see for
example
\cite{Bamba:2013fha,Barragan:2009sq,Farajollahi:2010pn,Escofet:2015gpa},
see also Refs.~\cite{Barragan:2010qb,Koivisto:2010jj,Komada:2014asa}
for bouncing cosmologies in the context of Palatini gravity.

The modified gravity description of our Universe cosmological
evolution is one physically appealing theoretical framework, which
can potentially explain the various evolution eras of the Universe,
for the simple reason that it can provide a unified and
theoretically consistent description. There exist a plethora of
modified gravity models, that can potentially describe our Universe
evolution and the most important criterion for the viability of a
modified gravity theory is the compatibility of the theory with
present time observations. The observations related to our Universe,
mainly consist of large scale observations and astrophysical
observations, related to compact gravitational objects or
gravitationally bound objects. In both cases, a successful modified
gravity theory, should pass all the tests related to observations.
But there is also another feature that may render a modified gravity
theoretical description as viable, namely the fact that the theory
will be able to predict new, yet undiscovered phenomena. We believe
that by studying alternative modified gravity models, even if we do
not succeed in finding the ultimate modified gravity theory, at
least we will pave the way towards finding the most successful
theory. In the context of modified gravity, the new Lagrangian
terms, introduce new degrees of freedom beyond the standard General
Relativity, and the Standard Model of particle physics. When these
new terms are applied at a cosmological level, the extra degrees of
freedom alter the evolution of the Universe, and may have as an
effect the desired behavior of the Universe. These theories may
however introduce certain pathologies or extra instabilities, as it
happens for example in the case of $F(R,\mathcal{G})$ gravity, where
superluminal extra modes appear, which are absent in the $F(R)$
gravity and the $f(\mathcal{G})$ gravity, or modified Gauss-Bonnet
gravity cases, and of course these are absent in General Relativity.

Also with regards to astrophysical solutions, like black holes,
neutron stars, and wormholes, modified gravity utterly changes the
conditions that are needed to be satisfied in order for the
solutions to be consistent. For example, in the case of wormholes in
the ordinary Einstein gravity, an exotic matter fluid needs
to be present, in order for the wormhole solution to be
self-consistent. On the contrary, in the modified gravity case, the
modified gravity part can offer a theoretically appealing and simple
remedy to this problem, see for example \cite{Lobo:2009ip}.

Our aim with this review is to present the latest developments in
the description of the inflationary era, dark energy and also in
bouncing cosmology, in the context of modified gravity. Our
motivation is the fact that up to date the $\Lambda$CDM model is
successful but does not provide a complete description of the
Universe. Also we shall provide all the necessary information for
the models we shall study, that make these models consistent both at
astrophysical and at cosmological level. Our presentation will have
a pedagogical and introductive character, in order to make this
review a pedagogical tool available to experts and non-experts. It
is conceivable that our work does not cover all the modified gravity
applications on inflation, dark energy and bouncing cosmologies,
this task would be very hard to be materialized, since the subject
is vast. However we shall provide the most important tools that will
enable one to study in some detail inflation and bounces with
modified gravity. Also for completeness and in order to render this
review an autonomous study on modified gravity, we shall discuss
some astrophysical applications of modified gravity and specifically
of $F(R)$ gravity, which is the most sound representative theory of
modified gravity. In most cosmological applications, the background
space-time geometry will be that of the flat Friedmann-Robertson-Walker
(FRW) geometry. We will start our presentation with chapter II,
where we shall present the most important modified gravity
descriptions. Specifically we shall provide the theoretical
framework of each modified gravity version, and also we also present
some illustrative examples in each case. Specifically, in section
II-A, we discuss the most important representative theory of
modified gravity, namely $F(R)$ gravity. We present in detail the
equations of motion of the FRW Universe and also we discuss the
criteria that render an $F(R)$ gravity theory a astrophysical and
cosmologically viable theory. We also discuss the stability, in
terms of the scalaron mass, and we present the corresponding
scalar-tensor description in the Einstein frame. Several viable
$F(R)$ gravities are presented, and the conditions of their
viability are also quoted. Also we discuss in brief the case that a
non-minimal coupling between the $F(R)$ gravity sector and the
matter Lagrangian exists. In section II-B, we present the
Gauss-Bonnet modified gravities. We present the basic equations of
motion, which can be used as a reconstruction method in order to
realize various cosmologies. Also we calculate the corrections to
the Newton law due to the Gauss-Bonnet gravity corrections. The
string inspired gravity follows in section II-C, along with several
characteristic examples. In section II-D, we discuss an important
modified gravity theory, namely that of an $F(T)$ gravity, with $T$
being the torsion scalar. Both the $f(\mathcal{G})$ and $F(T)$
gravity will be thoroughly discussed in this review article, since
these theories have appealing characteristics and are important
alternative theories to $F(R)$ gravity. In section II-E we discuss
massive gravity and bigravity theories, and in sections II-F and
II-G, we discuss mimetic $F(R)$ and mimetic $f(\mathcal{G})$
gravity. The mimetic framework has offered a quite interesting
appealing theoretical framework, so we present various versions of
the mimetic framework. In section II-H we present another $F(R)$
gravity extension, namely that of unimodular $F(R)$ gravity, and in
sections II-I, II-J, we discuss some combinations of unimodular and
mimetic $F(R)$ gravity, which are accompanied by illustrative
examples.

Chapter III is devoted on the study of inflationary dynamics in
various theoretical contexts. Traditionally, inflation was firstly
studied in the context of scalar-tensor theory in its simplest form,
namely that of a canonical scalar field, the inflaton, but in this
chapter we also present some modified gravity descriptions of the
inflationary era. For completeness in III-A we first study the
canonical scalar field inflationary paradigm, and in order to
illustrate the methods, we calculate the spectral index of
primordial curvature perturbations and also the scalar-to-tensor
ratio. In the same section we study the non-canonical scalar field
case and we also discuss in brief the form of the inflationary
dynamics in multi-scalar theories of gravity. In section III-B we
present the inflationary dynamics formalism for general $f(R,\phi)$
theories of gravity. We provide detailed calculations for the
slow-roll indices and also for the observational indices. Also we
focus on interesting subcases of $f(R,\phi)$ gravity, and
particularly $\phi f(R)$, mimetic $F(R)$ gravity and $F(R)$ gravity,
and we use various illustrative examples in order to better support
the theoretical formalism. In the same section we present a very
useful reconstruction technique which offers the possibility of
obtaining the $F(R)$ gravity from the Einstein frame theory with a
given potential. In section III-C we present the study of
inflationary dynamics in the context of the modified Gauss-Bonnet
gravity of the form $f(\mathcal{G})$, where $\mathcal{G}$ is the
Gauss-Bonnet scalar. We use a different approach in comparison to
the previous sections of this chapter, so we calculate directly the
power spectrum of primordial curvature perturbations, and from this
we calculate the spectral index of the primordial curvature
perturbations. As a peripheral study we discuss how the
perturbations evolve after the horizon crossing, by exploiting a
specifically chosen example. We also briefly present the formalism
of inflationary dynamics in the context of $F(T)$ and one loop
quantum gravity. Particularly, in the $F(T)$ case we realize the
intermediate inflation scenario in section III-D, while section
III-E is devoted to the singular inflation scenario and various
phenomenological implications of this cosmological evolution.
Finally,in section III-F we present a useful theoretical approach
for the graceful exit issue and in III-G the reheating era in the
context of modified gravity is studied. Usually in modified gravity
the graceful exit instance is identified with the moment that the
first slow-roll index becomes of the order one. However, the growing
perturbations that are caused by the unstable de Sitter points can also
provide a mechanism for graceful exit from inflation. So we present
in brief all the essential information for this alternative approach
to the graceful exit issue.

In chapter IV we address the dark energy issue in the context of
modified gravity. Particularly, we discuss how the dark energy era
can be realized by $F(R)$ and $f(\mathcal{G})$ gravity, and also we discuss
various phenomena related to the late-time era realization with
modified gravity. But firstly, we will show in sections IV-A and
IV-B, how the successful $\Lambda$CDM model can be realized by
$F(R)$ and $f(\mathcal{G})$ gravity respectively. In sections IV-C and IV-D we
will demonstrate how it is possible to provide a unified description
of early and late-time acceleration with $F(R)$ and $f(\mathcal{G})$ gravity.
In section IV-E we discuss how a phantom cosmological evolution can
be realized in the context of $F(R)$ gravity. Finally, in section
IV-F we will discuss an important feature of $F(R)$ gravity when the
late-time era is studied. Particularly, we present the dark energy
oscillations issue in $F(R)$ gravity, and we discuss how this may
affect the late-time era. We shall use various matter fluids
present, from perfect matter fluids to collisional, and we
critically discuss how the fluid viscosity may affect the dark
energy oscillations issue.

In chapter V we discuss some astrophysical applications of modified
gravity, emphasizing in $F(R)$ gravity applications. We shall
present some neutron stars and quark stars solutions in $F(R)$
gravity, and we briefly examine the astrophysical consequences of
these solutions, for general equations of state for the neutron
star. Also we discuss the possibility of anti-evaporation from
Reissner-Nordstr\"{o}m black holes in $F(R)$ gravity and finally we
present some wormhole solutions from $F(R)$ gravity.

In chapter VI we present the general reconstruction techniques which
can be used in order to realize bouncing cosmologies with $F(R)$,
$f(\mathcal{G})$ and $F(T)$ gravity. We shall present how to realize bounces
by vacuum modified gravity, but also in some cases we shall present
how the results are modified in the presence of perfect matter
fluids. In the beginning of the chapter we provide an informative
overview of bouncing cosmologies which is necessary in order to
understand the fundamental characteristics of a cosmological bounce.
In section VI-A we discuss how a specific bounce ca be realized with
vacuum $F(R)$ gravity, but we also present the case that perfect
matter fluids are present. In section VI-B we discuss the $f(\mathcal{G})$
realization of the same bounce cosmology as before, and finally in
section VI-C we discuss the $F(T)$ gravity realization. In all three
cases, we examine the stability of the modified gravity solutions we
found, at the level of the equations of motion. Specifically, we
study the stability of the equations of motion, if these are viewed
as a dynamical system, and we demonstrate in which cases stability
can be ensured.

Finally the conclusions follow at the end of this review, were we
summarize the successes of modified gravity and we outline its
shortcomings as a theory. Also we discuss in brief various
challenging problems in cosmology, which still need to be
incorporated successfully to a future theory.

\section{Modified Gravities and Cosmology \label{section1}}

\subsection{$F(R)$ Gravity}

The theory of $F(R)$ gravity could be considered as the most popular
among modified gravity theories. In this section, a general review
of the $F(R)$ gravity theory is given. In the literature there are
various reviews also discussing this topic, see
\cite{Bamba:2015uma,Nojiri:2013zza,Nojiri:2006ri,Capozziello:2011et,
Capozziello:2010zz,Nojiri:2010wj,Clifton:2011jh}.

\subsubsection{General properties}

The action of the $F(R)$ gravity \cite{Nojiri:2006ri} is given by
replacing the scalar curvature $R$ in the Einstein-Hilbert
action which is,\footnote{We use the following convention for the
curvatures and connections:
\[
R=g^{\mu\nu}R_{\mu\nu} \, , \quad
R_{\mu\nu} = R^\lambda_{\ \mu\lambda\nu} \, , \quad
R^\lambda_{\ \mu\rho\nu} = -\Gamma^\lambda_{\mu\rho,\nu}
+ \Gamma^\lambda_{\mu\nu,\rho} - \Gamma^\eta_{\mu\rho}
\Gamma^\lambda_{\nu\eta}
+ \Gamma^\eta_{\mu\nu}\Gamma^\lambda_{\rho\eta} \, ,\quad
\Gamma^\eta_{\mu\lambda} = \frac{1}{2}g^{\eta\nu}\left(
g_{\mu\nu,\lambda} + g_{\lambda\nu,\mu} - g_{\mu\lambda,\nu}
\right)\, .
\]
}
\begin{equation}
\label{JGRG6}
S_\mathrm{EH}=\int d^4 x \sqrt{-g} \left(
\frac{R}{2\kappa^2} + \mathcal{L}_\mathrm{matter} \right)\, ,
\end{equation}
by
using some appropriate function of the scalar curvature, as follows,
\begin{equation}
\label{JGRG7}
S_{F(R)}= \int d^4 x \sqrt{-g} \left(
\frac{F(R)}{2\kappa^2} + \mathcal{L}_\mathrm{matter} \right)\, .
\end{equation}nd{equation}
In Eqs.~(\ref{JGRG6}) and (\ref{JGRG7}),
$\mathcal{L}_\mathrm{matter}$ is the matter Lagrangian density. We
now review in brief the general properties of $F(R)$ gravity.

For later convenience, we define the effective equation of state
(EoS) parameter for the $F(R)$ gravity theory. The expression can be
used in any other modified gravity context. We start with the FRW
equations, which in the Einstein gravity coupled with perfect fluid
are:
\begin{equation}
\label{JGRG11}
\rho_\mathrm{matter}=\frac{3}{\kappa^2}H^2
\, ,\quad p_\mathrm{matter}
= - \frac{1}{\kappa^2}\left(3H^2 + 2\dot H\right) \, .
\end{equation}
Then, the EoS parameter can be given by using the
Hubble rate $H$, in the following way,
\begin{equation}
\label{JGRG12}
w_\mathrm{eff}= - 1 - \frac{2\dot H}{3H^2} \, .
\end{equation}
In principle we
can use the expression given in Eq.~(\ref{JGRG12}) even for modified
gravity theories, since it is useful for a generalized fluid
description of modified gravity.

By varying the action (\ref{JGRG7}) with respect to the metric, we
obtain the equation of motion for the $F(R)$ gravity theory as
follows,
\begin{equation}
\label{JGRG13}
\frac{1}{2}g_{\mu\nu} F(R) - R_{\mu\nu} F'(R) - g_{\mu\nu} \Box F'(R)
+ \nabla_\mu \nabla_\nu F'(R)
= - \frac{\kappa^2}{2}T_{\mathrm{matter}\,\mu\nu}\, .
\end{equation}
For the spatially flat FRW Universe, in which case the metric is given by,
\begin{equation}
\label{JGRG14}
ds^2 = - dt^2 + a(t)^2 \sum_{i=1,2,3}
\left(dx^i\right)^2\, ,
\end{equation}
Eq.~(\ref{JGRG13}) gives the FRW equations,
\begin{align}
\label{JGRG15}
0 =& -\frac{F(R)}{2} + 3\left(H^2 + \dot H\right)
F'(R) - 18 \left( 4H^2 \dot H + H \ddot H\right) F''(R)
+ \kappa^2 \rho_\mathrm{matter}\, ,\\
\label{Cr4b}
0 =& \frac{F(R)}{2} - \left(\dot H + 3H^2\right)F'(R)
+ 6 \left( 8H^2 \dot H + 4 {\dot H}^2 + 6 H \ddot H + \dddot H\right)
F''(R) + 36\left( 4H\dot H + \ddot H\right)^2 F'''(R) \nn &
+ \kappa^2 p_\mathrm{matter}\, ,
\end{align}
where the Hubble rate $H$ is equal to $H=\dot a/a$. In terms of the
Hubble rate $H$, the scalar curvature $R$ is equal to $R=12H^2
+ 6\dot H$.

We can find several (in many cases exact) solutions of
Eq.~(\ref{JGRG15}). Without the presence of matter, a simple
solution is given by assuming that the Ricci tensor is covariantly
constant, that is, $R_{\mu\nu}\propto g_{\mu\nu}$. Then
Eq.~(\ref{JGRG13}) is simplified to the following algebraic equation
\cite{Nojiri:2003ft} (see also \cite{Cognola:2005de}):
\begin{equation}
\label{JGRG16}
0 = 2 F(R) - R F'(R)\, .
\end{equation}
If Eq.~(\ref{JGRG16})
has a solution the (anti-)de Sitter and/or Schwarzschild- (anti-)de
Sitter space
\begin{equation}
\label{SdS}
ds^2 = - \left( 1 - \frac{2MG}{r} \mp
\frac{r^2}{L^2} \right) dt^2 + \left( 1 - \frac{2MG}{r} \mp
\frac{r^2}{L^2} \right)^{-1} dr^2 + r^2 d\Omega^2\, ,
\end{equation}
or the
Kerr - (anti-)de Sitter space is an exact solution in vacuum. In
Eq.~(\ref{SdS}), the minus and plus signs in $\pm$ correspond to the
de Sitter and anti-de Sitter space, respectively. In
Eq.~(\ref{SdS}), $M$ is the mass of the black hole, $G =
\frac{\kappa^2}{8\pi}$, and $L$ is the length parameter of (anti-)de
Sitter space, which is related to the curvature as follows $R=\pm
\frac{12}{L^2}$ (the plus sign corresponds to the de Sitter space
and the minus sign to the anti-de Sitter space).

We now consider the perfect fluid representation and scalar-tensor
representation of the $F(R)$ gravity. For convenience, we write
$F(R)$ as the sum of the scalar curvature $R$ and the part which
expresses the difference from the Einstein gravity case,
\begin{equation}
\label{FRf}
F(R) = R + f(R) \, .
\end{equation}
Eqs.~(\ref{JGRG15}) and (\ref{Cr4b}) indicate that we can express
the effective energy density $\rho_\mathrm{eff}$ and also the
effective pressure $p_\mathrm{eff}$ including the contribution from
$f(R)$ gravity as follows, (see, for instance, \cite{Nojiri:2009xw})
\begin{align}
\label{Cr4}
\rho_\mathrm{eff} =&
\frac{1}{\kappa^2}\left(-\frac{1}{2}f(R) + 3\left(H^2
+ \dot H\right) f'(R) - 18 \left(4H^2 \dot H
+ H \ddot H\right)f''(R)\right) + \rho_\mathrm{matter}\, ,\\
\label{Cr4bb}
p_\mathrm{eff} =&
\frac{1}{\kappa^2}\left(\frac{1}{2}f(R) - \left(3H^2
+ \dot H \right)f'(R) + 6 \left(8H^2 \dot H + 4{\dot H}^2 + 6 H \ddot H
+ \dddot H \right)f''(R) + 36\left(4H\dot H + \ddot H\right)^2f'''(R)
\right) \nn
& + p_\mathrm{matter}\, ,
\end{align}
which enables us to rewrite the equations (\ref{JGRG15}) and
(\ref{Cr4b}) as in the Einstein gravity case (\ref{JGRG11}),
\begin{equation}
\label{JGRG11B}
\rho_\mathrm{eff}=\frac{3}{\kappa^2}H^2 \, , \quad
p_\mathrm{eff}= - \frac{1}{\kappa^2}\left(3H^2 + 2\dot H\right)\, .
\end{equation}
The fluid representation for the FRW equations, however, often
leads to an unjustified treatment. For example, it is often ignored
that the generalized gravitational fluid contains higher-derivative
curvature invariants. The viable dark energy models of $F(R)$
gravity are discussed in
Refs.~\cite{Nojiri:2003ni,Nojiri:2006be,Starobinsky:2007hu,Nojiri:2007jr,
Nojiri:2003wx},
while the unified description of inflation with dark energy are
discussed in Refs.~\cite{Nojiri:2006gh,Nojiri:2003ft}, for a review
see \cite{Nojiri:2010wj}.

\subsubsection{Scalar-tensor description}

We should note that we can also rewrite $F(R)$ gravity in a
scalar-tensor form. We introduce an auxiliary field $A$ and rewrite
the action (\ref{JGRG7}) of the $F(R)$ gravity in the following
form:
\begin{equation}
\label{JGRG21}
S=\frac{1}{2\kappa^2}\int d^4 x \sqrt{-g}
\left\{F'(A)\left(R-A\right) + F(A)\right\}\, .
\end{equation}
We obtain $A=R$
by the variation of the action with respect to $A$ and by
substituting the obtained equation $A=R$ into the action
(\ref{JGRG21}), we find that the action in (\ref{JGRG7}) is
reproduced. If we rescale the metric by a kind of a scale
transformation (canonical transformation),
\begin{equation}
\label{JGRG22}
g_{\mu\nu}\to \e^\sigma g_{\mu\nu}\, ,\quad \sigma = -\ln F'(A)\, ,
\end{equation}
we obtain the action in the Einstein frame \footnote{Note that
the difference between the $F(R)$ (Jordan) frame and the
scalar-tensor (Einstein) frame description, may lead to a number of
issues, for example the Universe in one frame may be accelerating,
while decelerating in the other frame
\cite{Capozziello:2006dj,Bahamonde:2017kbs}, or the singularity
types changes from frame to frame
\cite{Briscese:2006xu,Bahamonde:2016wmz}.},
\begin{align}
\label{JGRG23}
S_E =& \frac{1}{2\kappa^2}\int d^4 x \sqrt{-g} \left(
R - \frac{3}{2}g^{\rho\sigma}
\partial_\rho \sigma \partial_\sigma \sigma - V(\sigma)\right) \, ,\nn
V(\sigma) =& \e^\sigma g\left(\e^{-\sigma}\right)
 - \e^{2\sigma} f\left(g\left(\e^{-\sigma}\right)\right)
=\frac{A}{F'(A)} - \frac{F(A)}{F'(A)^2}\, .
\end{align}
Here $g\left(\e^{-\sigma}\right)$ is given by solving the equation
$\sigma = -\ln\left( 1 + f'(A)\right)=- \ln F'(A)$ as
$A=g\left(\e^{-\sigma}\right)$. Due to the scale transformation
(\ref{JGRG22}), a coupling of the scalar field $\sigma$ with usual
matter is introduced. The mass of the scalar field $\sigma$ is given
by
\begin{equation}
\label{JGRG24}
m_\sigma^2 \equiv \frac{3}{2}\frac{d^2
V(\sigma)}{d\sigma^2} =\frac{3}{2}\left\{\frac{A}{F'(A)}
 - \frac{4F(A)}{\left(F'(A)\right)^2} + \frac{1}{F''(A)}\right\}\, ,
\end{equation}
and if the mass $m_\sigma$ is not very large, there appears a
large correction to the Newton law. Since we would like to explain
the accelerating expansion of the current Universe by using the
$F(R)$ gravity, we may naively expect that the order of the mass
$m_\sigma$ should be that of the Hubble rate, that is, $m_\sigma
\sim H \sim 10^{-33}\,\mathrm{eV}$. Since the mass is very small,
very large corrections to the Newton law could appear. In order to
avoid the above problem in the Newton law, a so-called ``realistic''
model was proposed in \cite{Hu:2007nk}. In the model, the mass
$m_\sigma$ becomes large enough in the regions where the curvature
$R=A$ is large or under the presence of matter fluids, as in the
Solar System, or in the Earth. Therefore the force mediated by the
scalar field becomes short-ranged, which also introduces a screening
effect in which only the surface of the massive objects like the
planets can contribute to the correction to the Newton law even in
the vacuum. This is called the Chameleon mechanism
\cite{Khoury:2003rn}, which prevents the large correction to the
Newton law.

For example, we may consider the following model,
\cite{Cognola:2007zu} (see also \cite{Linder:2009jz,Bamba:2010ws}),
\begin{equation}
\label{expModel}
f(R)=2\Lambda_\mathrm{eff} \left( \e^{-bR}-1
\right)\, .
\end{equation}
In the Solar System, where $R\sim 10^{-61}\,\mathrm{eV}^2$,
if we choose $1/b \sim R_0 \sim \left(10^{-33}\,
\mathrm{eV}\right)^2$, we obtain $m_\sigma^2 \sim
10^{1,000}\,\mathrm{eV}^2$, which is ultimately heavy. In the
atmosphere of the Earth, where $R \sim 10^{-50}\,\mathrm{eV}^2$, and
if we choose $1/b \sim R_0 \sim
\left(10^{-33}\,\mathrm{eV}\right)^2$, again, we obtain $m_\sigma^2
\sim 10^{10,000,000,000}\,\mathrm{eV}^2$. Then, the correction to
the Newton law looks to be extremely small in this kind of model.
The corresponding Compton wavelength, however, is very small and
much shorter than the distance between the atoms. Since the region
between the atoms can be regarded as a vacuum except several quantum
corrections and we cannot use the approximation to regard the matter
fluids as continuous fluid, the Chameleon mechanism does not apply.
Therefore, the Compton length cannot be shorter than the distance
between the atoms and the scalar field cannot become so large,
although this is not in conflict with any observation nor
experiment.

We now also need to mention the problem of antigravity.
Eq.~(\ref{JGRG21}) tells that the effective gravitational coupling
is given by $\kappa_\mathrm{eff}^2 = \frac{\kappa^2}{F'(R)}$.
Therefore when $F'(R)$ is negative, it is possible to have
antigravity regions \cite{Nojiri:2003ft}. Then, we need to require
\begin{equation}
\label{FR1}
F'(R) > 0 \, .
\end{equation}
We should note that from the
viewpoint of the field theory, the graviton becomes ghost in the
antigravity region.

It should be noted that the de Sitter or anti-de Sitter space
solution in (\ref{JGRG16}) corresponds to the extremum of the
potential $V(\sigma)$. In fact, we find,
\begin{equation}
\label{FRV1}
\frac{dV(\sigma)}{dA} = \frac{F''(A)}{F'(A)^3} \left( - AF'(A)
+ 2F(A) \right)\, .
\end{equation}
Therefore, if Eq.~(\ref{JGRG16}) is satisfied,
the scalar field $\sigma$ should be on the local maximum or local
minimum of the potential and $\sigma$ can be a constant. When the
condition (\ref {JGRG16}) is satisfied, the mass given by
(\ref{JGRG24}) has the following form:
\begin{equation}
\label{FRV3}
m_\sigma^2 =
\frac{3}{2 F'(A)} \left( - A + \frac{F'(A)}{F''(A)} \right)\, .
\end{equation}
Therefore, in the case that the condition (\ref{FR1}) for avoiding
the antigravity holds true, the mass squared $m_\sigma^2$ is
positive, showing that the scalar field is on the local minimum if,
\begin{equation}
\label{FRV4}
 - A + \frac{F'(A)}{F''(A)} > 0\, .
\end{equation}
On the other hand, the scalar field is on the local maximum of
the potential if,
\begin{equation}
\label{FRV5}
 - A + \frac{F'(A)}{F''(A)} < 0\, .
\end{equation}
In this case the mass squared $m_\sigma^2$ is negative. The
condition (\ref{FRV4}) is nothing but the stability condition of the
de Sitter space.

Note that in the Einstein frame, the Universe is always in the
non-phantom phase, where the effective EoS $w_\mathrm{eff}$ in
(\ref{JGRG12}) is larger than $-1$ although in the Jordan frame, the
Universe can be, in general, in a phantom phase. This is because the
scale transformation in (\ref{JGRG22}) changes the time coordinate.
We should note that we observe the expansion of the Universe via
matter. In the Einstein frame, the metric for the matter is not
given by $g_{\mu\nu}$ but by $\e^\sigma g_{\mu\nu}$, which is
nothing but the metric in the Jordan frame. Therefore the
observations in the Einstein frame is not changed from the
observation in the Jordan frame. This indicates that in the Einstein
frame, the metric $g_{\mu\nu}$ is not physical but $\e^\sigma
g_{\mu\nu}$ is the physical metric \cite{Capozziello:2006dj}.

We now shortly mention on the matter instability issue pointed out
in \cite{Dolgov:2003px}. The instability may appear when the energy
density or the scalar curvature is large compared with that in the
Universe, as for example on the Earth. Let $R_b$ the background
scalar curvature and separate the scalar curvature $R$ as a sum of
$R_b$ and the perturbed part $R_p$ as $R=R_b + R_p$
$\left(\left|R_p\right|\ll \left|R_b\right|\right)$. Then we obtain
the following perturbed equation:
\begin{align}
\label{JGRG28}
0 =& \Box R_b + \frac{F^{(3)}(R_b)}{F^{(2)}(R_b)}
\nabla_\rho R_b \nabla^\rho R_b
+ \frac{F'(R_b) R_b}{3F^{(2)}(R_b)} \nn
& - \frac{2F(R_b)}{3 F^{(2)}(R_b)} - \frac{R_b}{3F^{(2)}(R_b)} + \Box R_p
+ 2\frac{F^{(3)}(R_b)}{F^{(2)}(R_b)}\nabla_\rho R_b \nabla^\rho R_p
+ U(R_b) R_p\, , \\
U(R_b) \equiv& \left(\frac{F^{(4)}(R_b)}{F^{(2)}(R_b)}
 - \frac{F^{(3)}(R_b)^2}{F^{(2)}(R_b)^2}\right) \nabla_\rho R_b
\nabla^\rho R_b + \frac{R_b}{3} \nn & - \frac{F^{(1)}(R_b)
F^{(3)}(R_b) R_b}{3 F^{(2)}(R_b)^2}
 - \frac{F^{(1)}(R_b)}{3F^{(2)}(R_b)} + \frac{2 F(R_b) F^{(3)}(R_b)}{3
F^{(2)}(R_b)^2}
 - \frac{F^{(3)}(R_b) R_b}{3 F^{(2)}(R_b)^2} \, .
\end{align}
If we assume $R_b$ and $R_p$ are uniform, we can replace the
d'Alembertian the second derivative with respect to the time
coordinate and therefore Eq.~(\ref{JGRG28}) has the following
structure:
\begin{equation}
\label{JGRG30}
0=-\partial_t^2 R_p + U(R_b) R_p + \mathrm{const.}\, .
\end{equation}
Then, if $U(R_b)>0$, $R_p$ becomes
exponentially large with time $t$, we have $R_p\sim
\e^{\sqrt{U(R_b)} t}$ and the system is rendered unstable. We should
note, however, that the scalar field in (\ref{JGRG23}) is nothing
but the scalar curvature and therefore the above matter instability
occurs when the mass of the scalar field is negative,
$m_\sigma^2<0$. Conversely, if we choose $F(R)$ for large $R$ in
such a way, so that the mass becomes positive, $m_\sigma^2>0$, then
the instability does not occur.

\subsubsection{Viable $F(R)$ gravities}

Using the previous arguments, we summarize the conditions that need
to hold true in order for an $F(R)$ gravity to be a viable
cosmological model, which unifies the accelerating expansion of the
present Universe and the early-time acceleration of the Universe.
The first unified inflation dark energy $F(R)$ gravity model
appeared in Ref.~\cite{Nojiri:2003ft}, and in
Refs.~\cite{Nojiri:2007as,Nojiri:2007cq,Cognola:2007zu,Cognola:2008zp,
Sokolowski:2007pk,Brookfield:2006mq,
Abdelwahab:2007jp,Oikonomou:2013rba}, various viable models of
$F(R)$ gravity unifying both of the late-time and the early-time
acceleration were proposed by requiring several conditions, which we
list here:
\begin{enumerate}
\item A condition to generate the inflationary era is given by,
\begin{equation}
\label{Uf1}
\lim_{R\to\infty} f (R) = - \Lambda_i\, .
\end{equation}
Here,
$\Lambda_i$ is an effective cosmological constant characterizing the
early Universe.
\item In order to generate the accelerating expansion of the Universe at present
time, the current value of $f(R)$ should be a small constant,
\begin{equation}
\label{Uf3}
f(R_0)= - 2\tilde R_0\, ,\quad f'(R_0)\sim 0\, .
\end{equation}
Here, $R_0$ expresses the present time curvature $R_0\sim
\left(10^{-33}\mathrm{eV}\right)^2$. Note that $R_0> \tilde R_0$
because we need to take into account the contribution from matter,
since the trace part of Eq.~(\ref{JGRG13}) indicates that $R_0\sim
\tilde R_0 - \kappa^2 T_\mathrm{matter}$. Here, $T_\mathrm{matter}$
is the trace part of the energy-momentum tensor of all matter
fluids. We should note that the quantity $f'(R_0)$ needs not to
vanish completely but instead it should satisfy $\left| f'(R_0)
\right| \ll \left(10^{-33}\,\mathrm{eV}\right)^4$. This is due to
the fact that we consider the time scale to be $10^{12-13}$ years.
\item The last condition is given by
\begin{equation}
\label{Uf4}
\lim_{R\to 0} f(R) = 0\, ,
\end{equation}
that is, a flat
space-time solution (Minkowski space-time) should exist.
\end{enumerate}

Let us consider the power-law model where $F(R)$ behaves as
\begin{equation}
\label{X}
f(R) \sim F_0 + F_1 R^\epsilon\, ,
\end{equation}
when $R$ is large.
Here $F_0$ and $F_1$ are arbitrary constants. The constant $F_0$ may
vanish but $F_1$ should not, $F_1\neq 0$. Then, the trace equation,
which is the trace part of Eq.~(\ref{JGRG13}),
\begin{equation}
\label{Scalaron}
3\Box f'(R)= R+2f(R)-Rf'(R)-\kappa^2 T_\mathrm{matter}\, ,
\end{equation}
indicates that,
\begin{equation}
\label{XII}
3 F_1 \Box R^{\epsilon -1} = \left\{
\begin{array}{ll} R & \ \mbox{when $\epsilon<0$ or $\epsilon=2$} \\
\left(2-\epsilon\right) F_1 R^\epsilon & \ \mbox{when $\epsilon>1$
or $\epsilon\neq 2$}
\end{array} \right. \, .
\end{equation}
We now assume that the Hubble rate has a structural singularity
as follows,
\begin{equation}
\label{XIII}
H \sim \frac{h_0}{\left(t_s - t\right)^\beta}\, ,
\end{equation}
where $h_0$ and $\beta$ are arbitrary
constants suitably chosen so that the Hubble rate is real. Then the
scalar curvature $R=6\dot H + 12 H^2$ behaves as follows,
\begin{equation}
\label{XIV}
R \sim \left\{
\begin{array}{ll}
\frac{12h_0^2}{\left(t_s - t\right)^{2\beta}} & \ \mbox{when $\beta>1$} \\
\frac{6 h_0 + 12 h_0^2}{\left(t_s - t\right)^2} & \ \mbox{when $\beta=1$} \\
\frac{6\beta h_0}{\left(t_s - t\right)^{\beta + 1}} & \ \mbox{when
$\beta<1$}
\end{array} \right. \, .
\end{equation}
In Eqs.~(\ref{XIII}) and (\ref{XIV}), the case $\beta\geq 1$
corresponds to a Type I (Big Rip) singularity, see
Refs.~\cite{Caldwell:1999ew,McInnes:2001zw,Nojiri:2003vn,Nojiri:2003ag,
Faraoni:2001tq,GonzalezDiaz:2003rf,GonzalezDiaz:2004as,
Singh:2003vx,Csaki:2004ha,Wu:2004ex,Nesseris:2004uj,Sami:2003xv,
Stefancic:2003rc,Chimento:2003qy,Hao:2004ky,Babichev:2004qp,
Zhang:2005eg,Elizalde:2005ju,Dabrowski:2004hx,Lobo:2005us,Cai:2005ie,
Aref'eva:2004vw,Aref'eva:2005fu,Lu:2005qy,Godlowski:2005tw,Guberina:2005mp,Dabrowski:2006dd,Barbaoza:2006hf}.
The case $1>\beta>0$ corresponds to a Type III singularity, the case
$0>\beta>-1$ corresponds to a Type II, and finally the case
$\beta<-1$ (but $\beta\neq\mbox{integer}$) corresponds to a Type IV
singularity.

The above classification of the finite-time future singularities was
proposed in Ref.~\cite{Nojiri:2005sx}. Particularly, the finite-time
singularity classification is the following:
\begin{itemize}
\item Type I (``Big Rip'') : This type of singularity occurs for $t \to t_s$,
$a \to \infty$, $\rho_\mathrm{eff} \to \infty$ and
$\left|p_\mathrm{eff}\right| \to \infty$. Its manifestations in
various models and theoretical contexts, have been studied in
Ref.~\cite{Nojiri:2005sx}.
\item Type II (``sudden'') \cite{Barrow:2004he,Barrow:2004hk,
Barrow:2004xh,FernandezJambrina:2004yy,Dabrowski:2004bz,Lake:2004fu,
Nojiri:2004ip,deHaro:2012wv,BouhmadiLopez:2006fu,BouhmadiLopez:2007qb,BeltranJimenez:2016dfc}:
This type of singularity occurs for $t \to t_s$, $a \to a_s$,
$\rho_\mathrm{eff} \to \rho_s$ and $\left|p_\mathrm{eff}\right| \to
\infty$.
\item Type III : This type of singularity occurs for $t \to t_s$, $a \to a_s$,
$\rho_\mathrm{eff} \to \infty$ and $\left|p_\mathrm{eff}\right| \to
\infty$.
\item Type IV : This type of singularity occurs for $t \to t_s$, $a \to a_s$,
$\rho_\mathrm{eff} \to 0$, $\left|p_\mathrm{eff}\right| \to 0$ and
higher derivatives of $H$ diverge. This also includes the case in
which $p_\mathrm{eff}$ ($\rho_\mathrm{eff}$) or both
$p_\mathrm{eff}$ and $\rho_\mathrm{eff}$ tend to some finite values,
whereas higher derivatives of the Hubble rate $H$ diverge. This type
of singularity was proposed in \cite{Nojiri:2005sx}.
\end{itemize}
Here, $\rho_\mathrm{eff} $ and $p_\mathrm{eff}$ are defined in
Eq.~(\ref{JGRG11}).

Substituting Eq.~(\ref{XIV}) into Eq.~(\ref{XII}),
one finds that there are two classes of consistent solutions. One
solution is given by $\beta =1$ and $\epsilon>1$ (but $\epsilon\neq
2$), which corresponds to the Big Rip ($h_0>0$ and $t<t_s$) or Big
Bang ($h_0<0$ and $t>t_s$) singularity at $t=t_s$. Another solution
is $\epsilon<1$, and $\beta = - \epsilon/\left(\epsilon - 2\right)$
($-1<\beta<1$), which corresponds to the Type II future singularity.
We should note that when $\epsilon=2$, that is, $f(R) \sim R^2$,
there is no singular solution.

Therefore, if we add the above term $R^2 \tilde f(R)$, where
$\lim_{R\to 0} \tilde f(R) = c_1$, $\lim_{R\to \infty} \tilde f(R) =
c_2$ to the action of $f(R)$ and if the added term dominates for
large $R$, the modified gravity part of the $F(R)$ gravity, namely
$f(R)$, behaves as $f(R)\sim R^2$ and the future singularity could
disappear. On the other hand, if we add the term it dominates as it
behaves as an $R^n$-term with $n=3,4,5,\cdots$ for large $R$. Then
the singularity appears because this case corresponds to
$\epsilon=n>1$, that is, the case of the Big Rip singularity. In
case of $0<\epsilon<2$, the future singularity does not appear
\cite{Nojiri:2008fk}.

In order to avoid the occurrence of a finite-time future
singularity, an additional $R^2$-term is needed to be added in the
$F(R)$ gravity Lagrangian. The addition of this term was first
proposed first in Ref.~\cite{Abdalla:2004sw} in order for the Big
Rip singularity to disappear. Furthermore, the $R^2$-term, which
effectively eliminates the future singularity, generates the
early-time acceleration simultaneously. In other words, adding the
$R^2$ term to a gravitational dark energy model, may lead to the
emergence of an inflationary phase in the model as first observed in
Ref.~\cite{Nojiri:2003ft} and also the singularity is removed. In
Refs.~\cite{Nojiri:2008fk,Bamba:2008ut,Capozziello:2009hc,Bamba:2009uf},
it has been investigated that the $R^2$-term could cancel all types
of future singularities. The $R^2$-term often solves other
phenomenological problems
\cite{Kobayashi:2008wc,Sami:2009jx,Thongkool:2009js,Babichev:2009fi,
Appleby:2010dx} of $F(R)$ dark energy. Even for the models of dark
energy besides the $F(R)$ gravity, e.g., the models using the
perfect fluid approach or in a scalar field context, there could
often appear a singularity in the finite future Universe, but by the
addition of the $R^2$-term, the singularity can be eliminated
\cite{Nojiri:2009pf}.

Summarizing the above analysis, in order to obtain a realistic and
viable $F(R)$ gravity model:
\begin{enumerate}
\item\label{req1} In the limit of $R\to 0$, the Einstein gravity should
be recovered,
\begin{equation}
\label{E1}
F(R) \to R \quad \mbox{that is,} \quad
\frac{F(R)}{R^2} \to \frac{1}{R}\, .
\end{equation}
If this condition is
satisfied, a flat space (Minkowski) is also an solution as in
(\ref{Uf4}).
\item\label{req2} As we will discuss later, there should appear the stable
de Sitter solution,
which corresponds to the late-time acceleration, where the curvature
is small $R\sim R_L \sim \left( 10^{-33}\, \mathrm{eV}\right)^2$.
This requires that, when $R\sim R_L$,
\begin{equation}
\label{E2}
\frac{F(R)}{R^2} = f_{0L} - f_{1L} \left( R - R_L \right)^{2n+2}
+ {o} \left( \left( R - R_L \right)^{2n+2} \right)\, .
\end{equation}
Here, $f_{0L}$
and $f_{1L}$ are positive constants and $n$ is a positive integer.
In some cases this condition may not always be necessary.
\item\label{req3} As we will also discuss later on,
the quasi-stable de Sitter solution that corresponds to the
inflationary era, in the early Universe should appear. In this case,
the curvature is large $R\sim R_I \sim \left( 10^{16 \sim 19}\,
\mathrm{GeV}\right)^2$. The de Sitter space should not be exactly
stable so that the curvature decreases very slowly. This requires
the following condition,
\begin{equation}
\label{E3}
\frac{F(R)}{R^2} = f_{0I} - f_{1I} \left( R - R_I \right)^{2m+1}
+ {o} \left( \left( R - R_I
\right)^{2m+1} \right)\, .
\end{equation}
Here, $f_{0I}$ and $f_{1I}$ are
positive constants and $m$ is a positive integer.
\item\label{req4} In order to avoid the curvature singularity when
$R\to \infty$, we may assume that,
\begin{equation}
\label{E4}
F(R) \to
f_{\infty} R^2 \quad \mbox{that is} \quad \frac{F(R)}{R^2} \to
f_{\infty} \, .
\end{equation}
Here, $f_{\infty}$ is a positive and
sufficiently small constant. Instead of (\ref{E4}), we may choose
\begin{equation}
\label{E5}
F(R) \to f_{\tilde\infty} R^{2 - \epsilon} \quad
\mbox{that is} \quad \frac{F(R)}{R^2} \to
\frac{f_{\tilde\infty}}{R^\epsilon} \, .
\end{equation}
Here,
$f_{\tilde\infty}$ is a positive constant and the parameter
$\epsilon$ satisfies $0< \epsilon <1$. With the condition (\ref{E4})
or (\ref{E5}), the future singularity can be eliminated.
\item\label{req5} As shown in (\ref{FR1}), in order to avoid the antigravity
regime, we require the following condition,
\begin{equation}
\label{E6}
F'(R)>0\, ,
\end{equation}
which can be rewritten as follows,
\begin{equation}
\label{E7}
\frac{d}{dR}
\left( \ln \left( \frac{F(R)}{R^2} \right)\right) > - \frac{2}{R}\, .
\end{equation}
\item\label{req6}
By combining conditions (\ref{E1}) and (\ref{E6}), we find
\begin{equation}
\label{E8}
F(R)>0\, .
\end{equation}
\end{enumerate}
The conditions (\ref{req1}) and (\ref{req2}) indicate that an extra,
unstable solution describing the de Sitter space-time always appears
at $R=R_e$ $\left( 0< R_e
< R_L \right)$. Due to the fact that the de Sitter solution $R=R_L$
is stable, the evolution of the Universe will stop at $R=R_L$ and
therefore the curvature will not become smaller than $R_L$, which
indicates that the extra de Sitter solution is not realized. An
example of such $F(R)$ gravity is given in \cite{Nojiri:2010ny} (the
models in \cite{Nojiri:2007as,Nojiri:2007cq,Cognola:2007zu}
partially satisfy the above conditions). Also for a study on the
constraints of $F(R)$ gravity coming from large scale structures,
see \cite{Lombriser:2010mp,Schmidt:2009am}.

We may consider the non-standard non-minimal coupling of modified
gravity with the matter Lagrangian
\cite{Nojiri:2004bi,Allemandi:2005qs,Bertolami:2008zh} (see also
\cite{Capozziello:1999xt}),
\begin{equation}
\label{LR1}
S=\int d^4 x
\sqrt{-g}\left\{ \frac{R}{2\kappa^2} + f \left(R\right) L_d
\right\}\, .
\end{equation}
Here, $L_d$ is the Lagrangian density similar to
that of a standard matter fluid. In a more generalized theoretical
context, we may extend the model of Eq.~(\ref{LR1}) in the form of
an $F(R)$ gravity,
\begin{equation}
\label{LR1FR}
S=\int d^4 x \sqrt{-g}\left\{
\frac{F(R)}{2\kappa^2} + f \left(R\right) L_d \right\}\, .
\end{equation}
Then
by varying of the action (\ref{LR1FR}) with respect to the metric we
obtain,
\begin{align}
\label{LR2FR}
0=\frac{1}{2}g_{\mu\nu} F(R) - R_{\mu\nu}
\frac{\partial F(R)}{\partial R}
 - g_{\mu\nu} \Box \frac{\partial F(R)}{\partial R}
+ \nabla_\mu \nabla_\nu \frac{\partial F(R)}{\partial R}
+ \frac{\kappa^2}{2}\tilde T_{\mathrm{matter}\,\mu\nu}\, ,
\end{align}
where $\tilde T_{\mu\nu}$ is the effective energy momentum tensor
defined as follows,
\begin{align}
\label{w5}
\tilde T^{\mu\nu} \equiv& - f'(R) R^{\mu\nu} L_d
+ \left(\nabla^\mu \nabla^\nu
 - g^{\mu\nu}\nabla^2 \right)\left( f'(R) L_d\right) + f'(R) T^{\mu\nu}\, ,\nn
T^{\mu\nu} \equiv& \frac{1}{\sqrt{-g}} \frac{\delta}{\delta
g_{\mu\nu}} \left(\int d^4x\sqrt{-g} L_d\right) \, .
\end{align}
We should note that although $\tilde T^{\mu\nu}$ is conserved, that
is, $\nabla_\mu \tilde T^{\mu\nu}=0$, in contrast the tensor
$T^{\mu\nu}$ is not conserved in general. We can define this model
by specifying the Lagrangian density $L_d$ to be that of a free
massless scalar,
\begin{equation}
\label{LR4}
L_d = - \frac{1}{2}g^{\mu\nu}\partial_\mu \phi
\partial_\nu \phi\, .
\end{equation}
In the FRW Universe (\ref{JGRG14}), the
$(t,t)$ component and $(i,j)$ component in Eq.~(\ref{LR2FR}) give
the following equations,
\begin{align}
\label{JGRG15LR}
0 =& -\frac{F(R)}{2} + 3\left(H^2 + \dot H\right)
F'(R) - 18 \left( 4H^2 \dot H + H \ddot H\right) F''(R)
+ \kappa^2 \tilde\rho\, ,\\
\label{Cr4bLR} 0 =& \frac{F(R)}{2} - \left(\dot H + 3H^2\right)F'(R)
+ 6 \left( 8H^2 \dot H + 4 {\dot H}^2 + 6 H \ddot H +
\dddot H\right) F''(R) + 36\left( 4H\dot H + \ddot H\right)^2 F'''(R)
+ \kappa^2 \tilde p\, .
\end{align}
Here $\tilde \rho$ and $\tilde p$ have the following expressions,
\begin{align}
\label{rhoLR}
\tilde\rho \equiv & 3\left(H^2 + \dot H\right) f'(R)
L_d -3 H \frac{d \left( f'(R) L_d\right)}{dt}
+ f(R) \rho\, , \\
\label{pLR} \tilde p \equiv & - \left(\dot H + 3H^2\right) f'(R) L_d
+ \frac{d^2 \left( f'(R) L_d\right)}{dt^2} + 4 H \frac{d \left(
f'(R) L_d\right)}{dt} + f(R) p \, ,
\end{align}
where $\rho$ and $p$ are the energy density and the pressure given
by $T_{\mu\nu}$. This model may generate the accelerating expansion
of the Universe, due to the non-trivial coupling of the curvature,
although there are problems with geodesics \cite{Bertolami:2008ab}.

As a variation of the $F(R)$ gravity, we may also consider $F(R,T)$
gravity \cite{Harko:2011kv}, whose action is given by,
\begin{equation}
\label{FRT1}
S_{F(R,T)}= \int d^4 x \sqrt{-g} \left(
\frac{F(R,T)}{2\kappa^2} + \mathcal{L}_\mathrm{matter} \right)\, .
\end{equation}
Here $T$ stands for the trace of some ``energy-momentum'' tensor
$T_{\mu\nu}$ in some sense. Then, by varying the action with respect
to the metric, we obtain,
\begin{equation}
\label{FRT2}
\frac{1}{2}g_{\mu\nu}
F(R,T) - R_{\mu\nu} \frac{\partial F(R,T)}{\partial R}
 - g_{\mu\nu} \Box \frac{\partial F(R,T)}{\partial R}
+ \nabla_\mu \nabla_\nu \frac{\partial F(R,T)}{\partial R}
+ \frac{\partial T}{\partial g_{\mu\nu}} \frac{\partial
F(R,T)}{\partial T} = - \frac{\kappa^2}{2}T_{\mathrm{matter}\,\mu\nu}\, .
\end{equation}
Due to the presence of the term including
$\frac{\partial T}{\partial g_{\mu\nu}}$,
the theory cannot be correctly defined without
specifying the metric dependence of $T$. Usually, the energy
momentum tensor $T_{\mu\nu}$ is given by the variation of some
action $S_T$ with respect to the metric. If the action $S_T$ is
given in terms of the fields, the action (\ref{FRT1}) can be
expressed in terms of the fields directly. We should also note that
due to the existence of the $F(R,T)$ term in the action of
Eq.~(\ref{FRT1}), the conservation of $T_{\mu\nu}$ is violated in
general. For the FRW Universe (\ref{JGRG14}), the $(t,t)$ component
and $(i,j)$ component in Eq.~(\ref{FRT2}) give the following
equations,
\begin{align}
\label{FRT3}
0 =& -\frac{F(R,T)}{2} + 3\left(H^2 + \dot H\right)
\frac{\partial F(R,T)}{\partial R}
 - 18 \left( 4H^2 \dot H + H \ddot H\right)
\frac{\partial^2 F(R,T)}{\partial R^2}
+ \frac{\partial T}{\partial g_{tt}} \frac{\partial F(R,T)}{\partial T}
+ \kappa^2 \rho_\mathrm{matter}\, ,\\
\label{FRT4}
0 =& \frac{F(R,T)}{2} - \left(\dot H
+ 3H^2\right)\frac{\partial F(R,T)}{\partial R} + 6 \left( 8H^2 \dot H
+ 4 {\dot H}^2 + 6 H \ddot H + \dddot H\right) \frac{\partial^2
F(R,T)}{\partial R^2} \nn & + 36\left( 4H\dot H + \ddot H\right)^2
\frac{\partial^3 F(R,T)}{\partial R^3} + \frac{1}{3} g^{ij}
\frac{\partial T}{\partial g_{ij}} \frac{\partial F(R,T)}{\partial T}
+ \kappa^2 p_\mathrm{matter}\, .
\end{align}
If we require the conservation laws for $T_{\mu\nu}$, the additional
constraints must be imposed on the model. Furthermore by imposing
additional assumptions just for the solvability of the theory,
several solutions have been studied
\cite{Momeni:2011am,Sharif:2012zzd,Houndjo:2011tu,Alvarenga:2013syu,
Haghani:2013oma,Odintsov:2013iba,Houndjo:2011fb}.


\subsection{Modified Gauss-Bonnet Gravity}

We consider another class of models in modified gravity in which an
arbitrary function of the topological Gauss-Bonnet invariant is
added to the action of General Relativity. We call this class of
modified gravity theory ``modified Gauss-Bonnet gravity''. This
class of modified gravity could be closely related with
(super)string theory.

\subsubsection{General properties}

The starting action is given by
\cite{Nojiri:2005jg,Nojiri:2005am,Cognola:2006eg,Elizalde:2010jx,Izumi:2014loa}:
\begin{equation}
\label{GB1b}
S=\int d^4x\sqrt{-g} \left(\frac{1}{2\kappa^2}R
+ f(\mathcal{G}) + \mathcal{L}_\mathrm{matter}\right)\, .
\end{equation}
By
varying the action with respect to the metric $g_{\mu\nu}$, we
obtain the following equations of motion,
\begin{align}
\label{GB4b}
0=& \frac{1}{2\kappa^2}\left(- R^{\mu\nu} + \frac{1}{2}
g^{\mu\nu} R\right) + T_\mathrm{matter}^{\mu\nu}
+ \frac{1}{2}g^{\mu\nu} f(\mathcal{G}) -2 f'(\mathcal{G}) R R^{\mu\nu}
\nn
& + 4f'(\mathcal{G})R^\mu_{\ \rho} R^{\nu\rho} -2
f'(\mathcal{G}) R^{\mu\rho\sigma\tau} R^\nu_{\ \rho\sigma\tau} - 4
f'(\mathcal{G}) R^{\mu\rho\sigma\nu}R_{\rho\sigma} + 2 \left(
\nabla^\mu \nabla^\nu f'(\mathcal{G})\right)R \nn & - 2 g^{\mu\nu}
\left( \nabla^2 f'(\mathcal{G})\right) R - 4 \left( \nabla_\rho
\nabla^\mu f'(\mathcal{G})\right) R^{\nu\rho} - 4 \left( \nabla_\rho
\nabla^\nu f'(\mathcal{G})\right)R^{\mu\rho} \nn
& + 4 \left(
\nabla^2 f'(\mathcal{G}) \right)R^{\mu\nu} + 4g^{\mu\nu} \left(
\nabla_{\rho} \nabla_\sigma f'(\mathcal{G}) \right) R^{\rho\sigma}
 - 4 \left(\nabla_\rho \nabla_\sigma f'(\mathcal{G}) \right)
R^{\mu\rho\nu\sigma} \, .
\end{align}
We should note that Eq.~(\ref{GB4b}) does not contain
higher-derivative terms, that is, the terms including third or
higher derivatives.

For the FRW metric with spatially flat part (\ref{JGRG14}), we obtain
the equation corresponding to the first FRW equation as follows,
\begin{equation}
\label{GB7b}
0=-\frac{3}{\kappa^2}H^2 - f(\mathcal{G})
+ \mathcal{G}f'(\mathcal{G}) - 24 \dot{\mathcal{G}}f''(\mathcal{G})
H^3 + \rho_\mathrm{matter}\, .
\end{equation}
In the FRW Universe
(\ref{JGRG14}), the Gauss-Bonnet invariant $\mathcal{G}$ has the
following form: $\mathcal{G} = 24 \left( H^2 \dot H + H^4 \right)$.

If the EoS parameter $w\equiv
p_\mathrm{matter}/\rho_\mathrm{matter}$ is a constant, by using the
conservation of energy equation:
\begin{equation}
\label{CEm}
\dot \rho_\mathrm{matter} + 3H\left(\rho_\mathrm{matter}
+ p_\mathrm{matter}\right)=0\, ,
\end{equation}
we find that $\rho=\rho_0 a^{-3(1+w)}$. If the function
$f(\mathcal{G})$ is given by,
\begin{equation}
\label{mGB1}
f(\mathcal{G})=f_0\left|\mathcal{G}\right|^\beta\, .
\end{equation}
in case $\beta<1/2$ and in the small curvature regime, the
$f(\mathcal{G})$ term becomes dominant and we can neglect the
Einstein term. Then Eq.~(\ref{GB7b}) has the following solution,
\begin{align}
\label{mGB3}
& a= a_0 t^{h_0} \quad \mbox{when}\ h_0>0\, ,
\quad a = a_0 \left(t_s - t\right)^{h_0}
\quad \mbox{when}\ h_0<0\, , \nn
& h_0=\frac{4\beta}{3(1+w)}\, ,\quad a_0=\left[ -\frac{f_0(\beta - 1)}
{\left(h_0 - 1\right)\rho_0}\left\{24 \left|h_0^3 \left(- 1
+ h_0\right) \right|\right\}^\beta \left( h_0 - 1
+ 4\beta\right)\right]^{-\frac{1}{3(1+w)}}\, .
\end{align}
Then we find that the EoS parameter $w_\mathrm{eff}$ in
(\ref{JGRG12}) is smaller than $-1$ if $\beta<0$, and for $w>-1$,
$w_\mathrm{eff}$ is given by,
\begin{equation}
\label{mGB3b}
w_\mathrm{eff}=-1 + \frac{2}{3h_0}=-1 + \frac{1+w}{2\beta}\, ,
\end{equation}
which is again
smaller than $-1$ if $\beta<0$. Thus, in the case $\beta<0$, we
obtain an effective phantom evolution with negative $h_0$, even if
$w>-1$. One could presume that in the phantom phase, a Big Rip
singularity could occur at $t=t_s$. Near this Big Rip, however, the
curvature becomes large and the Einstein term dominates, hence the
$f(\mathcal{G})$ term can be neglected. This indicates that the
Universe behaves as $a=a_0t^{2/3(w+1)}$ and the Big Rip singularity
does not appear. Therefore the phantom era is transient regime.

We may consider the case that $0<\beta<1/2$. In this case, the
phantom phase does not appear because $\beta$ is positive. When the
curvature is large, since the $f(\mathcal{G})$ term in the action
(\ref{GB1b}) can be neglected, the theory can be approximated by the
Einstein gravity solely. Then, if $w$ is positive, the matter energy
density $\rho_\mathrm{matter}$ should behave as
$\rho_\mathrm{matter}\sim t^{-2}$, but $f(\mathcal{G})$ goes as
$f(\mathcal{G})\sim t^{-4\beta}$. At late times, so at large $t$
values, the $f(\mathcal{G})$ term may become dominant compared with
the Lagrangian density of matter. Then by neglecting the
contribution from matter, Eq.~(\ref{GB7b}) has the de Sitter Universe
solution. Even if we start from the deceleration phase with
$w>-1/3$, the asymptotically de Sitter Universe emerges.
Correspondingly, a transition from deceleration to acceleration of
the Universe occurs. We may also construct a model which can
describe both the inflation and a late-time accelerating expansion
of the Universe in a unified way.

We now consider the correction to the Newton law. Let $g_{(0)}$ be a
solution of (\ref{GB4b}) and we express the perturbation of the
metric as $g_{\mu\nu}=g_{(0)\mu\nu} + h_{\mu\nu}$. First, we
consider the perturbation around the de Sitter background, whose
metric is taken as $g_{(0)\mu\nu}$. Then, the Riemann tensor is
equal to
\begin{equation}
\label{GB35}
R_{(0)\mu\nu\rho\sigma}=H_0^2\left(g_{(0)\mu\rho} g_{(0)\nu\sigma}
 - g_{(0)\mu\sigma}g_{(0)\nu\rho}\right)\, .
\end{equation}
We may impose the
following gauge condition: $g_{(0)}^{\mu\nu}
h_{\mu\nu}=\nabla_{(0)}^\mu h_{\mu\nu}=0$. Then by using
Eq.~(\ref{GB4b}), we find
\begin{equation}
\label{GB38b}
0=\frac{1}{4\kappa^2}
\left(\nabla^2 h_{\mu\nu} - 2H_0^2 h_{\mu\nu}\right)
+ T_{\mathrm{matter}\, \mu\nu}\, .
\end{equation}
Then we find that the
contribution from the Gauss-Bonnet term does not appear except in
the length parameter $1/H_0$ of the de Sitter space, which is
determined by taking into account the Gauss-Bonnet term. This could
be due to the topological structure of the Gauss-Bonnet invariant.
 From Eq.~(\ref{GB38b}), we find that there is no correction to
the Newton law in the de Sitter space and also in the flat background,
which corresponds
to the limit of $H_0\to 0$. This property does not depend on the
form of $f$. In the case of $F(R)$ gravity, significant corrections
appear in the Newton law in $1/R$ gravity, as was shown in
Refs.~\cite{Capozziello:2004sm}, which conflicts with the
observations coming from the solar system.

We now consider the relation between the modified Gauss-Bonnet
gravity and the scalar-Einstein-Gauss-Bonnet gravity. By introducing
two auxiliary scalar fields, $A$ and $B$, we can rewrite the action
(\ref{GB1b}) as follows,
\begin{equation}
\label{GB3}
S = \int d^4
x\sqrt{-g}\left(\frac{1}{2\kappa^2}R + B\left(\mathcal{G}-A\right)
+ f(A) + \mathcal{L}_\mathrm{matter} \right)\, .
\end{equation}
By the variation
of the action (\ref{GB3}) with respect to $B$, we obtain
$A=\mathcal{G}$. By substituting the expression into (\ref{GB3}),
the action (\ref{GB1b}) is reobtained. On the other hand, the
variation of the action (\ref{GB3}) with respect to $A$ in
(\ref{GB3}), we obtain $B=f'(A)$. Then we rewrite the action
(\ref{GB1b}) as follows,
\begin{equation}
\label{GB6}
S=\int d^4
x\sqrt{-g}\left(\frac{1}{2\kappa^2}R + f'(A)\mathcal{G} - A f'(A)
+ f(A)\right)\, .
\end{equation}
By varying the action (\ref{GB6}) with respect
to $A$, we reobtain the relation $A=\mathcal{G}$. The scalar field
$A$ in (\ref{GB6}) is not dynamic because there is no kinetic term
for $A$. We may add, however, a kinetic term for $A$ to the action
by hand as follows,
\begin{equation}
\label{GB6b}
S=\int d^4 x\sqrt{-g}\left(
\frac{1}{2\kappa^2}R - \frac{\epsilon}{2}\partial_\mu A \partial^\mu A
+ f'(A)\mathcal{G} - A f'(A) + f(A)\right)\, .
\end{equation}
Then the
resulting theory is a dynamical scalar theory coupled with the
Gauss-Bonnet invariant and with a potential, which is the
scalar-Einstein-Gauss-Bonnet gravity. In general, the
scalar-Einstein-Gauss-Bonnet gravity has no ghosts and is stable.
This model is related to string-inspired dilaton gravity, which was
also proposed as an alternative model for the dark energy problem
\cite{Nojiri:2005vv,Sami:2005zc,Calcagni:2005im}. Then, since we can
obtain the limit of $\epsilon\to 0$ smoothly, the corresponding
$f(\mathcal{G})$ theory could not have any ghost and could be
stable.

\subsection{String-inspired Gravity}

Superstring theory is a ten dimensional space-time theory. Since the
observed space-time looks four dimensional, superstring theory was
equipped with a scenario which describes the compactification from
ten dimensions to four. The compactification technique makes use of
many kinds of scalar fields, which are known under the names
``moduli'' or ``dilaton'', and these scalar fields are coupled with
the curvature invariants. We now neglect the moduli fields
associated with the radii of the internal space and we only consider
the following action of the low-energy effective dilaton string
theories. The general action of the theory is given by,
\begin{equation}
\label{eq:action}
S = \int d^4 x \sqrt{-g}\left[\frac{R}{2}+\mathcal{L}_{\phi}
+ \mathcal{L}_{c}+\ldots\right]\, .
\end{equation}
In the above equation, $\phi$
is the dilaton field, which is related to the string coupling,
$\mathcal{L}_{\phi}$ is the Lagrangian of $\phi$, and
$\mathcal{L}_c$ expresses the string curvature correction terms to
the Einstein-Hilbert action,
\begin{equation}
\label{stcor}
\mathcal{L}_{\phi} = -\partial_{\mu}\phi\partial^{\mu}\phi-V(\phi) \, ,
\quad \mathcal{L}_c = c_1 {\alpha'}
\e^{2\frac{\phi}{\phi_0}}\mathcal{L}_c^{(1)}
+ c_{\alpha'}^2\e^{4\frac{\phi}{\phi_0}}\mathcal{L}_c^{(2)}
+ c_3{\alpha'}^3\e^{6\frac{\phi}{\phi_0}}\mathcal{L}_c^{(3)}\, .
\end{equation}
In addition, $1/\alpha'$ is the string tension,
$\mathcal{L}_c^{(1)}$, $\mathcal{L}_c^{(2)}$, and
$\mathcal{L}_c^{(3)}$ express the leading-order (the Gauss-Bonnet
term $\mathcal{G}$), the second-order, and the
third-order curvature corrections, respectively.
The Gauss-Bonnet invariant $\mathcal{G}$, which
is defined by
\begin{equation}
\label{GB}
\mathcal{G}=R^2 -4 R_{\mu\nu}
R^{\mu\nu} + R_{\mu\nu\xi\sigma} R^{\mu\nu\xi\sigma}\, .
\end{equation}
The explicit forms
of the terms $\mathcal{L}_c^{(1)}$, $\mathcal{L}_c^{(2)}$ and
$\mathcal{L}_c^{(3)}$ are given by,
\begin{equation}
\label{ccc}
\mathcal{L}_c^{(1)} = \Omega_2\, , \quad
\mathcal{L}_c^{(2)} = 2 \Omega_3
+ R^{\mu\nu}_{\alpha \beta} R^{\alpha\beta}_{\lambda\rho}
R^{\lambda\rho}_{\mu\nu}\, , \quad \mathcal{L}_c^{(3)}
= \mathcal{L}_{31} - \delta_\mathrm{H} \mathcal{L}_{32}
 -\frac{\delta_{B}}{2} \mathcal{L}_{33}\, .
\end{equation}
Here, $\delta_B$ and
$\delta_H$ equal to $0$ or $1$ and
\begin{align}
\Omega_2 =& \mathcal{G} \, , \nn
\Omega_3 \propto &
\epsilon^{\mu\nu\rho\sigma\tau\eta}
\epsilon_{\mu'\nu'\rho'\sigma'\tau'\eta'} R_{\mu\nu}^{\ \ \mu'\nu'}
R_{\rho\sigma}^{\ \ \rho'\sigma'} R_{\tau\eta}^{\ \ \tau'\eta'} \, , \nn
\mathcal{L}_{31} =& \zeta(3)
R_{\mu\nu\rho\sigma}R^{\alpha\nu\rho\beta} \left( R^{\mu\gamma}_{\ \
\delta\beta} R_{\alpha\gamma}^{\ \ \delta\sigma} - 2
R^{\mu\gamma}_{\ \ \delta\alpha} R_{\beta\gamma}^{\ \ \delta\sigma}
\right)\, , \nn
\mathcal{L}_{32} =& \frac{1}{8} \left(
R_{\mu\nu\alpha\beta} R^{\mu\nu\alpha\beta}\right)^2 + \frac{1}{4}
R_{\mu\nu}^{\ \ \gamma\delta} R_{\gamma\delta}^{\ \ \rho\sigma}
R_{\rho\sigma}^{\ \ \alpha\beta} R_{\alpha\beta}^{\ \ \mu\nu}
 -\frac{1}{2} R_{\mu\nu}^{\ \ \alpha\beta} R_{\alpha\beta}^{\ \
\rho\sigma} R^\mu_{\ \sigma\gamma\delta} R_\rho^{\ \nu\gamma\delta}
 - R_{\mu\nu}^{\ \ \alpha\beta} R_{\alpha\beta}^{\ \ \rho\nu}
R_{\rho\sigma}^{\ \ \gamma\delta} R_{\gamma\delta}^{\ \ \mu\sigma}
\, , \nn
\mathcal{L}_{33} =& \left( R_{\mu\nu\alpha\beta}
R^{\mu\nu\alpha\beta}\right)^2 - 10 R_{\mu\nu\alpha\beta}
R^{\mu\nu\alpha\sigma} R_{\sigma\gamma\delta\rho}
R^{\beta\gamma\delta\rho} - R_{\mu\nu\alpha\beta} R^{\mu\nu\rho}_{\
\ \ \sigma} R^{\beta\sigma\gamma\delta} R_{\delta\gamma\rho}^{\ \ \
\alpha} \, .
\end{align}
The values of $\delta_B$ and $\delta_H$ depend on the type of string
theory. The dependence is encoded in the curvature invariants, in
the coefficients $(c_1,c_2,c_3)$ and finally in the parameters
$\delta_H$, $\delta_B$, as follows,
\begin{itemize}
\item For the Type II superstring theory: $(c_1,c_2,c_3) = (0,0,1/8)$
and $\delta_H=\delta_B=0$.
\item For the heterotic superstring theory: $(c_1,c_2,c_3) = (1/8,0,1/8)$
and $\delta_H=1,\delta_B=0$.
\item For the bosonic superstring theory: $(c_1,c_2,c_3) = (1/4,1/48,1/8)$
and $\delta_H=0,\delta_B=1$.
\end{itemize}
Motivated by the string considerations, we now consider the theories
of the scalar-Einstein-Gauss-Bonnet gravity\footnote{For pioneering
work on the scalar-Einstein-Gauss-Bonnet gravity, see
\cite{Boulware:1986dr}.} based on
\cite{Nojiri:2005vv,Nojiri:2006je}. It was first proposed in
Ref.~\cite{Nojiri:2005vv} to consider the
scalar-Einstein-Gauss-Bonnet gravity as a gravitational model of the
dark energy.

The starting action is given by,
\begin{equation}
\label{ma22}
S=\int d^4 x
\sqrt{-g}\left[ \frac{R}{2\kappa^2} - \frac{1}{2}
\partial_\mu \phi
\partial^\mu \phi - V(\phi) - \xi(\phi) \mathcal{G}\right]\, .
\end{equation}
Here, we do not specify the forms of the potential $V(\phi)$ and
the Gauss-Bonnet coupling $\xi(\phi)$, which could be determined by
the non-perturbative effects in string theory (\ref{stcor}). We
should note that the action (\ref{ma22}) is given by adding the
kinetic term for the scalar field $\phi$ to the action of the
$f(\mathcal{G})$ gravity in the scalar-tensor form (\ref{GB6}) or by
putting $A=\phi$ and $\epsilon=1$ in (\ref{GB6b}).

By varying the action (\ref{ma22}) with respect to the metric
$g_{\mu\nu}$, we obtain the following equation,
\begin{align}
\label{ma23CC}
0=& \frac{1}{\kappa^2}\left(- R^{\mu\nu}
+ \frac{1}{2} g^{\mu\nu} R\right) + \frac{1}{2}\partial^\mu \phi
\partial^\nu \phi - \frac{1}{4}g^{\mu\nu}
\partial_\rho \phi \partial^\rho \phi
 - \frac{1}{2}g^{\mu\nu} V(\phi) \nn
& + 2 \left( \nabla^\mu \nabla^\nu \xi(\phi)\right)R - 2 g^{\mu\nu}
\left( \nabla^2\xi(\phi)\right)R - 4 \left( \nabla_\rho \nabla^\mu
\xi(\phi)\right)R^{\nu\rho} - 4 \left( \nabla_\rho \nabla^\nu
\xi(\phi)\right)R^{\mu\rho} \nn & + 4 \left( \nabla^2 \xi(\phi)
\right)R^{\mu\nu} + 4g^{\mu\nu} \left( \nabla_{\rho} \nabla_\sigma
\xi(\phi) \right) R^{\rho\sigma} + 4 \left(\nabla_\rho \nabla_\sigma
\xi(\phi) \right) R^{\mu\rho\nu\sigma} \, ,
\end{align}
where we have used the following identity which holds true in four
dimensions: $0 = \frac{1}{2}g^{\mu\nu} \mathcal{G} -2 R R^{\mu\nu}
 - 4 R^\mu_{\ \rho} R^{\nu\rho}
 -2 R^{\mu\rho\sigma\tau}R^\nu_{\ \rho\sigma\tau}
+4 R^{\mu\rho\nu\sigma}R_{\rho\sigma}$. We should note that in
Eq.~(\ref{ma23CC}), the derivatives of curvature such as $\nabla R$, do
not appear, which tells that the derivatives higher than two do not
appear.

This situation is contrasted with a general $\alpha R^2 + \beta
R_{\mu\nu}R^{\mu\nu} + \gamma
R_{\mu\nu\rho\sigma}R^{\mu\nu\rho\sigma}$ gravity, where fourth
derivatives of $g_{\mu\nu}$ appear in general. This tells that for
the classical theory, if we specify the values of $g_{\mu\nu}$ and
$\dot g_{\mu\nu}$ on a spatial surface as an initial condition, the
time dependence is uniquely determined. This situation is similar to
the case in classical mechanics, where we only need to specify the
values of the position and of the velocity of particle as initial
conditions. On the other hand, in general $\alpha R^2 + \beta
R_{\mu\nu}R^{\mu\nu} + \gamma
R_{\mu\nu\rho\sigma}R^{\mu\nu\rho\sigma}$ gravity, as initial
conditions we need to specify the values of $\ddot g_{\mu\nu}$ and
$\dddot g_{\mu\nu}$ in addition to $g_{\mu\nu}$, $\dot g_{\mu\nu}$
in order to obtain a unique time evolution. In the Einstein gravity,
we only need to specify the values of $g_{\mu\nu}$ and $\dot
g_{\mu\nu}$. This may tell that the scalar-Gauss-Bonnet gravity
could be a natural extension of the Einstein gravity.

In the FRW Universe (\ref{JGRG14}), Eqs.~(\ref{ma23CC}) have the
following form,
\begin{align}
\label{ma24}
0 =& - \frac{3}{\kappa^2}H^2 + \frac{1}{2}{\dot\phi}^2 + V(\phi)
+ 24 H^3 \frac{d \xi(\phi(t))}{dt}\, ,\\
\label{GBany5}
0 =& \frac{1}{\kappa^2}\left(2\dot H + 3 H^2 \right)
+ \frac{1}{2}{\dot\phi}^2 - V(\phi) - 8H^2 \frac{d^2 \xi(\phi(t))}
{dt^2} - 16H \dot H \frac{d\xi(\phi(t))}{dt} - 16 H^3 \frac{d
\xi(\phi(t))}{dt} \, .
\end{align}
In particular when we consider the following model inspired by
string theory \cite{Nojiri:2005vv},
\begin{equation}
\label{NOS1}
V=V_0\e^{-\frac{2\phi}{\phi_0}}\, , \quad \xi(\phi)=\xi_0
\e^{\frac{2\phi}{\phi_0}}\, ,
\end{equation}
we find that the solution which describes the de Sitter space is,
\begin{equation}
\label{NOS2}
H^2 = H_0^2 \equiv - \frac{\e^{-\frac{2\varphi_0}{\phi_0}}} {8\xi_0 \kappa^2} \, ,
\quad \phi = \varphi_0 \, ,
\end{equation}
where $\varphi_0$ is an arbitrary
constant. When $\varphi_0$ is larger, the Hubble rate $H=H_0$ is
smaller. Then in the case $\xi_0\sim \mathcal{O}(1)$, if we choose
$\varphi_0/\phi_0\sim 140$, the value of the Hubble rate $H=H_0$
becomes consistent with the observations of the current Universe.
There is also another exact solution for the model (\ref{NOS1})
which is,
\begin{align}
\label{NOS3}
H=&\frac{h_0}{t}\, ,\quad \phi=\phi_0
\ln \frac{t}{t_1}\ \mbox{when}\ h_0>0\, ,\nn
H=&-\frac{h_0}{t_s - t}\, ,\quad \phi=\phi_0 \ln \frac{t_s - t}{t_1}\
\mbox{when}\ h_0<0\, .
\end{align}
Here the constant $h_0$ is determined by solving the
following algebraic equations:
\begin{equation}
\label{NOS4}
0 = -\frac{3h_0^2}{\kappa^2} + \frac{\phi_0^2}{2}
+ V_0 t_1^2 - \frac{48 \xi_0 h_0^3}{t_1^2}\, ,\quad
0 = \left( 1 - 3h_0 \right)\phi_0^2 + 2V_0 t_1^2
+ \frac{48 \xi_0 h_0^3}{t_1^2}\left(h_0 - 1\right)\, .
\end{equation}
We can realize an arbitrary value of $h_0$ by properly choosing
the values of $V_0$ and $\xi_0$. By appropriately choosing $V_0$ and
$\xi_0$, it is possible to obtain a negative $h_0$, which
corresponds to an effective phantom regime, where the EoS parameter
(\ref{JGRG12}) is less than $-1$, $w_\mathrm{eff} < -1$. We should
note that the phantom era cannot be realized by a canonical scalar
in the usual scalar-tensor theory without the Gauss-Bonnet term.

\subsection{$F(T)$ Gravity}

As a gravitational theory besides Einstein's general relativity, we
may consider ``teleparallelism'' with the Weitzenb\"{o}ck
connection, where the torsion $T$ is used as a fundamental
ingredient instead of the curvature $R$, defined by the Levi-Civita
connection \cite{Hehl:1976kj,Hayashi:1979qx,Cai:2015emx,
Flanagan:2007dc,Maluf:2013gaa,Garecki:2010jj,Ferraro:2008ey}. In
order to explain both of the inflationary era and the accelerating
expansion of the current Universe, we may consider a model whose
Lagrangian density is extended to be a function of $T$ as $F(T)$
\cite{Bengochea:2008gz,Maluf:2013gaa,
Linder:2010py,Cai:2011tc,Chen:2010va,Bamba:2010wb,
Zhang:2011qp,Nashed:2014uta,Geng:2011aj,Bohmer:2011si,
Nashed:2014lva,Gonzalez:2011dr,Karami:2012fu,Bamba:2012vg,
Rodrigues:2012qua,Capozziello:2012zj,Chattopadhyay:2012eu,
Izumi:2012qj,Li:2013xea,Ong:2013qja,Otalora:2013tba,
Nashed:2013bfa,Kofinas:2014owa,Harko:2014sja,Hanafy:2014bsa,
Junior:2015bva,Ruggiero:2015oka,Hanafy:2014ica,Nunes:2016plz}, from
the teleparallel Lagrangian density given by the torsion scalar $T$
instead of $R$ in the Einstein gravity.

In contrast to other modified gravity theories studied in this
review, the $F(T)$ gravity theory does not have the local Lorentz
symmetry in general. Particularly, the torsion is not invariant
under the local Lorentz transformation. This is due to the fact that
we start with globally defined vierbein fields and in effect, the
theory depends on the choice of the vierbein even if the metric
induced from the vierbein is identical. In the case of general
relativity for example, we cannot define the vierbein field globally
in general, because there is a curvature for the local Lorentz
symmetry. The exception to this rule is the case that the action is
linear to the torsion $T$.  The difference from the Einstein-Hilbert
action, when we write the action in terms of the vierbein, is a
total derivative and therefore the symmetry is enhanced in this
case, and the action becomes invariant under the local Lorentz
transformation. Although the torsion is not invariant under the
local Lorentz transformation, the torsion is invariant under the
global Lorentz symmetry. The global Lorentz symmetry is the symmetry
of the tangent space of the space-time manifold. Therefore, if we
choose the vierbein to be diagonal, the space-time has the Lorentz
symmetry, which is inherited from the global Lorentz symmetry. The
lack of the local Lorentz symmetry, however, generates some
difficulties. For example, we cannot consistently couple the spinor
fields with $F(T)$ gravity, and therefore the model cannot be
realistic.

We need to note that the lack of the local Lorentz invariance is a
crucial difference between $F(R)$ gravity and $F(T)$ gravity, as it
was explained in \cite{Cai:2015emx}. Particularly, when the standard
formulation of $F(T)$ gravity is used, the Lorentz invariance is
absent or strongly restricted, since the spin connection vanishes.
The latter assumption makes the theory simpler for the derivation of
solutions, at the expense of having local Lorentz invariance
violation and therefore the theory is frame dependent. The Lorentz
violation issue can cause serious complications in the theory, which
may become apparent when spherically symmetric solutions are
discussed. The only way to construct a Lorentz invariant $F(T)$
theory is to impose covariance in the formulation, where both the
vierbein and the spin connection are used, in which case the
resulting connection makes the theory covariant, see for example
\cite{Krssak:2015oua}. However, for the FRW space-time, the absence of
local Lorentz invariance is not an important issue, so for the sake
of simplicity we shall use the standard formulation of $F(T)$
gravity.

In the context of $F(T)$ gravity, it is possible to explain the
late-time acceleration of the Universe
\cite{Bamba:2013jqa,Bengochea:2008gz,Linder:2010py,Geng:2011aj,
Otalora:2013tba,Chattopadhyay:2012eu,Dent:2011zz,Yang:2010hw,
Bamba:2010wb,Capozziello:2011hj,Geng:2011ka,Farajollahi:2011af,
Cardone:2012xq,Bahamonde:2015zma}, but there exist also local
astrophysical solutions, and various metric solutions, see for
example
\cite{Capozziello:2012zj,Paliathanasis:2014iva,Gonzalez:2011dr,
Bohmer:2011si,Nashed:2013bfa,Ruggiero:2015oka}. Finally for
inflationary, bouncing cosmology and perturbation evolution studies,
see \cite{Cai:2011tc,Chen:2010va,Izumi:2012qj,Nashed:2014lva,
Hanafy:2014bsa,Hanafy:2014ica,Ferraro:2006jd,Oikonomou:2017isf,Oikonomou:2017brl}.

We shall use the orthonormal tetrad components $e_A (x^{\mu})$ for
the teleparallelism theory analysis. We assume that the index $A$
runs over $0$, $1$, $2$, $3$ for the tangent space at each point
$x^{\mu}$ of the manifold and $e_A^\mu$ form the tangent vector of
the manifold. The relation with the metric $g^{\mu\nu}$ is given by
$g_{\mu\nu}=\eta_{A B} \e^A_{\mu} \e^B_\nu$. The torsion
$T^\rho_{\verb| |\mu \nu}$ and contorsion $K^{\mu\nu}_{\verb||\rho}$
tensors are defined as,
\begin{equation}
\label{eq:2.2}
T^\rho_{\ \mu\nu} \equiv
\e^\rho_A \left( \partial_\mu \e^A_\nu - \partial_\nu \e^A_\mu
\right)\, , \quad K^{\mu\nu}_{\ \ \rho} \equiv -\frac{1}{2}
\left(T^{\mu\nu}_{\ \ \rho} - T^{\nu \mu}_{\ \ \rho}
 - T_\rho^{\ \ \mu\nu}\right)\, .
\end{equation}
Furthermore the torsion scalar $T$ is defined
by \cite{Hayashi:1979qx,Maluf:2013gaa},
\begin{equation}
\label{eq:2.4}
T \equiv
S_\rho^{\ \mu\nu} T^\rho_{\ \mu\nu}\, , \quad S_\rho^{\ \mu\nu}
\equiv \frac{1}{2} \left(K^{\mu\nu}_{\ \ \rho} + \delta^\mu_\rho
T^{\alpha \nu}_{\ \ \alpha}
 - \delta^\nu_\rho T^{\alpha \mu}_{\ \ \alpha}\right)\, .
\end{equation}
Then the action of the modified teleparallel gravity describing
$F(T)$ gravity \cite{Linder:2010py} is given by,
\begin{equation}
\label{eq:2.6}
S = \int d^4x |e| \left[ \frac{F(T)}{2{\kappa}^2}
+{\mathcal{L}}_{\mathrm{matter}} \right]\, .
\end{equation}
Here $|e|= \det \left(\e^A_\mu \right)=\sqrt{-g}$.
Upon variation of the action
Eq.~(\ref{eq:2.6}) with respect to the vierbein vector field
$e_A^\mu$ we obtain, \cite{Bengochea:2008gz}
\begin{equation}
\label{eq:2.7}
\frac{1}{e}
\partial_\mu \left( eS_A^{\ \mu\nu} \right) f'
 - e_A^\lambda T^\rho_{\ \mu \lambda} S_\rho^{\ \nu\mu} f'
+S_A^{\ \mu\nu} \partial_\mu T f'' +\frac{1}{4} e_A^\nu f
= \frac{{\kappa}^2}{2} e_A^\rho {T_\mathrm{matter}}_{\rho}^{\ \nu}\, .
\end{equation}
We consider the FRW metric with flat spacial part (\ref{JGRG14})
and then we find the tetrad components $\e^A_\mu = \mathrm{diag}
\left(1,a,a,a \right)$, which yields $g_{\mu \nu}= \mathrm{diag}
\left(1, -a^2, -a^2, -a^2 \right)$. Then we find that the torsion
scalar $T$ is given by $T=-6H^2$. In the flat FRW background
(\ref{JGRG14}) , Eq.~(\ref{eq:2.7}) takes the following form,
\begin{equation}
\label{FT}
\frac{3}{\kappa^2} H^2 = \rho_\mathrm{matter} + \rho_\mathrm{DE}\, ,
\quad \frac{1}{\kappa^2} \left( H^2 + \dot H
\right) = p_\mathrm{matter} + p_\mathrm{DE} \, ,
\end{equation}
where,
\be
\label{eq:4.3}
\rho_{\mathrm{DE}} = \frac{1}{2{\kappa}^2}
\left( -T -f +2T f' \right) \, , \quad p_{\mathrm{DE}}
= -\frac{1}{2{\kappa}^2} \left( \left(4 - 4 f' - 8 T f''\right) \dot H
 -T -f +2T f' \right)\, .
\end{equation}We should note that $\rho_{\mathrm{DE}}$
and $p_{\mathrm{DE}}$
in Eq.~(\ref{eq:4.3}) satisfy the standard continuity equation
\begin{equation}
\label{eq:4.5}
0 = \dot{\rho}_{\mathrm{DE}}+3H \left(
\rho_{\mathrm{DE}} + p_{\mathrm{DE}} \right) \, .
\end{equation}
In the vacuum
case, that is when $\rho_\mathrm{matter} = p_\mathrm{matter} = 0$,
since $T=-6H^2$, the first Eq.~(\ref{FT}) gives
\begin{equation}
\label{FT2}
0 =- f + 2 T f' \, , \quad 0=f' + 2Tf''\, ,
\end{equation}
which can be integrated and we obtain,
\begin{equation}
\label{FT3}
f (T) = f_0 \left( T \right)^{\frac{1}{2}} \, ,
\end{equation}
where $f_0$ is a constant. Therefore,
the only consistent $F(T)$ is uniquely given by (\ref{FT2}). On the
other hand, if $F(T)$ is given by (\ref{FT3}), Eqs.~(\ref{FT}) can
be satisfied regardless of what the choice of $H$ is, which
indicates that an arbitrary cosmological evolution can be realized
by (\ref{FT3}). There are many works related with $F(T)$ gravity and
several solutions have been found, see for example,
\cite{Cai:2015emx,Iorio:2012cm,Capozziello:2012zj,Rodrigues:2012qua,
Paliathanasis:2014iva,Amoros:2013nxa,Jamil:2012ju,
Basilakos:2013rua,Bamba:2013fta,
Bamba:2014eea,Dent:2011zz}.

\subsection{Massive $F(R)$ Gravity and Massive Bigravity}

In the gravity models we presented in the previous sections, the
accelerating expansion of the Universe is achieved by the
condensation of scalar quantities. If there could occur the
condensation of a vector field, and especially of the spatial
components of the vector field, the isotropy of space could be
broken and therefore the theory would be rendered inconsistent.
Besides the case of a scalar quantity condensation, the condensation
of a rank-two symmetric tensor field, does not always break the
isotropy of space, if the spatial components of the tensor field are
proportional to the spatial components of the metric tensor. Usually
the rank-two symmetric tensor field corresponds to spin-two particles.
Since the massless mode cannot condense, we need to consider a
theory of massive spin-two particles, which could be regarded as
massive gravitons. Since the typical scale of the accelerating
expansion of the present Universe is about $10^{-33}\, \mathrm{eV}$,
it is natural to expect that the mass of the massive graviton would
be almost $10^{-33}\, \mathrm{eV}$. Since the mass is so small, we
may naively expect that the interaction of the massive graviton
could not be changed from that of the standard massless graviton,
although the situation is not so simple because the degrees of
freedom of the massless spin-two particle are two, which corresponds
to the helicity, but the degrees of freedom of the massive spin-two
field are five. There is a very long history in the study of the
massive spin-two particles. The free theory was found by  Fierz and
Pauli in 1939 \cite{Fierz:1939ix}. It was soon found that if the
massive spin-two field couples with the matter fields as in the
Einstein gravity, the extra three degrees of freedom do not
decouple, because their coupling with matter diverges in the
massless limit, which is called as vDVZ discontinuity
\cite{vanDam:1970vg,Zakharov:1970cc}. It was found, however, that
the extra three degrees of freedom can be screened by the Vainshtein
mechanism \cite{Vainshtein:1972sx} if we introduce a non-linearity
as in the Einstein gravity case. The non-linearity, however,
generates an extra degree of freedom, which is a ghost called the
Boulware-Deser ghost \cite{Boulware:1974sr}. The symmetric tensor
field has ten components in four dimensions. In case of the Einstein
gravity, the re-parametrization invariance removes four degrees of
freedom and one lapse function and three shift functions give in
total four constraints and therefore only two degrees of freedom,
corresponding to the helicity remain. In the case of the massive
graviton model by Fierz and Pauli, the three components
corresponding to the shift functions can be solved with respect to
other components and the component corresponding to the lapse
function becomes the Lagrange multiplier field, which is not
dynamical and gives one constraint, and therefore there remain five
degrees of freedom. If we introduce the non-linearity in the theory,
however, the shift function does not become the Lagrange multiplier
field but the shift function is solved with respect to other
components and does not give any further constraint.

Then there appears one extra degree of freedom, which corresponds to
the Boulware-Deser ghost. After that, there had not been much
progress in the study of the massive gravity for a long time until
the dRGT model \cite{deRham:2010ik,deRham:2010kj} was found. In the
formulation of the dRGT model, it becomes clear how we can introduce
the non-linearity to keep the shift function to be the Lagrange
multiplier field and therefore there does not appear any ghost. It
is possible if the massive spin-two particle coupled with the standard
gravity, which is called as bigravity and even in this model, the
ghost does not appear \cite{Hassan:2011zd}. For the general review
on the massive gravity, see \cite{Hinterbichler:2011tt} and for a
more recent review, see \cite{deRham:2014zqa}.

In the massive gravity, the same cosmological constant is replaced
by the small mass of the massive graviton. Therefore, the problem of
the cosmological constant is never solved, but we need to explain
the unnaturally small mass of the massive graviton. The massive
gravity model could be, however, a low energy effective theory,
because the unitarity is broken in high energy regime, even for the
dRGT model. Therefore, the small mass might be explained by a more
fundamental theory. Another problem could be that it is difficult to
generate the accelerating expansion of the Universe only by the
massive gravity theory or the bigravity theory. Then we need to
extend the models to the $F(R)$ gravity form  in the way we
demonstrate in the present section. By such an extension, however,
we need to introduce more small parameters beside the graviton mass
and therefore the situation of the cosmological constant problem
becomes much more complicated than in the standard $\Lambda$CDM
model. In spite of such difficulties, we may expect that a more
fundamental theory could solve the problem and the study of the
$F(R)$ extension we present in this section may provide new hints
for the more fundamental theory.

Now we consider a model of massive gravity which is an extension of
$F(R)$ gravity, whose action is given by,
\begin{equation}
\label{mFR1}
S_\mathrm{mg} = M_g^2\int d^4x\sqrt{-\det g}\,F \left(R^{(g)}
+2m^2 \int d^4x\sqrt{-\det g}\sum_{n=0}^{4} \beta_n\,
e_n\left(\sqrt{g^{-1} f} \right) \right)+ S_\mathrm{matter}\, .
\end{equation}
Here $R^{(g)}$ is the scalar curvature for $g_{\mu \nu}$ and $f_{\mu
\nu}$ is a non-dynamical reference metric. We define the tensor
$\sqrt{g^{-1} f}$ by the square root of $g^{\mu\rho} f_{\rho\nu}$,
$\left(\sqrt{g^{-1} f}\right)^\mu_{\ \rho} \left(\sqrt{g^{-1}
f}\right)^\rho_{\ \nu} = g^{\mu\rho} f_{\rho\nu}$. For a general
tensor $X^\mu_{\ \nu}$, we define the quantities $e_n(X)$'s as
follows,
\begin{align}
\label{ek}
& e_0(X)= 1 \, , \quad e_1(X)= [X] \, , \quad e_2(X)=
\tfrac{1}{2}([X]^2-[X^2])\, ,\nn & e_3(X)=
\tfrac{1}{6}([X]^3-3[X][X^2]+2[X^3]) \, ,\nn & e_4(X)
=\tfrac{1}{24}([X]^4-6[X]^2[X^2]+3[X^2]^2 +8[X][X^3]-6[X^4])\, ,\nn
& e_k(X) = 0 ~~\mbox{for}~ k>4 \, .
\end{align}
Here $[X]$ expresses the trace of arbitrary tensor $X^\mu_{\ \nu}$,
$[X] \equiv X^\mu_{\ \mu}$.

In the following, just for simplicity, we only consider the minimal
case, where,
\begin{equation}
\label{mFR2}
2m^2 \sum_{n=0}^{4} \beta_n\, e_n
\left(\sqrt{g^{-1} f} \right) = 2m^2 \left( 3 - \tr \sqrt{g^{-1} f}
+ \det \sqrt{g^{-1} f} \right)\, .
\end{equation}
Then the equation motion
which are obtained from the action when it is varied with respect to
the metric, have the following form:
\begin{align}
\label{mFR3}
0 =& M_g^2 \left( \frac{1}{2} g_{\mu\nu} F\left(
{\tilde R}^{(g)} \right) - R^{(g)}_{\mu\nu} F'\left( {\tilde
R}^{(g)} \right) + \nabla_\nu \nabla_\mu F'\left( {\tilde R}^{(g)}
\right) - g_{\mu\nu} \nabla^2 F'\left( {\tilde R}^{(g)} \right)
\right) \nn & + m^2 M_g^2 F'\left( {\tilde R}^{(g)} \right)\left\{
\frac{1}{2} f_{\mu\rho} \left( \sqrt{ g^{-1} f } \right)^{-1\,
\rho}_{\qquad \nu} + \frac{1}{2} f_{\nu\rho} \left( \sqrt{ g^{-1} f
} \right)^{-1\, \rho}_{\qquad \mu} - g_{\mu\nu} \det \sqrt{g^{-1}
f}
\right\} 
+ \frac{1}{2} T_{\mathrm{matter}\, \mu\nu} \, .
\end{align}
Here,
\begin{equation}
\label{mFR4}
{\tilde R}^{(g)} \equiv R^{(g)} + 2m^2 \left(
3 - \tr \sqrt{g^{-1} f}+ \det \sqrt{g^{-1} f} \right)\, .
\end{equation}
Now
the covariant derivative $\nabla_\mu$ is defined, as usual, in terms
of the Levi-Civita connection defined by the metric $g_{\mu\nu}$. In
this section, we do not use the covariant derivative with respect to
the metric $f_{\mu\nu}$. First we observe that,
\begin{align}
\label{mFbis1}
& \nabla^\mu \left( \frac{1}{2} g_{\mu\nu} F\left(
{\tilde R}^{(g)} \right) - R^{(g)}_{\mu\nu} F'\left( {\tilde
R}^{(g)} \right) + \nabla_\nu \nabla_\mu F'\left( {\tilde R}^{(g)}
\right) - g_{\mu\nu} \nabla^2 F'\left( {\tilde R}^{(g)} \right)
\right) \nn =& m^2 F'\left( {\tilde R}^{(g)} \right)
\partial_\nu \left( - \tr \sqrt{g^{-1} f}+ \det \sqrt{g^{-1} f} \right) \, ,
\end{align}
which can be obtained by using the Bianchi identity
$0=\nabla^\mu\left( \frac{1}{2} g_{\mu\nu} R^{(g)}
 - R^{(g)}_{\mu\nu} \right)$. By multiplying the covariant derivative
$\nabla^\mu$ with respect to the metric $g$ with Eq.~(\ref{mFR3})
and using the conservation law $0=\nabla^\mu T_{\mathrm{matter}\,
\mu\nu} $, we find the following equation,
\begin{align}
\label{mFR5}
0 = & F'\left( {\tilde R}^{(g)} \right)
\partial_\nu \left( - \tr \sqrt{g^{-1} f}+ \det \sqrt{g^{-1} f} \right) \nn
&+ \left( \partial^\mu F'\left( {\tilde R}^{(g)} \right)
\right)\left\{ \frac{1}{2} f_{\mu\rho} \left( \sqrt{ g^{-1} f }
\right)^{-1\, \rho}_{\qquad \nu} + \frac{1}{2} f_{\nu\rho} \left(
\sqrt{ g^{-1} f } \right)^{-1\, \rho}_{\qquad \mu}
 - g_{\mu\nu} \det \sqrt{g^{-1} f}
\right\} \nn & + F'\left( {\tilde R}^{(g)} \right) \nabla^\mu
\left\{ \frac{1}{2} f_{\mu\rho} \left( \sqrt{ g^{-1} f }
\right)^{-1\, \rho}_{\qquad \nu} + \frac{1}{2} f_{\nu\rho} \left(
\sqrt{ g^{-1} f } \right)^{-1\, \rho}_{\qquad \mu}
 - g_{\mu\nu} \det \sqrt{g^{-1} f} \right\} \, .
\end{align}
We consider that the metric $g_{\mu\nu}$ is the flat FRW Universe and
flat Minkowski space-time for $f_{\mu\nu}$. We also use the
conformal time $t$ with metric $g_{\mu\nu}$ as follows,
\begin{equation}
\label{Fbi10}
ds_g^2 = \sum_{\mu,\nu=0}^3 g_{\mu\nu} dx^\mu dx^\nu
= a(t)^2 \left( - dt^2 + \sum_{i=1}^3 \left( dx^i \right)^2\right)
\, , \quad ds_f^2 = \sum_{\mu,\nu=0}^3 f_{\mu\nu} dx^\mu dx^\nu
= - dt^2 + \sum_{i=1}^3 \left( dx^i \right)^2 \, .
\end{equation}
Then, although
the $\nu=i$ component in (\ref{mFR5}) is trivially satisfied, the
$\nu=t$ component gives the following equation,
\begin{equation}
\label{mFR6}
0 =
\partial_t \left( - 4 a^{-1} + a^{-4} \right) F'\left( {\tilde R}^{(g)} \right)
+ \left( a^{-1} - a^{-4} \right) \partial_t
F'\left( {\tilde R}^{(g)} \right) = a^{-4} \partial_t \left\{ F'
\left({\tilde R}^{(g)} \right) \left( a^3 - 1 \right) \right\}\, ,
\end{equation}
which further gives a constraint,
\begin{equation}
\label{mFR7}
F'\left( {\tilde R}^{(g)} \right) \left( a^3 - 1 \right) = C\, .
\end{equation}
Here $C$
is a constant. By using Eq.~(\ref{mFR7}), we can determine the form
of $F'\left( {\tilde R}^{(g)} \right)$. For a given evolution of the
scale factor $a=a(t)$ with respect to the time, we can find the
$t$-dependence of ${\tilde R}^{(g)}$,
${\tilde R}^{(g)}={\tilde R}^{(g)}(t)$.
This equation can be solved with respect to $t$ as a
function of ${\tilde R}^{(g)}$, $t = t \left( {\tilde R}^{(g)}
\right)$. By using the obtained expression and Eq.~(\ref{mFR7}), we
find the following form of $F'\left( {\tilde R}^{(g)} \right)$,
\begin{equation}
\label{mFR8}
F'\left( {\tilde R}^{(g)} \right) = \frac{C}{ a\left(t
\left( {\tilde R}^{(g)} \right)\right)^3 - 1 } \, ,
\end{equation}
We should
note that as we show shortly, the time evolution of the scale factor
$a=a(t)$ cannot be arbitrary. Furthermore $F'\left( {\tilde R}^{(g)}
\right)$ diverges when the scale factor $a$ goes to unity.

In the FRW metric with conformal time in (\ref{Fbi10}), the
$(\mu,\nu)=(t,t)$ component in (\ref{mFR3}) has the following form:
\begin{equation}
\label{mFR9}
0 = - \frac{1}{2} a^{-2} F\left( {\tilde R}^{(g)}
\right) + 3 \dot H F'\left( {\tilde R}^{(g)} \right)
 - 3 H \partial_t F'\left( {\tilde R}^{(g)} \right)
+ \left( - a + a^{-3} \right) F'\left( {\tilde R}^{(g)} \right)
+ \frac{1}{2 M_g^2}\rho_\mathrm{matter}\, ,
\end{equation}
and the
$(\mu,\nu)=(i,j)$ component gives,
\begin{equation}
\label{mFR10}
0 = \frac{1}{2}
a^{-2} F\left( {\tilde R}^{(g)} \right)
 -\left( \dot H + 2 H^2 \right) F'\left( {\tilde R}^{(g)} \right)
+ \left( \partial_t^2 + H \partial_t \right) F'\left( {\tilde R}^{(g)} \right)
 - \left( - a + a^{-3} \right) F'\left( {\tilde R}^{(g)} \right)
+ \frac{1}{2 M_g^2}p_\mathrm{matter}\, .
\end{equation}
By combining
Eqs.~(\ref{mFR9}) and (\ref{mFR10}), we obtain,
\begin{equation}
\label{mFR11}
0 = 2 \left( \dot H - H^2 \right) F'\left( {\tilde R}^{(g)} \right)
+ \left( \partial_t^2 - 2 H \partial_t \right)
F'\left( {\tilde R}^{(g)} \right) + \frac{1}{2 M_g^2}
\left( \rho_\mathrm{matter} + p_\mathrm{matter} \right)\, .
\end{equation}
In contrast to the Einstein
gravity case, Eqs.~(\ref{mFR9}), (\ref{mFR10}), and the conservation
law (\ref{CEm}) are independent equations. Note that the form of the
conservation law in terms of the conformal time is not changed from
that in terms of the cosmological time. We now treat
Eqs.~(\ref{mFR7}), (\ref{mFR11}), and the conservation law
(\ref{CEm}) as independent equations. By using Eq.~(\ref{mFR7}), we
can rewrite Eq.~(\ref{mFR11}) as,
\begin{equation}
\label{mFR14}
0 = \frac{\left\{ \dot H \left( - a^6 - a^3 + 2 \right)
+ H^2 \left( 13 a^6 + 7 a^3 - 2 \right) \right\}C}{\left( a^3 - 1 \right)^3}
+ \frac{1}{2 M_g^2}\left( \rho_\mathrm{matter} + p_\mathrm{matter}
\right)\, .
\end{equation}
This equation (\ref{mFR14}) describes the dynamics
of the Universe, which does not depend on the form of $F\left(
{\tilde R}^{(g)} \right)$.

Since it is difficult to solve (\ref{mFR14}) explicitly, we consider
the following three cases, that is, a) The $a\to 1$ case, b) The
$a\gg 1$ case, c) The $a\ll 1$ case. In the following, for
simplicity, we consider the case that the matter has a constant EoS
parameter $w$.

\noindent
a) The $a\to 1$ case. By putting $a= 1 + \delta a$ and by
using (\ref{mFR14}), we find,
\begin{equation}
\label{mFR15}
0 \sim - 9 \dot H + 18 H^2
\sim - 9 \delta\ddot a \delta a + 18 \left( \delta \dot a
\right)^2 = - 9 \left( \delta a \right)^3 \frac{d}{dt} \left(
\frac{\delta \dot a}{\left( \delta a \right)^2 } \right)\, ,
\end{equation}
which can be solved as,
\begin{equation}
\label{mFR16}
\delta a = \frac{C_1}{t + C_2}\, .
\end{equation}
Here we denote the constants of integration by $C_1$
and $C_2$. By using Eq.~(\ref{mFR16}), we find that the limit $a\to
1$ $\left( \delta a \to 0 \right)$ is realized in the infinite past
or future in conformal time, $t \to \pm \infty$.

\noindent
b) The $a\gg 1$ case. In this case, we can approximate
Eq.~(\ref{mFR14}) as,
\begin{equation}
\label{mFR17}
0 \sim C a^{-3} \left( - \dot H + 13 H^2 \right)
+ \frac{1+w}{2} \rho_0 a^{- 3 \left( w + 1 \right)}\, .
\end{equation}
\begin{enumerate}
\item The $\rho_0 = 0$ case.
In this case, we find that the solution of (\ref{mFR17}) is,
\begin{equation}
\label{mFR18}
H = \frac{1}{13 \left( t_0 - t \right)}\, .
\end{equation}
This
solution describes a phantom Universe which has a Big Rip
singularity at $t=t_0$ because we assume $a\gg 1$.
\item The $\rho_0 \neq 0$ case. In case $w\neq 0$,
we find the following power law solution,
\begin{equation}
\label{mFR19}
a = a_0 t^{\frac{2}{3w}}\, .
\end{equation}
Here $a_0$ is given by solving the
following equation,
\begin{equation}
\label{mFR20}
0 = \frac{2C}{3w} \left( 1 + \frac{26}{3w} \right)
+ \frac{w + 1}{2} a_0^{-3w} \rho_0\, .
\end{equation}
On
the other hand, for the case $w=0$, we obtain a solution describing
the de Sitter Universe:
\begin{equation}
\label{FR20b}
H^2 = \frac{\rho_0}{26C} \, .
\end{equation}
This could be interesting because the accelerating expansion
of the present Universe can be realized by dust, which could be
identified with cold dark matter. Then we find
\begin{equation}
\label{FR20c}
\frac{1}{26C} \rho_0 \sim \left( 10^{-33}\, \mathrm{eV} \right)^2
\, ,\quad \rho_0 a^3 \sim \left( 10^{-3}\, \mathrm{eV} \right)^4\, .
\end{equation}
\end{enumerate}

\noindent
c) The $a\ll 1$ case. We find the approximated form of
Eq.~(\ref{mFR14}) as follows,
\begin{equation}
\label{mFR21}
0 = - 2 C \left(
\dot H - H^2 \right) + \frac{1+w}{2} \rho_0 a^{- 3 \left( w + 1
\right)}\, .
\end{equation}
\begin{enumerate}
\item The $\rho_0 = 0$ case. The solution is given by,
\begin{equation}
\label{mFR22}
H = \frac{1}{t_0 - t}\, ,
\end{equation}
with $t_0$ being a
constant of integration . The expression (\ref{mFR22}) is valid when
$t \to \pm \infty$ because we are assuming $a\ll 1$. Then,
regardless of the form (\ref{mFR22}), there does not always occur a
Big Rip singularity.
\item The $\rho_0 \neq 0$ case. The solution of Eq.~(\ref{mFR22})
is given by,
\begin{equation}
\label{mFR23}
a = a_0 t^{\frac{2}{3\left( w + 1 \right)}}\, .
\end{equation}
The constant $a_0$ is given by solving the
following equation,
\begin{equation}
\label{mFR24}
0 = - \frac{4C}{{3\left( w + 1 \right)}} \left( 1
 - \frac{2}{3\left( w + 1 \right)} \right) + \frac{w + 1}{2}
a_0^{-3w} \rho_0\, .
\end{equation}
Then we find that the qualitative behavior
is similar to the Einstein gravity coupled with matter having a
constant EoS parameter $w$.
\end{enumerate}

Just for comparison reasons, instead of the model in
Eq.~(\ref{mFR1}), we may consider an $F(R)$ gravity
extension of massive
gravity proposed in \cite{Cai:2013lqa}, whose action is given by
\begin{equation}
\label{massivegravity}
S_\mathrm{mg} = M_g^2\int d^4x\sqrt{-\det
g}\,F \left(R^{(g)}\right) +2m^2 M_g^2 \int d^4x\sqrt{-\det
g}\sum_{n=0}^{4} \beta_n\, e_n \left(\sqrt{g^{-1} f} \right)
+ S_\mathrm{matter}\, .
\end{equation}
For simplicity, we only consider the
minimal case in the following, so as in (\ref{mFR2}) we have,
\begin{equation}
\label{mg3}
S_\mathrm{mg} = M_g^2\int d^4x\sqrt{-\det g}\,F \left(
R^{(g)} \right) +2m^2 M_g^2 \int d^4x\sqrt{-\det g} \left( 3 - \tr
\sqrt{g^{-1} f} + \det \sqrt{g^{-1} f} \right) + S_\mathrm{matter}\, ,
\end{equation}
Then upon variation with respect to the metric $g_{\mu\nu}$,
we find,
\begin{align}
\label{Fbi8}
0 =& M_g^2 \left( \frac{1}{2} g_{\mu\nu} F\left(
R^{(g)} \right) - R^{(g)}_{\mu\nu} F'\left( R^{(g)} \right)
+ \nabla_\nu \nabla_\mu F'\left( R^{(g)} \right)
 - g_{\mu\nu} \nabla^2 F'\left( R^{(g)} \right) \right) \nn
& + m^2 M_g^2 \left\{ g_{\mu\nu} \left( 3 - \tr \sqrt{g^{-1} f}
\right) + \frac{1}{2} f_{\mu\rho} \left( \sqrt{ g^{-1} f }
\right)^{-1\, \rho}_{\qquad \nu} + \frac{1}{2} f_{\nu\rho} \left(
\sqrt{ g^{-1} f } \right)^{-1\, \rho}_{\qquad \mu}
\right\} 
+ \frac{1}{2} T_{\mathrm{matter}\, \mu\nu} \, .
\end{align}
Then, instead of Eq.~(\ref{mFbis1}), we have the following
equations,
\begin{equation}
\label{Fbis1}
0 = \nabla^\mu \left( \frac{1}{2}
g_{\mu\nu} F\left( R^{(g)} \right)
 - R^{(g)}_{\mu\nu} F'\left( R^{(g)} \right)
+ \nabla_\nu \nabla_\mu F'\left( R^{(g)} \right)
 - g_{\mu\nu} \nabla^2 F'\left( R^{(g)} \right) \right)\, ,
\end{equation}
and
\begin{equation}
\label{identity1}
0 = - g_{\mu\nu} \nabla^\mu \left( \tr
\sqrt{g^{-1} f} \right) + \frac{1}{2} \nabla^\mu \left\{ f_{\mu\rho}
\left( \sqrt{ g^{-1} f } \right)^{-1\, \rho}_{\qquad \nu}
+ f_{\nu\rho} \left( \sqrt{ g^{-1} f } \right)^{-1\, \rho}_{\qquad
\mu} \right\} \, ,
\end{equation}
which corresponds to Eq.~(\ref{mFR5}). We
choose the $g_{\mu\nu}$ metric to describe the FRW Universe and the
metric $f_{\mu\nu}$ describing a flat Minkowski space-time and also
we use the conformal time $t$ as in Eq.~(\ref{Fbi10}). Then the
$(t,t)$ component of (\ref{Fbi8}) has the following form,
\begin{equation}
\label{Fbi11}
0 = - 3 M_g^2 H^2 - 3 m^2 M_g^2 \left( a^2 - a \right)
+ \rho_\mathrm{matter} \, ,
\end{equation}
and $(i,j)$ components are given by,
\begin{equation}
\label{Fbi12}
0 = M_g^2 \left( 2 \dot H + H^2 \right) + 3 m^2
M_g^2 \left( a^2 - a \right) + p_\mathrm{matter}\, .
\end{equation}
On the
other hand, Eq.~(\ref{identity1}) gives the following constraint:
\begin{equation}
\label{identity3}
\frac{\dot a}{a} = 0\, .
\end{equation}
In contrast to
Eq.~(\ref{mFR7}), the identity (\ref{identity3}) shows that $a$
should be a constant $a=a_0$. This indicates that the only
consistent solution for $g_{\mu\nu}$ is the flat Minkowski space.
Therefore, we cannot obtain the expanding Universe without extra
fields and/or fluids.

Before we close this section, we also review in brief the
construction of ghost-free $F(R)$ bigravity by following
Ref.~\cite{Nojiri:2012zu}. The consistent model of bimetric gravity,
which includes two metric tensors $g_{\mu\nu}$ and $f_{\mu\nu}$, was
proposed in Ref.~\cite{Hassan:2011zd}. This model contains a
massless spin-two field, corresponding to graviton, and also a
massive spin-two field. The gravity model which only contains the
massive spin-two field is called massive gravity but now we consider
the model including both the massless and massive spin-two field,
which is called bigravity. The Boulware-Deser ghost
\cite{Boulware:1974sr} does not appear in such a theory.

We start with the following action,
\begin{align}
\label{bimetric}
S_\mathrm{bi} =&M_g^2\int d^4x\sqrt{-\det
g}\,R^{(g)}+M_f^2\int d^4x
\sqrt{-\det f}\,R^{(f)} \nn
&+2m^2 M_\mathrm{eff}^2 \int d^4x\sqrt{-\det g}\sum_{n=0}^{4}
\beta_n\, e_n \left(\sqrt{g^{-1} f} \right) \, .
\end{align}
Here $R^{(g)}$ is the scalar curvature for $g_{\mu \nu}$ again and
$R^{(f)}$ is the scalar curvature for $f_{\mu \nu}$. The massive
parameter $M_{\mathrm{eff}}$ is defined by,
\begin{equation}
\label{Meff}
\frac{1}{M_{\mathrm{eff}^2}} = \frac{1}{M_g^2} + \frac{1}{M_f^2}\, .
\end{equation}
Furthermore, the tensor $\sqrt{g^{-1} f}$ is defined by the
square root of $g^{\mu\rho} f_{\rho\nu}$ again.

In order to construct a consistent $F(R)$ bigravity, we add the
following terms to the action (\ref{bimetric}):
\begin{align}
\label{Fbi1}
S_\varphi =& - M_g^2 \int d^4 x \sqrt{-\det g} \left\{
\frac{3}{2} g^{\mu\nu} \partial_\mu \varphi \partial_\nu \varphi
+ V(\varphi) \right\} + \int d^4 x \mathcal{L}_\mathrm{matter}
\left( \e^{\varphi} g_{\mu\nu}, \Phi_i \right)\, ,\\
\label{Fbi7b}
S_\xi =& - M_f^2 \int d^4 x \sqrt{-\det f} \left\{
\frac{3}{2} f^{\mu\nu} \partial_\mu \xi \partial_\nu \xi + U(\xi)
\right\} \, .
\end{align}
By using the conformal transformations $g_{\mu\nu} \to \e^{-\varphi}
g^{\mathrm{J}}_{\mu\nu}$ and $f_{\mu\nu}\to \e^{-\xi}
f^{\mathrm{J}}_{\mu\nu}$, the total action $S_{F} = S_\mathrm{bi} +
S_\varphi + S_\xi$ is transformed as follows,
\begin{align}
\label{FF1}
S_{F} =& M_f^2\int d^4x\sqrt{-\det f^{\mathrm{J}}}\,
\left\{ \e^{-\xi} R^{\mathrm{J}(f)} - \e^{-2\xi} U(\xi) \right\}
\nn & +2m^2 M_\mathrm{eff}^2 \int d^4x\sqrt{-\det
g^{\mathrm{J}}}\sum_{n=0}^{4} \beta_n \e^{\left(\frac{n}{2} -2
\right)\varphi - \frac{n}{2}\xi} e_n
\left(\sqrt{{g^{\mathrm{J}}}^{-1} f^{\mathrm{J}}} \right) \nn &
+ M_g^2 \int d^4 x \sqrt{-\det g^{\mathrm{J}}} \left\{ \e^{-\varphi}
R^{\mathrm{J}(g)} - \e^{-2\varphi} V(\varphi) \right\} + \int d^4 x
\mathcal{L}_\mathrm{matter} \left( g^{\mathrm{J}}_{\mu\nu}, \Phi_i
\right)\, .
\end{align}
Then the kinetic terms for $\varphi$ and $\xi$ vanish. By the
variations of the action with respect to $\varphi$ and $\xi$, as in
the case of convenient $F(R)$ gravity \cite{Nojiri:2003ft}, we
obtain,
\begin{align}
\label{FF2}
0 =& 2m^2 M_\mathrm{eff}^2 \sum_{n=0}^{4} \beta_n
\left(\frac{n}{2} -2 \right) \e^{\left(\frac{n}{2} -2 \right)\varphi
- \frac{n}{2}\xi} e_n \left(\sqrt{{g^{\mathrm{J}}}^{-1}
f^{\mathrm{J}}}\right) + M_g^2 \left\{ - \e^{-\varphi}
R^{\mathrm{J}(g)} + 2 \e^{-2\varphi} V(\varphi)
+ \e^{-2\varphi} V'(\varphi) \right\}\, ,\\
\label{FF3}
0 =& - 2m^2 M_\mathrm{eff}^2 \sum_{n=0}^{4}
\frac{\beta_n n}{2} \e^{\left(\frac{n}{2} -2 \right)\varphi -
\frac{n}{2}\xi} e_n \left(\sqrt{{g^{\mathrm{J}}}^{-1}
f^{\mathrm{J}}}\right) + M_f^2 \left\{ - \e^{-\xi} R^{\mathrm{J}(f)}
+ 2 \e^{-2\xi} U(\xi) + \e^{-2\xi} U'(\xi) \right\}\, .
\end{align}
Then, Eqs.~(\ref{FF2}) and (\ref{FF3}) can be solved algebraically
with respect to $\varphi$ and $\xi$ as $\varphi = \varphi \left(
R^{\mathrm{J}(g)}, R^{\mathrm{J}(f)}, e_n
\left(\sqrt{{g^{\mathrm{J}}}^{-1} f^{\mathrm{J}}}\right) \right)$
and $\xi = \xi \left( R^{\mathrm{J}(g)}, R^{\mathrm{J}(f)}, e_n
\left(\sqrt{{g^{\mathrm{J}}}^{-1} f^{\mathrm{J}}}\right) \right)$.
Substituting the $\varphi$ and $\xi$ we found above, into
(\ref{FF1}), one gets the $F(R)$ bigravity which is described as
follows:
\begin{align}
\label{FF4}
S_{F} =& M_f^2\int d^4x\sqrt{-\det f^{\mathrm{J}}}
F^{(f)}\left( R^{\mathrm{J}(g)}, R^{\mathrm{J}(f)}, e_n
\left(\sqrt{{g^{\mathrm{J}}}^{-1} f^{\mathrm{J}}}\right) \right) \nn
& +2m^2 M_\mathrm{eff}^2 \int d^4x\sqrt{-\det g}\sum_{n=0}^{4}
\beta_n \e^{\left(\frac{n}{2} -2 \right) \varphi\left(
R^{\mathrm{J}(g)}, e_n \left(\sqrt{{g^{\mathrm{J}}}^{-1}
f^{\mathrm{J}}}\right) \right)} e_n
\left(\sqrt{{g^{\mathrm{J}}}^{-1} f^{\mathrm{J}}} \right) \nn &
+ M_g^2 \int d^4 x \sqrt{-\det g^{\mathrm{J}}} F^{\mathrm{J}(g)}\left(
R^{\mathrm{J}(g)}, R^{\mathrm{J}(f)}, e_n
\left(\sqrt{{g^{\mathrm{J}}}^{-1} f^{\mathrm{J}}}\right) \right)
+ \int d^4 x \mathcal{L}_\mathrm{matter} \left(
g^{\mathrm{J}}_{\mu\nu}, \Phi_i \right)\, ,
\end{align}
\begin{align}
\label{FF4BBB}
F^{\mathrm{J}(g)}\left( R^{\mathrm{J}(g)},
R^{\mathrm{J}(f)}, e_n \left(\sqrt{{g^{\mathrm{J}}}^{-1}
f^{\mathrm{J}}}\right) \right) \equiv & \left\{ \e^{-\varphi\left(
R^{\mathrm{J}(g)}, R^{\mathrm{J}(f)}, e_n
\left(\sqrt{{g^{\mathrm{J}}}^{-1} f^{\mathrm{J}}}\right) \right)}
R^{\mathrm{J}(g)} \right. \nn & \left.
 - \e^{-2\varphi\left( R^{\mathrm{J}(g)}, R^{\mathrm{J}(f)},
e_n \left(\sqrt{{g^{\mathrm{J}}}^{-1} f^{\mathrm{J}}}\right)
\right)} V \left(\varphi\left( R^{\mathrm{J}(g)}, R^{\mathrm{J}(f)},
e_n \left(\sqrt{{g^{\mathrm{J}}}^{-1} f^{\mathrm{J}}}\right)
\right)\right) \right\} \, ,\nn F^{(f)}\left( R^{\mathrm{J}(g)},
R^{\mathrm{J}(f)}, e_n \left(\sqrt{{g^{\mathrm{J}}}^{-1}
f^{\mathrm{J}}}\right) \right) \equiv & \left\{ \e^{-\xi\left(
R^{\mathrm{J}(g)}, R^{\mathrm{J}(f)}, e_n
\left(\sqrt{{g^{\mathrm{J}}}^{-1} f^{\mathrm{J}}}\right) \right)}
R^{\mathrm{J}(f)} \right. \nn & \left.
 - \e^{-2\xi\left( R^{\mathrm{J}(g)}, R^{\mathrm{J}(f)},
e_n \left(\sqrt{{g^{\mathrm{J}}}^{-1} f^{\mathrm{J}}}\right)
\right)} U \left(\xi\left( R^{\mathrm{J}(g)}, R^{\mathrm{J}(f)}, e_n
\left(\sqrt{{g^{\mathrm{J}}}^{-1} f^{\mathrm{J}}}\right)
\right)\right) \right\} \, .
\end{align}
Note that it is difficult to solve Eqs.~(\ref{FF2}) and (\ref{FF3})
with respect to $\varphi$ and $\xi$ explicitly. Therefore, it might
be easier to define the model in terms of the auxiliary scalars
$\varphi$ and $\xi$ as in Eq.~(\ref{FF1}) (for details see
\cite{Nojiri:2012re}).

In Ref.~\cite{Kluson:2013yaa}, another kind of non-linear massive
$F(R)$ gravity has been proposed. The action is given by,
\begin{equation}
\label{mFR1BB}
S_\mathrm{mg} = M_g^2\int d^4x\sqrt{-\det g}\,F\left(R^{(g)}
+2m^2 \int d^4x\sqrt{-\det g} \sum_{n=0}^{4} \beta_n\,
e_n \left(\sqrt{g^{-1} f} \right) \right)+ S_\mathrm{matter}\, .
\end{equation}
In this action, $f_{\mu \nu}$ is a non-dynamical reference metric.
By using the Hamiltonian formulation in the scalar-tensor frame, it
has been shown that this kind of $F(R)$ theory is ghost-free.

\subsection{Mimetic $F(R)$ gravity }

Mimetic gravity is a modified gravity theory for which the conformal
symmetry is respected as an internal degree of freedom
\cite{Chamseddine:2013kea} (for its generalizations, see
\cite{Chamseddine:2014vna,Golovnev:2013jxa,Momeni:2014qta,
Deruelle:2014zza,Chaichian:2014qba}). The terminology mimetic
gravity firstly appeared in Ref.~\cite{Chamseddine:2013kea}, and
since then, several works were devoted on mimetic gravity in various
contexts, see the recent review \cite{Sebastiani:2016ras}. For
example in Ref.~\cite{Nojiri:2014zqa}, see also
Refs.~\cite{Odintsov:2015wwp,Odintsov:2015ocy,
Odintsov:2015cwa,Leon:2014yua} the generalization of mimetic gravity
in the context of $F(R)$ gravity was performed and in
Refs.~\cite{Myrzakulov:2016hrx,Momeni:2015gka,
Astashenok:2015haa,Myrzakulov:2015qaa} the mimetic gravity
extensions of various modified gravity models were studied. In
Refs.~\cite{Cognola:2016gjy,Arroja:2015wpa,Ijjas:2016pad,
Saadi:2014jfa,Matsumoto:2015wja,Myrzakulov:2015nqa,Rabochaya:2015haa}
several cosmological applications of mimetic gravity were studied in
various theoretical contexts. In addition, in
Refs.~\cite{Myrzakulov:2015kda,Oikonomou:2015lgy,Myrzakulov:2015sea,Oikonomou:2016fxb,
Astashenok:2015qzw} various astrophysical solutions in mimetic
gravity were presented. Finally for some late-time applications of
mimetic gravity, see \cite{Odintsov:2016oyz,Oikonomou:2016pkp} and
for some more involved mimetic unimodular gravity models see
\cite{Nojiri:2016ppu,Odintsov:2016imq}.

Although we usually regard the metric $g_{\mu\nu}$ as a fundamental
variable of gravity, in the case of mimetic gravity, the metric is
expressed in a different way by using new degrees of freedom. We may
consider the variation of the action with respect to the new degrees
of freedom and the obtained equation may admit a new or wider class
of solutions compared with the equations given by the variation with
respect to the metric $g_{\mu\nu}$. In the case of mimetic gravity,
we use the following parametrization of the metric,
\begin{equation}
\label{Mi1}
g_{\mu\nu}= - \hat g^{\rho\sigma} \partial_\rho \phi
\partial_\sigma \phi \hat g_{\mu\nu} \, ,
\end{equation}
and we consider the variation with respect to $\hat g_{\mu\nu}$
and $\phi$. Since the parametrization is invariant under the Weyl
transformation $\hat g_{\mu\nu} \to \e^{\sigma(x)} \hat g_{\mu\nu}$,
the variation over $\hat g_{\mu\nu}$ gives the traceless part of the
equation. In fact, in case of the Einstein gravity, whose action is
given by,
\begin{equation}
\label{EHaction}
S=\int d^4 x \sqrt{-g\left({\hat
g}_{\mu\nu}, \phi \right)} \left( \frac{R\left({\hat g}_{\mu\nu},
\phi \right)}{2\kappa^2} + \mathcal{L}_\mathrm{matter}\right)\, ,
\end{equation}
the variation with respect to $\hat g_{\mu\nu}$ gives the
traceless part of the Einstein equation:
\begin{equation}
\label{Mi2}
0 = - R\left({\hat g}_{\mu\nu}, \phi \right)_{\mu\nu}
+ \frac{1}{2} g\left({\hat g}_{\mu\nu}, \phi \right)_{\mu\nu}
R\left({\hat g}_{\mu\nu}, \phi \right) + \kappa^2 T_{\mu\nu}
+ \partial_\mu \phi \partial_\nu \phi \left( R\left({\hat g}_{\mu\nu},
\phi \right)
+ \kappa^2 T \right)\, .
\end{equation}
Here $T$ is the trace of the matter
energy-momentum tensor $T_{\mu\nu}$, $T=g\left({\hat g}_{\mu\nu},
\phi \right)^{\mu\nu}T_{\mu\nu}$. Eq.~(\ref{Mi1}) shows
\begin{equation}
\label{Mi2B}
g\left({\hat g}_{\mu\nu}, \phi \right)^{\mu\nu}
\partial_\mu \phi
\partial_\nu \phi = - 1\, .
\end{equation}
Usually, the variation with respect to the Weyl factor in the
metric gives the equation for the trace part, that is, in the case
of the Einstein gravity, $R + \kappa^2 T =0$, but due to the
parametrization in Eq.~(\ref{Mi1}), the variation with respect to
$\phi$ gives,
\begin{equation}
\label{Mi3}
0 = \nabla\left(g\left({\hat g}_{\mu\nu},
\phi \right)_{\mu\nu}\right)^\mu \left(\partial_\mu
\phi \left( R\left({\hat g}_{\mu\nu}, \phi \right)
+ \kappa^2 T \right) \right) \, .
\end{equation}
Note that $\nabla_\mu$ is the covariant
derivative with respect to $g_{\mu\nu}$. Eq.~(\ref{Mi3}) shows that
there can be a wider class of the solutions in the mimetic model,
compared with the Einstein gravity. In fact, Eq.~(\ref{Mi3})
effectively induces dark matter \cite{Chamseddine:2013kea}, which
will be discussed in the following section for the case of the
$F(R)$ extension of the mimetic model. We should note that in
Eqs.~(\ref{Mi2}) and (\ref{Mi3}), ${\hat g}_{\mu\nu}$ appears only
in the combination of $g_{\mu\nu}$ in (\ref{Mi1}) and therefore
${\hat g}_{\mu\nu}$ does not appear explicitly.

Now we discuss the mimetic $F(R)$ gravity which was developed in
Ref.~\cite{Nojiri:2014zqa}. This theory seems to be a ghost-free
theory, like the conventional $F(R)$ gravity, and in addition, it is
a conformal invariant theory. In the literature it was shown how
early-time and late-time acceleration of the Universe, can be
realized in the context of mimetic $F(R)$ gravity
\cite{Odintsov:2015wwp}. Also it was shown that an inflationary era
which can be consistent with the current observational data, may be
realized in the context of mimetic $F(R)$ gravity
\cite{Odintsov:2015ocy,Odintsov:2015cwa}. Furthermore, it was shown
in \cite{Odintsov:2015wwp}, that the reconstruction of $\Lambda$CDM
model is also possible, as well as the unification of early-time
acceleration with the late-time acceleration can be done, in the
spirit of first proposal in $F(R)$ gravity \cite{Nojiri:2003ft}.
Also, cosmological bounces can be realized in the context of mimetic
$F(R)$ gravity. We need to note that a specified cosmological
evolution which is realized by an $F(R)$ gravity, in principle it is
realized by a different mimetic $F(R)$ gravity. In addition, like in
the case of usual mimetic gravity, the mimetic $F(R)$ theory can be
extended to include a scalar potential and a Lagrange multiplier.

We start from $F(R)$ gravity, whose action is given by,
\begin{equation}
\label{FRaction}
S=\int d^4 x \sqrt{-g} \left( F\left(R\right)
+ \mathcal{L}_\mathrm{matter}\right)\, .
\end{equation}
The function $F(R)$ is
some function of the scalar curvature $R$ and
$\mathcal{L}_\mathrm{matter}$ is matter Lagrangian. If we
parameterize the metric as in Eq.~(\ref{Mi1}), upon variation of the
action with respect to the metric, we get,
\begin{equation}
\label{dg}
\delta
g_{\mu\nu} = \hat g^{\rho\tau} \delta \hat g_{\tau\omega}
\hat g^{\omega\sigma}
\partial_\rho \phi \partial_\sigma \phi \hat g_{\mu\nu}
 - \hat g^{\rho\sigma} \partial_\rho \phi \partial_\sigma \phi \delta
\hat g_{\mu\nu}
 - 2 \hat g^{\rho\sigma} \partial_\rho \phi \partial_\sigma \delta \phi
\hat g_{\mu\nu} \, .
\end{equation}
In effect, in the case of $F(R)$ gravity, by
using the parametrization of the metric as in Eq.~(\ref{Mi1}), the
action can be written as follows,
\begin{equation}
\label{miFRaction}
S=\int d^4
x \sqrt{-g\left({\hat g}_{\mu\nu}, \phi \right)} \left(
F\left(R\left({\hat g}_{\mu\nu}, \phi \right)\right)
+ \mathcal{L}_\mathrm{matter}\right)\, .
\end{equation}
Note that in principle
any solution of the standard $F(R)$ gravity, is also a solution of
the mimetic $F(R)$ gravity.

Assume that the background geometry is the flat FRW space-time
(\ref{JGRG14}), where $\phi$ depends only on the cosmic time $t$,
and due to Eq.~(\ref{Mi2B}), we find that the cosmic time can be
identified with $\phi$, that is $\phi=t$. For the FRW space-time, we
obtain,
\begin{align}
\label{Mi5b}
\frac{C_\phi}{a^3} = & 2 F(R) - R F'(R) - 3 \Box F'(R)
+ \frac{1}{2} T \nn = & 2 F(R) - 6 \left( \dot H + 2 H^2 \right)
F'(R) + 3 \frac{d^2 F'(R)}{dt^2} + 9H \frac{d F'(R)}{dt}
+ \frac{1}{2} \left( - \rho + 3 p \right) \, ,
\end{align}
where $C_\phi$ is an arbitrary integration constant. Since the term
containing $C_\phi$ behaves as $a^{-3}$, the case $C_\phi\neq 0$
describes mimetic dark matter. We also obtain the following
equation,
\begin{equation}
\label{Mi6}
0 = \frac{d^2 F'(R)}{dt^2} + 2H
\frac{dF'(R)}{dt} - \left(\dot H + 3 H^2 \right) F'(R) + \frac{1}{2}
F(R) + \frac{1}{2} p \, .
\end{equation}
By combining Eqs.~(\ref{Mi5b}) and
(\ref{Mi6}), we obtain,
\begin{equation}
\label{Mi7}
0 = \frac{d^2 F'(R)}{dt^2} - H \frac{d F'(R)}{dt}
+ 2 \dot H F'(R) + \frac{1}{2} \left( p + \rho
\right) + \frac{4C_\phi}{a^3}\, .
\end{equation}
In the case $C_\phi=0$, the
FRW equations above are identical to the ones corresponding to the
standard $F(R)$ gravity, or in other words, in the case that
$C_\phi\neq 0$, the FRW equations and therefore the corresponding
solutions, are in principle different from those corresponding to
the standard $F(R)$ gravity.

Let us briefly illustrate how it is possible to realize an arbitrary
cosmological evolution, in the context of mimetic $F(R)$ gravity.
Consider an arbitrary cosmological evolution with scale factor
$a(t)$ and assume that the explicit $\rho$ and $p$ depend explicitly
on the scale factor, as it happens usual perfect fluids. Consider
the differential equation:
\begin{equation}
\label{Mi8}
0 = \frac{d^2 f(t)}{dt^2}
 - H(t) \frac{d f(t)}{dt} + 2 \dot H(t) f(t)\, .
\end{equation}
We denote the
two solutions of Eq.~(\ref{Mi8}) as $f_1(t)$ and $f_2(t)$. Then
$F'(R)$, which is the solution of Eq.~(\ref{Mi7}), is equal to,
\begin{align}
\label{Mi9}
F'\left(R \left(t\right)\right) =& - f_1 (t) \int^t dt'
\frac{ \gamma(t') f_2(t') }{W\left(f_1(t'),f_2(t')\right)} + f_2 (t)
\int^t dt' \frac{ \gamma(t') f_1(t') }{
W\left(f_1(t'),f_2(t')\right)} \, , \nn
\gamma(t) \equiv & - \frac{1}{2} \left( p
+ \rho \right) - \frac{4 C_\phi}{a^3}\, , \nn
W\left(f_1,f_2\right) \equiv & f_1(t) f_2'(t) - f_2(t) f_1'(t)\, .
\end{align}
By using the explicit $t$-dependence of the scalar curvature
$R=R(t)$, we find the cosmic time $t$ as a function of the
scalar curvature $R$, that is $t=t(R)$. Therefore, one gets
the explicit form of the expression $F'(R)$.

In Ref.~\cite{Chamseddine:2014vna}, it has been proposed that
instead of parameterizing the metric tensor as in Eq.~(\ref{Mi1}),
the condition of Eq.~(\ref{Mi2B}) can be imposed in the theory, by
adding the Lagrange multiplier auxiliary field $\lambda$. In effect,
instead of Eq.~(\ref{miFRaction}), we may consider the following
gravitational action,
\begin{equation}
\label{FRphilambda0}
S=\int d^4 x
\sqrt{-g} \left( F \left(R\left(g_{\mu\nu}\right)\right)
 + \lambda \left( g^{\mu\nu} \partial_\mu \phi \partial_\nu \phi + 1
\right) + \mathcal{L}_\mathrm{matter}\right)\, .
\end{equation}
In addition, a
scalar potential $V(\phi)$ can be added as follows,
\begin{equation}
\label{FRphilambda000}
S=\int d^4 x \sqrt{-g} \left( F \left(R\left(g_{\mu\nu}\right)\right)
 - V(\phi) + \lambda \left( g^{\mu\nu} \partial_\mu \phi \partial_\nu \phi
+ 1 \right) + \mathcal{L}_\mathrm{matter}\right)\, .
\end{equation}
This
gravitational action is a modified gravity with Lagrange multiplier
constraints. These Lagrange multiplier techniques were developed in
\cite{Lim:2010yk,Capozziello:2013xn,Capozziello:2010uv,Gao:2010gj},
see also \cite{Makarenko:2016jsy}. Then by appropriately choosing
the scalar potential $V(\phi)$, in principle any arbitrary
cosmological evolution can be realized. Since the standard $F(R)$
gravity has no ghost issues, the explicit addition of the Lagrange
multiplier constraint, in principle, does not violate the ghost-free
property that the standard $F(R)$ version has. Therefore, it can be
expected that the mimetic $F(R)$ gravity theory under consideration,
it is also a ghost-free theory, although this should be verified
explicitly by using the Hamiltonian formulation of the theory. In a
similar way, it can be expected that the same applies for the Newton
law, so in principle it should be the same as in standard $F(R)$
gravity.

\subsection{Mimetic $f(\mathcal{G})$ Gravity}

The mimetic framework can be extended in the context of the modified
Gauss-Bonnet gravity, and in this section we present the essential
features of this framework based on Ref.~\cite{Astashenok:2015haa}.
Consider the following $f(\mathcal{G})$ gravity action,
\begin{equation}
\label{action}
S=\int
d^4x\sqrt{-g}\left[\frac{R}{2\kappa^2}+f(\mathcal{G})\right]+S_\mathrm{matter}\, .
\end{equation}
with $S_\mathrm{matter}$ denoting the action for the matter fluids present. The
Gauss-Bonnet scalar is given in (\ref{GB}).
The signature for the
Riemannian metric is $(-+++)$, and also $\kappa^2=8\pi G/c^4=1$,
with $G$ being the Newtonian constant.

In order to introduce the mimetic formalism in the Gauss-Bonnet
gravity, we make the following parametrization of the metric
\cite{Chamseddine:2013kea,Golovnev:2013jxa}:
\begin{equation}
\label{100}
g_{\mu\nu}=-\hat{g}^{\rho\sigma}\partial_{\rho}\phi
\partial_{\sigma}\phi \hat{g}_{\mu\nu}\, .
\end{equation}
Upon variation of the metric, we obtain,
\begin{align}
\label{dgggg}
\delta g_{\mu\nu}=&\hat{g}^{\rho\tau}\delta\hat{g}_{\tau\omega}
\hat{g}^{\omega\sigma}\partial_{\rho}\phi \partial_{\sigma}\phi
\hat{g}_{\mu\nu} -\hat{g}^{\rho\sigma}\partial_{\rho}\phi
\partial_{\sigma}\phi \delta\hat{g}_{\mu\nu} \nn
& -2\hat{g}^{\rho\sigma}\partial_{\rho}\phi
\partial_{\sigma}\delta\phi \hat{g}_{\mu\nu}\, .
\end{align}
Moreover, upon variation of the action (\ref{action}), with respect
to the metric $\hat{g}_{\mu\nu}$, and also with respect to the
scalar field $\phi$, the gravitational equations are equal to,
\begin{align}
\label{eom}
& R_{\mu\nu}- \frac{1}{2}R g_{\mu\nu} \nn
& +8\left[R_{\mu\rho\nu\sigma}+R_{\rho\nu}g_{\sigma\mu}
 -R_{\rho\sigma}g_{\nu\mu}-R_{\mu\nu}g_{\sigma\rho}
+R_{\mu\sigma}g_{\nu\rho}
+\frac{R}{2}\left(g_{\mu\nu}g_{\sigma\rho}
 -g_{\mu\sigma}g_{\nu\rho}\right)\right]
\nabla^{\rho}\nabla^{\sigma}f_\mathcal{G}
+ \left(f_\mathcal{G}\mathcal{G}-f(\mathcal{G})\right)g_{\mu\nu} \nn
& +\partial_{\mu}\phi\partial_{\nu}\phi\left(-R+8\left(-R_{\rho\sigma}
+\frac{1}{2}R
g_{\rho\sigma}\right)\nabla^{\rho}\nabla^{\sigma}f_\mathcal{G}
+4(f_\mathcal{G}\mathcal{G}-f(\mathcal{G}))\right) \nn
=&T_{\mu\nu}+\partial_{\mu}\phi\partial_{\nu}\phi T\, ,
\end{align}
with $f_\mathcal{G}$ being equal to
$f_{\mathcal{G}}=d f(\mathcal{G}) / d\mathcal{G}$.
Also the
covariant derivative $\nabla_{\mu}$, acts on vectors as follows,
\begin{equation}
\label{asxetious}
\nabla_{\mu}V_{\nu}=\partial_{\mu}V_{\nu}
 -\Gamma_{\mu\nu}^{\lambda}V_{\lambda}\, .
\end{equation}
Also, upon variation of the action (\ref{action}), with respect
to the mimetic scalar degree of freedom $\phi$, we obtain,
\begin{equation}
\label{FG}
\nabla^{\mu}\left(\partial_{\mu}\phi\left(-R+8\left(-R_{\rho\sigma}
+\frac{1}{2} R
g_{\rho\sigma}\right)\nabla^{\rho}\nabla^{\sigma}f_\mathcal{G}
+4(f_\mathcal{G}\mathcal{G}-f(\mathcal{G})\right)-T)\right)=0\, .
\end{equation}
For the flat FRW metric, with line element (\ref{JGRG14}), the
scalar curvature and the Gauss-Bonnet invariant are given by
$R=6\left(\frac{\ddot{a}}{a}+\frac{\dot{a}^2}{a^2}\right)=6(\dot{H}+2H^2)$,
and
$\mathcal{G}=24\frac{\ddot{a}\dot{a}^2}{a^3}=24H^2(\dot{H}+H^2)$.
with $H(t)$ being the Hubble rate $H(t)=\dot{a}(t)/a(t)$. From
Eq.~(\ref{100}) we obtain that,
\begin{equation}
\label{introeqns}
g^{\mu\nu}\partial_{\mu}\phi\partial_{\nu}\phi=-1\, ,
\end{equation}
and due to
the fact that the mimetic scalar $\phi$ has an explicit dependence
solely on the cosmic time $t$, from Eq.~(\ref{introeqns}), we have
$\phi=t$. In effect, the $(t,t)$ component of Eq.~(\ref{eom}), is
equal to,
\begin{equation}
\label{FRIED-1}
2\dot{H}+3H^{2}+16H(\dot{H}+H^{2})\frac{d
f_{\mathcal{G}}}{dt}+8H^{2}\frac{d^2 f_{\mathcal{G}} }{dt^2}
 -(f_{\mathcal{G}}\mathcal{G}-f(\mathcal{G}))=-p\, ,
\end{equation}
and as it
can be checked, the same equation results if one considers the
$(r,r)$ component. Upon integration, Eq.~(\ref{FG}) yields,
\begin{equation}
\label{RRRRR}
 -R+8\left(-R_{\rho\sigma}+\frac{1}{2}R
g_{\rho\sigma}\right)\nabla^{\rho}\nabla^{\sigma}f_{\mathcal{G}}
+4(f_{\mathcal{G}}\mathcal{G}-f(\mathcal{G}))+\rho-3p
=-\frac{C}{a^3} \, ,
\end{equation}
and after some algebra we obtain,
\begin{equation}
\label{FRIED-2}
\dot{H}+2H^{2}+4H^{2}\frac{d^{2}f_{\mathcal{G}}}{dt^2}
+ 4H\left(2\dot{H}+3H^{2}\right)\frac{d
f_{\mathcal{G}}}{dt}+\frac{2}{3}(f_{\mathcal{G}}\mathcal{G}
 -f(\mathcal{G}))+ \frac{\rho}{6}-\frac{p}{2}=-\frac{C}{a^3}\, .
\end{equation}
with $C$ being an arbitrary integration constant. By combining
Eqs.~(\ref{FRIED-1}) and (\ref{FRIED-2}), we get,
\begin{equation}
\label{fried3}
\dot{H}+4H^{2}\frac{d^{2}f_{\mathcal{G}}}{dt^2}
+4H(2\dot{H}-H^{2})\frac{df_{\mathcal{G}}}{dt}
=-\frac{1}{2}\left(\rho+p\right)-\frac{C}{a^{3}}\, .
\end{equation}
where the constant $C$ is redefined for notational simplicity. We
introduce the function $g(t)$,
$g(t)=\frac{df_{\mathcal{G}}}{dt}$,
which satisfies the following equation:
\begin{equation}
\label{FRIED4}
4H^{2}\frac{dg(t)}{dt}+4H(2\dot{H}-H^2)g(t)
=-\dot{H}-\frac{1}{2}\left(\rho+p\right)-\frac{C}{a^{3}}\, .
\end{equation}
For the value $C=0$, both Eqs.~(\ref{FRIED-1}) and (\ref{fried3})
are identical with the Friedmann equations of ordinary $f(\mathcal{G})$
gravity. If the right hand side of the above equation is equal to
zero, that is,
\begin{equation}
\label{asxetiouss2}
B(t)=-\dot{H}-\frac{1}{2}\left(\rho+p\right)-\frac{C}{a^{3}}= 0\, ,
\end{equation}
then we easily obtain the following solution,
\begin{equation}
g(t)=g_{0}\left(\frac{H_{0}}{H}\right)^{2}\exp\left(\int_{{0}}^{t}Hdt\right)
\, ,
\end{equation}
where $g_{0}$ is an integration constant and also $H_{0}=H({0})$.
For non-vanishing right hand side of Eq.~(\ref{FRIED4}), which means
that $B(t)\neq 0$, then we obtain the following solution,
\begin{equation}
\label{genericbouncesol}
g(t)=g_{0}\left(\frac{H_{0}}{H}\right)^{2}\exp\left(\int_{{0}}^{t}Hdt\right)
+\frac{1}{4H^{2}}\int_{0}^{t}dt_{1}B(t_{1})\exp\left(\int_{t_{1}}^{t}
H(\tau)d\tau\right)\, .
\end{equation}
Hence, given a cosmic evolution $H(t)$, then we can find the
function $g(t)$, by applying the formulas we quoted above.
Therefore one can specify the cosmic evolution of the Universe, in
terms of the Hubble rate $H(t)$ and then obtain the function $g(t)$
and we may obtain the function $f(t)$,
$f_{\mathcal{G}}(t)=\int g(t) dt$.
By also finding the function $t(\mathcal{G})$, we may substitute it in the
above equation, and then we can find the function
$f_{\mathcal{G}}(\mathcal{G})$.
From this a simple integration with respect to the Gauss-Bonnet
scalar, yields the function $f(\mathcal{G})$.

Let us apply the method we just presented in order to realize the
following cosmological evolution,
\begin{equation}
\label{apopano}
H(t)=\frac{H_{0}}{1+\alpha t}, \quad \alpha>0 \, .
\end{equation}
The
Gauss-Bonnet invariant for the evolution (\ref{apopano}) is equal to
$\mathcal{G}=24\frac{H_{0}^{2}(\alpha-H_{0})}{(1+\alpha t)^{3}}$.
For $B=0$, the solution to Eq.~(\ref{FRIED4}),
is, $g(t)=g_{0}\left(1+\alpha t\right)^{H_{0}/\alpha+2}$. For a
Universe that does not contain any matter or mimetic dark matter,
which means that $\rho=p=C=0$, the function $g(t)$ is,
$g(t)=g_{0}\left(1+\alpha t\right)^{H_{0}/\alpha+2}-\frac{1}{4H_{0}}
\frac{1+\alpha t}{\alpha+H_{0}}$. Hence, the function $f_{\mathcal{G}}(t)$ is,
$f_{\mathcal{G}}(t)=f_{0}\left(1+\alpha t\right)^{H_{0}/\alpha+3}
 -\frac{1}{8\alpha H_{0}}\frac{(1+\alpha t)^{2}}{\alpha+H_{0}}$.
By inverting the Gauss-Bonnet scalar, we can
find the function $t=t(\mathcal{G})$, and eventually the function
$f_{\mathcal{G}}(G)$ is equal to,
$f_{\mathcal{G}}=A\mathcal{G}^{-1-H_{0}/3\alpha}+D\mathcal{G}^{-2/3}$,
$D=-\frac{1}{8\alpha
H_{0}(\alpha+H_{0})}\left(24H_{0}^{2}(\alpha-H_{0})\right)^{2/3}$.
By integrating with respect to $\mathcal{G}$, we obtain the
non-mimetic $f(\mathcal{G})$ gravity,
\begin{equation}
\label{nomtc}
f(\mathcal{G})=F_{0}
\mathcal{G}^{-H_{0}/\alpha}+3D\mathcal{G}^{1/3}\, .
\end{equation}
In the case
that matter and mimetic dark matter is present, in which case we
have $C\neq 0$, $\rho= \rho_{0} a^{-3}$, $p=0$, the function $g(t)$
is,
\begin{equation}
\label{axris}
g(t)=g_{0}\left(1+\alpha t\right)^{H_{0}/\alpha+2}-\frac{1}{4H_{0}}
\frac{1+\alpha t}{\alpha+H_{0}}+\frac{\tilde{C}}{4H_{0}-\alpha}
\frac{1}{4H_{0}^{2}}\left(1+\alpha t\right)^{3(1-H_{0}/\alpha)}, \quad
\tilde{C}=C+\rho_{0}\, ,
\end{equation}
and
therefore $f_{\mathcal{G}}(t)$ is equal to,
\begin{equation}
\label{fffff}
f_{\mathcal{G}}(t)=A\mathcal{G}^{-1-H_{0}/3\alpha}+D\mathcal{G}^{-2/3}
+E\mathcal{G}^{-4/3+H_{0}/\alpha}\, ,
\end{equation}
with $E$ being equal to,
\begin{equation}
\label{EEEEEEE}
E=\frac{\tilde{C}}{4H_{0}^{2}}\frac{\alpha}{(4H_{0}-\alpha)(4\alpha-H_{0})}
\left(24H_{0}^{2}(\alpha-H_{0})\right)^{4/3-H_{0}/\alpha}\, .
\end{equation}
Hence, the mimetic $f(\mathcal{G})$ gravity is equal to,
\begin{equation}
\label{answ}
f(\mathcal{G})=F_{0}
\mathcal{G}^{-H_{0}/\alpha}+3D\mathcal{G}^{1/3}
+\frac{3\alpha}{3H_{0}-\alpha}E\mathcal{G}^{-1/3+H_{0}/\alpha}\, .
\end{equation}
By comparing Eqs.~(\ref{answ}) and (\ref{nomtc}), we can easily
spot the differences between mimetic and non-mimetic
$f(\mathcal{G})$ gravity.

There is another formalism for mimetic $f(\mathcal{G})$ gravity, which employs
a Lagrange multiplier function, in order to realize the constraint
for the mimetic scalar we found in the previous formalism, namely
Eq.~(\ref{introeqns}). In this case, the gravitational action is,
\begin{equation}
\label{SSSSSS}
S_{\phi}=\int d^{4}x\sqrt{-g}\left(-\epsilon
g^{\mu\nu}\partial_{\mu}\phi\partial_{\nu}\phi-V(\phi)\right)\, ,
\end{equation}
where a mimetic potential is included too. The values of the
parameter $\epsilon={\pm 1}$, correspond to a canonical and a
phantom scalar and also for $\epsilon=0$, the theory is conformally
invariant, and the above action is rendered equivalent to an $f(\mathcal{G})$
gravity, with $U=1$ and $V=0$. We consider the following generalized
action with Lagrange multiplier and mimetic potential,
\begin{equation}
\label{laction}
S=\int d^{4}x\sqrt{-g}\left(\frac{R}{2}+f(\mathcal{G})-\epsilon
g^{\mu\nu}\partial_{\mu}\phi\partial_{\nu}\phi-V(\phi)
+\lambda(g^{\mu\nu}\partial_{\mu}\phi\partial_{\nu}\phi+U(\phi))
+\mathcal{L}_\mathrm{matter}\right)\, ,
\end{equation}
and upon variation with respect to $g_{\mu\nu}$, we get,
\begin{align}
\label{FGFR-1}
& 3H^2+24H^{3}\frac{df_{\mathcal{G}}(\mathcal{G})}{dt}+f(\mathcal{G})
 -f_{\mathcal{G}}(\mathcal{G})\mathcal{G}
=\rho+\epsilon\dot{\phi}^{2}+V(\phi)
 -\lambda(\dot{\phi}^{2}+U(\phi))\, , \\
\label{FGFR-2}
& -2\dot{H}-3H^2-8H^2\frac{d^{2}f_{\mathcal{G}}(\mathcal{G})}{dt^{2}}
 -16H(\dot{H}+H^2)\frac{df_{\mathcal{G}}(\mathcal{G})}{dt}
+f_{\mathcal{G}}(\mathcal{G})\mathcal{G}-f(\mathcal{G})
=p+\epsilon\dot{\phi}^{2}-V(\phi)-\lambda(\dot{\phi}^{2}-U(\phi))\, .
\end{align}
for the flat FRW metric. In addition, upon variation with respect to
$\lambda$ , we obtain.
\begin{equation}
\label{phiphiphi}
\dot{\phi}^{2}-U(\phi)=0\, ,
\end{equation}
which is a subcase of the
constraint (\ref{introeqns}). Moreover, by varying with respect to
the mimetic scalar, we get,
\begin{equation}
\label{SF}
2\partial_{t}((\lambda-\epsilon)
\dot{\phi})+6H(\lambda-\epsilon)\dot{\phi}-V'(\phi)+\lambda
U'(\phi)=0\, .
\end{equation}
In the case that non-relativistic matter is
present ($p=0$), the mimetic potential $V(t)$ and the corresponding
Lagrange multiplier $\lambda(t)$, are equal to,
\begin{align}
\label{p1}
V(t)=&2\dot{H}+3H^2+\epsilon\dot{\phi}^{2}
+8H^2\frac{d^{2}f_{\mathcal{G}}(\mathcal{G})}{dt^{2}}
+16H(\dot{H}+H^2)\frac{df_{\mathcal{G}}(\mathcal{G})}{dt}
 -f_{\mathcal{G}}(\mathcal{G})\mathcal{G}+f(\mathcal{G})\, , \\
\label{lambdat}
\lambda(t)=&\dot{\phi}^{-2}\left(\frac{\rho}{2}+\dot{H}
+4H(2\dot{H}-H^{2})\frac{df_{\mathcal{G}}(\mathcal{G})}{dt}
+4H^{2}\frac{d^{2}f_{\mathcal{G}}(\mathcal{G})}{dt^{2}}\right)
+\epsilon\, .
\end{align}
Hence, given the function $f(\mathcal{G})$, we can find the cosmological
evolution and the scalar potential that can realize such an
evolution. The method can work in the inverse way, so for a given
potential and also for an arbitrary cosmic evolution, we can find
the $f(\mathcal{G})$ gravity which may realize such an evolution. We refer to
Ref.~\cite{Astashenok:2015haa} for some illustrative applications of
this reconstruction method.

\subsection{Unimodular $F(R)$ Gravity }

In principle the cosmological constant can be regarded as the vacuum
energy, but in the context of quantum field theory, the magnitude of
the vacuum energy is around $60$-$120$ orders higher in comparison
to that of the observed cosmological constant. For this problem, the
unimodular gravity theory
\cite{Anderson:1971pn,Buchmuller:1988wx,Henneaux:1989zc,
Unruh:1988in,Ng:1990xz,Finkelstein:2000pg,Alvarez:2005iy,
Alvarez:2006uu,Abbassi:2007bq,Ellis:2010uc,Jain:2012cw,Singh:2012sx,
Kluson:2014esa,Padilla:2014yea,Barcelo:2014mua,Barcelo:2014qva,
Burger:2015kie,Alvarez:2015sba,Jain:2012gc,Jain:2011jc,Cho:2014taa,
Basak:2015swx,Gao:2014nia,Eichhorn:2015bna,Saltas:2014cta,
Chaturvedi:2016fea}
may offer an interesting and conceptually simple theoretical
proposal. However, this theory cannot provide a full solution to the
problem of the cosmological constant \cite{Padilla:2014yea}. In this
section, we review the extension of unimodular gravity, in the
context of $F(R)$ gravity. This was proposed and studied in detail
in \cite{Nojiri:2015sfd}.

For the standard Einstein-Hilbert unimodular gravity case, the
determinant of the metric is fixed, and therefore a constraint of
the metric exists, $g_{\mu \nu}\delta g^{\mu \nu}=0$. We assume,
without loss of generality, that the determinant of the metric is
constrained to be equal to one,
\begin{equation}
\label{Uni1}
\sqrt{-g}=1\, .
\end{equation}
Since the above unimodular constraint of Eq.~(\ref{Uni1}) is not
satisfied by the metric (\ref{JGRG14}), we need to redefine the
cosmic time coordinate to be $d\tau = a(t)^3 dt$ and we rewrite the
FRW metric (\ref{JGRG14}) in the following way,
\begin{equation}
\label{UniFRW}
ds^2 = - a\left(t\left(\tau\right)\right)^{-6} d\tau^2
+ a\left(t\left(\tau\right)\right)^{2} \sum_{i=1}^3 \left( dx^i
\right)^2 \, .
\end{equation}
In effect, the metric of Eq.~(\ref{UniFRW})
satisfies the unimodular constraint (\ref{Uni1}). By using the
coordinate $\tau$, the de Sitter Universe, for which $a(t) = \e^{H_0
t}$, is expressed as follows,
\begin{equation}
\label{Uni5}
ds^2 = - \left( 3 H_0
\tau \right)^{-2} d\tau^2 + \left( 3 H_0 \tau \right)^\frac{2}{3}
\sum_{i=1}^3 \left( dx^i \right)^2 \, .
\end{equation}
On the other hand, for
the Universe with a power-law scale factor, $a(t) = \left(
\frac{t}{t_0} \right)^{h_0}$, we find that,
\begin{equation}
\label{Uni8}
ds^2 =
- \left( \frac{\left( 3 h_0 + 1 \right) \tau}{t_0}
\right)^\frac{-6h_0}{3h_0 + 1}d\tau^2 + \left( \frac{\left( 3 h_0
+ 1 \right) \tau}{t_0} \right)^\frac{2h_0}{3h_0 + 1} \sum_{i=1}^3
\left( dx^i \right)^2 \, .
\end{equation}
We should note that the metric of
Eq.~(\ref{Uni5}) becomes identical with that of the de Sitter
Universe, when $h_0, t_0 \to \infty$, by keeping $\frac{\left( 3 h_0
+ 1 \right) \tau}{t_0}$ to be finite, that is, $\frac{\left( 3 h_0 +
1 \right) \tau}{t_0} \to 3 H_0$. Hence, we can unify the form of the
metric in the following way,
\begin{equation}
\label{Uni10}
ds^2 = - \left(
\frac{\tau}{\tau_0} \right)^{-6f_0}d\tau^2 + \left(
\frac{\tau}{\tau_0} \right)^{2f_0} \sum_{i=1}^3 \left( dx^i
\right)^2 \, ,
\end{equation}
where the parameters $\tau_0$ and $f_0$ are
arbitrary constant real numbers. The case for which
$f_0=\frac{1}{3}$ corresponds to the de Sitter Universe. On the other
hand, if the condition $\frac{1}{4}\leq f_0 < \frac{1}{3}$ holds
true, which occurs when, $h_0\geq 1$, the Universe is described by a
quintessential evolution. In the case that $0<f_0<\frac{1}{4}$,
which implies, $0<h_0<1$, the Universe is expanding but in a
decelerating way. Finally, when $f_0<0$ or if $f_0>\frac{1}{3}$,
which in turn implies, $h_0<0$, the Universe's evolution is a
phantom evolution
\cite{Bamba:2008hq,Caldwell:2003vq,Elizalde:2004mq,Haro:2011zzb,Nesseris:2006er,Chen:2008ft,
Caldwell:1999ew,Leith:2007bu,Faraoni:2003jh,Saridakis:2010mf,
Sushkov:2004qy,Kahya:2009sz,Onemli:2004mb,Faraoni:2005gg,
Dabrowski:2003jm,deHaro:2012cj,Brevik:2006md}.

Let us now see how the Newton law becomes in unimodular $F(R)$
gravity. First, as in the case of standard $F(R)$ gravity theory, we
can rewrite the action in a scalar-tensor theory form,
\begin{equation}
\label{UF3}
S = \int d^4 x \left\{ \sqrt{-g} \left(
\frac{1}{2\kappa^2} \left( R
 - \frac {3}{2} g^{\mu\nu} \partial_\mu \phi \partial_\nu \phi - V(\phi)
\right)
 - \lambda \e^{2\phi} \right) + \lambda \right\}
+ S_\mathrm{matter} \left( \e^\phi g_{\mu\nu}, \Psi \right)\, ,
\end{equation}
where the scalar potential $V(\phi)$ is given in Eq.~(\ref{JGRG23}).
The unimodular constraint of Eq.~(\ref{Uni1}) is modified in the
following way $\e^{2\phi}\sqrt{-g} = 1$. By eliminating the scalar
field $\phi$, the action of Eq.~(\ref{UF3}) can be written in the
following way,
\begin{equation}
\label{UF7}
S = \int d^4 x \sqrt{-g} \left(
\frac{1}{2\kappa^2} \left( R - \frac{3}{32 g^2} g^{\mu\nu}
\partial_\mu g \partial_\nu g - V\left(\frac{1}{4}\ln \left( - g
\right) \right) \right) \right) + S_\mathrm{matter} \left( \left( -g
\right)^\frac{1}{4} g_{\mu\nu}, \Psi \right)\, .
\end{equation}
Now we consider
the perturbation of the metric tensor $g_{\mu\nu}$ around the
background metric $g_{\mu\nu}^{(0)}$, in the following way,
$g_{\mu\nu} = g_{\mu\nu}^{(0)} + h_{\mu\nu}$. In addition, we assume
that the background metric is a flat metric, hence we have,
$g_{\mu\nu}^{(0)} = \eta_{\mu\nu}$. In effect, we can obtain the
following equations,
\begin{equation}
\label{UF13}
\partial_\lambda \partial^\lambda h_{\mu\nu}
 - \partial_\mu \partial^\lambda h_{\lambda\nu}
 - \partial_\nu \partial^\lambda h_{\lambda\mu}
+ \partial_\mu \partial_\nu h + \eta_{\mu\nu} \partial^\rho
\partial^\sigma h_{\rho\sigma}
 - \frac{13}{16} \eta_{\mu\nu} \partial_\lambda \partial^\lambda h - m^2
\eta_{\mu\nu} h = \kappa^2 \left( T_{\mu\nu} - \frac{1}{4}
\eta_{\mu\nu} T \right) \, ,
\end{equation}
with $T_{\mu\nu}$ being the
energy-momentum tensor of the matter fluids present, and in addition
$T$ stands for the trace of $T_{\mu\nu}$, $T \equiv
\eta^{\rho\sigma} T_{\rho\sigma}$. In order to study the Newton law,
we may consider a point source at the origin of the coordinate
system, so that the components of the energy-momentum tensor have
the following form, $T_{00} = M \delta \left( \bm{r} \right)$,
$T_{ij} = 0$ $\left(i,j=1,2,3\right)$, so we seek a static solution
of Eq.~(\ref{UF13}). In the case of the Einstein gravity, four
gauge degrees of freedom exist, but in the unimodular $F(R)$ gravity
case, only three gauge degrees of freedom exist, since the
unimodular constraint (\ref{Uni1}) needs to be satisfied. In
effects, we additionally impose three gauge conditions, $\partial^i
h_{ij} = 0$. Moreover, by using appropriate boundary conditions, and
by defining the Newtonian potential $\Phi$ as follows, $h_{00} = 2
\Phi$, we may obtain the Poisson equation that the Newtonian
potential $\Phi$ satisfies, which is, $\partial_i
\partial^i \Phi = \frac{3 \kappa^2}{8} M \delta \left( \bm{r}
\right)$. Hence, by redefining the gravitational constant $\kappa$
by $\frac{3 \kappa^2}{4} \to \kappa^2 = 8\pi G$, we easily obtain
the standard Poisson equation satisfied by the Newtonian potential
$U$, which is, $\partial_i \partial^i \Phi = 4\pi G M \delta \left(
\bm{r} \right)$, the solution of which is given by,
\begin{equation}
\label{UF29}
\Phi = - \frac{GM}{r}\, .
\end{equation}
The result we found above is quite
different from the solution in the context of the standard $F(R)$
gravity (for reviews, see
\cite{Capozziello:2011et,Nojiri:2010wj,Nojiri:2006ri,Capozziello:2010zz}),
in which case the propagation of the scalar mode $\phi = - \ln
F'(A)$ yields a non-trivial correction to the standard Newton law of
gravity. In the case of unimodular $F(R)$ gravity, due to the fact
that the unimodular constraint (\ref{Uni1}) is rewritten in the
following way, $\e^{2\phi}\sqrt{-g} = 1$, the degree of the freedom
corresponding to the scalar mode $\phi$, is eventually eliminated
from the resulting field equations, and in effect, the scalar $\phi$
does not propagate. In effect, no correction to the Newton law of
gravity occurs.

We need to note that since the unimodular constraint (\ref{Uni1})
needs to be satisfied, the unimodular $F(R)$ gravity is not a fully
covariant theory. A covariant formulation of the unimodular
Einstein gravity was proposed in
Ref.~\cite{Henneaux:1989zc}. By using this formulation, we may start
from the following action,
\begin{equation}
\label{LUF1}
S = \int d^4 x \left\{
\sqrt{-g} \left( \mathcal{L}_\mathrm{gravity}
 - \lambda \right) + \lambda
\epsilon^{\mu\nu\rho\sigma} \partial_\mu a_{\nu\rho\sigma} \right\}
+ S_\mathrm{matter} \left( g_{\mu\nu}, \Psi \right)\, ,
\end{equation}
with
$\mathcal{L}$ being the Lagrangian density of any gravitational
theory, and $a_{\nu\rho\sigma}$ is a three-form field. Upon
variation with respect to $a_{\nu\rho\sigma}$, we obtain the
equation $0 =
\partial_\mu \lambda$, so in effect, $\lambda$ is a constant. On the
other hand, upon variation with respect to $\lambda$, yields,
\begin{equation}
\label{LLUF19}
\sqrt{-g} = \epsilon^{\mu\nu\rho\sigma} \partial_\mu
a_{\nu\rho\sigma}\, ,
\end{equation}
instead of the unimodular constraint.
Since Eq.~(\ref{LLUF19}) can be solved with respect to
$a_{\mu\nu\rho}$, there is no constraint imposed on the metric
$g_{\mu\nu}$.

We should note that the quantity $a_{\mu\nu\rho}$, has four degrees
of freedom. The gravitational action is invariant under the
following gauge transformation $\delta a_{\mu\nu\rho} = \partial_\mu
b_{\nu\rho} +
\partial_\nu b_{\rho\mu} + \partial_\rho b_{\mu\nu}$, with
$b_{\mu\nu}$ being an anti-symmetric tensor field, that is,
$b_{\mu\nu} = - b_{\nu\mu}$, which contributes in total six degrees
of freedom. We should note that the gauge transformation is actually
invariant under the following gauge transformation, $\delta
b_{\mu\nu} = \partial_\mu c_\nu - \partial_\nu c_\mu$. The quantity
$c_\mu$ is a vector field, which contributes four degrees of
freedom. Note that the gauge transformation of the original gauge
transformation is actually invariant under the following gauge
transformation $\delta c_\mu = \partial_\mu \varphi$. The field
$\varphi$ is a scalar field which contributes a degree of freedom.
Therefore, the total number of degrees of freedom in the gauge
transformation is $6-4+1=3$ and the number of degrees of freedom of
the quantity $a_{\mu\nu\rho}$ is $4-3=1$. In effect, we may choose
the following gauge condition,
\begin{equation}
\label{LUF5}
a_{tij} \left( =
a_{jti} = a_{ijt} \right) = 0\, , \quad i,j=1,2,3\, ,
\end{equation}
and we
find that the only remaining degree of freedom is given by $a_{ijk}$
$\left( i,j,k = 1,2,3 \right)$. We can rewrite the action as
follows,
\begin{equation}
\label{LUF6}
S_{\lambda\alpha} = \int d^4x \lambda
\left( - \sqrt{-g} + \epsilon^{\mu\nu\rho\sigma}
\partial_\mu a_{\nu\rho\sigma} \right) = \int d^4x \lambda
\left( - \sqrt{-g} + \partial_t \alpha \right)\, ,
\end{equation}
where $\alpha \equiv \frac{1}{3!}
a_{123}$. The system described by Eq.~(\ref{LUF6}) might yield a
non-trivial correction to the Newton law, or it might be a ghost and
it can contribute negative norm states in the corresponding quantum
field theory. We can use an analogy with the quantum mechanical
system with Lagrangian,
\begin{equation}
\label{L1}
L= B y \dot x\, ,
\end{equation}
which
describes the massless limit of a charged particle in a uniform
magnetic field $B$. Then, we find the following commutation
relations satisfied by the fields $\lambda$ and $\alpha$,
\begin{equation}
\label{LUF7}
\left[ \alpha, \lambda \right] = i \delta \left( \bm{x}
\right)\, ,
\end{equation}
with $\bm{x} = \left( x^1, x^2, x^3 \right)$ and the
Hamiltonian $H$ is given by,
\begin{equation}
\label{LUF8}
H = \int_S d S \sqrt{-g} \lambda \, ,
\end{equation}
with $S$ being an arbitrary space-like
surface.

In effect, regardless that $\lambda$ is a constant, the time
evolution of the quantity $\alpha$ is given equal to,
\begin{equation}
\label{LUF9}
\frac{d\alpha}{dt} = i \left[ H, \alpha \right] = \sqrt{-g}\, ,
\end{equation}
which is compatible with the classical equation
which is obtained by the variation of the action of
Eq.~(\ref{LUF6}) with respect to the quantity $\lambda$. The eigenstate
of the Hamiltonian $H$, could be provided by the eigenstate of
$\lambda$. In the representation of the quantum states by using
$\alpha$, the commutation relation of Eq.~(\ref{LUF7}), yields
$\lambda = i \frac{\delta}{\delta \alpha}$. In effect, the
eigenstate $\Psi_{\lambda_0}(\alpha)$ of $\lambda$ with the
eigenvalue $\lambda_0$ can be expressed as follows,
\begin{equation}
\label{LUF11b}
\Psi_{\lambda_0}(\alpha) = \exp \left( i \lambda_0
\int_S d S \alpha \left( \bm{x} \right) \right)\, .
\end{equation}
The
eigenvalue of the Hamiltonian (\ref{LUF8}) is infinite, due to the
fact that the volume of $S$ is infinite, and in addition unbounded
from below. Note however that no quantum transition between the
states occurs, and in effect, the states may be considered quantum
mechanically stable.

In the case that covariant unimodular $F(R)$ gravity is considered,
the action is,
\begin{equation}
\label{LUF10}
\mathcal{L}_\mathrm{gravity} = \frac{F(R)}{2\kappa^2}\, ,
\end{equation}
and as in the standard $F(R)$ gravity
case, we may rewrite the action in a scalar-tensor form as follows,
\begin{equation}
\label{LUF11}
S = \int d^4 x \left\{ \sqrt{-g} \left(
\frac{1}{2\kappa^2} \left( R
 - \frac{3}{2} g^{\mu\nu} \partial_\mu \phi \partial_\nu \phi - V(\phi)
\right)
 - \lambda \e^{2\phi} \right) + \lambda \epsilon^{\mu\nu\rho\sigma}
\partial_\mu a_{\nu\rho\sigma} \right\}
+ S_\mathrm{matter} \left( \e^\phi g_{\mu\nu}, \Psi \right)\, .
\end{equation}
In the action of Eq.~(\ref{LUF11}), one obtains $0 = \partial_\mu
\lambda$ and therefore $\lambda$ is a constant in this case too.
Thus, the scalar potential $V(\phi)$ is effectively changed as
follows,
\begin{equation}
\label{LUF12}
V(\phi) \to \tilde V(\phi) =
\frac{A(\phi)}{F'\left( A \left( \phi \right) \right)}
 - \frac{F\left( A \left( \phi \right) \right)}{F'\left( A \left( \phi
\right) \right)^2} + 2\kappa^2 \lambda \e^{2\phi}\, ,
\end{equation}
Then if
the mass of field $\phi$, which is defined to be $m_\phi^2 =
\frac{3}{2} \frac{d^2 \tilde V(\phi)}{d\phi^2}$ is small, a large
correction to the Newton law of gravity could appear. By using the
equation $\phi = - \ln F'(A)$ and Eq.~(\ref{LUF12}), the explicit
expression for the mass $m_\phi^2$ is given by,
\be
\label{ULF14}
m_\phi^2 = \frac{3}{2} \left\{
\frac{A(\phi)}{F'\left(A\left(\phi\right)\right)}
 - \frac{4F\left( A \left(\phi\right) \right) }{F' \left( A
\left(\phi\right) \right)^2 } + \frac{1}{F'' \left( A
\left(\phi\right) \right) } + \frac{8 \kappa^2 \lambda}{ F' \left( A
\left(\phi\right) \right)^2 } \right\} \, . \end{equation}The last term is a
characteristic feature of the unimodular $F(R)$ gravity theory but
since $\lambda$ is constant, the last term can be absorbed into the
redefinition of $F(R)$, that is, $F(R) \to F(R) + 2 \kappa^2
\lambda$. The expression for $m_\phi^2$ obtained by the above
redefinition, is identical to the expression in the standard $F(R)$
gravity. Therefore, there is no essential difference in the Newton
law corrections, between the covariant unimodular $F(R)$ gravity and
the standard $F(R)$ gravity
\cite{Capozziello:2011et,Nojiri:2010wj,Nojiri:2006ri,Capozziello:2010zz}.
This feature is quite different from the non-covariant unimodular
$F(R)$ gravity version, where the standard Newton law is recovered.

\subsection{Unimodular Mimetic Gravity}

The unimodular gravity formalism may be extended in the context of
mimetic gravity, and in this section we discuss the formalism of
mimetic unimodular (U-M) gravity, which was developed in
\cite{Nojiri:2016ppu}. The formalism we shall present constitutes a
reconstruction method, and we now we shall briefly present the
essential features of this
 reconstruction method. The standard Einstein gravity
 approach for unimodular gravity
\cite{Alvarez:2006uu,Ellis:2010uc,Kluson:2014esa,Barcelo:2014mua,
Burger:2015kie,Alvarez:2015sba,Jain:2012gc,Jain:2011jc,Cho:2014taa,
Basak:2015swx,Gao:2014nia},
is based on the basic assumption that the determinant of the metric
is a fixed number, so that the metric satisfies $g_{\mu \nu}\delta
g^{\mu \nu}=0$, which in effect implies that the various components
of the metric can be appropriately adjusted, so that the resulting
determinant of the metric $\sqrt{-g}$ is some fixed function of
space-time. Hence, it can be assumed, that the metric tensor
satisfies the unimodular constraint of Eq.~(\ref{Uni1}). In order to
satisfy the unimodular constraint, we shall employ the Lagrange
multiplier method, and in effect the unimodular constraint will
appear as a part of the resulting equations of motion. Bearing this
in mind, we can generalize the mimetic gravity with potential and
Lagrange multiplier \cite{Chamseddine:2014vna}, in order to take
into account the unimodular constraint of Eq.~(\ref{Uni1}). Hence,
the generalization is the following,
\begin{equation}
\label{UM1}
S = \int d^4 x
\left\{ \sqrt{-g} \left( \frac{R}{2\kappa^2}+f(R) - V(\phi)
 - \eta \left( \partial_\mu \phi \partial^\mu \phi + 1 \right)
 - \lambda \right) + \lambda \right\}
+ S_\mathrm{matter} \, ,
\end{equation}
with $\phi$ being the real mimetic
scalar field, and $R$ is the scalar curvature. Moreover,
$S_\mathrm{matter}$ stands for the action for the matter fluids
present. We need to note that the action of Eq.~(\ref{UM1})
describes the action in (\ref{FRf}), but we shall take $f(R)=0$. The
case with $f(R)\neq 0$ will be studied in a later section. The
Lagrange multipliers $\eta$ and $\lambda$ correspond to the mimetic
gravity constraint and to the unimodular constraint respectively. We
can easily see that by varying the action of Eq.~(\ref{UM1}) with
respect to $\eta$, in which case we obtain,
\begin{equation}
\label{UM2}
\partial_\mu \phi \partial^\mu \phi = - 1 \, ,
\end{equation}
which is the mimetic constraint we discussed earlier. Moreover,
by varying the action of Eq.~(\ref{UM1}), with respect to $\lambda$,
we obtain the unimodular constraint appearing in Eq.~(\ref{Uni1}).

Both the functions $\eta $ and $\lambda$ are functions of the cosmic
time variable, which can be identified with the field $\phi$, as we
show shortly. The last identification is owing to the mimetic
constraint. Upon variation of the action (\ref{UM1}), with respect
to the metric, we obtain the equations of motion,
\begin{equation}
\label{UM2B}
0=\frac{1}{2}g_{\mu\nu} \left( \frac{R}{2\kappa^2} - V(\phi)
 - \eta \left( \partial_\mu \phi \partial^\mu \phi + 1 \right)
 - \lambda \right) - \frac{1}{2\kappa^2} R_{\mu\nu}
+ \eta \partial_\mu \phi \partial_\nu \phi + \frac{1}{2} T_{\mu\nu}
\, ,
\end{equation}
where $T_{\mu\nu}$ denotes the energy-momentum tensor of
the perfect matter fluids which are present. In addition, upon
variation of the action (\ref{UM1}), with respect to the field
$\phi$, we obtain,
\begin{equation}
\label{UM3}
0 = 2
\nabla^\mu \left( \lambda \partial_\mu \phi \right) - V' (\phi)\, .
\end{equation}
By assuming that the metric is the flat FRW metric, we can see
that the metric does not satisfy the unimodular constraint
(\ref{Uni1}), so the cosmic time variable needs to be redefined as
follows, $ d\tau = a(t)^3 dt$. In effect, the FRW metric of
Eq.~(\ref{JGRG14}) can be rewritten as in Eq.~(\ref{UniFRW}). We shall
refer to the metric of Eq.~(\ref{UniFRW}) as ``the unimodular
metric'' hereafter. For the unimodular metric of Eq.~(\ref{UniFRW}),
below we quote only the non-vanishing components of the Ricci
tensor and of the Levi-Civita connection, which are,
\begin{align}
\label{Uni13}
& \Gamma^t_{tt} = - 3 K\, , \quad \Gamma^t_{ij} = a^8
K \delta_{ij}\, , \quad \Gamma^i_{jt} = \Gamma^i_{tj} = K
\delta_j^{\ i} \, , \nn & R_{tt} = - 3 \frac{d K}{d\tau} - 12 K^2\, ,
\quad R_{ij} = a^8 \left( \frac{d K}{d\tau} + 6 K^2 \right)
\delta_{ij}\, .
\end{align}
In Eq.~(\ref{Uni13}), the function $K$ is a direct generalization of
the usual Hubble rate, in terms of the $\tau$ variable this time,
that is, $K\equiv \frac{1}{a} \frac{da}{d\tau}$. In addition, the
scalar curvature $R$ as a function of the $\tau$ variable is given
below,
\begin{equation}
\label{scalarunifrw}
R = a^6 \left( 6 \frac{d K}{d\tau} + 30 K^2
\right) \, .
\end{equation}
Due to our assumption that the auxiliary scalar field $\phi$, is
only dependent on the cosmic time $t$ (or $\tau$), we can rewrite
the mimetic constraint as follows,
\begin{equation}
\label{UM4}
a^{-6}\left( \frac{d\phi}{d\tau} \right)^2 = 1\, ,
\end{equation}
which can be
rewritten in terms of the cosmic time $t$ by employing $ d\tau =
a(t)^3 dt$, in the following way,
\begin{equation}
\label{UM5}
\left(
\frac{d\phi}{dt} \right)^2 = 1\, .
\end{equation}
Therefore, we may identify
the auxiliary scalar field $\phi$ with the cosmic time coordinate
$t$, that is $\phi=t$. The $(\tau,\tau)$ and $(i,j)$ components of
the equations appearing in Eq.~(\ref{UM2B}) are equal to,
\begin{align}
\label{UM6}
0 =& - \frac{3a^6}{2\kappa^2} K^2 + \frac{V(\phi)}{2}
+ \frac{\lambda}{2} + \eta + \frac{\rho}{2}\, , \\
\label{UM7}
0 =& \frac{a^6}{2\kappa^2} \left( 2 \frac{dK}{d\tau}
+ 9 K^2\right) - \frac{V(\phi)}{2} - \frac{\lambda}{2}
+ \frac{p}{2}\, ,
\end{align}
with $\rho$ and $p$ denoting the energy density and the pressure of
the matter fluids which are present. We need to note that in order
to get Eqs.~(\ref{UM6}) and (\ref{UM7}), we employed the constraint
of Eq.~(\ref{UM4}). Eqs.~(\ref{UM6}) and (\ref{UM7}) can be
rewritten by making use of the cosmological time $t$ variable, as
follows,
\begin{align}
\label{UM8}
0 = & - \frac{3 H^2}{2\kappa^2} + \frac{V(\phi)}{2} + \frac{\lambda}{2}
+ \eta + \frac{\rho}{2}\, , \\
\label{UM9}
0 = & \frac{1}{2\kappa^2} \left( 3 H^2 + 2 \frac{dH}{dt}
\right) - \frac{V(\phi)}{2} - \frac{\lambda}{2} + \frac{p}{2}\, .
\end{align}
In addition, Eq.~(\ref{UM3}) may be written in the following way,
\begin{equation}
\label{UM10}
0 = - 6H \lambda - 2 \frac{d\lambda}{dt}
 - V'(\phi)\, .
\end{equation}
By making use of Eqs.~(\ref{UM9}) and (\ref{UM10}),
we may eliminate $\lambda$ from the equations of motion, and hence
we have,
\begin{equation}
\label{UM11}
0 = 6 H V(\phi) - 3 V'(\phi) - 6 H p - 2
\frac{dp}{dt} + \frac{1}{\kappa^2} \left( - 18 H^3 - 6 H
\frac{dH}{dt} + 4 \frac{d^2 H}{dt^2} \right)\, .
\end{equation}
Since we
identified $\phi = t$, we can integrate Eq.~(\ref{UM12}) and we
obtain the scalar potential $V(\phi)$,
\begin{equation}
\label{UM12}
V(\phi) =
\frac{a \left( t = \phi \right)^2}{3} \int^\phi dt \, a(t)^{-2}
\left\{ - 6 H(t) p(t) - 2 \frac{dp(t)}{dt} + \frac{1}{\kappa^2}
\left( - 18 H(t)^3
 - 6 H(t) \frac{d H(t)}{dt} + 4 \frac{d^2 H(t)}{dt^2} \right) \right\} \, .
\end{equation}
Therefore, by specifying the cosmological evolution in terms of
its scale factor, and also by specifying the functional dependence
of the energy density and pressure in terms of the scale factor, by
employing Eq.~(\ref{UM12}), we can find the mimetic potential which
generates the cosmology with scale factor $a(t)$. In addition, by
employing Eqs.~(\ref{UM9}) and (\ref{UM8}), we can easily find the
functions $\lambda (t)$ and $\eta (t)$, and the complete unimodular
mimetic gravity which realizes $a(t)$ is determined. Actually, the
Eqs.~(\ref{UM12}), (\ref{UM9}) and (\ref{UM8}) constitute a
reconstruction method for finding the unimodular-mimetic theory
which can realize a specific cosmological evolution.

\subsection{Unimodular Mimetic $F(R)$ Gravity}

The unimodular mimetic $F(R)$ gravity framework
\cite{Odintsov:2016imq} uses the Lagrange multiplier formalism. So
we introduce two Lagrange multiplier functions $\eta$ and $\lambda$,
and the corresponding mimetic unimodular $F(R)$ gravity action with
mimetic potential and Lagrange multipliers is,
\begin{equation}
\label{actionmimeticfraction}
S=\int d x^4\left( \sqrt{-g}\left (
F(R)-V(\phi)+\eta \left(g^{\mu
\nu}\partial_{\mu}\phi\partial_{\nu}\phi +1\right)-\lambda \right
)+\lambda \right )\, .
\end{equation}
Upon variation of the above action
(\ref{actionmimeticfraction}), with respect to the metric, we obtain
the gravitational equations of motion,
\begin{align}
\label{eqnsofm1}
&\frac{g_{\mu \nu}}{2}\left(F(R)-V(\phi)+\eta \left(g^{\mu
\nu}\partial_{\mu}\phi\partial_{\nu}\phi +1\right)-\lambda
\right)-R_{\mu \nu}F'(R) -\eta
\partial_{\mu}\phi\partial_{\nu}\phi+\nabla_{\mu}\nabla_{\nu}F'(R)-g_{\mu
\nu}\square F'(R)=0\, .
\end{align}
Moreover, upon variation with respect to the mimetic scalar $\phi$,
we obtain,
\begin{equation}
\label{eqnsofm2}
 -2\nabla^{\mu}\left( \eta
\partial_{\mu}\phi\right)-V'(\phi)=0\, .
\end{equation}
By varying the action
(\ref{actionmimeticfraction}) with respect to the function $\eta$ we
get,
\begin{equation}
\label{mimeticconstraint}
g^{\mu \nu}(\hat{g}_{\mu \nu},\phi)\partial_{\mu}\phi\partial_{\nu}\phi=-1\, ,
\end{equation}
while
varying the action with respect to the function $\lambda (t)$, we
obtain,
\begin{equation}
\label{unimodularconstraint}
\sqrt{-g}=1\, .
\end{equation}
For the
flat FRW metric, the $(t,t)$ components of the expression appearing
in Eq.~(\ref{eqnsofm1}), are equal to,
\begin{equation}
\label{enm1}
 -F(R)+6(\dot{H}+H^2)F'(R)-6H\frac{d F'(R)}{d t}-\eta
(\dot{\phi}^2+1)+\lambda+V(\phi)=0\, ,
\end{equation}
while the corresponding $(i,j)$ components are equal to,
\begin{equation}
\label{enm2}
F(R)-2(\dot{H}+3H^2)+2\frac{d ^2F'(R)}{d t^2}
+4H\frac{d F'(R)}{d t}-\eta
(\dot{\phi}^2-1)-V(\phi)-\lambda=0\, ,
\end{equation}
For the flat FRW metric, Eq.~(\ref{eqnsofm2})can be written as
follows,
\begin{equation}
\label{enm3}
2\frac{d (\eta \dot{\phi})}{d t}+6H\eta
\dot{\phi}-V'(\phi)=0\, .
\end{equation}
Finally, from the mimetic constraint we have,
\begin{equation}
\label{enm4}
\dot{\phi}^2-1=0\, .
\end{equation}
We need to note that in Eqs.~(\ref{enm1}), (\ref{enm2}),
(\ref{enm3}), and (\ref{enm4}), the ``prime'' indicates
differentiation with respect to $R$, the scalar curvature, and also with
respect to the mimetic scalar. By taking into account
Eq.~(\ref{enm4}), the equation (\ref{enm1}), can be rewritten as
follows,
\begin{equation}
\label{enm1sim}
 -F(R)+6(\dot{H}+H^2)F'(R)-6H\frac{d F'(R)}{d t}
 -2\eta+\lambda+V(\phi)=0\, ,
\end{equation}
and in addition Eq.~(\ref{enm2}) is rewritten in the following
way,
\begin{equation}
\label{enm2sim}
F(R)-2(\dot{H}+3H^2)+2\frac{d ^2F'(R)}{d t^2}
+4H\frac{d F'(R)}{d t}-V(\phi)-\lambda=0\, .
\end{equation}
Furthermore, Eq.~(\ref{enm4}) becomes,
\begin{equation}
\label{onlali}
2\frac{d \eta}{d t}+6H\eta-V'(t)=0\, .
\end{equation}
If we eliminate $\lambda$ from Eqs.~(\ref{enm1sim}) and
(\ref{enm2sim}), we get,
\begin{equation}
\label{sone}
6(\dot{H}+H^2)F'(R)-2H\frac{d F'(R)}{d t}-2\eta
 -2(\dot{H}+3H^2)+2\frac{d ^2F'(R)}{d t^2}-V(t)=0\, .
\end{equation}
 From Eqs.~(\ref{sone}) and (\ref{onlali}), we get,
\begin{equation}
\label{auxeqn}
 -2V'(t)-3H V(t)+f_0(t)=0\, ,
\end{equation}
with the function $f_0(t)$ being equal to,
\begin{align}
\label{explicitf0cosmictime}
& f_0(t)=-\left[18 H(t)\left(
\dot{H}+H^2\right)F'(R)-6H^2\frac{d F'(R)}{d t}
 -6H\left(\dot{H}3H^2
\right)+6 H\frac{d ^3F'(R)}{d t^3} \right. \nn
& \left. 6(\ddot{H}+2\dot{H}H)F'(R)
+H\dot{H}\frac{d F'(R)}{d t}+6
H^2\frac{d F'(R)}{d t}
-2H\frac{d ^2F'(R)}{d t^2}-2\left(
\ddot{H}+6\dot{H}H\right)
+2\frac{d ^2F'(R)}{d t^2} \right] \, .
\end{align}
The differential equation of Eq.~(\ref{auxeqn}) can be solved and we
get the following general solution,
\begin{equation}
\label{generalsol1}
V(t)=\frac{a^{3/2}(t)}{2}\int a^{-3/2}(t)f_0(t)d t\, ,
\end{equation}
where $a(t)$ is the scale factor. Hence, given an arbitrary
cosmological evolution, quantified in terms of its scale factor, and
also the form of the $F(R)$ gravity, we can use
Eq.~(\ref{generalsol1}) in order to obtain the mimetic potential, and
with it, by using Eq.~(\ref{enm2sim}), we can obtain the function
$\lambda (t)$. Eventually, by using the resulting function $\lambda
(t)$ and also the potential $V(t)$, we can substitute these in
Eq.~(\ref{enm1sim}), and we can obtain the function $\eta (t)$. Several
illustrative examples on how this reconstruction method works, can
be found in \cite{Odintsov:2016imq}.

In the previous sections we provided a quick review for a number of
modified gravities. Definitely, the models we presented are not all
the models one can construct, different approaches and
generalizations of the above can be constructed. In the next chapter
we shall present cosmological applications of the above models in
more detail.

\section{Inflationary Dynamics in Modified Gravity}

The Standard Big Bang cosmology had several issues which rendered
the theory incomplete. Particularly, the flatness problems, the
monopole problem, the entropy problem and the horizon problem, found
a successful explanation in the context of inflation. In some sense
the horizon problem can represent all the above problems, since it
is related to all of the above. The horizon problem refers to the
question how two distinct parts of the Universe which are very far
away at present time, could be in causal connection in the past.
These two distinct parts of the Universe seem to have almost the
same density at present time, so this could not be explained at a
microphysics level, unless these two parts were in causal connection
in the past. If someone adopts the Big Bang cosmology evolution, and
solves the equations following the cosmic time backwards, then two
parts of the Universe which are now very far away and seem to have
almost the same density, cannot be causal connection in the past.
The inflationary paradigm solved this issue by adding a period of
nearly exponential accelerated expansion during early times.

The inflationary paradigm
\cite{Guth:1980zm,Linde:2007fr,Gorbunov:2011zzc,Lyth:1998xn,
Linde:1983gd,Linde:1985ub,Linde:1993cn,Sasaki:1995aw,Turok:2002yq,
Linde:2005dd,Kachru:2003sx,
Brandenberger:2016uzh,Bamba:2015uma,Martin:2013tda,
Martin:2013nzq,Baumann:2014nda,Baumann:2009ds,Linde:2014nna,
Pajer:2013fsa,Yamaguchi:2011kg,Byrnes:2010em} is up to date one of
the most successful descriptions of the early Universe, since in
most cases the outcome is a nearly scale invariant power spectrum.
It is quite common in the cosmology literature to use a single
canonical scalar field which slow-rolls on its potential, in order
to describe the inflationary era, and actually this was the original
approach of inflation
\cite{Linde:2007fr,Gorbunov:2011zzc,Lyth:1998xn,Linde:1983gd,Linde:1985ub,
Linde:1993cn,Sasaki:1995aw,Brandenberger:2016uzh,Bamba:2015uma,
Martin:2013tda,Martin:2013nzq,Baumann:2014nda,Baumann:2009ds,
Linde:2014nna,Pajer:2013fsa,Yamaguchi:2011kg,Byrnes:2010em,
Turok:2002yq,Linde:2005dd,Kachru:2003sx}. In the literature there
are many reviews on this topic and in this section we shall provide
a quick overview of the various approaches on inflation, emphasizing
in the modified gravity description. We shall not expand our
description going into many details, but we provide a concrete and
condensed description of the quantitative features of an
inflationary theory, that is, the calculation of the spectral index
of the primordial curvature perturbations and of the
scalar-to-tensor ratio. We start off with the single and
multi-scalar field descriptions of inflation and then we proceed to
the modified gravity description. For simplicity, in all the cases
we shall assume that the inflationary dynamics is controlled by the
vacuum modified gravity, but the same issue can be addressed in the
presence of perfect matter fluids. The original papers and reviews
on cosmological perturbations can be found in
Refs.~\cite{Gorbunov:2011zzc,Mukhanov:1990me}, but we do not discuss
perturbations in detail here.

The Hubble radius $R_H$, defined as $R_H=1/(aH)$, where $a$ is the
scale factor of the Universe and $H$ the corresponding Hubble rate,
quantifies perfectly the conditions that have to be satisfied in
order for inflation to occur. In terms of the Hubble radius, the
primordial modes before the inflationary era were well inside the
Hubble radius, and during the inflationary era these exited the
Hubble radius, since the Hubble radius shrunk exponentially. In
terms of the Hubble radius, the conditions for inflation are,
$\dot{R}_H<0$, which implies that $\ddot{a}>0$. In turn the latter
conditions indicate that the Universe described by the scale factor
$a$ is accelerating, and this is why the inflationary era is often
referred to as early-time accelerating era.

\subsection{Scalar Field Descriptions}

\subsubsection{Canonical Scalar Field Inflation}

Consider a Universe described by the FRW metric and a canonical
scalar field $\varphi$, with the following action, \be
\label{canonicalscalarfieldaction} \mathcal{S}=\int d^4x
\sqrt{-g}\left(\frac{R}{2\kappa^2}-\frac{1}{2}\partial_{\mu}\phi
\partial^{\mu}\phi -V(\varphi))\right)\, ,
\end{equation}where $V(\varphi)$ the canonical scalar field potential, $\kappa^2$ is related to the gravitational constant as $\kappa^2=8\pi
G=\frac{1}{M_p^2}$ and $M_p$ is the reduced Planck mass. The
energy momentum tensor $T_{\mu \nu}$ corresponding to the scalar
field action (\ref{canonicalscalarfieldaction}) is equal to,
\begin{equation}
\label{energymomentumcanonicalscalarfield}
T_{\mu\nu}=\partial_{\mu}\varphi\partial_{\nu}\varphi-g_{\mu \nu}
\left(\frac{1}{2}g^{\mu
\nu}\partial_{\mu}\varphi\partial_{\nu}\varphi-V(\varphi)\right)\, ,
\end{equation}
and therefore from this we obtain that the energy density is
$T^0_0=\rho$, that is,
\begin{equation}
\label{energydensitysinglescalar}
\rho=\frac{1}{2}\dot{\varphi}^2+V(\varphi)\, .
\end{equation}
Also from the
component $T^{i}_j=-P\delta^i_j$, we obtain that the total pressure
in a Universe filled with a canonical scalar field is,
\begin{equation}
\label{pressuresinglescalar}
P=\frac{1}{2}\dot{\varphi}^2-V(\varphi)\, .
\end{equation}
The Friedmann equation is,
\begin{equation}
\label{friedmaneqnsinglescalar}
H^2=\frac{\kappa^2}{3}\rho\, ,
\end{equation}
so by substituting the energy
density (\ref{energydensitysinglescalar}) and taking the first
derivative with respect to the cosmic time $t$, by also using the
expression for the pressure (\ref{pressuresinglescalar}) we obtain,
\begin{equation}
\label{dothsinglescalar}
\dot{H}=-\frac{\kappa^2}{2}\dot{\varphi}^2\, ,
\end{equation}
which combined
with the above leads to the Klein-Gordon equation for the canonical
scalar field $\varphi$,
\begin{equation}
\label{kleingordonsingle}
\ddot{\varphi}+3H\dot{\varphi}+V'=0\, ,
\end{equation}
where the prime denotes differentiation with respect to the scalar
field $\varphi$. The definition of the slow-roll indices for the
scalar theory are,
\begin{equation}\label{definitionscalarslowrollindices}
\epsilon_1=-\frac{\dot{H}}{H^2},\,\,\,\epsilon_2=\frac{\ddot{\varphi}}{H\dot{\varphi}}\,
,
\end{equation}
and the slow-roll conditions for the canonical scalar field
require that the slow-roll indices satisfy $\epsilon_1,\epsilon_2
\ll 1$, so in terms of the scalar field and the potential, this
means $\frac{1}{2}\dot{\varphi}^2\ll V(\varphi)$. The first
slow-roll condition $\epsilon_1\ll 1$ or equivalently $\dot{H}\ll
H^2$ ensures the occurrence of the inflationary era in the first
place, while the second ensures that inflation lasts for a
sufficient amount of time. In view of the slow-roll conditions,
the Friedmann equation (\ref{friedmaneqnsinglescalar}) is
simplified as follows,
\begin{equation}
\label{firstslowrollcond}
H^2\simeq \kappa^2\frac{V}{3}\, .
\end{equation}
The second slow-roll condition $\epsilon_2 \ll 1$ implies that
$\ddot{\varphi}\ll H\dot{\varphi}$, so the Klein-Gordon equation
(\ref{kleingordonsingle}) is simplified as follows,
\begin{equation}
\label{secondslowroll} 3H\dot{\varphi}\simeq -V'\, .
\end{equation}
The spectral index of the primordial scalar curvature
perturbations $n_s$ and the tensor-to-scalar ratio $r$ can be
written in terms of the slow-roll indices $\epsilon_1$ and
$\epsilon_2$ in the following way,
\begin{equation}\label{obsvindicespreliminary}
n_s=1-4\epsilon_1-2\epsilon_2,\,\,\,r=16\epsilon_1\, .
\end{equation}
We can express the slow-roll indices $\epsilon_1$ and $\epsilon_2$
as functions of the scalar field $\varphi$ using the following
approach: the slow-roll index $\epsilon_1$ in view of Eqs.
(\ref{dothsinglescalar}) and (\ref{firstslowrollcond}) is written
as follows,
\begin{equation}\label{epsilon1initial}
\epsilon_1=\frac{3\dot{\varphi}^2}{2V(\varphi)}\, ,
\end{equation}
which in view of Eq. (\ref{secondslowroll}) becomes,
\begin{equation}\label{secondstep}
\epsilon_1=\frac{V'(\varphi)^2}{6 H^2 V(\varphi)}\, ,
\end{equation}
so in view of Eq. (\ref{firstslowrollcond}) we finally have,
\begin{equation}\label{finalfirstslowrollindex}
\epsilon=\epsilon_1=\frac{1}{2\kappa^2}\left(\frac{V'(\varphi)}{V(\varphi)}
\right)^2\, ,
\end{equation}
where we changed the notation $\epsilon_1=\epsilon$ since for
single minimally coupled canonical scalar field theories the
notation for the first slow-roll index in the literature is
$\epsilon$. Let us now turn our focus on the second slow-roll
index $\epsilon_2$, and by differentiating Eq.
(\ref{secondslowroll}) with respect to the cosmic time, we get,
\begin{equation}\label{secondslowrollprelim}
\frac{d \dot{\varphi}}{d
t}=\ddot{\varphi}=-\frac{V''(\varphi)\dot{\varphi}H-V'(\varphi)\dot{H}}{3
H^2}\, ,
\end{equation}
hence in view of Eqs. (\ref{dothsinglescalar}) and
(\ref{firstslowrollcond}) we have after some simple algebra,
\begin{equation}\label{epsilon2final}
\epsilon_2=\frac{\ddot{\varphi}}{H\dot{\varphi}}=-\eta+\epsilon_1\,
,
\end{equation}
where $\eta$ is defined as follows,
\begin{equation}\label{eta}
\eta=\frac{1}{\kappa^2}\frac{V''(\varphi)}{V(\varphi)}\, .
\end{equation}
The most familiar expressions for the slow-roll indices in
minimally coupled canonical scalar theory are $\epsilon$ and
$\eta$, which as we showed in terms of the canonical scalar field
potential $V(\varphi)$ are,
\begin{equation}
\label{slowrollindicespotentialversion} \epsilon\simeq
\frac{1}{2\kappa^2}\left(
\frac{V'(\varphi)}{V(\varphi)}\right)^2\, ,\quad \eta\simeq
\frac{1}{\kappa^2}\frac{V''(\varphi)}{V(\varphi)}\, .
\end{equation}
Moreover, the graceful exit from the inflationary era occurs when
the first slow-roll index becomes of the order $\epsilon\sim
\mathcal{O}(1)$. Having the above at hand, one can easily calculate
the spectral index of primordial curvature perturbations $n_s$ and
the scalar-to-tensor ratio $r$, which for a slow-rolling canonical
scalar can be expressed in terms of the slow-roll indices as
follows,
\begin{equation}
\label{observationalindices} n_s\simeq 1-6\epsilon+2\eta\, ,\quad
r\simeq 16 \epsilon\, ,
\end{equation}
where we combined Eqs. (\ref{obsvindicespreliminary}),
(\ref{finalfirstslowrollindex}) and (\ref{epsilon2final}). For the
calculation of the spectral index $n_s$ and the tensor-to-scalar
ratio $r$, these must be evaluated at the first horizon crossing
time instance. Let us use the single scalar field inflation
formalism in order to calculate the slow-roll indices and the
corresponding observational indices for the Starobinsky model of
inflation \cite{Starobinsky:1982ee}, or Higgs inflation
\cite{Bezrukov:2007ep},
\begin{equation}
\label{starobinflpotential1}
V(\varphi)=\alpha \mu^2 \left(
1-\e^{-\sqrt{\frac{2}{3}}\varphi}\right)^{2}\, ,
\end{equation}
where for simplicity we use the physical units system in which
$\hbar=c=8\pi G=\kappa^2=1$, which is known as reduced Planck
units system. So let us calculate the slow-roll indices
(\ref{slowrollindicespotentialversion}) at the horizon crossing,
which we assume it occurs for $\varphi=\varphi_k$, and these are,
\begin{equation}
\label{slowrollstarobinsky}
\epsilon=\frac{4}{3\left(\e^{\sqrt{\frac{2}{3}} \varphi_k}-1\right)^2}\, ,
\quad \eta=-\frac{4 \left(\e^{\sqrt{\frac{2}{3}}
\varphi_k }-2\right)}{3 \left(\e^{\sqrt{\frac{2}{3}} \varphi_k}-1\right)^2}\, .
\end{equation}
Then, the observational indices $n_s$ and $r$,
calculated at the horizon crossing read,
\begin{equation}
\label{observstarobinsky}
n_s\simeq \frac{-14 \e^{\sqrt{\frac{2}{3}}
\varphi_k }+3 \e^{2 \sqrt{\frac{2}{3}} \varphi_k }-5}{3
\left(\e^{\sqrt{\frac{2}{3}} \varphi_k}-1\right)^2}\, , \quad
r\simeq \frac{64}{3 \left(\e^{\sqrt{\frac{2}{3}} \varphi_k
}-1\right)^2}\, .
\end{equation}
We can express the observational indices as
functions of the $e$-foldings number $N$, which for a slow-rolling
canonical scalar field is,
\begin{equation}
\label{nfoldingsstarob1}
N\simeq
\int_{\varphi_f}^{\varphi_{k}}\frac{V(\varphi)}{V'(\varphi)}d
\varphi \, ,
\end{equation}
where $\varphi_k$ is the value of the scalar field
at the horizon crossing and $\varphi_f$ is the value of the scalar
at the end of inflation. By using the potential
(\ref{starobinflpotential1}), the $e$-foldings number at leading
order is,
\begin{equation}
\label{leadingn}
N\simeq \frac{-3}{4}
\e^{\sqrt{\frac{2}{3}} \varphi_f }+\frac{3}{4}
\e^{\sqrt{\frac{2}{3}} \varphi_k}\, ,
\end{equation}
and the expression
containing the value of the scalar field at the end of the
inflationary era can be determined by the condition $\epsilon
(\varphi_f)\simeq 1$, which yields the condition,
\begin{equation}
\label{conditionendofinflation}
\frac{4}{3\left(\e^{\sqrt{\frac{2}{3}} \varphi_f }-1\right)^2}=1\, .
\end{equation}
Combining Eqs.~(\ref{observstarobinsky}), (\ref{leadingn}), and
(\ref{conditionendofinflation}), and also for large $e$-folding
values, we finally obtain the observational indices of the
Starobinsky model, which at leading-$N$ order  are,
\begin{equation}
\label{finalstarob}
n_s\simeq 1-\frac{2}{N}-\frac{3}{N^2}\, ,\quad
r\simeq \frac{12}{N^2}\, .
\end{equation}
Before we close this section, it is
worth discussing an interesting aspect with regard to the graceful
exit from inflation in the context of inflationary theories with a
single canonical scalar field. Actually, the slow-roll parameters
$\epsilon$ and $\eta$ are the lowest order terms in the Hubble
slow-roll expansion
\cite{Liddle:1994dx,Copeland:1993jj,Liddle:1992wi}, and there are
many more higher order parameters in the expansion. Then it is
possible that the graceful exit might occur even at higher order and
thus much more earlier than in the case where $\epsilon\sim
\mathcal{O}(1)$. The slow-roll expansion
\cite{Liddle:1994dx,Copeland:1993jj,Liddle:1992wi}, quantitatively
completes the slow-roll approximation, and it actually enables us to
securely find the final inflationary attractor of the theory. It is
worth recalling some fundamental features of the slow-roll
expansion, and for details we refer to
\cite{Liddle:1994dx,Copeland:1993jj,Liddle:1992wi}. According to the
Hubble slow-roll expansion, the physical system has an inflationary
attractor, to which all the slow-roll solutions tend to
asymptotically. Then, we may write the single canonical scalar field
Friedmann equation as follows,
\begin{equation}
\label{frqwe}
H^2(\varphi)=\frac{8 \pi\kappa^2}{3}V(\varphi)
\left( 1-\frac{1}{3}\epsilon (\varphi) \right)^{-1}\, ,
\end{equation}
where
we expressed the slow-roll parameter $\epsilon$ as a function of the
canonical scalar field $\varphi$. By employing the binomial theorem,
we can obtain the perturbative expansion of Eq.~(\ref{frqwe}) in
terms of the slow-roll parameters $\epsilon_V$, $\xi_V$ and $\eta_V$
\begin{equation}
\label{frqwe1}
H^2(\varphi )\simeq \frac{8 \pi\kappa^2}{3}V(\varphi )\left(
1+\epsilon-\frac{4}{3}\epsilon^2+\frac{2}{3}\epsilon\eta_V
+\frac{32}{9}\epsilon^3+\frac{5}{9}\epsilon\eta_V^2
 -\frac{10}{3}\epsilon^2\eta_V+\frac{2}{9}\epsilon\xi^2_V
+\mathcal{O}_4 \right)\, ,
\end{equation}
where in the expansion above we kept terms up to fourth order in the
parameters $\eta_V,\xi_V$, which in terms of the slow-roll
parameters $\epsilon$ and $\eta$, are defined below,
\begin{align}
\label{shelll}
& \eta_V=(3-\epsilon)^{-1}\left(
3\epsilon+3\eta-\eta^2-\xi_H^2\right)\, , \\ & \notag
\xi_V=(3-\epsilon)^{-1} \left(27\epsilon\eta+9\xi_H^2
 -9\epsilon\eta^2-12\eta\xi_H^2-3\sigma_H^3+3\eta^2\xi_H^2
+\eta\sigma_H^3 \right)\, .
\end{align}
In addition, the parameters $\xi_H$ and $\sigma_H$ appearing in
Eq.~(\ref{shelll}), are also functions of the slow-roll parameters
$\epsilon$ and $\eta$ and are defined as follows,
\begin{equation}
\label{insertcode123sledgehammerenteraccept}
\xi_H^2=\epsilon\eta-\sqrt{\frac{1}{4\pi\kappa^2}}\sqrt{\epsilon}
\eta'\, ,\quad \sigma_H^3=\xi_H^2(2\epsilon-\eta)
-\sqrt{\frac{1}{\kappa^2\pi}}\sqrt{\epsilon}\xi_H\xi_H'\, .
\end{equation}
Note
that in the above equations, the prime indicates differentiation
with respect to the canonical scalar field $\varphi$, and in
addition we assumed the Hubble rate solution $H(t)$ is the final
attractor of the theory.

Having the slow-roll expansion (\ref{frqwe1}) at hand, it is easy to
support our argument that the slow-roll expansion might actually
break down much more earlier than the slow-roll approximation does,
and this is due to the existence of the higher order terms in the
expansion. In the case of the slow-roll approximation, ones takes
into account only $\epsilon$ and $\eta$ which are simply the lowest
order terms. Moreover, as was also noted in
Ref.~\cite{Liddle:1994dx,Copeland:1993jj,Liddle:1992wi}, the Hubble
slow-roll expansion can break down for large values of the slow-roll
parameters or if a singularity occurs in the perturbative expansion.
For example, if the canonical scalar field potential has the
following form $V(\varphi)\sim \varphi^2$, then the Hubble slow-roll
expansion contains inverse powers of the canonical scalar field, and
the inflationary era might end before $\epsilon\sim \mathcal{O}(1)$.
In fact, although definitely the end of inflation occurs at
$\varphi=0$, where the absolute minimum of the potential is, the
Hubble slow-roll expansion might break down earlier, due to the fact
that the expansion contains terms $\varphi^{-n}$, with $n>0$ and
this is clearly an indication that inflation might end earlier. For
similar results on instabilities of the second slow-roll index
$\eta$ and its connection to the graceful exit from inflation, see
\cite{Odintsov:2015gba,Odintsov:2016plw,Odintsov:2015jca,
Odintsov:2015tka}.

\subsubsection{Non-Canonical Scalar Field Inflation}

Consider now a non-canonical scalar field, in which case the
gravitational action is,
\begin{equation}
\label{grvact1}
S=\int d^4 x \sqrt{-g}\left\{ \frac{1}{2\kappa^2}R
 - \frac{1}{2}\omega(\phi)\partial_\mu \phi
\partial^\mu\phi - V(\phi) + L_\mathrm{matter} \right\}\, .
\end{equation}
In this case the appearance of the kinetic term $\omega(\phi)$,
makes the scalar field $\phi$ non-canonical, so it is conceivable
that the slow-roll indices in this case, are related to this kinetic
function $\omega (\phi)$. In principle the kinetic function is
irrelevant and can be easily absorbed by making a redefinition of
the scalar field $\phi$, as follows,
\begin{equation}
\label{rdfne1}
\varphi
\equiv \int^\phi d\phi \sqrt{\omega(\phi)} \, ,
\end{equation}
where it is
assumed that $\omega(\phi)>0$. Then, one can obtain a canonical
scalar field action, since the kinetic scalar term becomes,
\begin{equation}
\label{kntrterm}
 - \omega(\phi) \partial_\mu \phi \partial^\mu\phi
= - \partial_\mu \varphi \partial^\mu\varphi\, ,
\end{equation}
Then, only in
the case for which the resulting expression from the integration
(\ref{rdfne1}), is an invertible function of $\phi$, then if we find
the function $\phi=\phi (\varphi)$, we may substitute $\phi$ in the
action (\ref{grvact1}) and then we have the canonical scalar theory
at hand. However, in the case that the result of the integral
(\ref{rdfne1}) is not an invertible function of $\phi$, then it is
not possible to obtain the canonical scalar theory, and therefore we
need an alternative way to calculate the slow-roll indices. This
case was studied in detail in
\cite{Liddle:1994dx,Copeland:1993jj,Liddle:1992wi}, and the
slow-roll indices as functions of the non-canonical kinetic term
$\omega (\phi)$ and of the scalar potential $V(\phi)$ are given
below \cite{Liddle:1994dx,Lyth:1998xn,Bamba:2014daa},
\begin{align}
\label{noncaninflationslowroollindices}
\epsilon =&
\frac{1}{2\kappa^2} \left(\frac{d\phi}{d\varphi} \right)^2 \left(
\frac{V'(\phi)}{V(\phi)} \right)^2 = \frac{1}{2\kappa^2}
\frac{1}{\omega(\phi)} \left( \frac{V'(\phi)}{V(\phi)} \right)^2 \, , \nn
\eta = & \frac{1}{\kappa^2 V(\phi)} \left[
\frac{d\phi}{d\varphi} \frac{d}{d\phi} \left( \frac{d\phi}{d\varphi}
\right) V'(\phi) + \left( \frac{d\phi}{d\varphi} \right)^2 V''(\phi)
\right] = \frac{1}{\kappa^2 V(\phi)}
\left[ -\frac{\omega'(\phi)}{2 \omega(\phi)^2} V'(\phi)
+ \frac{1}{\omega(\phi)} V''(\phi) \right] \, .
\end{align}
Then, one may study the analytic slow-roll of the non-canonical
scalar field $\phi$, following the research line we presented in the
previous section, but we will not go into details for brevity.

Before closing this section we need to note that in the case
$\omega(\phi)<0$, the non-canonical scalar $\phi$ is a phantom
field, and in this case instead of the integration performed in
Eq.~(\ref{rdfne1}), one should perform the following redefinition of the
scalar $\phi$,
\begin{equation}
\label{ma13preview}
\varphi \equiv \int^\phi
d\phi \sqrt{-\omega(\phi)} \, .
\end{equation}
In this case then, the kinetic term reads,
\begin{equation}
\label{kntrterm1}
 - \omega(\phi) \partial_\mu \phi \partial^\mu\phi
= \partial_\mu \varphi \partial^\mu\varphi\, ,
\end{equation}
instead of the
corresponding expression appearing in Eq.~(\ref{kntrterm}). However,
in the case of a ghost scalar field, the situation is rather
physically not appealing, since phantom inflation occurs. In this
case, the total energy density is unbounded from below when one
deals with the classical theory at least. However, in the quantum
theory, the energy becomes bounded from below, at the cost of having
a negative norm in the Hilbert space states.

\subsubsection{Multi-Scalar Field Inflation}

The slow-roll formalism for inflation we developed in the previous
sections for a single scalar field, can be generalized in the case
of multiple scalar fields. In this section we briefly outline the
fundamental features of this formalism and for details we refer to
the original papers that address this issue, see
Refs.~\cite{Sasaki:1998ug,Kaiser:2012ak,Turzynski:2014tza,Kaiser:2013sna}.

The gravitational action in the presence of multiple scalar fields
$\phi^I$ reads,
\be
\label{miltuiscale}
S=\int d ^4x\sqrt{-\hat{g}}\left(
\frac{\hat{R}}{2\kappa^2}-\frac{1}{2}G_{IJ}(\phi^I)\hat{g}^{\mu \nu
}\partial_{\mu }\phi^I\partial_{\nu }\phi^J -V(\phi^{I} ) \right)\, ,
\end{equation}
with $I,J=1,2,..N$, where $N$ is the total number of the scalar
fields present. The metric $G_{IJ}(\phi^I)$ depends solely on the
scalar fields and therefore it is the metric in the configuration
space formed by the scalars $\phi^I$.

The FRW equations corresponding to the gravitational action
(\ref{miltuiscale}) are given below,
\begin{equation}
\label{gfgdgd}
H^2=\frac{\kappa^2}{3}\left(
\frac{1}{2}G_{IJ}\dot{\varphi}^I\dot{\varphi}^J+V(\varphi^I)\right) \, , \quad
\dot{H}=-\frac{1}{\kappa^2}G_{IJ}\dot{\varphi}^I\dot{\varphi}^J \, , \quad
\square \phi^I+\hat{g}^{\mu \nu }\Gamma^I_{IK}\partial_{\mu}
\phi^J\partial_{\nu }\phi^K-G^{IK}V_{,K}=0\, ,
\end{equation}
where $V_{,K}=\partial V/\partial \phi^K$, and also the Christoffel
symbols $\Gamma^I_{IK}=\Gamma^I_{IK}(\varphi,\phi)$ are the
connections in the configuration space formed by the scalar fields
$\phi^I$, equipped with the metric $\Gamma^I_{IK}$.

We introduce the scalar field function $\sigma$ and the vector field
$\hat{\sigma}^{I}$, which are defined in terms of the scalar fields
and the metric as follows,
\begin{equation}
\label{dsigmadot}
\dot{\sigma}=\sqrt{G_{IJ}\dot{\phi}^{I}\dot{\phi}^{J}}\, ,\quad
\hat{\sigma}^{I}=\frac{\dot{\phi}^I}{\dot{\sigma}}\, .
\end{equation}
The FRW equations (\ref{gfgdgd}) can be rewritten in terms of the
variables $\dot{\sigma}$ and $\hat{\sigma}^{I}$, in the following
way,
\begin{equation}
\label{fieldhf}
H^2=\frac{\kappa^2}{3}\left(\frac{1}{2}\dot{\sigma}^2+V \right)\, , \quad
\dot{H}=-\frac{\kappa^2}{2}\dot{\sigma}^2\, , \quad
\ddot{\sigma}+3H\dot{\sigma}+V_{,\sigma}=0\, ,
\end{equation}
where $V_{,\sigma}=\hat{\sigma}^IV_{,I}$. Finally, the generalized
multi-field slow-roll indices, as functions of the scalar fields,
are equal to,
\begin{equation}
\label{000}
\epsilon=\frac{3\dot{\sigma}^2}{\dot{\sigma}^2+2V}\, ,\quad
\eta_{\sigma
\sigma}=\frac{1}{\kappa^2}\frac{\hat{\sigma}_I
\hat{\sigma}^JM^I_J}{V}\, ,\quad
\eta_{ss}=\frac{1}{\kappa^2}\frac{\hat{s}_I\hat{s}^JM^I_J}{V}\, ,
\end{equation}
where the vectors $\hat{\sigma}^I$ and $\hat{s}^I=\omega^I/\omega$
are the unit vectors of the adiabatic and iso-curvature directions
respectively, in the curved configurations space of the scalar
fields. In addition, the vector $\omega^I$ is the turn-rate vector
given by $\omega^I=\mathcal{D}_t\hat{\sigma}^I$, with
$\omega=|\omega^I|$ and $\mathcal{D}_t$ the covariant derivative
along the $t$-direction, which on an arbitrary vector field $A^I$
acts as follows,
\begin{equation}
\label{mathcaldcov}
\mathcal{D}_tA^I=\dot{A}^I+\Gamma^I_{JK}A^J\dot{\phi}^K\, .
\end{equation}
For a recent work with a detailed computation with two scalar
fields, we refer the reader to Ref.~\cite{Odintsov:2015tka}.

\subsection{Inflation in $f(\varphi,R)$ Theories of Gravity}

A quite general class of inflationary theories can be described by
the following action \cite{Noh:2004bc,Hwang:2002fp,Hwang:2001fb,
Hwang:2000jh,Hwang:1996xh,Sebastiani:2015kfa,Farajollahi:2010tc},
\begin{equation}
\label{actionfrphigeneral}
\mathcal{S}=\int d ^4x\sqrt{-g}\left
(\frac{f(R,\phi)}{2\kappa^2} -\frac{1}{2}\omega (\phi)
g^{\mu\nu}\partial_{\mu}\phi\partial_{\nu}\phi-V(\phi )\right)\, ,
\end{equation}
where $f(R,\phi)$ is a smooth function of the scalar curvature and
of the scalar field, which is in general a non-canonical scalar
field, if the kinetic term satisfies $\omega (\phi)\neq 1$. It is
conceivable that the $F(R)$ gravity theory, the canonical and
non-canonical scalar field gravity, and also theories of the form
$f(R,\phi)=f(\phi)R$, are subcases of the theory described by the
action (\ref{actionfrphigeneral}). In this section we present the
essential features of the inflationary dynamics for the general
action (\ref{actionfrphigeneral}) and later on we specify how the
dynamics are modified in each of the $F(R)$ and $f(R,\phi)=f(\phi)R$
cases, since the single scalar field case was studied in the
previous sections. The slow-roll indices for the action
(\ref{actionfrphigeneral}) have the following general form,
\begin{equation}
\label{slowrollgenerarlfrphi}
\epsilon_1=-\frac{\dot{H}}{H^2}\, ,\quad
\epsilon_2=\frac{\ddot{\phi}}{H\dot{\phi}}\, , \quad
\epsilon_3=\frac{\dot{f}_R}{2Hf_R}\, ,\quad
\epsilon_4=\frac{\dot{E}}{2HE}\, ,
\end{equation}
where $f_R=\frac{\partial f(R,\phi)}{\partial R}$.
Also the function $E$ appearing in
Eq.~(\ref{slowrollgenerarlfrphi}) is equal to,
\begin{equation}
\label{epsilonfnction}
E=f_R \omega+\frac{3\dot{f}_R^2}{2\kappa^2\dot{\phi}^2}\, .
\end{equation}
In
addition for later convenience, we define the following function,
$Q_s$,
\begin{equation}
\label{qsfunction}
Q_s=\dot{\phi}^2\frac{E}{f_RH^2(1+\epsilon_3)^2}\, ,
\end{equation}
which plays
an important role for the calculation of the scalar-to-tensor ratio
during the slow-roll era. Assuming the flat FRW background, then by
varying the action (\ref{actionfrphigeneral}) with respect to the
metric and also with respect to the scalar field, we obtain the
following equations of motion,
\begin{align}
\label{eqnsofmotionfrphinonslowroll}
& \frac{3f_R(R,\phi)H^2}{\kappa^2}=\frac{\omega
(\phi)\dot{\phi}^2}{2}+V(\phi)+\frac{1}{\kappa^2}\left(
Rf_R(R,\phi)-f(R,\phi)\right)-3H\dot{f}_R(R,\phi) \, , \nn
& -\frac{2f_R(R,\phi)}{\kappa^2}\dot{H}=\omega
(\phi)\dot{\phi}^2+\ddot{f}_R(R,\phi)-H\dot{f}_R(R,\phi) \, ,\nn
& \ddot{\phi}+3H\dot{\phi}
+\frac{1}{2\omega (\phi)}\left( \dot{\omega
(\phi)}\dot{\phi}-\frac{1}{\kappa^2}\frac{\partial
f(R,\phi)}{\partial \phi}+2\frac{\partial V}{\partial
\phi}\right)=0\, .
\end{align}
During the slow-roll era, the following relations hold true between
the Hubble rate, the scalar field and the functions $f(R,\phi)$,
$\omega (\phi)$,
\begin{equation}
\label{slowrollconditions}
\frac{\dot{\omega}\dot{\phi}}{H^2\dot{f }_R}\ll 1\, ,\quad
\frac{\dot{\omega}\dot{\phi}^2}{H^2\dot{f}_R}\gg \frac{2\omega
\dot{\phi}\epsilon_3}{H\dot{f}_R}\, ,
\end{equation}
and in view of the
relations (\ref{slowrollconditions}), the equations of motion
(\ref{eqnsofmotionfrphinonslowroll}) become,
\begin{align}
\label{slowrolleqnsofmotionfrphigeneral}
& \frac{3f_R(R,\phi)H^2}{\kappa^2}\simeq
V(\phi)+\frac{Rf_R(R,\phi)-f(R,\phi)}{2}\, , \nn
& 3H\dot{\phi}+\frac{1}{2\omega (\phi)}\left(\dot{\omega
(\phi)}\dot{\phi}-\frac{1}{\kappa^2}\frac{\partial
f(R,\phi)}{\partial \phi}+2\frac{\partial V}{\partial
\phi}\right)=0\, .
\end{align}
In the context of the slow-roll approximation, where the slow-roll
indices satisfy $\epsilon_i\ll 1$, $i=1,...4$, the spectral index of
primordial curvature perturbations has the following form
\cite{Noh:2004bc,Hwang:2002fp,Hwang:2001fb,
Hwang:2000jh,Hwang:1996xh,Sebastiani:2015kfa},
\begin{equation}
\label{nsfrphi}
n_s\simeq 1-4\epsilon_1-2\epsilon_2+2\epsilon_3-2\epsilon_4\, .
\end{equation}
Also, the general form of the scalar-to-tensor ratio $r$, as a
function of the function $Q_s$, which we introduced in
Eq.~(\ref{qsfunction}), is equal to,
\begin{equation}
\label{rgeneralform}
r=8\kappa^2\frac{Q_s}{f_R(R,\phi)}\, .
\end{equation}
For example, in the case
of a canonical scalar field, the function $Q_s$ is,
\begin{equation}
\label{qscannon}
Q_s=\frac{\dot{\phi}^2}{H^2}\, ,
\end{equation}
and the slow-roll approximation yields,
$\dot{\phi}^2=-\frac{2\dot{H}}{\kappa^2}$, so the resulting
scalar-to-tensor ratio is $r\simeq 16\epsilon_1$, as we saw in the
previous section. Also in the pure $F(R)$ gravity case, the
parameter $Q_s$ is,
\begin{equation}
\label{qsfrpreliminary}
Q_s=\frac{3\dot{f_R}^2}{2f_RH^2\kappa^2}\, ,
\end{equation}
and so the
resulting scalar-to-tensor ratio is $r=48\epsilon_3^2$. In the
following sections we shall present several examples of inflationary
theories, which belong to some types of the model
(\ref{actionfrphigeneral}).

\subsubsection{Inflation in $f(\phi)R$ Theories of Gravity}

As a first example of a theory belonging to the class of the model
in action (\ref{actionfrphigeneral}), we consider $f(\phi)R$
theories of gravity, which describe non-minimally coupled theories
of gravity, in which case the action is
\cite{Noh:2004bc,Hwang:2002fp,Hwang:2001fb,
Hwang:2000jh,Hwang:1996xh,Myrzakulov:2015qaa},
\begin{equation}
\label{graviactionnonminimal}
\mathcal{S}=\int d ^4x\sqrt{-g}\left(
\frac{f(\phi)R}{2\kappa^2}-\frac{1}{2}g^{\mu\nu}
\partial_{\mu}\phi\partial_{\nu}\phi-V(\phi) \right)\, ,
\end{equation}
where $f(\phi)$ is an analytic function of the scalar field $\phi$.
By varying the action (\ref{graviactionnonminimal}) with respect to
the metric and with respect to the scalar field $\phi$, we obtain
the following equations of motion,
\begin{equation}
\label{gravieqnsnonminimal}
\frac{3f}{\kappa^2}H^2=\frac{\dot{\phi}^2}{2}
+V(\phi)-3h\frac{\dot{f}}{\kappa^2}\, ,\quad
 -\frac{2f}{\kappa^2}\dot{H}=\dot{\phi}^2
+\frac{\ddot{f}}{\kappa^2}-H\frac{\dot{f}}{\kappa^2}\, ,\quad
\ddot{\phi}+3H\dot{\phi}
 -\frac{1}{2\kappa^2}R\frac{df}{d \phi}+\frac{d V}{d \phi}=0\, ,
\end{equation}
with the ``dot'' denoting as usual, differentiation with respect
to the cosmic time. The slow-roll indices in the case of a
non-minimally coupled scalar theory, are equal to,
\begin{equation}
\label{slowrollnonminimal}
\epsilon_1=-\frac{\dot{H}}{H^2}\, ,\quad
\epsilon_2=\frac{\ddot{\phi}}{H\dot{\phi}}\, ,\quad
\epsilon_3=\frac{\dot{f}}{2Hf} \, ,\quad
\epsilon_4=\frac{\dot{E}}{2HE}\, ,
\end{equation}
and in this case the function
$E$ is equal to,
\begin{equation}
\label{epsilonparameter}
E=f+\frac{3\dot{f}^2}{2\kappa^2\dot{\phi}^2}\, .
\end{equation}
The spectral
index of primordial curvature perturbations $n_s$ and the
scalar-to-tensor ratio in terms of the slow-roll indices, are equal
to,
\begin{equation}
\label{observatinalindices1}
n_s\simeq
1-4\epsilon_1-2\epsilon_2+2\epsilon_3-2\epsilon_4\, ,\quad
r=8\kappa^2\frac{Q_s}{f}\, ,
\end{equation}
and it is assumed that the
slow-roll indices satisfy the slow-roll condition $\epsilon_i\ll 1$,
$i=1,..,4$. In addition, the parameter $Q_s$ in the case at hand is
equal to,
\begin{equation}
\label{qs1nonslowroll}
Q_s=\dot{\phi}^2\frac{E}{fH^2(1+\epsilon_3)^2}\, .
\end{equation}
We can find
an approximate expression for the function $Q_s$, by using the
slow-roll approximation, so we find the slow-roll approximated
gravitational equations of motion, which take the following form,
\begin{align}
\label{gravieqnsslowrollapprx12}
& \frac{3fH^2}{\kappa^2}\simeq V(\phi)\, ,\quad
3H\dot{\phi}-\frac{6H^2}{\kappa^2}f'+V'\simeq 0\, , \\
\label{gravieqnsslowrollapprx3}
& \dot{\phi}^2\simeq
\frac{H\dot{f}}{\kappa^2}-\frac{2f\dot{H}}{\kappa^2}\, ,
\end{align}
with the ``prime'' denoting this time differentiation with respect
to the scalar $\phi$. In view of the slow-roll approximated
equations of motion, the parameter $Q_s$ becomes,
\begin{equation}
\label{qs2}
Q_s=\frac{\dot{\phi}^2}{H^2}+\frac{3\dot{f}^2}{2\kappa^2fH^2}\, ,
\end{equation}
and in conjunction with Eq.~(\ref{gravieqnsslowrollapprx3}), the
parameter $Q_s$ finally becomes,
\begin{equation}
\label{qs3}
Q_s\simeq
\frac{H\dot{f}}{H^2\kappa^2}-\frac{2f\dot{H}}{\kappa^2H^2}\, .
\end{equation}
Hence, by combining Eqs.~(\ref{observatinalindices1}) and
(\ref{qs3}), we can find a simplified expression for the
scalar-to-tensor ratio during the slow-roll era, which is,
\begin{equation}
\label{rscalartensornoniminal}
r\simeq 16 (\epsilon_1+\epsilon_3)\, .
\end{equation}
Moreover, we may express the spectral index of the primordial
curvature perturbations $n_s$ as a function of the slow-roll indices
in the slow-roll approximation, which takes the following form,
\begin{equation}
\label{nsintersmofslowroll}
n_s\simeq 1-2\epsilon_1\left (\frac{3H\dot{f}}{\dot{\phi}^2}+2
\right)-2\epsilon_2-6\epsilon_3\left(
\frac{H\dot{f}}{\dot{\phi}^2}-1\right)\, .
\end{equation}
The formalism we just developed, enables us to calculate
analytically the observational indices during the slow-roll era, for
any given function $f(\phi)$. As an example we shall consider
theories in which the function $f(\phi)$ has a special symmetry
$\beta\to \frac{1}{\beta}$, where $\beta>1$. These kind of
potentials were considered in \cite{Odintsov:2016jwr}. So suppose
that the function $f(\phi)$ is equal to,
\begin{equation}
\label{frphispec}
f(\phi)=\frac{1+\xi \left( \e^{-\beta
n\phi}+\e^{-\frac{1}{\beta}n\phi}\right)}{2}\, ,
\end{equation}
where the parameters $\xi$ and $n$ are positive real parameters. The
scalar potential $V(\phi)$ will be chosen in such a way so that it
does not affect the dynamics, in comparison to the function
$f(\phi)$, for example it can contain higher powers of the
exponential appearing in Eq.~(\ref{frphispec}). However, in the case
at hand we can simply choose $V(\phi)=\Lambda$, with $\Lambda$ being
a positive constant parameter. Notice that the function $f(\phi)$ of
Eq.~(\ref{frphispec}), is symmetric under the transformation $\beta
\to \frac{1}{\beta}$. By using the formalism we developed earlier,
we shall calculate analytically the slow-roll indices and the
corresponding observational indices for the case at hand. For
simplicity we shall choose a physical units system for which
$\kappa^2=1$, and in addition we choose $\xi=1$ in order to simplify
the intermediate expressions. It can be shown though that the
resulting expressions for $n_s$ and $r$ do not depend on the
parameter $\xi$. For $\beta>1$, the function $f(\phi)$ can be
approximated as follows,
\begin{equation}
\label{fphifirstmainapprox}
f(\phi)\simeq \frac{1+ \e^{-\frac{n}{\beta}\phi}}{2}\, ,
\end{equation}
and by using the slow-roll equations of motion, we get the following
expression,
\begin{equation}
\label{slowrollapproxeqnsofmotion}
\dot{\phi}\simeq -H\frac{n}{\beta}\e^{-\frac{n}{\beta}\phi}\, .
\end{equation}
Moreover, the function $\dot{f}$ during the slow-roll era, can be
written as follows,
\begin{equation}
\label{slowrollapprxofdot}
\dot{f}\simeq \frac{n^2\e^{-2\frac{n}{\beta}\phi }H}{2}\, .
\end{equation}
We can then calculate the slow-roll indices, with the first one
being approximately equal to,
\begin{equation}
\label{firstdslowrollapproxanalyt}
\epsilon_1\simeq \left(\frac{\dot{\phi}}{H}
\right)^2\frac{1}{2f}-\frac{\dot{f}}{2fH}\, .
\end{equation}
By combining Eqs.~(\ref{slowrollapproxeqnsofmotion}) and
(\ref{slowrollapprxofdot}), the index $\epsilon_1$ reads
\begin{equation}
\label{epsilon1afteranalysis}
\epsilon_1\simeq \frac{n^2\e^{-2\frac{n}{\beta}\phi }}{2\beta^2}\, .
\end{equation}
Accordingly, the slow-roll indices $\epsilon_2$ and $\epsilon_3$ can
be calculated, and these are equal to,
\begin{equation}
\label{slowrolllast}
\epsilon_2\simeq \frac{n^2}{\beta^2}\e^{-\frac{n}{\beta}\phi
}+\epsilon_1\, ,\quad \epsilon_3\simeq \epsilon_1\, .
\end{equation}
Finally, the observational indices $n_s$ and $r$ can be easily found
by combining Eqs.~(\ref{rscalartensornoniminal}),
(\ref{nsintersmofslowroll}), (\ref{epsilon1afteranalysis}), and
(\ref{slowrolllast}), so these read,
\begin{equation}
\label{nsandrprefinal}
n_s=1-2\frac{n^2}{\beta^2}\e^{-\frac{n}{\beta}\phi }\, ,\quad
r\simeq 16
\frac{n^2}{\beta^2}\e^{-2\frac{n}{\beta}\phi }\, .
\end{equation}
It is more appropriate to have the resulting expressions for $n_s$
and $r$ as functions of the $e$-foldings number, so we shall make
use of the following relation-definition,
\begin{equation}
\label{efolidngsanalyticresults}
N=\int_{t_k}^{t_f}H(t)d t
=\int_{\phi_k}^{\phi_f}\frac{H}{\dot{\phi}}d \phi\simeq
\frac{\beta^2}{n^2}\e^{\frac{n}{\beta}\phi }\, ,
\end{equation}
with $t_k$ being chosen as the horizon crossing time instance and in
addition, $\phi_k$ is the value of the scalar field at the horizon
crossing time instance. Finally the time instance $t_f$ is the time
that inflation ends, and also $\phi_f$ is the corresponding scalar
field. Not that for the derivation of relation
(\ref{efolidngsanalyticresults}), we made use of the approximation
$\phi_k\gg \phi_f$, which holds true during the slow-roll era.
 From Eq.~(\ref{efolidngsanalyticresults}) we easily obtain that
$\frac{n^2}{\beta^2}\e^{-\frac{n}{\beta}\phi}=\frac{1}{N}$, so the
observational indices can be expressed in terms of $N$ as follows,
\begin{equation}
\label{nsandrprefinal1}
n_s=1-\frac{2}{N}\, ,\quad r\simeq
\frac{16\beta^2}{n^2N^2}\, .
\end{equation}
A quite interesting choice for the
parameter $n$ is $n=\frac{2}{\sqrt{3}}$, and the observational
indices become,
\begin{equation}
\label{nsandrprefinal2}
n_s=1-\frac{2}{N}\, ,\quad r\simeq \frac{12\beta^2}{N^2}\, .
\end{equation}
The final expressions for the
observational indices (\ref{nsandrprefinal2}) are identical to the
ones corresponding to the $\alpha$-attractor models
\cite{Kallosh:2013hoa,Ferrara:2013rsa,Kallosh:2013yoa,
Galante:2014ifa,Carrasco:2015pla,Linde:2015uga,Roest:2015qya,
Kallosh:2014rga,Sebastiani:2013eqa,Cai:2014bda,Yi:2016jqr}, with the
difference being that $\beta\ll 1$ in the case of the
$\alpha$-attractor models.

\subsubsection{Inflation from $F(R)$ Theories}

As another example of a theory belonging to the general class of
models of Eq.~(\ref{actionfrphigeneral}), we shall consider $F(R)$
theories of gravity. As we showed in section \ref{section1}, the
$F(R)$ gravity has a unique Einstein frame canonical scalar theory,
so in many cases a viable cosmological model in the Einstein frame
has a unique $F(R)$ gravity in the Jordan frame, which is also
viable. In this section we shall present the theoretical study of
slow-roll inflation in the pure $F(R)$ gravity case. For simplicity
we shall consider only a vacuum theory, but the results can be
easily extended in the case that perfect matter fluids are present.

Consider the $F(R)$ gravity. The detailed study of inflation for
modified gravities was developed in
Refs.~\cite{Noh:2004bc,Hwang:2002fp,Hwang:2001fb,
Hwang:2000jh,Hwang:1996xh,Sebastiani:2015kfa,Sebastiani:2013eqa}
(see also \cite{Carloni:2007yv} for a study on the evolution of
density perturbations in $F(R)$ gravity), and the dynamics of
inflation are determined by four generalized slow-roll indices,
namely $\epsilon_1$ ,$\epsilon_2$, $\epsilon_3$, $\epsilon_4$, which
we shall intensively use in this section. The parameter $\epsilon_1$
is simply equal to $\epsilon_1=-\frac{\dot{H}}{H^2}$, and in the
case of the pure $F(R)$ gravity of Eq.~(\ref{JGRG7}), the slow-roll
parameters can be expressed in terms of the slow-roll index
$\epsilon_1$ as follows \cite{Noh:2004bc,Hwang:2002fp,Hwang:2001fb,
Hwang:2000jh,Hwang:1996xh,Sebastiani:2015kfa,Sebastiani:2013eqa},
\begin{equation}
\label{restofparametersfr}
\epsilon_2=0\, ,\quad
\epsilon_1\simeq
 -\epsilon_3\, ,\quad \epsilon_4\simeq
 -3\epsilon_1+\frac{\dot{\epsilon}_1}{H\epsilon_1}\, .
\end{equation}
The spectral index of primordial curvature perturbations and the
scalar-to-tensor ratio in the slow-roll limit, have the following
form,
\begin{equation}
\label{epsilonall}
n_s\simeq 1-6\epsilon_1-2\epsilon_4,\quad r=48\epsilon_1^2\, .
\end{equation}
The expressions (\ref{epsilonall}) are valid only in the slow-roll
limit where $\epsilon_1,\epsilon_4\ll 1$. Let us apply this
methodology in order to calculate the observational indices $n_s$
and $r$ for some special cases.

The $R^2$ model of inflation \cite{Starobinsky:1982ee}, is widely
known and also has very appealing features since it is compatible to
the latest (2015) Planck data \cite{Ade:2015lrj}. In this case, the
$F(R)$ gravity is $F(R)=R+\frac{R^2}{36H_i}$, where $H_i$ has
dimensions of mass$^2$. During the slow-roll era of inflation, it is
expected that $F'(R)\simeq R/(18H_i)$. For the FRW metric, the FRW
equations read,
\begin{equation}
\label{frweqnsr2}
\ddot{H}-\frac{\dot{H}^2}{2H}+3H_iH=-3H\dot{H}\, ,\quad
\ddot{R}+3HR+6H_iR=0\, ,
\end{equation}
and during the slow-roll era, the first
two terms in the first equation of Eq.~(\ref{frweqnsr2}) can be
neglected, hence the evolution during the inflationary era is the
quasi-de Sitter evolution of the form,
\begin{equation}
\label{quasievolbnexampler2}
H(t)\simeq H_0-H_i(t-t_k)\, .
\end{equation}
The
time instance can be arbitrarily chosen, but we assume that this is
the time instance that the primordial curvature modes exit the
horizon. Now let us proceed to the calculation of the indices, so at
first let us note that the slow-roll approximation ends when the
first slow-roll parameter becomes $\epsilon_1\simeq \mathcal{O}(1)$,
and by assuming that this occurs at the time instance $t=t_f$ at
which point $H(t_f)=H_f$, the condition $\epsilon_1(t_f)\simeq 1$
yields the relation $H_f\simeq \sqrt{H_i}$. Accordingly, by
substituting in Eq.~(\ref{quasievolbnexampler2}), we obtain,
\begin{equation}
\label{finalghf}
H_f-H_0\simeq -H_i (t_f-t_k)\, ,
\end{equation}
and by
substituting $H_f$ we obtain,
\begin{equation}
\label{timerelation}
t_f-t_k=\frac{H_0}{H_i}-\frac{\sqrt{H_i}}{H_i}\, .
\end{equation}
Both the
parameters $H_0$ and $H_i$ are expected to have large values during
the slow-roll era, so we may omit the second term in
Eq.~(\ref{timerelation}), and hence,
\begin{equation}
\label{timerelation1}
t_f-t_k\simeq \frac{H_0}{H_i}\, .
\end{equation}
In terms of the Hubble rate,
the $e$-foldings number $N$ is equal to
\begin{equation}
\label{efold1}
N=\int_{t_k}^{t_f}H(t)d t\, .
\end{equation}
By using the Hubble rate
(\ref{quasievolbnexampler2}), and performing the integral in
Eq.~(\ref{efold1}), we obtain,
\begin{equation}
\label{nefold1}
N=H_0(t_f-t_k)-\frac{H_i(t_f-t_k)^2}{2}\, ,
\end{equation}
and by using
(\ref{timerelation1}), we finally get,
\begin{equation}
\label{nfinal1}
N=\frac{H_0^2}{2H_i}\, .
\end{equation}
So at leading order we have,
\begin{equation}
\label{leadingtf}
t_f-t_k\simeq \frac{2N}{H_0}\, ,
\end{equation}
and now we
can calculate the observational indices by combining
Eqs.~(\ref{restofparametersfr}), (\ref{epsilonall}), and
(\ref{quasievolbnexampler2}), so we finally have,
\begin{equation}
\label{spectrscatotensor}
n_s\simeq 1-\frac{4 H_i}{\left(H_0-\frac{2
H_i N}{H_0}\right)^2}\, , \quad r=\frac{48 H_i^2}{\left(H_0-\frac{2
H_i N}{H_0}\right)^4}\, .
\end{equation}
By taking the large $N$ limit, the
observational indices take the following form,
\begin{equation}
\label{spectrscatotensor1}
n_s\simeq 1-\frac{H_0^2}{H_i N^2}\, ,
\quad r=\frac{3 H_0^4}{H_i^2 N^4}\, ,
\end{equation}
so by using
Eq.~(\ref{nfinal1}) we finally obtain,
\begin{equation}
\label{jordanframeattract}
n_s\simeq 1-\frac{2}{N}\, ,\quad r\simeq
\frac{12}{N^2}\, .
\end{equation}
As it can be seen by comparing
Eqs.~(\ref{finalstarob}) and (\ref{jordanframeattract}), the
Einstein frame $R^2$ model and the Jordan frame $R^2$ model yield
the same observational indices at leading order, as it was possible
expected. This equivalence of the indices in the Jordan and Einstein
frame is not accidental, see for example
\cite{Kaiser:1995nv,Faraoni:2007yn,Domenech:2016yxd,
Brooker:2016oqa,Kaiser:1994vs}. It should be noted that the graceful
exit from inflation in these theories occurs due to the $R^2$ term.
Indeed this term induces graceful exit from inflation, since the de
Sitter point becomes unstable. Particularly, as it was shown in
\cite{Bamba:2014jia}, the final attractor of the theory is a de
Sitter vacuum and by perturbing this solution, it is shown that the
perturbations grow. Thus the final de Sitter attractor is unstable,
and hence the graceful exit comes as a result of growing curvature
perturbations.

\subsubsection{Inflation in mimetic-$F(R)$ Theories of Gravity}

Another interesting class of models that belongs to the models with
action (\ref{actionfrphigeneral}), is the mimetic $F(R)$ gravity
models, with Lagrange multiplier and potential. These models were
introduced in \cite{Nojiri:2014zqa} and further studied in
\cite{Nojiri:2016vhu}. In this section we shall be interested mainly
on the study of inflationary dynamics in the context of these
theories. Also, another issue that is related to these theories is
the study of graceful exit, which in general can be a tedious study.
We will address this issue at the end of this chapter.

The theoretical framework of mimetic $F(R)$ gravity in principle
provides us with much freedom in realizing various cosmological
scenarios, in a viable way, meaning that the theories are compatible
with the recent observational data. In the literature there are
various approaches for these kind of theories, with one approach
being based on the perfect fluid description
\cite{Odintsov:2015wwp}, and the other description deals with the
mimetic $F(R)$ gravity as it is an $F(R)$ gravity in the presence of
a non-canonical scalar field. In this section we discuss the second
paradigm, in order to apply the formalism of the previous sections,
since the mimetic $F(R)$ gravity can be viewed as a special case of
an $F(R,\phi)$ scalar-tensor theory. The slow-roll conditions are in
this case,
\begin{equation}
\label{slowrollerarealtions}
\dot{H}\ll H^2\, ,\quad \ddot{H}\ll
H\dot{H}\, ,
\end{equation}
with $H$ being the Hubble rate. The mimetic $F(R)$ gravity action
describes a special class of $F(R,\phi)$ scalar-tensor theory, since
the mimetic $F(R)$ action can be written in the following way,
\begin{equation}
\label{MF1sc}
S = \int \sqrt{-g}
\left\{ \frac{F(R)}{2\kappa^2} + \lambda (\phi) \partial_\mu \phi
\partial^\mu \phi + \lambda(\phi ) - V(\phi) \right\} \, ,
\end{equation}
and therefore, the kinetic term is, $\omega (\phi)=-2 \lambda
(\phi )$, while the scalar potential is, $\mathcal{V}(\phi)=\lambda
(\phi)-V(\phi)$. The slow-roll indices for the general $F(R,\phi)$
gravity are \cite{Noh:2001ia,Hwang:2001qk,Hwang:2001pu},
\begin{equation}
\label{slowrollindices}
\epsilon_1=-\frac{\dot{H}}{H^2}\, ,\quad
\epsilon_2=\frac{\ddot{\phi}}{H\dot{\phi}}\, , \quad
\epsilon_3=\frac{\dot{F'}(R,\phi)}{2HF'(R,\phi)}\, , \quad
\epsilon_4=\frac{\dot{E}}{2HE}\, ,
\end{equation}
where the function $E(R,\phi)$ is,
\begin{equation}
\label{epsilonfunction}
E(R,\phi)=F(R,\phi)\omega
(\phi)+\frac{3\dot{F'}(R,\phi)^2}{2\dot{\phi}^2}\, .
\end{equation}
In the case
of mimetic $F(R)$ gravity, owing to the fact that $\phi=t$ and
$\omega (\phi)=-2\lambda (\phi)$, the slow-roll indices take the
following form,
\begin{equation}
\label{slowrollindicesmim}
\epsilon_1=-\frac{\dot{H}}{H^2}\, ,\quad \epsilon_2=0\, ,\quad
\epsilon_3=\frac{\dot{F'}(R,\phi)}{2HF'(R,\phi)}\, ,\quad
\epsilon_4=\frac{\dot{E}}{2HE}\, ,
\end{equation}
with the function $E(R,\phi)$
being equal to,
\begin{equation}
\label{epsilonfunction1}
E(R,\phi)=2F(R,\phi)\lambda (\phi)+\frac{3\dot{F'}(R,\phi)^2}{2}\, .
\end{equation}
Then, given the form of the $F(R)$ gravity, it is possible to
calculate analytically the slow-roll indices and the corresponding
observational indices. However, note that without the slow-roll
approximation, the calculation of the slow-roll indices can be
tedious, so we assume that the slow-roll approximation holds true.
Therefore, if the slow-roll indices $\epsilon_i$, $i=1,..,4$ satisfy
$\epsilon_i\ll 1$, the observational indices can be written in terms
of the slow-roll indices as follows,
\cite{Noh:2001ia,Hwang:2001qk,Hwang:2001pu},
\begin{equation}
\label{obsindices}
n_s\simeq 1-4\epsilon_1-2\epsilon_2+2\epsilon_3-2\epsilon_4\, ,\quad
r=16(\epsilon_1+\epsilon_3)\, .
\end{equation}
Then, by specifying the explicit
form of the $F(R)$ gravity and also the Hubble rate, we can realize
various cosmological scenarios which are compatible with the
observational data coming from Planck \cite{Ade:2015lrj} and the
most recent BICEP2/Keck Array data \cite{Array:2015xqh}. For
convenience we quote here the Planck constraints, for the spectral
index of the primordial curvature perturbations $n_s$ and for the
scalar-to-tensor ratio $r$, which are constrained as follows,
\begin{equation}
\label{planckdata}
n_s=0.9644\pm 0.0049\, , \quad r<0.10\, .
\end{equation}
In
addition, the latest BICEP2/Keck-Array data \cite{Array:2015xqh}
further constrain the scalar-to-tensor ratio to satisfy,
\begin{equation}
\label{scalartotensorbicep2}
r<0.07\, ,
\end{equation}
at $95\%$ confidence
level. For the purposes of our analysis, we shall consider two quite
popular $F(R)$ gravity models, first the $R^2$ inflation model,
\begin{equation}
\label{rsquare}
F(R)=R+\frac{R^2}{6M^2}\, ,
\end{equation}
and secondly the power-law model,
\begin{equation}
\label{frpowerlaw}
F(R)=\alpha R^n\, ,
\end{equation}
with
$n>0$. With regards to the non-mimetic $R^2$ model, this model is
compatible with the observational data, but the non-mimetic version
of the model (\ref{frpowerlaw}) is not compatible with the data. As
we shall demonstrate, the mimetic version of the power-law model
(\ref{frpowerlaw}), can be compatible with observations, if the
parameters of the model are suitably chosen.

Let us first consider the case that the $F(R)$ gravity is the $R^2$
model of Eq.~(\ref{rsquare}), so by using the mimetic $F(R)$
formalism, we shall realize the cosmological evolution,
\begin{equation}
\label{hubblersquare}
H(t)=H_0-\frac{d}{6}(t-t_0)+c (t-t_0)^2\, ,
\end{equation}
with $H_0$, $d$, $c$ and $t_0$ being arbitrary free parameters
of the theory. These parameters must be chosen in the following way,
in order for the slow-roll condition to hold true,
\begin{equation}
\label{relationsrquare}
H_0>d,c,t_0\, .
\end{equation}
In addition, the
parameter $t_0$ must be $t_0\ll 1$, since this time instance
corresponds to the beginning of the inflationary era. Moreover, we
need to note that the slow-roll approximation is valid for the
Hubble rate (\ref{hubblersquare}), due to the fact that the cosmic
time take values in the interval $(10^{-35},10^{-15})$sec, and
therefore the time-dependent terms of Eq.~(\ref{hubblersquare}),
have extremely small values. The mimetic potential and the Lagrange
multiplier have to be chosen appropriately in order to realize the
cosmological evolution (\ref{hubblersquare}). Due to the slow-roll
conditions (\ref{slowrollerarealtions}), the $F'(R)$ can be
simplified as $F'(R)\simeq R/(3M^2)$, and also in the same way,
$\dot{F'}\simeq 8 H \dot{H}/M^2$. Therefore, the mimetic $R^2$
gravity gravitational equations are simplified as follows,
\begin{equation}
\label{rsquareprofinal}
\ddot{H}-\frac{\dot{H}^2}{2 H}
+\frac{M^2H}{2}=-3H\dot{H}+M^2\frac{V(\phi)-2\lambda (\phi)}{12 H}\, ,
\end{equation}
and by neglecting the first two terms in
Eq.~(\ref{rsquareprofinal}), we obtain,
\begin{equation}
\label{rsquarefinal}
\frac{M^2H}{2}=-3H\dot{H}+M^2\frac{V(\phi)-2\lambda (\phi)}{12 H} \, .
\end{equation}
Consequently, for the cosmological evolution
(\ref{hubblersquare}), the Lagrange multiplier and the potential
satisfy,
\begin{equation}
\label{vlambdacombo}
V(t)-2\lambda (t)=\frac{\left(-d+M^2+12 c (t-t_0)\right) (-6 H_0
+(t-t_0) (d+6 c (-t+t_0)))^2}{6 M^2}\, .
\end{equation}
We can find the exact forms of the
Lagrange multiplier and of the potential as follows,
\begin{equation}
\label{lambdap}
\lambda (\phi)=4 c \phi -\frac{10 c d \phi}{M^2}-\frac{120 c^2 t_0 \phi }{M^2}
+\frac{60 c^2 \phi^2}{M^2}+\mathcal{A}_1\, ,
\end{equation}
with the constant $\mathcal{A}_1$ being,
\begin{equation}
\label{constanta}
\mathcal{A}_1=-\frac{d}{3}+\frac{d^2}{3 M^2} +\frac{12 c H_0}{M^2}-4 c t_0
+\frac{10 c d t_0}{M^2} +\frac{60 c^2 t_0^2}{M^2}\, .
\end{equation}
In the same way we can easily find the
mimetic potential, but we omit this for brevity. The slow-roll
indices $\epsilon_3$ and $\epsilon_4$ can be simplified during the
slow-roll era, in the following way,
\begin{equation}
\label{slowrollapprox}
\epsilon_3\simeq -2\epsilon_1\, ,\quad \epsilon_4\simeq
\frac{\dot{\lambda}}{2H\lambda}\, ,
\end{equation}
and the slow-roll index
$\epsilon_1$ remains the same. By combining Eqs.~(\ref{lambdap}) and
(\ref{hubblersquare}), we can find the observational indices, which
are,
\begin{align}
\label{obsfinalrsquare}
& \epsilon_1=\frac{\frac{d}{6}-2 c
(t-t_0)}{\left(H_0-\frac{1}{6} d (t-t_0)+c (t-t_0)^2\right)^2}
\, ,\quad \epsilon_3\simeq -\frac{2 \left(\frac{d}{6}-2 c
(t-t_0)\right)}{\left(H_0-\frac{1}{6} d (t-t_0)+c
(t-t_0)^2\right)^2} \nn & \epsilon_4\simeq \frac{18 c \left(-5 d+2
\left(M^2+30 c (t-t_0)\right)\right)}{\left(d^2-d \left(M^2 +30 c
(t-t_0)\right)+12 c \left(3 H_0+\left(M^2+15 c (t-t_0)\right)
(t-t_0)\right)\right) (6 H_0-(t-t_0) (d+6 c (-t+t_0)))}\, .
\end{align}
Accordingly, by using the $e$-foldings number $N$, we can express
the spectral index $n_s$ as a function of $N$, as follows,
\begin{equation}
\label{spectralrsquare}
n_s\simeq 1-\frac{48 (d-12 c z)}{(-6 H_0+z
(d-6 c z))^2}-\frac{36 c \left(-5 d+2 M^2+60 c z\right)}{(6 H_0+z
(-d+6 c z)) \left(d^2-d \left(M^2+30 c z\right)+12 c \left(3 H_0+z
\left(M^2+15 c z\right)\right)\right)}\, ,
\end{equation}
and the
$N$-dependence of $n_s$ is given in terms of the parameter $z$, the
explicit form of which is,
\begin{align}
\label{parameterz}
z=& \frac{M^2}{12 c}-\frac{144 c H_0-M^4}{6\
2^{2/3} c \left(-432 c H_0 M^2+2 M^6+5184 c^2 N+\sqrt{4 \left(144 c
H_0-M^4\right)^3+\left(-432 c H_0 M^2+2 M^6+5184 c^2
N\right)^2}\right)^{1/3}} \nn & +\frac{\left(-432 c H_0 M^2+2
M^6+5184 c^2 N+\sqrt{4 \left(144 c H_0-M^4\right)^3+\left(-432 c H_0
M^2+2 M^6+5184 c^2 N\right)^2}\right)^{1/3}}{12\ 2^{1/3} c}\, .
\end{align}
In the same way, the scalar-to-tensor ratio is equal to,
\begin{equation}
\label{sctotensorr}
r\simeq -\frac{16 \left(\frac{d}{6}-2 c
z\right)}{\left(H_0-\frac{d z}{6}+c z^2\right)^2}\, .
\end{equation}
By
assuming that the free parameters have the following values,
\begin{equation}
\label{valuesparmeters}
N=50\, ,\quad c=0.0000009\, ,\quad M=0.085
\, ,\quad H_0=0.26\, ,\quad d=0.002\, ,
\end{equation}
the observational
indices $n_s$ and $r$, take the following values,
\begin{equation}
\label{observindicesfinalresults}
n_s\simeq 0.967429\, ,\quad r\simeq 0.0134796\, ,
\end{equation}
which are in concordance with the Planck
data (\ref{planckdata}) and also with the BICEP2/Keck Array data
(\ref{scalartotensorbicep2}). It is conceivable that the results we
just presented, strongly depend on the choice of the free
parameters, and specifically on the parameter $H_0$, and the
compatibility with the observational data is achieved at the expense
of having fine-tuning. For more details on this model, see
\cite{Nojiri:2016vhu} for a detailed analysis.

As we just showed, the mimetic $R^2$ model is compatible with the
observational data, or at least can be by appropriately choosing the
free parameters of the theory. As we now show, the mimetic version
of the power law model of Eq.~(\ref{frpowerlaw}) can also be
compatible with the observations, although the ordinary power-law
model is not compatible with the observational data. Particularly,
in the ordinary power-law case, the slow-roll indices of
Eq.~(\ref{slowrollindices}) are equal to,
\begin{equation}
\label{slowrollfrordinary}
\epsilon_1=\frac{2-n}{(n-1)(2 n-1)}\, ,\quad
\epsilon_2=0\, ,\quad \epsilon_3=\frac{(1-n) (2-n)}{(-1+n)
(-1+2 n)}\, ,\quad \epsilon_4=\frac{n-2}{n-1}\, ,
\end{equation}
and the resulting spectral index $n_s$ is,
\begin{equation}
\label{nsordfr}
n_s\simeq \frac{-7+5 n}{1-3 n+2 n^2}\, ,
\end{equation}
and in addition, scalar-to-tensor ratio $r$ is,
\begin{equation}
\label{sttrationfrord}
r\simeq \frac{16 (-2+n)^2}{(-1+n) (-1+2 n)}\, .
\end{equation}
In the ordinary power-law $F(R)$ gravity model, the only free
parameter is $n$, so by choosing $n=1.8105$, the spectral index and
the scalar-to-tensor ratio become,
\begin{equation}
\label{obsrveatinindordfr}
n_s\simeq 0.966191\, ,\quad r\simeq
0.27047\, ,
\end{equation}
so only the spectral index can be compatible with observations. As
we show, the mimetic power-law $F(R)$ model can produce results
compatible with observations. We first consider the quasi-de Sitter
evolution of the form,
\begin{equation}
\label{hubblefrpowerlaw}
H(t)=H_0-\frac{M^2}{6}(t-t_0)\, ,
\end{equation}
with $H_0>M^2$. Since $t\ll 1$, then the slow-roll conditions of
Eq.~(\ref{slowrollerarealtions}) hold true for the evolution of
Eq.~(\ref{hubblefrpowerlaw}). The gravitational equations during the
slow-roll era in this case become,
\begin{equation}
\label{powerlawfrgravequations}
\alpha n=2\alpha (n-1)-\alpha n
(n-1)24 \dot{H}+\frac{V(t)-2\lambda (t)}{6(12H^2)^{n-1}H^2}\, ,
\end{equation}
and therefore the Lagrange multiplier is,
\begin{equation}
\label{lagrafrsimple}
\lambda (\phi)=-2^{-3+2 n} 3^{-3+n} M^2 n
\left(2-n+4 M^2 (-1+n) n\right) \alpha \left(H_0+\frac{1}{6} M^2
(t_0-\phi )\right)^{2 (-1+n)}\, ,
\end{equation}
while the mimetic potential becomes,
\begin{equation}
\label{mimeticpotfrpowerlaw}
V(\phi)=3^{-3+n} 4^{-1+n} \left(2-n+4
M^2 (-1+n) n\right) \alpha \left(9-\frac{36 M^2 n}{\left(6 H_0+M^2
(t_0-\phi )\right)^2}\right) \left(H_0+\frac{1}{6} M^2 (t_0-\phi
)\right)^{2 n}\, .
\end{equation}
We can easily find the slow-roll indices in the slow-roll
approximation, which are,
\begin{equation}
\label{slowrollgeneralrelations}
\epsilon_3\simeq -(n-1)\epsilon_1\, ,\quad
\epsilon_4\simeq \frac{\dot{\lambda}}{2H\lambda}\, ,
\end{equation}
so by using Eqs.~(\ref{hubblefrpowerlaw}), (\ref{lagrafrsimple}), and
(\ref{mimeticpotfrpowerlaw}), we obtain,
\begin{equation}
\label{poorlone}
\epsilon_1=\frac{M^2}{6 \left(H_0-\frac{1}{6} M^2
(t-t_0)\right)^2}\, ,\quad \epsilon_3=\frac{M^2 (1-n)}{6
\left(H_0-\frac{1}{6} M^2 (t-t_0)\right)^2}\, ,\quad
\epsilon_4=-\frac{6 M^2 (-1+n)}{\left(6 H_0+M^2 (-t+t_0)\right)^2}
\, .
\end{equation}
As in the $R^2$ case, we shall express the observational indices as
functions of the $e$-foldings number, so we have,
\begin{equation}
\label{obsrindifin}
n_s=\frac{3 H_0^2-M^2 (2+N)}{3 H_0^2-M^2 N}\, ,\quad
r=\frac{8 M^2 (-2+n)}{-3 H_0^2+M^2 N}\, .
\end{equation}
By assuming that the free parameters and $N$ have the following
values,
\begin{equation}
\label{powerlawfreeparametersvalues}
N=50\, ,\quad H_0=12.04566\, ,\quad
M=2\, ,\quad n=1.8\, ,
\end{equation}
the spectral index $n_s$ and the scalar-to-tensor ration become,
\begin{equation}
\label{obsrvfinalvalues}
n_s\simeq 0.966\, ,\quad r\simeq 0.0272\, ,
\end{equation}
and therefore compatibility with the observational data is achieved.
It can be shown that compatibility with the observations is always
achieved if $M$ and $H_0$ satisfy $H_0=6.0228\times M$, and if $n$
is appropriately chosen in each case. As a second example, consider
the following cosmological evolution,
\begin{equation}
\label{newinfldesitter}
H(t)=H_0 - d/6 (t - t_0) + c (t - t_0)^2\, ,
\end{equation}
and the Lagrange multiplier that generates (\ref{newinfldesitter}),
for the power-law gravity (\ref{frpowerlaw}), is at leading order,
\begin{equation}
\label{leadinglagra}
\lambda (\phi)\simeq -2^{1+2 n} 3^{-1+n} c
H_0^{-1+2 n} (-1+n) n \alpha\, ,
\end{equation}
Accordingly, the leading order mimetic potential is,
\begin{equation}
\label{corrspotelagra}
V(\phi)= 12^{-1+n} H_0^{-1+2 n} \alpha
\left(-16 c (-1+n) n+H_0 \left(2+4 n^2 (d+12 c (t_0-\phi ))-n (1+4
d+48 c (t_0-\phi ))\right)\right)\, .
\end{equation}
Then, the spectral index $n_s$ and the scalar-to-tensor ratio can be
easily found, but we do not quote their explicit form here for
brevity. It can be shown \cite{Nojiri:2016vhu} that if we choose the
parameters as follows,
\begin{equation}
\label{finalvaluesforprm}
N=50\, ,\quad c=0.014\, ,\quad d=-0.1\, ,\quad
n=11\, ,\quad H_0=11,
\end{equation}
the observational indices take the following values,
\begin{equation}
\label{finalvaluesnsr}
n_s\simeq 0.966121\, ,\quad r\simeq
0.0176679\, ,
\end{equation}
which of course are compatible with the current observational data.
As we also mentioned in the $R^2$ model, compatibility is achieved
at the expense of having to fine-tune the parameters, so we refer
the reader to \cite{Nojiri:2016vhu} for more details on the allowed
ranges of the parameters.

Also, in the context of mimetic $F(R)$ gravity and for similar
theories, it is very difficult to address the graceful exit issue by
using the methods we already described in this chapter. So at the
end of this chapter, we devote a whole section for this issue.

\subsubsection{Reconstruction of $F(R)$ gravity from Einstein Frame
Scalar Potential}

In chapter \ref{section1} we demonstrated how to obtain the Einstein
frame canonical scalar theory, from the Jordan frame $F(R)$ gravity,
so in this section we shall demonstrate how it is possible to obtain
the $F(R)$ gravity which corresponds to an inflationary potential.
As we showed in chapter \ref{section1}, the Einstein frame potential
of the canonical scalar field is,
\begin{align}
\label{potentialvsigma}
V(\varphi)
=\frac{1}{2}\left(\frac{A}{F'(A)}-\frac{F(A)}{F'(A)^2}\right)=\frac{1}{2}\left
( \e^{-\sqrt{2/3}\varphi }R\left (\e^{\sqrt{2/3}\varphi} \right )
 - \e^{-2\sqrt{2/3}\varphi }F\left [ R\left (\e^{\sqrt{2/3}\varphi}
\right ) \right ]\right )\, .
\end{align}
The Einstein and Jordan frames are connected via the following
canonical transformation,
\begin{equation}
\label{can}
\varphi =\sqrt{\frac{3}{2}}\ln (F'(A))\, ,
\end{equation}
so by solving Eq.~(\ref{can}) with respect to $A$, we may obtain the
scalar curvature $R$ as a function of the scalar field, since $A=R$.
Then, finding the $F(R)$ gravity from a given canonical scalar
potential is a straightforward task, since we have only to combine
Eqs.~(\ref{potentialvsigma}) and (\ref{can}). Particularly, taking
the first derivative of Eq.~(\ref{potentialvsigma}), with respect to
$R$, that is with respect to the scalar curvature, and also by using
$\frac{d \varphi}{d R}
=\sqrt{\frac{3}{2}}\frac{F''(R)}{F'(R)}$,
we get the following differential equation,
\begin{equation}
\label{solvequation}
RF_R=2\sqrt{\frac{3}{2}}\frac{d }{d \varphi}
\left(\frac{V(\varphi)}{\e^{-2\left(\sqrt{2/3}\right)\varphi}}\right)
\, ,
\end{equation}
where $F_R=\frac{d F(R)}{d R}$. Given the above
differential equation, in conjunction with the canonical
transformation Eq.~(\ref{can}), will give us the resulting $F(R)$
gravity, given the potential $V(\varphi)$. Let us demonstrate how
the method works by using two simple examples. Firstly let us assume
that the potential in the Einstein frame is the Starobinsky
potential of Eq.~(\ref{starobinflpotential1}), so by substituting
the Starobinsky potential in Eq.~(\ref{solvequation}), and also due
to the fact that $F_R=\e^{\sqrt{\frac{2}{3}}\varphi}$, we finally get
the following algebraic equation,
\begin{equation}
\label{staroalgebreqn}
F_R R-\left(4 F_R^2 \mu ^2-4 F_R \mu ^2\right)=0\, .
\end{equation}
By solving the algebraic equation (\ref{staroalgebreqn}) with
respect to $F_R$ , we easily obtain,
\begin{equation}
\label{starsol1}
F_R=\frac{4 \mu ^2+R}{4 \mu ^2}\, ,
\end{equation}
and in turn, by integrating with respect to the scalar curvature $R$, we
get the corresponding $F(R)$ gravity, which is
$F(R)=R+\frac{R^2}{8 \mu^2}$.

As another example consider the following limiting attractor
potential studied in Ref.~\cite{Odintsov:2016jwr},
\begin{equation}
\label{limit1}
V(\varphi)\simeq \mu \alpha
\left(1-2n\e^{-\frac{1}{\alpha}\sqrt{\frac{2}{3}}\varphi}\right)\, .
\end{equation}
The Starobinsky model tends asymptotically, that is for large
field values, to the potential (\ref{limit1}) and also the
$E$-attractor models and $T$-attractor models
\cite{Kallosh:2013hoa,Ferrara:2013rsa,Kallosh:2013yoa,Galante:2014ifa},
also tend to the potential (\ref{limit1}). Moreover, a special class
of attractors, the inverse symmetric attractors
\cite{Odintsov:2016jwr}, also tend to the potential (\ref{limit1}).
The differential equation (\ref{solvequation}), for the potential
(\ref{limit1}), becomes,
\begin{equation}
\label{algegeneralalpha}
F_R =\frac{R}{4\alpha\mu}-F_R^{1-\frac{1}{\alpha}}(\frac{n}{\alpha}+2n)\, .
\end{equation}
The
algebraic equation (\ref{algegeneralalpha}) has a particularly
simple solution during the slow-roll era, since $F_R\gg 1$ during
this era. Therefore, for $\alpha>1$, the second term in the right
hand side of Eq.~(\ref{algegeneralalpha}), satisfies the inequality
$1-\frac{1}{\alpha}<1$, and therefore the term $\sim
F_R^{1-\frac{1}{\alpha}}$, becomes subdominant at leading order.
Therefore, during the slow-roll era we have $F_R\simeq
\frac{R}{4\alpha \mu}$, and in effect, upon substitution in
Eq.~(\ref{algegeneralalpha}), we finally obtain,
\begin{equation}
\label{fr}
F(R)\simeq \frac{R^2}{8\alpha\mu}-\frac{\frac{n}{\alpha}
+2n}{(2-\frac{1}{\alpha})(8\alpha\mu)^{}}R^{2-\frac{1}{\alpha}}+\Lambda\, ,
\end{equation}
where the parameter
$\Lambda$ is an arbitrary integration constant.

An extensive analysis of various phenomenologically interesting
canonical scalar field potentials was developed in
Ref.~\cite{Sebastiani:2013eqa}. In Table~\ref{dosimo} we have gathered
the most interesting outcomes of the research carried out in
Ref.~\cite{Sebastiani:2013eqa}.
\begin{table*}[h]
\small \caption{\label{dosimo}Scalar Potential and the Corresponding
$F(R)$ Gravity}
\begin{tabular}{@{}crrrrrrrrrrr@{}}
\tableline \tableline \tableline
 Scalar Potential $\quad$ &$\quad $ $F(R)$
Gravity$\quad $$\quad $$\quad $$\quad $$\quad $$\quad $
$\quad $$\quad $$\quad $$\quad $
\\\tableline
$V(\varphi )=c_0+c_1\e^{-\sqrt{\frac{2}{3}}\kappa
\varphi}+c_2\e^{-2\sqrt{\frac{2}{3}}\kappa \varphi}$ $\quad $ &
$F(R)=-\frac{c_1}{2c_0}R+\frac{R^2}{4c_0}+\frac{c_1^2}{4c_0}-c_2$
\\\tableline
$V(\varphi )=c_0\e^{-n\sqrt{\frac{2}{3}}\kappa \varphi}$ $\quad $ &
$F(R)\simeq c_0\left( \frac{n+1}{n+2}\right)\left(
\frac{1}{4(n+2)}\right)^{\frac{1}{n+1}}\left(
\frac{R}{c_0}\right)^{\frac{n+2}{n+1}}$$\quad R\gg c_1$
\\\tableline
$V(\varphi
)=\frac{c_0}{\kappa^2}-\frac{c_1}{\kappa^2}\e^{-\sqrt{\frac{2}{3}}\frac{\kappa}{2}
\varphi}$ $\quad $ & $F(R)\simeq
\frac{R}{2}+\frac{R^2}{6c_1}+\frac{\sqrt{3}}{36}\left(4 R/c_1+3
\right)^{3/2}+\frac{c_1}{4}$$\quad R\gg c_1$
\\\tableline
$V(\varphi
)=\frac{c_1(2-n)}{2\kappa^2}-\frac{c_1}{\kappa^2}\e^{-n\sqrt{\frac{2}{3}}\kappa
\varphi}$ $\quad $ & $F(R)\simeq
\frac{R^2}{2c_1(2-n)}+\frac{1}{2-n}\left(\frac{1}{2c_1(2-n)}\right)^{1-n}R^{2-n}$$\quad R\gg
c_1$
\\\tableline
\tableline \tableline
\end{tabular}
\end{table*}
As it can be seen in Table~\ref{dosimo}, we present the canonical
scalar field potential in the Einstein frame and the corresponding
Jordan frame $F(R)$ gravity. The parameters $c_0$, $c_1$ and $n$ in
Table~\ref{dosimo}, are positive numbers.

\subsection{Inflation from Gauss-Bonnet Theories of Gravity}

\subsubsection{Explicit Calculation of the Power Spectrum}

In the previous section, we calculated the spectral index of the
primordial curvature perturbations by finding the explicit form of
the slow-roll indices. However, there are more direct ways towards
the calculation of the spectral index, for example by calculating
directly the power spectrum of curvature perturbations. In this
section we shall present the general calculation for a vacuum $f(\mathcal{G})$
theory of gravity \cite{Bamba:2007ef,Bamba:2009uf,Odintsov:2015uca},
with $\mathcal{G}$ being the Gauss-Bonnet invariant.

We shall realize a quite popular inflationary scenario, namely that
of intermediate inflation
\cite{Barrow:1990td,Barrow:1993zq,Rezazadeh:2014fwa,
Barrow:2014fsa,Herrera:2014mca,Jamil:2013nca,Herrera:2010vv,
Rendall:2005if} in which case the scale factor is,
\begin{equation}
\label{bambabounce}
a(t)=\e^{f_0t^{\alpha+1}}\, ,
\end{equation}
and the
corresponding Hubble rate is,
\begin{equation}
\label{hubratepresentpaper}
H(t)=f_0(\alpha+1) t^{\alpha }\, ,
\end{equation}
with $-1<\alpha<0$ and also
$f_0>0$. Also for notational simplicity we set
$\beta=f_0(\alpha+1)$. Note that we use a somewhat strange notation
for the intermediate inflation scale factor and there is a specific
reason for this (later on we shall change this ``heavy'' notation).
The reason is that the intermediate inflation scenario is a type of
singular inflation, and specifically a Type III singular inflation,
as it can be seen by looking at the scale factor
(\ref{hubratepresentpaper}) and the singularity classification after
Eq.~(\ref{singstarobhub}). The intermediate inflation scenario is
quite popular, and in Refs.~\cite{Barrow:2014fsa,Herrera:2014mca}
the scalar-tensor description of this scenario was confronted with
observational data. Here we shall realize the intermediate inflation
with $f(\mathcal{G})$ gravity. The background metric will be assumed
to be the flat FRW metric. Before we calculate the power spectrum we
need to find which $f(\mathcal{G})$ gravity can realize the
intermediate inflation Hubble rate (\ref{hubratepresentpaper}),
emphasizing on early cosmic times. To this end we shall use some
very well known $f(\mathcal{G})$ gravity reconstruction schemes
\cite{Bamba:2007ef,Bamba:2009uf,Odintsov:2015uca}. Also a detailed
presentation of the issues that follow can be found in
\cite{Oikonomou:2015qha}. The starting point of our analysis is the
Jordan frame vacuum $f(\mathcal{G})$ gravity, with the action being,
\begin{equation}
\label{actionfggeneral}
\mathcal{S}=\frac{1}{2\kappa^2}\int d^4x
\sqrt{-g}\left( R+f(\mathcal{G})\right)\, ,
\end{equation}
with
$\kappa^2=1/M_\mathrm{pl}^2$, and $M_\mathrm{pl}=1.22\times 10^{19}$GeV, and $g$
is the trace of the background metric $g_{\mu \nu}$. If we vary the
action with respect to the metric tensor $g_{\mu \nu}$, we easily
obtain the following gravitational equations of motion,
\begin{align}
\label{fgr1}
& R_{\mu \nu}-\frac{1}{2}g_{\mu \nu}f(\mathcal{G})-\left( -2RR_{\mu
\nu}+4R_{\mu \rho}R_{\nu}^{\rho}-2R_{\mu}^{\rho \sigma \tau}R_{\nu
\rho \sigma \tau}+4g^{\alpha \rho}g^{\beta \sigma}R_{\mu \alpha \nu
\beta}R_{\rho \sigma}\right) F'(\mathcal{G})\nn
& -2 \left(\nabla_{\mu}\nabla_{\nu}F'(\mathcal{G})\right )R+2g_{\mu \nu}
\left(\square
F'(\mathcal{G}) \right )R-4 \left (\square F'(\mathcal{G}) \right )R_{\mu \nu }
+4 \left(\nabla_{\mu}\nabla_{\nu}F'(\mathcal{G})\right )R_{\nu}^{\rho }\nn
&+4
\left (\nabla_{\rho}\nabla_{\nu}F'(\mathcal{G})\right ) R_{\mu}^{\rho}
 -4g_{\mu \nu} \left (\nabla_{\rho}\nabla_{\sigma }F'(\mathcal{G})
\right)R^{\rho \sigma }+4 \left (\nabla_{\rho}\nabla_{\sigma }F'(\mathcal{G})
\right)g^{\alpha \rho}g^{\beta \sigma }R_{\mu \alpha \nu \beta }=0\, .
\end{align}
Recall from the previous sections that for the flat FRW metric, the
Gauss-Bonnet invariant can be expressed as a function of the Hubble
rate as follows,
\begin{equation}
\label{gausbonehub}
\mathcal{G}=24H^2\left
(\dot{H}+H^2 \right )\, ,
\end{equation}
and also for the flat FRW metric, the
equations of motion (\ref{fgr1}), can be written as follows,
\begin{align}
\label{eqnsfggrav}
& 6H^2+f(\mathcal{G})-\mathcal{G}F'(\mathcal{G})
+24H^3\dot{\mathcal{G}}F''(\mathcal{G})=0 \nn
& 4\dot{H}+6H^2+f(\mathcal{G})-\mathcal{G}F'(\mathcal{G})
+16H\dot{\mathcal{G}}\left( \dot{H}+H^2\right) F''(\mathcal{G}) \nn
& +8H^2\ddot{\mathcal{G}}F''(\mathcal{G})
+8H^2\dot{\mathcal{G}}^2F'''(\mathcal{G})=0\, .
\end{align}
A vital element of the reconstruction method we shall use, is the
introduction of an auxiliary scalar field $\phi$, which can be
identified eventually with the cosmic time, as it was shown in
\cite{Bamba:2007ef,Bamba:2009uf,Odintsov:2015uca}. Then, by using
two proper functions of the cosmic time, namely $P(t)$ and $Q(t)$,
the Jordan frame $f(\mathcal{G})$ gravity action (\ref{actionfggeneral}), can
be rewritten as follows,
\begin{align}
\label{actionfrg}
& \mathcal{S}=\frac{1}{2\kappa^2}\int d ^4x\sqrt{-g}\left (
R+P(t)\mathcal{G}+Q(t)\right )\, ,
\end{align}
so by varying the above action with respect to the cosmic time $t$,
we get the following differential equation,
\begin{equation}
\label{auxeqnsvoithitiko}
\frac{d P(t)}{d t}\mathcal{G}
+\frac{d Q(t)}{d t}=0 \, .
\end{equation}
Then, one may solve Eq.~(\ref{auxeqnsvoithitiko}) with respect to
$t=t(\mathcal{G})$, and eventually by substituting the result in the following
equation,
\begin{equation}
\label{ebasc}
f(\mathcal{G})=P(t)\mathcal{G}+Q(t)\, ,
\end{equation}
one may obtain the resulting $f(\mathcal{G})$ gravity. Hence, the functions
$P(t)$ and $Q(t)$ are very important for the calculation of the
resulting $f(\mathcal{G})$ gravity which may realize a specific cosmic
evolution, so now we demonstrate how to find these, given a specific
cosmological evolution. Combining Eqs.~(\ref{ebasc}) and
(\ref{eqnsfggrav}), we get the following differential equation,
\begin{equation}
\label{ak}
Q(t)=-6H^2(t)-24H^3(t)\frac{d P}{d t}\, ,
\end{equation}
and in addition, by combining (\ref{ebasc}) and (\ref{ak}), we get,
\begin{equation}
\label{diffept}
2H^2(t)\frac{d ^2P}{d t^2}
+2H(t)\left(2\dot{H}(t)-H^2(t) \right)
\frac{d P}{d t}+\dot{H}(t)=0\, .
\end{equation}
By solving the differential equation (\ref{diffept}), we can obtain
the analytic form of the function $P(t)$, and eventually the
function $Q(t)$. So by substituting these in
Eq.~(\ref{auxeqnsvoithitiko}), we obtain the resulting form of the
function $t=t(\mathcal{G})$, and having this at hand, enables us to obtain the
$f(\mathcal{G})$ gravity, by simply substituting $t=t(\mathcal{G})$ in
Eq.~(\ref{ebasc}). Let us apply this method for the intermediate
inflation cosmology (\ref{hubratepresentpaper}), at early times, in
which case the differential equation (\ref{diffept}) becomes,
\begin{equation}
\label{diffept2}
\frac{2 \beta }{\alpha }t^{1+\alpha}\frac{d ^2P}{d t^2}
+4 t^{\alpha } \beta+1=0\, ,
\end{equation}
and by analytically solving this we obtain,
\be
\label{dfesolu}
P(t)=-\frac{t^{1-2 \alpha } \left(t^{\alpha }-2 t^{\alpha } \alpha
 -2 \beta C_1+2 \alpha \beta C_1\right)}{2 (-1+\alpha ) (-1+2
\alpha ) \beta }+C_2\, . \end{equation}By combining Eqs.~(\ref{dfesolu}) and
(\ref{ak}), we get the function $Q(t)$, which is,
\begin{align}
\label{analyticqt}
Q(t)=&-2 t^{-1+2 \alpha } \alpha \beta ^2 \nn &
-72 t^{-1+3 \alpha } \alpha \beta ^3 \left(-\frac{t^{1-2 \alpha }
\left(t^{-1+\alpha } \alpha -2 t^{-1+\alpha } \alpha ^2\right)}{2
(-1+\alpha ) (-1+2 \alpha ) \beta }
 -\frac{t^{-2 \alpha }
(1-2 \alpha ) \left(t^{\alpha } -2 t^{\alpha } \alpha -2 C_1 \beta
+2 C_1 \alpha \beta \right)}{2 (-1+\alpha ) (-1+2 \alpha ) \beta
}\right) \nn & + 24 t^{3 \alpha } \beta ^3\left( -\frac{t^{-2\alpha
} (1-2 \alpha ) \left(t^{-1+\alpha } \alpha -2 t^{-1+\alpha } \alpha
^2\right)}{(-1+\alpha ) (-1+2 \alpha ) \beta }
 -\frac{t^{1-2 \alpha } \left(t^{-2+\alpha } (-1+\alpha ) \alpha -2
t^{-2+\alpha } (-1+\alpha ) \alpha ^2\right)}{2 (-1+\alpha ) (-1+2
\alpha ) \beta } \right. \nn & \left. +\frac{t^{-1-2 \alpha } (1-2
\alpha ) \alpha \left(t^{\alpha }-2 t^{\alpha } \alpha -2 C_1 \beta
+2 C_1 \alpha \beta \right)}{(-1+\alpha ) (-1+2 \alpha ) \beta
}\right) \, ,
\end{align}
Finally, by using the resulting $Q(t)$ and $P(t)$, we can obtain the
final form of Eq.~(\ref{auxeqnsvoithitiko}), which is,
\begin{equation}
\label{finaformofauxiliaryeq}
\frac{t^{-1-2 \alpha } \left(4 t^{3
\alpha } \alpha \beta ^3 \left(11 t^{\alpha }-12 C_1 \beta \right)
 - \mathcal{G} x \left(t^{\alpha }-2 C_1 \beta \right)\right)}{2 \beta }=0\, .
\end{equation}
We shall be interested for early times, so we solve the algebraic equation
above for $t\to 0$, in which case, the algebraic equation
(\ref{finaformofauxiliaryeq}) becomes,
\begin{equation}
\label{dgfshsyyss}
C_1
\mathcal{G} t^{-2 \alpha }-24 C_1 t^{-1+\alpha } \alpha \beta ^3=0
\, ,
\end{equation}
which results to,
\begin{equation}
\label{onyfe}
t=\frac{\mathcal{G}^{\frac{1}{(3\alpha -1)}}}{\left(24 \alpha
\beta^3\right)^{\frac{1}{(3\alpha -1)}}}\, ,
\end{equation}
Substituting
(\ref{onyfe}) in $P(t)$ and $Q(t)$ and also by making use of
Eq.~(\ref{ebasc}), the resulting $f(\mathcal{G})$ gravity which
realizes the intermediate inflation scenario at early times reads,
\begin{equation}
\label{actaulfg}
f(\mathcal{G})=C_2\mathcal{G} +A
G^{\frac{2\alpha }{-1+3 \alpha }} +B \mathcal{G}^{\frac{\alpha
}{-1+3 \alpha}}\, ,
\end{equation}
where the parameters $A$ and $B$ are equal
to,
\begin{align}
\label{dgdgdg}
A=&11\ 24^{\frac{2 \alpha }{1-3 \alpha }} \beta ^2
\left(\alpha \beta ^3\right)^{\frac{2 \alpha }{1-3 \alpha }}\nn B=&
\frac{\left(24^{\frac{1-2 \alpha }{1-3 \alpha }} C_1-24^{\frac{1-2
\alpha }{1-3 \alpha }} C_1 \alpha -\frac{2^{-1+\frac{3 (1-2 \alpha
)}{1-3 \alpha }} 3^{\frac{1-2 \alpha }{1-3 \alpha }}}{\beta
}+\frac{24^{\frac{1-2 \alpha }{1-3 \alpha }} \alpha }{\beta }\right)
\left(\alpha \beta ^3\right)^{\frac{1-2 \alpha }{1-3 \alpha }}}{1-3
\alpha +2 \alpha ^2}\nn & -24^{1+\frac{\alpha }{1-3 \alpha }} C_1
G^{\frac{\alpha }{-1+3 \alpha }} \beta ^3 \left(\alpha \beta
^3\right)^{\frac{\alpha }{1-3 \alpha }} \, .
\end{align}
We can further simplify the resulting $f(\mathcal{G})$ gravity, by
using the fact that we are interested in early times. Particularly,
the Gauss-Bonnet invariant is equal to,
\begin{equation}
\label{argausbbbe}
\mathcal{G} =24 t^{-1+3 \alpha } \alpha \beta ^3+24 t^{4 \alpha }
\beta ^4\, ,
\end{equation}
so for $t\to 0$, we also have $\mathcal{G}\to 0$.
Hence we can simplify the $f(\mathcal{G})$ gravity by keeping only
the leading order terms in the small-$\mathcal{G}$ limit, so we
finally have,
\begin{equation}
\label{fgsmalglimit}
f(\mathcal{G})\simeq C_2
\mathcal{G}+B \mathcal{G}^{\frac{\alpha }{-1+3 \alpha }}\, .
\end{equation}
Having the resulting $f(\mathcal{G})$ gravity at hand, we can easily
calculate the power spectrum of the primordial curvature
perturbations, which will determine how the perturbations evolve.

To proceed to the calculation of the power spectrum, consider scalar
linear perturbations of the flat FRW metric of the following form,
\begin{equation}
\label{scalarpertrbubations}
d s^2=-(1+\psi)d
t^2-2a(t)\partial_i\beta d td x^i+a(t)^2\left(
\delta_{ij}+2\phi\delta_{ij}+2\partial_i\partial_j\gamma \right) d
x^id x^j\, ,
\end{equation}
where $\psi$, $\phi$, $\gamma$ and $\beta$ are
smooth scalar perturbations. The primordial perturbations are
analyzed by studying a gauge invariant quantity, so we shall use the
comoving curvature perturbation, which is equal to,
\begin{equation}
\label{confedf}
\Phi=\phi-\frac{H\delta \xi}{\dot{\xi}}\, ,
\end{equation}
where $\xi=\frac{d F}{d G}$. In the case of $f(\mathcal{G})$
gravity, there are no $k^4$ scalar propagating modes, where $k$ is
the wavelength, so in this case there is no superluminal
propagation. Hence the propagation of the scalar perturbations
contains only $k^2$ terms, and the evolution of perturbations is
governed by the following equation,
\begin{equation}
\label{perteqnmain}
\frac{1}{a(t)^3Q(t)}\frac{d }{d t}\left(a(t)^3Q(t)
\dot{\Phi}\right)+B_1(t)\frac{k^2}{a(t)^2}\Phi=0\, ,
\end{equation}
and it can
be seen that the $k^2$ terms dominate the evolution. The speed of
the propagating modes is determined by $B_1(t)$, which for an
$f(\mathcal{G})$ theory is,
\begin{equation}
\label{b1}
B_1(t)=1+\frac{2\dot{H}}{H^2}\, .
\end{equation}
In addition, the term $Q(t)$
which appears in Eq.~(\ref{perteqnmain}), for a general
$f(\mathcal{G})$ gravity is,
\begin{equation}
\label{gfgdhbhhyhjs}
Q(t)=\frac{6\left( \frac{d ^2F}{d \mathcal{G}^2}
\right)^2\dot{\mathcal{G}}^2\left(1
+4F''(\mathcal{G})\dot{\mathcal{G}}H\right)}{\left(1
+6HF''(\mathcal{G})\dot{\mathcal{G}}\right)^2}\, ,
\end{equation}
where we used
$\dot{\xi}=\frac{d F^2}{d \mathcal{G}^2} {\dot{\mathcal{G}}}$. Note
that the ``prime'' in Eq.~(\ref{gfgdhbhhyhjs}) denotes
differentiation with respect to the Gauss-Bonnet scalar, while the
``dot'' denotes differentiation with respect to the cosmic time. We
shall try to find an approximate solution of the differential
equation (\ref{gfgdhbhhyhjs}) corresponding to early times. The
differential equation (\ref{perteqnmain}) can be recast as follows,
\begin{equation}
\label{difnew}
a(t)^3Q(t)\ddot{\Phi}+\left(3a(t)^2\dot{a}Q(t)+a(t)^3
\dot{Q}(t)\right)\dot{\Phi}+B_1(t)Q(t)a(t)k^2\Phi=0\, .
\end{equation}
Then
after some calculations, the leading order differential equation
which governs the evolution of the perturbations at early times
reads,
\begin{equation}
\label{simplifiedeqn1}
t^{1+\alpha }
\Omega_4\ddot{\Phi}-t^{\alpha }\Omega_2\dot{\Phi}+\Omega_1\Phi=0\, ,
\end{equation}
where the explicit form of the parameters $\Omega_i$ ($i=1,2,4$)
is,
\begin{align}
\label{longago}
\Omega_1=&\frac{2^{-4+\frac{6 \alpha }{-1+3 \alpha
}} 3^{-1+\frac{2 \alpha }{-1+3 \alpha }} B^2 k^2\text{ }(1-2 \alpha
)^2 \alpha \left(\alpha \beta ^3\right)^{\frac{2 \alpha }{-1+3
\alpha }} }{(1-3 \alpha )^4 \beta ^7}\, , \nn \Omega_2=&
-\frac{2^{-3+\frac{6 \alpha }{-1+3 \alpha }} 3^{-1+\frac{2
\alpha}{-1+3 \alpha }} B^2\text{ }(1-2 \alpha )^2 \alpha
\left(\alpha \beta ^3\right)^{\frac{2 \alpha }{-1+3 \alpha }}}{(1-3
\alpha )^4 \beta ^6}\, , \nn \Omega_3=&\frac{2^{-5+\frac{6 \alpha
}{-1+3 \alpha }} 3^{\frac{2 \alpha }{-1+3 \alpha }} B^2\text{ }(1-2
\alpha )^2 \alpha \left(\alpha \beta ^3\right)^{\frac{2 \alpha
}{-1+3 \alpha }}}{(1-3 \alpha )^4 (1+\alpha ) \beta ^5}\, , \nn
\Omega_4=&\frac{2^{-5+\frac{6 \alpha }{-1+3 \alpha }} 3^{-1+\frac{2
\alpha }{-1+3 \alpha }} B^2 (1-2 \alpha )^2 \left(\alpha \beta
^3\right)^{\frac{2 \alpha }{-1+3 \alpha }} }{(1-3 \alpha )^4 \beta
^6}\, .
\end{align}
We need to note that for deriving the resulting differential
equation, we omitted a term $\sim (\Omega_3 \dot{\Phi}t)^{2 \alpha
}$, since the term $t^{\alpha}\Omega_2\dot{\Phi}$ dominates, and the
explicit form of the parameter $\Omega_3$ can be found in Eq.
(\ref{longago}). The differential equation (\ref{difnew}) can be
solved and the solution is,
\begin{equation}
\label{solutionevolution}
\Phi(t)=
\Delta_1 t^{\frac{\mu }{2(-1+\alpha )}} J_{\mu}(\zeta
t^{\frac{1-\alpha }{2}})+\Delta_2 t^{\frac{\mu }{2 (-1+\alpha
)}}J_{-\mu}(\zeta t^{\frac{1-\alpha }{2}})\, ,
\end{equation}
where the
parameters $\mu$ and $\zeta$, are equal to,
\begin{equation}
\label{app}
\mu =
\frac{\Omega_2+\Omega_4}{(-1+\alpha ) \Omega_4}\, , \quad \zeta =
\frac{2\text{ }\sqrt{\Omega_1}}{\left(-1+\frac{1}{\alpha } \right)
\alpha \sqrt{\Omega_4}} \, ,
\end{equation}
and $J_{\mu}(y)$ is the Bessel
function of the first kind. Also the explicit form of the parameters
$\Delta_i$, $i=1,2$ appears below,
\begin{align}
\label{app1}
& \Delta_1=\left(-1+\frac{1}{\alpha }\right)^{\mu } \alpha ^{\mu }
\Omega_1^{-\frac{\mu }{2}} \Omega_4^{\frac{\mu }{2}}\text{ }C_3
\Gamma\left[\frac{\alpha }{-1+\alpha }+\frac{\Omega_2}{(-1+\alpha )
\Omega_4}\right],\nn & \Delta_2=\left(-1+\frac{1}{\alpha
}\right)^{3\mu } \alpha ^{\mu } \Omega_1^{-\frac{\mu }{2}}
\Omega_4^{\frac{\mu }{2}}C_4
\Gamma\left[\frac{\Omega_2}{\Omega_4-\alpha \Omega_4}+\frac{2
\Omega_4}{\Omega_4-\alpha \Omega_4}-\frac{\alpha
\Omega_4}{\Omega_4-\alpha \Omega_4}\right]\, .
\end{align}
We can further simplify the solution by using the small argument
limit of the Bessel function,
\begin{equation}
\label{besselapprox}
J_{\mu}(y)\simeq \frac{y^{\mu }2^{-\mu }}{\Gamma[1+\mu ]}\, ,
\end{equation}
so
the solution (\ref{solutionevolution}) is simplified as follows,
\begin{equation}
\label{approximatebehavior}
\Phi(t)\simeq \Delta_2 \frac{2^{-\mu }
\zeta ^{\mu }}{\Gamma[1 +\mu ]}x^{\frac{\left(2-2 \alpha
+\alpha^2\right) \mu }{2 (-1+\alpha )}}\, .
\end{equation}
Hence, from
Eq.~(\ref{approximatebehavior}) it is obvious that the evolution of
the primordial curvature scalar perturbations depends linearly on
the cosmic time $t$. Now we investigate whether the power spectrum
is nearly scale invariant and also whether this can be in
concordance with the observational data. The gauge invariant
function $\Phi$ of Eq.~(\ref{confedf}), satisfies the following
equation \cite{DeFelice:2009ak},
\begin{equation}
\label{secondoredperturbas}
\mathcal{S}_p=\int d x^4a(t)^3Q_s\left(
\frac{1}{2}\dot{\Phi}-\frac{1}{2}\frac{c_s^2}{a(t)^2}(\nabla \Phi
)^2\right)\, ,
\end{equation}
with $Q_s$ being this time,
$Q_s=\frac{4}{\kappa^2}Q(t)$ and the function $Q(t)$ was defined in
Eq.~(\ref{gfgdhbhhyhjs}). The power spectrum of the primordial
curvature perturbations for the gauge invariant variable $\Phi$, can
be calculated by using standard approaches in cosmological
perturbation theory \cite{Linde:2007fr,Gorbunov:2011zzc,Lyth:1998xn,
Linde:1983gd,Linde:1985ub,Linde:1993cn,
Brandenberger:2016uzh,Bamba:2015uma,Martin:2013tda,
Martin:2013nzq,Baumann:2014nda,Baumann:2009ds,Linde:2014nna,
Pajer:2013fsa,Yamaguchi:2011kg,Byrnes:2010em,Turok:2002yq,
Linde:2005dd,Kachru:2003sx}, and it is equal to,
\begin{equation}
\label{powerspecetrumfgr}
\mathcal{P}_R=\frac{4\pi
k^3}{(2\pi)^3}|\Phi|_{k=aH}^2\, .
\end{equation}
If we look at the analytic
forms of the parameters $\Omega_i$, $\Delta_2$ and $\zeta$ above,
one may easily conclude that the power spectrum is not scale
invariant, and we demonstrate that now in some detail. The
parameters $\Omega_1$, $\Delta_2$, $\zeta$ contain the wavenumber
$k$ implicitly via the parameter $\Omega_1$, and in detail we have,
\begin{equation}
\label{snddfr}
\Omega_1\sim k^2,\quad \zeta\sim
\sqrt{\Omega_1},\quad \Delta_2\sim \Omega_1^{-\frac{\mu}{2}}\, .
\end{equation}
Due to the fact that the power spectrum depends on
$\Delta\zeta^{\mu}$, the from Eq.~(\ref{snddfr}), we get,
\begin{equation}
\label{powerspectrajb}
\mathcal{P}_R\sim k^3 \left| C_4
(t-t_s)^{\frac{\left(2-2 \alpha +\alpha ^2\right) \mu }{2
(-1+\alpha)}}\right|^2_{k=aH}\, .
\end{equation}
However it is not obvious if
the above spectrum is scale invariant, since the constant of
integration contained in the parameter $\Delta_2$ in
Eq.~(\ref{app1}), namely $C_4$, depends on $k$, and its exact form
can be found if the initial conditions on the gauge invariant
quantity $\Phi (t)$ are given. Moreover, a hidden $k$-dependence
exists in the term $t^{\frac{\left(2-2 \alpha +\alpha^2\right) \mu
}{2 (-1+\alpha )}}$, since the power spectrum is calculated at the
horizon crossing time instance, where $k=aH$. So let us calculate
these in detail. Since we are working for $t\to 0$ cosmic times, the
conformal time $\tau$, which is defined as $d \tau =a^{-1}(t)d t$,
satisfies $\tau \sim t$, since for $t\to 0$, the scale factor in
Eq.~(\ref{bambabounce}), is approximately equal to $a\sim 1$. In
addition, for $t\to 0$, we approximately have $k\simeq H$ at the
horizon crossing time instance, and therefore the following relation
holds true,
\begin{equation}
\label{explicitequation}
\beta t^{\alpha}\simeq k\, .
\end{equation}
By solving the above equation with respect to the cosmic time
$t$, we obtain,
\begin{equation}
\label{solvedwithrespecttotts}
t\simeq \left(\frac{k}{\beta}\right )^{\frac{1}{\alpha}}\, .
\end{equation}
It is obvious
that Eq.~(\ref{solvedwithrespecttotts}) determines the
$k$-dependence of the cosmic time $t$, at the horizon crossing.
Having found this, let us now proceed to find $C_4$ as a function of
$k$. We introduce the canonical variable $u=z_s \Phi$, where,
$z_s=Q(t)a(t)$, and owing to the fact that $a(t)\simeq 1$ for $t\to
0$, we finally conclude that $z_s\simeq Q(t)$, and this implies
that,
\begin{equation}
\label{safakebelieve}
u\sim \Phi Q(t)\, ,
\end{equation}
with $Q(t)$
being defined in Eq.~(\ref{gfgdhbhhyhjs}). Then we can write the
action of Eq.~(\ref{secondoredperturbas}), in terms of the variable
$u$, at early times, as follows,
\begin{equation}
\label{actiaonerenearthebounce}
\mathcal{S}_u\simeq \int d ^3d \tau \left[
\frac{u'}{2}-\frac{1}{2}(\nabla u)^2+\frac{z_s''}{z_s}u^2\right ]\, ,
\end{equation}
with the prime this time indicating differentiation with respect
to $\tau\sim t$. Assuming a Bunch-Davies vacuum state for the
canonical scalar field $u$ \cite{Brandenberger:2016uzh}, before the
inflationary era starts, then we have $u\sim
\frac{\e^{-ik\tau}}{\sqrt{k}}$. Note that the imaginary phase plays
no role, since we are interested in calculating $|\Phi (t=t_s)|^2$.
Due to Eq.~(\ref{safakebelieve}), we get,
\begin{equation}
\label{imanrun}
\Phi(t_0)\sim C_4\sim \frac{1}{\sqrt{k}Q(t)}\, ,
\end{equation}
where $t_0$
indicates the start of the inflationary era. In addition, the
function $Q(t)$ is equal to,
\begin{equation}
\label{isssoripiametaksytrelas}
Q(t)\simeq \frac{2^{-5+\frac{6 \alpha }{-1+3 \alpha }} 3^{-1+\frac{2
\alpha }{-1+3 \alpha }} B^2 (t-t_s)^{-4 \alpha } (1-2 \alpha )^2
\left(\alpha \beta ^3\right)^{\frac{2 \alpha }{-1+3 \alpha }}}{(1-3
\alpha )^4 \beta ^6}\, .
\end{equation}
Then, by using
Eq.~(\ref{solvedwithrespecttotts}), we obtain,
\be
\label{newsrevisison3}
Q(t(k)) \simeq \mathcal{Z}_1 k^{\frac{-4
\alpha }{\alpha}} =\mathcal{Z}_1 k^{-4}\, ,
\end{equation}
with the parameter
$\mathcal{Z}_1$ being equal to,
\begin{equation}
\label{dfjedjsuestrsureend}
\mathcal{Z}_1= \frac{2^{-5+\frac{6 \alpha }{-1+3 \alpha }}
3^{-1+\frac{2 \alpha }{-1+3 \alpha }} B^2 (1-2 \alpha )^2
\left(\alpha \beta ^3\right)^{\frac{2 \alpha }{-1+3 \alpha
}}}{(1-3\alpha )^4 \beta ^6\beta^{1/{\alpha}}}\, .
\end{equation}
By combining
Eqs.~(\ref{newsrevisison3}) and (\ref{imanrun}), we obtain that,
\begin{equation}
\label{equationforc4final}
C_4\simeq
\frac{1}{\mathcal{Z}_1}\frac{1}{\sqrt{k}k^{-4}}\sim
k^{\frac{7}{2}}\, .
\end{equation}
Finally combining
Eqs.~(\ref{equationforc4final}), (\ref{solvedwithrespecttotts}), and
(\ref{powerspectrajb}), the power spectrum $\mathcal{P}_R$ as a
function of $k$ is,
\begin{equation}
\label{powerspectrumfinal}
\mathcal{P}_R\sim
k^{\frac{7}{2}+3 +\frac{\left(2-2 \alpha +\alpha ^2\right) \mu
}{2(-1+\alpha )}}\, ,
\end{equation}
and it can be seen that the resulting
power spectrum is not scale invariant. We can compute the spectral
index of primordial curvature perturbations if we combine
Eqs.~(\ref{app}) and (\ref{longago}) and in effect we get,
$\mu=11/(1-\alpha)$. Then, the spectral index of the primordial
curvature perturbations is,
\begin{equation}
\label{pcp}
n_s-1\equiv\frac{d\ln\mathcal{P}_{\mathcal{R}}}{d\ln k}
=\frac{7}{2}+3+\frac{2-2\alpha+\alpha^2}{2(\alpha-1)}\mu
=1-\frac{11}{2(\alpha-1)^2}\, .
\end{equation}
There can be agreement of $n_s$
with the observational data for two values of $\alpha$, namely for
$\alpha=13.9$ and for $\alpha=-12$. The latter is not allowed by the
intermediate inflation scenario, and the same applies  for the
former, so we can see that the intermediate inflation scenario, when
realized by an $f(\mathcal{G})$ gravity, does not result to a power
spectrum which scale or nearly scale invariant and it is not
compatible with observational data.

\subsubsection{Evolution of Perturbations After the Horizon Crossing}

As another task relevant to the dynamics of the inflationary era, we
shall now investigate whether the cosmological perturbations remain
constant or not after the horizon crossing. We are mainly presenting
this issue in order to have a complete idea of all the topics
related to the inflationary era.

In many scenarios, the conservation of the comoving curvature
perturbations does not occur, like for example in the matter bounce
scenario \cite{Brandenberger:2016uzh}, so this issue should be
properly investigated. The focus is on cosmological times for which
the wavenumber satisfies $k\ll a(t)H(t)$. Due to the last relation,
the differential equation (\ref{perteqnmain}) can be simplified
since the last term of it can be safely neglected, so it becomes,
\begin{equation}
\label{perteqnmainportoriko}
\frac{1}{a(t)^3Q(t)}\frac{d }{d t}
\left(a(t)^3Q(t)\dot{\Phi}\right)=0\, ,
\end{equation}
and by analytically solving this we obtain,
\begin{equation}
\label{soldiffeqn}
\Phi (t)=\mathcal{C}_1+\mathcal{C}_2\int
\frac{1}{a(t)^3Q(t)}d t\, ,
\end{equation}
where $Q(t)$ is defined in Eq.~(\ref{gfgdhbhhyhjs}). Clearly the
integral in Eq.~(\ref{soldiffeqn}), determines the evolution of the
comoving curvature perturbations after the horizon crossing. It is
therefore crucial to find the function $Q(t)$ and also to determine
the $f(\mathcal{G})$ gravity which governs the evolution for $k\ll a(t)H(t)$.
Since the condition $t\to 0$ holds true after the horizon crossing
and for $t\ll 1$sec, the $f(\mathcal{G})$ gravity which generates the
evolution is still the one appearing in Eq.~(\ref{fgsmalglimit}),
and in effect, the function $Q(t)$ is,
\begin{equation}
\label{qtfunctionscenarioi}
Q(t)\simeq \mathcal{Z}_2 t^{-4\alpha}\, .
\end{equation}
Hence, due to the fact that $a(t)\simeq \e^{f_0 t^{\alpha+1}}$ , the
term $\frac{1}{a(t)^3Q(t)}$, has the following behavior,
\begin{equation}
\label{integraterm1}
\frac{1}{a(t)^3Q(t)}\sim
\frac{t^{4\alpha}}{\e^{f_0 t^{\alpha+1}}}\, .
\end{equation}
Since the exponential term dominates the evolution, the integral
term eventually decays, as the cosmic time increases, so in effect
we have,
\begin{equation}
\label{shotshottohea}
\int \frac{1}{a(t)^3Q(t)}\rightarrow 0\, .
\end{equation}
Therefore, the comoving curvature perturbation can be approximated
as follows,
\begin{equation}
\label{scomobafter}
\Phi (t)=\mathcal{C}_1\, ,
\end{equation}
and therefore the comoving curvature perturbation is conserved after
the horizon crossing, since $\Phi (t)$ is constant.

\subsection{Inflation from $F(T)=T+f(T)$ Theories in the Jordan Frame}

In this section we shall present in brief the formalism for
obtaining the power spectrum of primordial curvature perturbations
in the context of $F(T)=T+f(T)$ gravity. For detailed analysis on
these issues we refer to Refs.~\cite{Bengochea:2008gz,
Linder:2010py,Cai:2011tc,Chen:2010va,Bamba:2010wb,Zhang:2011qp,
Geng:2011aj,Bohmer:2011si,Gonzalez:2011dr,Karami:2012fu,Bamba:2012vg,
Rodrigues:2012qua,Capozziello:2012zj,Chattopadhyay:2012eu,Izumi:2012qj,
Li:2013xea,
Ong:2013qja,Otalora:2013tba,
Nashed:2013bfa,Kofinas:2014owa,Harko:2014sja,Hanafy:2014bsa,
Junior:2015bva,Ruggiero:2015oka}, and we adopt the notation of
Ref.~\cite{Cai:2011tc}. The analysis that follows has been presented
in detail in Refs.~\cite{Oikonomou:2017isf,Oikonomou:2017brl}. We
consider only the longitudinal gauge, so only scalar-type metric
fluctuations are considered, with the perturbed metric being of the
following form,
\begin{equation}
\label{metricscalar}
ds^2=(1+2\Phi)d
t^2-a(t)^2(1-2\Psi) \sum_id x^2_i\, ,
\end{equation}
and the scalar
fluctuations in the metric are quantified in terms of the functions
$\Phi$ and $\Psi$. The perturbation of the torsion scalar in terms
of the scalar perturbations $\Psi$ and $\Phi$, at leading order, is
equal to,
\begin{equation}
\label{trosionpertubrscalar}
\delta T=12
H(\dot{\Phi}+H\Psi)\, ,
\end{equation}
with $H$ being the Hubble rate. The
$F(T)=T+f(T)$ gravity perturbed gravitational equations
corresponding to the metric of Eq.~(\ref{metricscalar}), are equal
to,
\begin{align}
\label{ftgravieqnspertrb}
& (1+F_{,T})\frac{\nabla^2}{a^2}\Psi-3(1+f_{,T})H\dot{\Psi}
 -3(1+f_{,T})H^2\Phi+36f_{,TT}H^3(\dot{\Psi}+H\Phi)
=4\pi G \delta \rho\, ,\nn &
(1+f_{,T}-12H^2f_{,TT})(\dot{\Psi}+H\Phi)=4\pi G\delta q\, ,\nn &
(1+f_{,T})(\Psi-\Phi)=8\pi G\delta s\, , \nn &
(1+f_{,T}-12H^2f_{,TT})\ddot{\Psi}+3H(1+f_{,T}-12H^2f_{,TT}
 -12\dot{H}f_{,TT}+48H^2\dot{H}f_{,TTT})\dot{\Psi}\nn
& \left[
3H^2(1+f_{,T}-12H^2f_{,TT})+2\dot{H}(1+f_{,T}-30H^2f_{,TT}+72H^4f_{,TTT})
\right]\Phi+\frac{1+f_{,T}}{2a^2}\nabla^2(\Psi-\Phi)=4\pi G \delta p
\, ,
\end{align}
where $f_{,T}$, stands for $\partial_T f(T)$, and the rest of the
derivatives, namely $f_{,TT}$ and $f_{,TTT}$ are defined
correspondingly. Also, the functions $\delta \rho$, $\delta q$,
$\delta s$, $\delta p$, are the fluctuations of the total energy
density, of the fluid velocity, of the anisotropic stress and of the
total pressure respectively. By assuming that the matter fluids
present consist of a canonical scalar field with potential
$V(\phi)$, we obtain the following relations,
\begin{equation}
\label{deltarhodrelations}
\delta
\rho=\dot{\phi}(\delta\dot{\phi}-\dot{\phi}\Phi) +V_{,\phi}\delta
\phi\, ,\quad \delta q=\dot{\phi}\delta \phi\, ,\quad \delta s=0\, ,
\quad \delta p=\dot{\phi}(\delta
\dot{\phi}-\dot{\phi}\Phi)-V_{,\phi}\delta \phi\, .
\end{equation}
Due to the
above equations, it can be shown \cite{Cai:2011tc} that we have
$\Psi=\Phi$ and also that the gravitational potential $\Phi$ can be
completely determined by the scalar fluctuation $\delta \phi$. Hence
it is conceivable that the minimally coupled $F(T)$ gravity to a
scalar field, has only one degree of freedom.

The evolution of scalar perturbations is governed by the following
differential equation,
\begin{equation}
\label{masterequation}
\ddot{\Phi}_k+\alpha
\dot{\Phi}_k+\mu^2\Phi_k+c_s^2\frac{k^2}{a^2}\Phi_k=0\, ,
\end{equation}
where $\Phi_k$ is the scalar Fourier mode of the scalar potential
$\Phi$, and the functions $\alpha$, $\mu^2$ and $c_s^2$ are the
frictional term, the effective mass and the speed of sound parameter
for the potential $\Phi$ respectively. These functions are equal to,
\begin{align}
\label{functionsanalytical}
& \alpha=7H+\frac{2V_{,\phi}}{\dot{\phi}}-\frac{36H\dot{H}(f_{,TT}
 -4H^2f_{,TTT})}{1+f_{,T}-12H^2f_{,TT}}\, ,\nn &
\mu^2=6H^2+2\dot{H}+\frac{2HV_{,\phi}}{\dot{\phi}}
 -\frac{36H^2\dot{H}(f_{,TT}-4H^2f_{,TTT})}{1+f_{,T}-12H^2f_{,TT}}
\, , \nn & c_s^2=\frac{1+f_{,T}}{1+f_{,T}-12H^2f_{,TT}}\, .
\end{align}
By taking into account that the gravitational equation of motion for
the canonical scalar field is,
\begin{equation}
\label{eqnforscaux}
\ddot{\phi}+3H\dot{\phi}+V_{,\phi}=0\, ,
\end{equation}
and also by rewriting
the second $F(T)$ gravity Friedmann equation as follows,
\begin{equation}
\label{secondfriedmann}
(a+f_{,T}-12H^2f_{,TT})\dot{H}=-4\pi G
\dot{\phi}^2\, ,
\end{equation}
the evolution of scalar perturbations
differential equation (\ref{masterequation}), becomes,
\begin{equation}
\label{evolutionequationfinal}
\ddot{\Phi}_k+\left(H-\frac{\ddot{H}}{\dot{H}}\right)\dot{\Phi}_k
+\left(2\dot{H}-\frac{H\ddot{H}}{\dot{H}}
\right)\Phi_k+\frac{c_s^2k^2}{a^2}\Phi_k=0\, .
\end{equation}
By looking at the
evolution equation (\ref{evolutionequationfinal}), we can see that
it is identical to the one corresponding to the Einstein-Hilbert
case, except for the speed of sound parameter.

The physical quantity that quantifies perfectly the cosmological
inhomogeneities is the comoving curvature fluctuation $\zeta$,
expressed in comoving coordinates, which in this case is,
\begin{equation}
\label{comovcurv}
\zeta =\Phi-\frac{H}{\dot{H}}\left(\dot{\Phi}+H\Phi \right)\, .
\end{equation}
The parameter $\zeta$ is a gauge
invariant quantity, and it can be used to simplify the calculation
of the power spectrum and of the corresponding spectral index of the
primordial curvature perturbation. Particularly, we can introduce
the variable $v$ which is equal to,
\begin{equation}
\label{varintr1}
v=z\zeta\, ,
\end{equation}
with $z$ being equal to,
\begin{equation}
\label{fgr}
z=a\sqrt{2\epsilon}\, ,
\end{equation}
and with $\epsilon$ being the first slow-roll index
$\epsilon=-\frac{\dot{H}}{H^2}$. Eventually, the differential
equation that determines the evolution of the primordial
perturbations is \cite{Cai:2011tc},
\begin{equation}
\label{mastereqn2}
v_k''+\left( c_s^2k^2-\frac{z''}{z}\right) v_k=0\, ,
\end{equation}
where the
sound speed parameter was defined in
Eq.~(\ref{functionsanalytical}). The prime in Eq.~(\ref{mastereqn2})
denotes differentiation with respect to the conformal time, which is
defined as follows,
\begin{equation}
\label{conformtime}
\tau=\int d
t\frac{1}{a}\, .
\end{equation}
In order to demonstrate how the calculation of
the power spectrum works in practise, we shall realize the
intermediate inflation scenario
\cite{Barrow:1993zq,Rezazadeh:2014fwa,
Barrow:2014fsa,Herrera:2014mca,Jamil:2013nca,
Herrera:2010vv,Rendall:2005if}, in which case, the scale factor and
the Hubble rate are,
\begin{equation}
\label{intermed}
a(t)=\e^{At^n}\, ,\quad
H(t)=A n t^{n-1}\, ,
\end{equation}
where $0<n<1$ and also $A>0$ (here we use
for notational convenience, the notation $n$ for the exponent in the
scale factor). In
Refs.~\cite{Barrow:1993zq,Rezazadeh:2014fwa,Barrow:2014fsa}, the
intermediate inflation scenario was realized in the context of
scalar-tensor theory, but here we shall realize the intermediate
inflation in the context of vacuum $F(T)$ gravity, and after that we
shall calculate the power spectrum in detail. For a similar
presentation to ours, see \cite{Jamil:2013nca}. Notice that the
$F(T)$ gravity function appears implicitly in
Eq.~(\ref{mastereqn2}), via the sound speed parameter $c_s$. For the
FRW background, the first Friedmann equation of $F(T)$ gravity in
vacuum is,
\begin{equation}
\label{fteqn1}
H^2=-\frac{f(T(t))}{6}-2f_{,T}(t)H^2\, ,
\end{equation}
and also $T=-6H^2$, which in the case of the intermediate
inflation Hubble rate, we get,
\begin{equation}
\label{explicitttrelation}
T=-6
A^2 n^2 t^{2 n-2}\, ,
\end{equation}
which can be solved with respect to $t$
and we get the function $t(T)$,
\begin{equation}
\label{ttexplic}
t(T)=6^{-\frac{1}{2 n-2}} \left(-\frac{T}{A^2
n^2}\right)^{\frac{1}{2 n-2}}\, .
\end{equation}
By inserting the intermediate
inflation Hubble rate from Eq.~(\ref{intermed}) in the Friedmann
equation (\ref{fteqn1}), and also by substituting $t(T)$ from
Eq.~(\ref{ttexplic}), we can easily solve it to get the approximate
$F(T)$ gravity, which is,
\begin{equation}
\label{ftgravitytyfn1}
f(T)=c_1
T^{\frac{A n}{2}}-\frac{T}{2\left(1-\frac{A n}{2}\right)}\, ,
\end{equation}
where $c_1$ is an arbitrary integration constant. Since we are
interested in the inflationary era, which corresponds to early
times, when the intermediate inflation scenario is considered, the
conformal time $\tau$ defined in Eq.~(\ref{conformtime}) can be
approximately equal with the cosmic time $t$, since the exponential
for small values of $t$ is approximately equal to $\e^{At^n}\sim 1$.
Therefore, since $t\equiv \tau$, we can calculate $z(t)$, which is,
\begin{equation}
\label{zrt}
z(t)=\sqrt{\frac{2 (1-n) t^{-n}}{A n}}\, ,
\end{equation}
and
in addition, by using Eq.~(\ref{ftgravitytyfn1}), and also the
explicit form of the Hubble rate for the intermediate inflation
scenario, we can calculate the sound speed parameter which is,
\begin{equation}
\label{cscalculation}
c_s^2(t)=\frac{A c_1 n 6^{\frac{A n}{2}} (A
n-2) \left(-A^2 n^2 t^{2 n-2} \right)^{\frac{A n}{2}}-12 A^2 n^2 (A
n-1) t^{2 n-2}}{2 (A n-1) \left(c_1 6^{\frac{A n}{2}} (A n-2)
\left(-A^2 n^2 t^{2 n-2}\right)^{\frac{A n}{2}}-6 A^2 n^2 t^{2
n-2}\right)}\, ,
\end{equation}
and since the dominant term in the above
expression during the inflationary era, is $\sim t^{2 n-2}$, the
sound speed parameter can be approximated as $c_s^2\simeq 1$.
Therefore, the evolution equation (\ref{mastereqn2}) becomes as
follows,
\begin{equation}
\label{mastereqn23}
v''_k(t)+\left(
k^2-\frac{\left(\frac{n}{2} +1\right) n}{2 t^2}\right) v_k(t)=0\, ,
\end{equation}
which can be analytically solved to yield,
\be
\label{evoltpertbsol}
v_k(t)=C_1\sqrt{t}J_{\frac{n+1}{2}}(k
t)+C_2\sqrt{t} Y_{\frac{n+1}{2}}(kt)\, ,
\end{equation}
where $J_n(z)$ and
$Y_n(z)$ are the Bessel functions of first and second kind
respectively, and also $C_1$, $C_2$ are arbitrary constants. We can
approximate the expression in Eq.~(\ref{evoltpertbsol}) for small
values of the argument of the Bessel function, so we have,
\begin{align}
\label{evoltpertbsol12}
v_k(t)=& \frac{C_1 \sqrt{t} (k t)^{n/2} \left(2^{\frac{1}{2}
(-n-1)} \sqrt{k t}\right)}{\Gamma \left(\frac{n+1}{2}+1\right)}
\nn
& + C_2 \sqrt{t}
\left(-\frac{2^{\frac{n}{2}+\frac{1}{2}} \Gamma
\left(\frac{n+1}{2}\right) (k t)^{-\frac{n}{2}-\frac{1}{2}}}{\pi}
 -\frac{2^{-\frac{n}{2}-\frac{1}{2}} \cos \left(\frac{1}{2} \pi
(n+1)\right) \Gamma \left(\frac{1}{2} (-n-1)\right)
(k t)^{\frac{n}{2}+\frac{1}{2}}}{\pi }\right)\, ,
\end{align}
so by keeping the most dominant term, we finally have,
\begin{equation}
\label{dominanthubbleev}
v_k(t)=\frac{C_2 \sqrt{t} \left(2^{\frac{n}{2}+\frac{1}{2}} \Gamma
\left(\frac{n+1}{2}\right) (k
t)^{-\frac{n}{2}-\frac{1}{2}}\right)}{\pi }\, .
\end{equation}
The power spectrum of the primordial curvature perturbations can be
written in terms of the function $v_k$ and $z$ as follows,
\begin{equation}
\label{powerspectrmrelation}
\mathcal{P}_{\zeta}=\frac{k^3}{2\pi^2}\left|
\frac{v_k}{z} \right|_{k=a H}^2\, ,
\end{equation}
which is evaluated at the horizon crossing $k=aH$. So our aim is to
express the quantity $|\frac{v_k}{z}|^2$ in terms of the wavenumber
$k$. We have already calculated the function $v_k(t)$ which appears
in Eq.~(\ref{dominanthubbleev}), however the integration constant
$C_2$ has a hidden $k$-dependence, which can be determined if we
assume that at some initial time $t_0$, the function $v_k(t_0)$ is
described by a Bunch-Davies vacuum, that is $v_k\simeq
\frac{\e^{-ikt}}{\sqrt{2k}}$, so in this way we obtain that the
constant $C_2$ is equal to,
\begin{equation}
\label{constantc4}
C_2\simeq \frac{\pi 2^{-\frac{n}{2}-\frac{1}{2}} k^{n/2}
t^{n/2}}{\Gamma \left(\frac{n+1}{2}\right)}\, .
\end{equation}
Also the cosmic time in Eqs.~(\ref{dominanthubbleev}) and
(\ref{constantc4}) can be expressed in terms of the wavenumber $k$,
by using the fact that the power spectrum is evaluated at the
horizon crossing, so when $k=aH$, and since $a\sim 1$ for the
intermediate inflation scenario, we obtain,
\begin{equation}
\label{inthorcross}
t\simeq \frac{k^{\frac{1}{n-1}}}{(A n)^{\frac{1}{n-1}}}\, .
\end{equation}
Therefore, by combining Eqs.~(\ref{zrt}), (\ref{dominanthubbleev}),
(\ref{constantc4}), and (\ref{inthorcross}), the power spectrum of
Eq.~(\ref{powerspectrmrelation}) becomes,
\begin{equation}
\label{powerspectrumfinalrelationforft}
\mathcal{P}_{\zeta}\simeq \frac{\left(n^{1-\frac{n}{n-1}}
A^{1-\frac{n}{n-1}}\right) k^{\frac{1}{n-1}+3}}{4 \pi ^2 (1-n)}\, ,
\end{equation}
which is not scale invariant. However, the spectral index can be
compatible with the 2015 Planck data of Eq.~(\ref{planckdata}), by
suitably choosing the parameter $n$. Indeed, the spectral index of
the power spectrum is,
\begin{equation}
\label{spcrelation}
n_s-1=\frac{d \ln \mathcal{P}_{\zeta}}{d \ln k}\, ,
\end{equation}
so in the case at hand, the spectral index is,
\begin{equation}
\label{nsfinalforft}
n_s=\frac{1}{n-1}+4\, ,
\end{equation}
so by choosing $n=0.670576$, the spectral index becomes
$n_s=0.9644$, which is identical with the 2015 Planck constrains on
the spectral index.

\subsection{Singular Inflation}

Recently, another type of inflationary dynamics was introduced in
the relevant literature, under the name ``singular inflation''
\cite{Barrow:2015ora,Nojiri:2015fra}, where the terminology singular
refers to a soft singularity type which occurs during the
inflationary era. The most interesting type of singularity is the
Type IV singularity, which according to the classification of
singularities we did in section II-A-3, the Type IV singularity is
not of crushing type and the Universe may smoothly pass through it
without having catastrophic consequences on physical observable
quantities. These singularities were studied in the literature in
various contexts, for example in the context of scalar-tensor
gravity, see
Refs.~\cite{Barrow:2015ora,Odintsov:2015jca,Odintsov:2015tka,
Nojiri:2015fia,Nojiri:2015fra},
while in the context of modified gravity in general, see
\cite{Odintsov:2015gba,Odintsov:2016plw,Oikonomou:2015qfh,
Nojiri:2015qyc,Odintsov:2015zza,Kleidis:2016vmd}.
Also when inhomogeneous equation of state fluids are considered, see
\cite{Nojiri:2015wsa}.

In this section, we shall present how a Type IV singularity can be
incorporated to the $R^2$ model, and the details on this issue can
be found in Ref.~\cite{Odintsov:2015gba}. As it was shown in
\cite{Odintsov:2015gba}, and we now briefly discuss, the
incorporation of a Type IV singularity during the slow-roll era of
the $R^2$ model, leads to dynamical instabilities in the second
slow-roll index, which indicates that the exit from inflation might
occur due to these instabilities. We can easily incorporate the Type
IV singularity in the $R^2$ model if the quasi-de Sitter Hubble rate
of the $R^2$ model is modified as follows,
\begin{equation}
\label{singstarobhub}
H(t)\simeq H_i-\frac{M^2}{6}\left ( t-t_i\right )+f_0\left (t-t_s
\right)^{\alpha} \, ,
\end{equation}
where the singularity occurs at $t=t_s$. The singularity structure
which the Hubble rate (\ref{singstarobhub}) implies is the
following,
\begin{itemize}\label{lista}
\item $\alpha<-1$ corresponds to the Type I singularity.
\item $-1<\alpha<0$ corresponds to Type III singularity.
\item $0<\alpha<1$ corresponds to Type II singularity.
\item $\alpha>1$ corresponds to Type IV singularity.
\end{itemize}
Therefore, in order for a Type IV singularity to occur, we must
require that $\alpha>1$. Moreover, we assume that $H_i\gg f_0$,
$M\gg f_0$ and in addition that $f_0\ll 1$. When these conditions
hold true, the singularity term in Eq.~(\ref{singstarobhub}), is
much more smaller when it is compared the first two terms. In
effect, near the singularity, the $F(R)$ gravity that generates the
Hubble rate (\ref{singstarobhub}) is the following,
\begin{equation}
\label{starobfr}
F(R)=R+\frac{1}{6M^2}R^2\, ,
\end{equation}
with $M\gg 1$. By calculating the Hubble flow parameters
\cite{Odintsov:2015gba} for the Hubble rate (\ref{singstarobhub}),
these read,
\begin{align}
\label{singstarohubflow}
\epsilon_1=& -\frac{-\frac{M^2}{6}+f_0 (t-t_s)^{-1+\alpha }
\alpha}{\left(H_i-\frac{1}{6} M^2 (t-t_i)+f_0 (t-t_s)^{\alpha }\right)^2}\, , \nn
\epsilon_3=& \frac{f_0 (t-t_s)^{-2+\alpha } (-1+\alpha )
\alpha +4 \left(H_i-\frac{1}{6} M^2 (t-t_i)+f_0 (t-t_s)^{\alpha
}\right) \left(-\frac{M^2}{6}+f_0 (t-t_s)^{-1+\alpha } \alpha
\right)}{M^2 \left(1+\frac{2 \left(-\frac{M^2}{6}+2
\left(H_i+\frac{1}{6} M^2 (-t+t_i)\right)^2\right)}{M^2}\right)
\left(H_i-\frac{1}{6} M^2 (t-t_i)+f_0 (t-t_s)^{\alpha }\right)} \, ,\nn
\epsilon_4=&\frac{4 f_0 H(t) (t-t_s)^{-2+\alpha } (-1+\alpha
) \alpha +f_0 (t-t_s)^{-3+\alpha } (-2+\alpha ) (-1+\alpha ) \alpha
+4 \left(-\frac{M^2}{6}+f_0 (t-t_s)^{-1+\alpha } \alpha
\right)^2}{H(t) \left(f_0 (t-t_s)^{-2+\alpha } (-1+\alpha ) \alpha
+4 H(t) \left(-\frac{M^2}{6}+f_0 (t-t_s)^{-1+\alpha } \alpha
\right)\right)}\, .
\end{align}
and near the Type IV singularity at $t=t_s$, these become,
\begin{align}
\label{singstarohubflow1}
\epsilon_1=&\frac{M^2}{6 \left(H_i-\frac{1}{6} M^2 (t-t_i)\right)^2}\, , \nn
\epsilon_3=& \frac{f_0 (t-t_s)^{-2+\alpha } (-1+\alpha )
\alpha +4 \left(H_i-\frac{1}{6} M^2 (t-t_i)\right)
\left(-\frac{M^2}{6}\right)}{M^2 \left(1+\frac{2
\left(-\frac{M^2}{6}+2 \left(H_i+\frac{1}{6} M^2
(-t+t_i)\right)^2\right)}{M^2}\right) \left(H_i-\frac{1}{6} M^2
(t-t_i)\right)} \, , \nn
\epsilon_4=&\frac{\frac{M^4}{9}+4 f_0
\left(H_i-\frac{1}{6} M^2 (t-t_i)\right) (t-t_s)^{-2+\alpha }
(-1+\alpha ) \alpha +f_0 (t-t_s)^{-3+\alpha } (-2+\alpha )
(-1+\alpha ) \alpha }{\left(H_i-\frac{1}{6} M^2 (t-t_i)\right)
\left(-\frac{2}{3} M^2 \left(H_i-\frac{1}{6} M^2 (t-t_i)\right)+f_0
(t-t_s)^{-2+\alpha } (-1+\alpha ) \alpha \right)}\, .
\end{align}
As it can be seen from the equations above, the parameter
$\epsilon_1$ is identical to the one corresponding to the $R^2$
model, and the difference can be found when the parameters
$\epsilon_3$ and $\epsilon_4$ are considered. In the case of
$\epsilon_3$ the difference is due to the presence of the term $\sim
(t-t_s)^{-2+\alpha }$ which may render the parameter $\epsilon_3$
singular, only in the case $1<\alpha<2$, at $t=t_s$. In the case of
$\epsilon_4$, the terms $\sim (t-t_s)^{-3+\alpha }$ and
$(t-t_s)^{-2+\alpha }$, may become singular. So we have three
possible scenarios which may occur, and now we discuss these in
brief.

\subsubsection{Scenario I}

If the singularity occurs before inflation ends, which occurs at
$t=t_f$, then we have $t_s<t_f$, and if $2<\alpha <3$, the slow-roll
parameter $\epsilon_4$ has a singularity at $t=t_s$, and it is
simplified as follows,
\begin{equation}
\label{epsilon4sdforscenar}
\epsilon_4\simeq -\frac{3 \left(\frac{M^4}{9}+f_0 (t-t_s)^{-3+\alpha
} (-2+\alpha ) (-1+\alpha ) \alpha \right)}{2 M^2
\left(H_i-\frac{1}{6} M^2 (t-t_i)\right)^2}\, ,
\end{equation}
while the rest slow-roll parameters are finite. As it was argued in
Ref.~\cite{Odintsov:2015gba}, this singularity indicates that the
slow-roll expansion breaks at a higher order, so this can be viewed
as a dynamical instability of the system. Therefore, this mechanism
may serve as a mechanism for generating the graceful exit from
inflation. Also by calculating the spectral index of primordial
curvature perturbations $n_s$ and the scalar-to-tensor ratio $r$, in
the case at hand, we get,
\begin{equation}
\label{hubslowatts}
\epsilon_H(t_s)=\frac{M^2}{6 \left(H_i-\frac{1}{6} M^2
(t_s-t_i)\right)^2}\, ,\quad \eta_H(t_s)=0\, ,
\end{equation}
where $N$ is calculated in the interval $[t_i,t_s]$
\begin{equation}
\label{nforsingscenarioi}
N=\int_{t_i}^{t_s}H(t)d t=H_i(t_s-t_i)-\frac{M^2}{12}\left(
t_s-t_i\right)^2\, ,
\end{equation}
and upon solving with respect to the difference $(t_s-t_i)$, we
obtain,
\begin{equation}
\label{tsti}
t_s-t_i=\frac{2 \left(3 H_i+\sqrt{3} \sqrt{3 H_i^2-M^2
N}\right)}{M^2}\, .
\end{equation}
Finally by substituting in Eq.~(\ref{hubslowatts}), we get,
\begin{equation}
\label{hubslowatt123s}
\epsilon_H(t_s)=\frac{M^2}{6 H_i^2-2 M^2 N}\, ,\quad
\eta_H(t_s)=0\, ,
\end{equation}
which are identical to $R^2$ inflation model ones.

\subsubsection{Scenario II}

In the case $t_s<t_f$, and if $1<\alpha<2$, the second Hubble
slow-roll parameter $\eta_H$ becomes,
\begin{equation}
\label{gdhdtf}
\eta_H\simeq -\frac{f_0 (t-t_s)^{-2+\alpha } (-1+\alpha ) \alpha }{2
\left(H_i-\frac{1}{6} M^2 (t-t_i)+f_0 (t-t_s)^{\alpha }\right)
\left(-\frac{M^2}{6}+f_0 (t-t_s)^{-1+\alpha } \alpha \right)}\, .
\end{equation}
In the present case, the parameters $\epsilon_3$, $\epsilon_4$ and
$\eta_H$ diverge at $t=t_s$, which indicates a strong instability at
this time instance, and therefore inflation ends abruptly at that
point.

\subsubsection{Scenario III}

In the case that $t_s<t_f$ and $\alpha>3$, all the parameters
$\epsilon_3$, $\epsilon_4$ are not singular at $t=t_s$, and
therefore inflation ends when $\epsilon_1$ becomes of order one. The
difference of this scenario with the ordinary $R^2$ model is that in
the case at hand, the second slow-roll parameter is not exactly
zero, but it is equal to,
\begin{equation}
\label{fgeddsecondhubinde}
\eta_H\simeq -\frac{f_0 (t_f-t_s)^{-2+\alpha } (-1+\alpha ) \alpha
}{2 \left(H_i-\frac{1}{6} M^2 (t_f-t_i)+f_0 (t_f-t_s)^{\alpha
}\right) \left(-\frac{M^2}{6}+f_0 (t_f-t_s)^{-1+\alpha } \alpha
\right)}\, .
\end{equation}
Hence, in this case, the spectral index $n_s$ is equal to,
\begin{equation}
\label{secanrioiiobservindex}
n_s=-\frac{9 f_0 (t_f-t_s)^{-1+\alpha } (-1+\alpha ) \alpha
}{\left(-\sqrt{9 H_i^2-3 M^2 N}+3 f_0 (t_f-t_s)^{\alpha }\right)
\left(M^2 (-t_f+t_s)+6 f_0 (t_f-t_s)^{\alpha } \alpha \right)} \, .
\end{equation}
However, the observational differences between this scenario and of
the Starobinsky model are minor, as was shown in
\cite{Odintsov:2015gba}.

\subsubsection{The Scenarios with $t_s=t_f$}

Finally, let us consider in brief the case that the singularity
occurs at the end of inflation, so that we have $t_s=t_f$. If
$2<\alpha<3$, the Hubble flow parameter $\epsilon_4$ diverges at
$t=t_f$. This case is similar to Scenario I, so inflation ends when
$\epsilon_1\sim 1$, but since $\epsilon_4$ diverges at $t=t_f=t_s$,
the graceful exit from inflation becomes more pronounced in
comparison to Scenario I. The same situation occurs in the case
$1<\alpha<2$, since the slow-roll indices $\epsilon_3$, $\epsilon_4$
and $\eta_H$ diverge at $t=t_s$. In the case that $\alpha>3$, the
situation is similar to Scenario III, with the only difference being
that the second Hubble slow-roll index is zero in the case at hand.
Hence this case is indistinguishable for the ordinary $R^2$ model.

In conclusion, the Type IV singular $R^2$ model has the advantage
that inflation ends more abruptly in comparison to the ordinary
$R^2$ model. Also, although the singularity term is insignificant at
early times, it becomes dominant at late times, so this term can be
responsible for late-time acceleration, see \cite{Odintsov:2015gba}
for details. Moreover we need to note that the $F(R)$ gravity of
Eq.~(\ref{starobfr}) generates the singular Hubble rate of
Eq.~(\ref{singstarobhub}) only at early times, so at late times this
description no longer holds true, and therefore a different
description applies, as was shown in detail in
Ref.~\cite{Odintsov:2015gba}.

\subsection{A Qualitative Study for the Graceful Exit from Inflation Issue}

In the previous sections we discussed the inflationary dynamics in
various modified gravities, and as we showed, in many cases, the
graceful exit from inflation quantitatively occurs when the first
slow-roll index becomes of the order $\mathcal{O}(1)$. In the cases
of $F(R)$ gravity and also in the case of a single scalar field this
is easy to address in a quantitative way. However, in more
complicated cases, like for example in the mimetic $F(R)$ gravity,
this problem is difficult to handle with, by using conventional
approaches. In this section we shall discuss the graceful exit issue
in the context of complicated theories. The formalism we shall
develop is quite general and can be applied in various theories of
gravity. In order to make the presentation more clear and for
illustrative purposes, we shall discuss the graceful exit problem in
the context of the mimetic $F(R)$ gravity we discussed earlier in
this chapter. Specifically, we consider the power-law model of
Eq.~(\ref{frpowerlaw}), so the Lagrange multiplier and the mimetic
potential are the ones appearing in Eqs.~(\ref{lagrafrsimple}) and
(\ref{mimeticpotfrpowerlaw}). As we showed, cosmologically viable
results for this specific model are obtained when $n=1.8$, and as we
show this crucially affects the graceful exit from inflation.

The formalism for the graceful exit from inflation that we now
discussed, is based on the existence of the unstable de Sitter
attractors in the theory. These unstable de Sitter vacua generate
curvature perturbations which grow with time, so this mechanism
triggers the graceful exit from inflation. More theoretical details
on this issue can be found in
Refs.~\cite{Odintsov:2015wwp,Bamba:2014jia}. To proceed with the
method, the gravitational equation for mimetic $F(R)$ gravity can be
written as follows,
\begin{equation}
\label{diffgrexit}
6 \alpha n H(t)^2-(\alpha n -\alpha )\left(12
H(t)^2+6H'(t)\right) +6 \alpha n (n-1)(24 H(t) H'(t)+6
H''(t))-\frac{V(t)-2\lambda (t)}{12 H(t)^2+6H'(t)}=0\, .
\end{equation}
We investigate whether the de Sitter solutions to the above equations
exist, so we seek solutions of the form $H(t)=H_d$, and by
substituting this in (\ref{diffgrexit}), we get the following
algebraic equation,
\begin{equation}
\label{gr1}
6 H_d^2 n \alpha -12 H_d^2 (-\alpha +n \alpha
)-12^{-2+n} H_d^{-2+2 n} (-2 (-1+n) \alpha +n \alpha )=0\, ,
\end{equation}
which when solved, yields,
\begin{equation}
\label{desittersolution}
H_d=\left(2^{-5+2 n}
3^{-3+n}\right)^{\frac{1}{4-2 n}}\, .
\end{equation}
Having found the de Sitter vacuum for the theory, we need to
investigate whether this is unstable or not, towards linear
perturbations, so we set $H(t)=H_d+\Delta H(t)$ in
Eq.~(\ref{diffgrexit}, and we keep only linear terms in $\Delta H(t)$,
$\Delta H'(t)$ and $\Delta H''(t)$. After expanding and keeping only
linear terms in the aforementioned variables, we obtain,
\begin{align}
\label{resultingdiffeqn}
& \frac{1}{288 H_d^2}\alpha \left(-4 H_d
\left(864 H_d^2+12^n H_d^{2 n}\right) (-2+n) \Delta H(t)\right. \nn
& +\left(-12^n H_d^{2 n} (-2+n)-1728 H_d^2 (-1+n)+41472 H_d^3 (-1+n)
n+3^{1+n} 4^{2+n} H_d^{1+2 n} (-1+n) n\right) \Delta H'(t) \nn
& \left.+2 H_d \left(\left(-864 H_d^2+12^n H_d^{2 n}\right)
(-2+n)+5184 H_d (-1+n) n \Delta H''(t)\right)\right)=0\, .
\end{align}
This differential equation can be solved analytically, so we obtain,
\begin{equation}
\label{desitterunstablesol}
\Delta H(t)=-\frac{864 H_d^2-2^{2 n}
3^n H_d^{2 n}}{2 \left(864 H_d^2+2^{2 n} 3^n H_d^{2 n}\right)} +c_1
\e^{\mu_1t}+c_2 \e^{\mu_2 t}\, ,
\end{equation}
where the detailed form of the
parameters $\mu_1$ and $\mu_2$ is,
\begin{equation}
\label{mu1andmu2}
\mu_1=\frac{q_1+\sqrt{q_2}}{2 \left(-10368 H_d^2 n+10368 H_d^2
n^2\right)}\, ,\quad \mu_2=\frac{q_1-\sqrt{q_2}}{2 \left(-10368
H_d^2 n+10368 H_d^2 n^2\right)}\, ,
\end{equation}
where $q_1$ stands for,
\begin{align}
\label{q1}
q_1=& -1728 H_d^2-2^{1+2 n} 3^n H_d^{2 n}+1728 H_d^2
n+41472 H_d^3 n \nn & +2^{2 n} 3^n H_d^{2 n} n+2^{4+2 n} 3^{1+n}
H_d^{1+2 n} n-41472 H_d^3 n^2-2^{4+2 n} 3^{1+n} H_d^{1+2 n} n^2\, ,
\end{align}
while $q_2$ stands for,
\begin{align}
\label{q2}
q_2 = & -4 \left(6912 H_d^3+2^{3+2 n} 3^n H_d^{1+2
n}-3456 H_d^3 n-2^{2+2 n} 3^n H_d^{1+2 n} n\right) \left(-10368
H_d^2 n+10368 H_d^2 n^2\right) \nn & +\left(1728 H_d^2+2^{1+2 n} 3^n
H_d^{2 n}-1728 H_d^2 n-41472 H_d^3 n-2^{2 n} 3^n H_d^{2 n} n-2^{4+2
n} 3^{1+n} H_d^{1+2 n} n \right. \nn & \left. +41472 H_d^3
n^2+2^{4+2 n} 3^{1+n} H_d^{1+2 n} n^2\right)^2 \, .
\end{align}
and also $c_1$, $c_2$ are arbitrary integration constants. So by
combining Eqs.~(\ref{desittersolution}) and
(\ref{desitterunstablesol}), for $n=1.8$, we have,
\begin{equation}
\label{explicitsol1}
\Delta H(t)=-0.499989+c_1 \e^{0.0775432t}+c_2
\e^{0.00195451 t}\, ,
\end{equation}
which tells us that the perturbation
$\Delta H(t)$ is exponentially growing with time. It is more
convenient and useful to have the solution (\ref{explicitsol1}) in
terms of the $e$-foldings number, so now we demonstrate how this can
be done formally. Assume that the graceful exit occurs at the time
instance $t_f$, so from the condition $\epsilon_1(t_f)\sim
\mathcal{O}(1)$ and from (\ref{hubblefrpowerlaw}), we obtain
$H_f^2\simeq \frac{M^2}{6}$. Therefore, at $t=t_f$, the quasi-de
Sitter evolution of (\ref{hubblefrpowerlaw}), yields,
\begin{equation}
\label{hubblecrevis}
H_f=H_0-\frac{M^2}{6}(t_f-t_0)\, .
\end{equation}
By
solving with respect to $t_f-t_0$, and also by substituting
$H_f^2\simeq \frac{M^2}{6}$ we obtain,
\begin{equation}
\label{fgrdsol}
t_f-t_0\simeq \frac{6}{M^2}\left(H_0-\frac{M}{\sqrt{6}} \right)\, .
\end{equation}
The definition of the $e$-foldings number $N$ is,
\begin{equation}
\label{defrev}
N=\int_{t_0}^{t_f}H(t)d t\, .
\end{equation}
By choosing the
initial time $t_0$ to be the time instance that the horizon crossing
occurs, then we have from (\ref{defrev}),
\begin{equation}
\label{nrev}
N=H_0(t_f-t_0)-\frac{M^2}{12}(t_f-t_0)^2\, ,
\end{equation}
and by substituting
$(t_f-t_0)$ given in Eq.~(\ref{fgrdsol}), we obtain,
\begin{equation}
\label{finalalmostrev}
N\simeq \frac{3 H_0^2}{M^2}-\frac{1}{2}\, .
\end{equation}
Combining Eqs.~(\ref{fgrdsol}) and (\ref{finalalmostrev}), we
get,
\begin{equation}
\label{dfrfefinal}
t_f-t_i\simeq
\frac{2(N+\frac{1}{2})}{H_0}-\frac{6}{M\sqrt{6}}\, ,
\end{equation}
and by
substituting $(t_f-t_0)$ in Eq.~(\ref{explicitsol1}), we get,
\begin{equation}
\label{explicitsol1new}
\Delta H(N)=-0.499989+c_1 \e^{0.0775432
\left(\frac{2(N+\frac{1}{2})}{H_0}-\frac{6}{M\sqrt{6}}\right)}+c_2
\e^{0.00195451\left(\frac{2(N+\frac{1}{2})}{H_0}
 -\frac{6}{M\sqrt{6}}\right)}\, .
\end{equation}
Having the resulting perturbation $\Delta H$ as a function of $N$,
enables us to decide if enough inflation is produced for at least
$N\sim 50$. For the example at hand, if we choose
$H_0=12.04566$ and $M=2$, we obtain,
\begin{align}
\Delta
H(N)=&-0.499989+c_1\exp\left\{0.0775432\left[-\sqrt{\frac{3}{2}}
+0.166035\left(\frac{1}{2}+N\right)\right]\right\} \nn
&+c_2\exp\left\{0.00195451\left[-\sqrt{\frac{3}{2}}
+0.166035\left(\frac{1}{2}+N
\right)\right]\right\}\, . \nonumber
\end{align}
In effect, even for small values of $N$, we find $\Delta H\sim
-0.5+0.9c_1+1.0c_2$, and therefore unless $c_1$ and $c_2$, are
chosen to be extremely small, the perturbation $\Delta H$ is of the
order $\mathcal{O}(1)$ at the beginning of inflation, so not enough
inflation is produced. As a conclusion, we need to note that the
qualitative results of the method we presented, are quite model
dependent, however, the quantitative approach is quite general, and
can be applied in quite complicated modified gravity theories.

Before closing it is worth discussing an important question related
to the graceful exit issue in different frames. The question is
whether the Universe can exit from the inflationary era in both of the
Jordan and Einstein frames, and if yes what is the physical
reasoning. The answer to this question is not easy to address
because this issue is strongly model dependent. It is known that
conformal invariant quantities have the same description in both of the
Jordan and Einstein frames, for example the observational indices of
inflation \cite{Domenech:2016yxd,Kaiser:1995nv,Faraoni:2007yn}, that
is the spectral index $n_s$ and the scalar-to-tensor ratio $r$, so
only the conformal invariant quantities are the same in the Jordan and
Einstein frame.

The point is that for slow-roll inflation, the Einstein and the
Jordan frame give the same indexes. Then, the graceful exit occurs
in both of the Einstein and Jordan frames. For example in the vacuum $R^2$
gravity the graceful exit occurs in the $F(R)$ frame and at the same
time it occurs in the Einstein frame \cite{Bezrukov:2007ep}.
However, the two frames are effectively equivalent only in the
vacuum case only. When we have matter fluids present, due to the
fact that $F(R)$ gravity with matter is not equivalent to the
Einstein frame with same matter fluids (rather it is the Einstein frame
with matter coupled to a scalar field), then we expect that there is
a difference between the two frames, and the impact of the
difference can be seen in the graceful exit issue too. In this
section we assumed that the physical theory is the $F(R)$ gravity
one, so we focused on the $F(R)$ gravity frame only.


Differences between the Jordan and Einstein frames were pointed out in
the literature. As it was shown in \cite{Bahamonde:2017kbs}, the
acceleration in one frame corresponds to deceleration in the other
frame, so the physical interpretation in the two frames is totally
different. However the graceful exit issue is more involved and we
remain skeptical on answering the question with a simple yes or no,
in the case that matter fluids are present. Traditionally, the exit
is triggered when the first slow-roll index becomes of the order
one, however this ends the slow-roll era but not inflation to our
opinion. Practically, in order for the inflationary era to come to
an end, the unstable de Sitter solution must exist in the theory and
if this exists, it has to be the final attractor of the cosmological
dynamical system. The de Sitter instabilities generate
uncontrollable curvature perturbations which cease the inflationary
era. According to Refs.~\cite{Bahamonde:2017kbs}, the de Sitter
solution in the Jordan frame may not be the de Sitter solution in the
other frame, when matter fluids are present, and in fact it may
describe a totally new physical evolution. So we refrain from going
into deeper details for this intriguing issue and we defer the study
to a focused future research paper.

Thus we presented the various models in which inflation can be
realized in the context of modified gravity. Basically speaking, all
modified gravities we presented may describe the inflationary era of
the sort appearing in this chapter. What now remains is to briefly
discuss the reheating issue in the context of $F(R)$ gravity. This
is the subject of the next section.

\subsection{Reheating in $F(R)$ Gravity}

Describing successfully the inflationary era is one step towards the
understanding of early-time dynamics, however the inflationary era
must be followed by the Friedmann evolution, as this is described by
standard Big-Bang cosmology. The reheating period fills the gap
between inflation and the radiation and matter domination era, and
it is actually the most important mechanism in order to thermalize
the cold Universe that results after the inflationary era. However,
it is conceivable that the full understanding of how the Universe
transits from a supercooled state at the end of inflation, to a
thermal radiation dominated epoch, is a very complicated task.
Actually, the conceptual understanding of the Universe's evolution
in the era between the ending of the inflationary era and Big Bang
Nucleosynthesis, is still an active research stream
\cite{Amin:2014eta}. In some sense, during this era, the energy that
drives the quasi-de Sitter inflationary evolution should be endowed
to the matter fields present, which are Standard Model particles.
Now in most descriptions, the inflaton directly transfers its energy
to the Standard Model particles
\cite{Kofman:1997yn,Greene:1997fu,Kofman:1994rk}, however there are
alternative ways in order for this reheating process to occur, such
as the modified gravity description of reheating
\cite{Mijic:1986iv}.

Although the reheating study can go quite deep in theoretical
descriptions, for example considering non-perturbative resonances as
preheating sources \cite{Amin:2014eta}, in this section we shall
discuss in brief this thermalization era, and we shall try to
outline the basic facts about reheating, always in the context of
modified gravity. Particularly, we shall focus on the $F(R)$ gravity
theory description of reheating, since this is the prominent and
most representative modified gravity theory. The modified gravity
description offers a more refined reheating theory since the
reheating occurs as a back-reaction effect on the metric, due to
curvature fluctuations acting as matter sources in the field
equations. This description is more refined in comparison to
standard inflaton approaches, since in the modified case there is no
need to fine-tune the couplings of the inflation to the Standard
Model fields, in order for these to be extremely suppressed during
the inflationary era.

We shall focus on the Starobinsky $R^2$ model we used earlier,
\begin{equation}
\label{starobinskyfr}
F(R)=R+\frac{R^2}{36H_i}\, ,
\end{equation}
where $H_i$ has dimensions of mass$^2$, and as we will see will play
an important role in determining the reheating and Friedman
temperature. We shall base our analysis on Ref.~\cite{Mijic:1986iv}.
Let us note here that by using the terminology ``Friedman
evolution'' or ``Friedman temperature'', we refer to the era that
the Big Bang description for the Universe evolution apply, so this
era refers to the radiation domination era and after. During the
reheating era, the first term of the second differential equation in
Eq.~(\ref{frweqnsr2}), namely $\ddot{R}$ cannot be neglected
anymore. Therefore, the evolution of the scalar curvature looks like
a damped harmonic oscillation, as the differential equation in Eq.
(\ref{frweqnsr2}) suggests, with the restoring force being $\sim
3H_i$. The parameter $H_i$ has a prominent role in the theory for
three reasons, firstly since the observational indices of inflation
strongly depend on this as we showed earlier, secondly since it
affects the reheating temperature and thirdly since it affects the
Friedman temperature and evolution. With regard to the reheating
temperature, it has to be high enough in order for the baryogenesis
to occur, but also it has to be low enough in order for the grand
unified theories phase transitions not to occur, in which case the
monopoles will be absent. Hence the parameter $H_i$ is strongly
constrained by the above requirements as we show later on in this
section.

The evolution of the Hubble rate is given in the first differential
equation of Eq.~(\ref{frweqnsr2}), which we quote here too for
reading convenience,
\begin{equation}
\label{maindiffhubble}
\ddot{H}-\frac{\dot{H}^2}{2H}+3H_iH=-3H\dot{H}\, .
\end{equation}
During the slow-roll era, the first and second term in
Eq.~(\ref{maindiffhubble}) can be neglected, due to the slow-roll
conditions and since the following condition holds true,
\begin{equation}
\label{condition1}
\left| \frac{\dot{H}^2}{2H} \right| \ll \left| 3H\dot{H} \right|\, ,
\end{equation}
so as we showed earlier, the evolution during the slow-roll era is
the quasi-de Sitter evolution given in
Eq.~(\ref{quasievolbnexampler2}). When the condition (\ref{condition1})
ceases to hold true, then the reheating phase begins, which is an
oscillatory era, in which case the terms $\sim \ddot{H}$ and $\sim
\frac{\dot{H}^2}{2H}$ dominate, but the term $\sim \dot{H}H$ becomes
subdominant. Then by solving the differential equation
(\ref{condition1}), we obtain the following solution,
\begin{equation}
\label{htsolutionreheating}
H(t)\simeq \frac{\cos^2\omega
(t-t_r)}{\frac{3}{\omega}+\frac{3}{4}(t-t_r)+\frac{3}{8\omega}\sin
2\omega (t-t_r)}\, ,
\end{equation}
where $t=t_r$ is the time instance that the quasi-de Sitter phase
stops and the reheating era starts, which occurs when the following
equality holds true,
\begin{equation}
\label{condition1equal}
\left|\frac{\dot{H}^2}{2H}\right| = \left| 3H\dot{H} \right|\, .
\end{equation}
Also the scale factor during reheating is equal to,
\begin{equation}
\label{reheatingscalefactor}
a(t)=a_r\left(1+\frac{\omega (t-t_r)}{4} \right)^{2/3}\, ,
\end{equation}
where $a_r$ is equal to
$a_r=a_0 \e^{\frac{H_0^2}{2H_i}-\frac{1}{12}}$, and $a_0$ is the scale
factor corresponding to the onset of inflation. The parameter
$\omega$ appearing in Eq.~(\ref{htsolutionreheating}) can be found
by using the equation (\ref{condition1equal}) and matching the quasi
de-Sitter solution with the solution (\ref{htsolutionreheating}) at
$t=t_r$, so the parameter $\omega$ and the reheating time $t=t_r$
are,
\begin{equation}
\label{omegareheating}
\omega=\sqrt{\frac{3H_i}{2}}\, \quad t_r\simeq H_i H_0\, ,
\end{equation}
where $H_0$ is the Hubble rate at the onset of inflation. During the
reheating phase, the scalar curvature is approximately equal to
$R\simeq 6\dot{H}$, and this is to be contrasted with the slow-roll
era, in which case the scalar curvature is approximately equal to
$R\sim 12H^2$. Hence, during the reheating era, the scalar curvature
is approximately equal to,
\begin{equation}
\label{approxscalarcurvaturereheating}
R(t)\simeq -\frac{6\omega \sin 2\omega
(t-t_r)}{\left(\frac{3}{\omega}+\frac{3}{4}(t-t_r)+\frac{3}{8\omega}\sin
2\omega (t-t_r) \right)}\, .
\end{equation}
It is conceivable that the scalar curvature appearing in Eq.
(\ref{approxscalarcurvaturereheating}) is an approximation and not
an exact solution of the differential equation (\ref{frweqnsr2}).
Also notice how the parameter $H_i$ affects the resulting scalar
curvature and Hubble rate via the parameter $\omega$, and it affects
the shape and size of the reheating phase.

During the reheating phase, the matter fields present will be
excited and the Universe will be reheated. Let us consider the
simplest case of a matter field by considering the effects of
curvature oscillations on a scalar field $\phi$, with equation of
motion $g^{\mu \nu}\phi_{;\mu \nu}=0$. The energy density of this
scalar field strongly depends on the averaged square of the scalar
curvature as it was shown in \cite{Mijic:1986iv}, and the evolution
of the energy density is determined by the following differential
equation,
\begin{equation}
\label{liatdiff}
\frac{ d \rho}{ d t}=-4\rho H+\frac{\omega
\bar{R}^2}{1152\pi}\, .
\end{equation}
The differential equation above can determine the way that matter
fields are affected by the curvature oscillations, and when the
reheating era ends, the last term in (\ref{liatdiff}) can be
neglected, so the differential equation becomes
$\frac{ d \rho a^4}{ d t}=0$, and the corresponding
equation of state of the matter fields is $p=\frac{1}{3}$, which
perfectly describes a radiation domination era. However, during the
reheating era, the equation of state of the matter field becomes,
\begin{equation}
\label{eqnofstatereheating}
p=\frac{1}{3}\rho-\frac{\omega \bar{R}^2}{1152\pi H}\, .
\end{equation}
Hence it is obvious that the curvature oscillations affect
significantly the matter content during the reheating era, and this
in turn has a back-reaction effect in the equations of motion.
Indeed, by using (\ref{eqnofstatereheating}), the curvature
evolution during reheating in the presence of matter is,
\begin{equation}
\label{frwregheating}
\ddot{R}+3HR+6H_iR=\frac{G N H_i}{8H}\omega \bar{R}^2\, ,
\end{equation}
where the factor $N$ denotes the number of matter fields that are
involved in the reheating process, excluding the massless conformal
fields which are not excited during the reheating era. Also the
$(t,t)$ component of the field equations is,
\begin{equation}
\label{eqnmotionreheatingmain}
H^2+\frac{1}{18H_i}\left(HR-\frac{R^2}{12}RH^2
\right)=\frac{8\pi}{3}G\rho_\mathrm{matter}\, ,
\end{equation}
where the term $\rho_\mathrm{matter}$ is equal to,
\begin{equation}
\label{rhomatter}
\rho_\mathrm{matter}=\frac{N}{a^4}\int_{t_r}^t\frac{\omega}{1152\pi}
\bar{R}^2a^4 d t\, .
\end{equation}
After discussing the back-reaction of the reheated matter on the
equations of motion, we can have a qualitative estimate of the
reheating temperature by using Eqs.~(\ref{reheatingscalefactor}) and
(\ref{approxscalarcurvaturereheating}), so for $t\simeq t_r+10\omega
$, we obtain $\rho \simeq 7776\times 10^{-7}H_i^2$, and the
corresponding reheating temperature is,
\begin{equation}
\label{reheatingtemperature1}
T_r\simeq 4\times 10^{17}\left(\frac{1}{36H_il_\mathrm{pl}^2}
\right)^{-1/2}\mathrm{GeV}\, ,
\end{equation}
where $l_\mathrm{pl}$ is the Planck length. The reheating temperature $T_r$
imposes further restrictions on the parameter $H_i$, which recall
that is restricted from the inflationary observational indices.
Particularly, the reheating temperature has to be large enough in
order for the baryogenesis to take place, and at the same time it
has to be smaller than the grand unified theory phase transition
temperatures, so that the theory remains monopole free.
Specifically, the reheating temperature imposes the following
constraint on the parameter $H_i$,
\begin{equation}
\label{hiconstraints}
36^{-1}10^{-15}-10^{-12}l_\mathrm{pl}^{-2}<H_i<10^{-3}l_\mathrm{pl}^{-2}\, .
\end{equation}
For cosmic times $t\gg t_r+\frac{1}{\omega}$, the energy density
$\rho_\mathrm{matter}$ becomes approximately,
\begin{equation}
\label{rhomatterasymptotically}
\rho_\mathrm{matter}\simeq
\frac{3}{5}\frac{32}{1152\pi}\frac{N\omega^3}{t-t_r}\, ,
\end{equation}
and the matter energy density term $\rho_\mathrm{matter}$ tends to zero, so
the solution of (\ref{eqnmotionreheatingmain}) tends to the
radiation domination solution. Finally, it can be shown that the
Friedman temperature is equal to,
\begin{equation}
\label{newpaok}
T_F\sim 10^{17}\left(\frac{1}{36H_il_\mathrm{pl}^2}
\right)^{-3/4}N^{-3/4}\mathrm{GeV}\, ,
\end{equation}
but no important constraint can imposed on this parameter.

\section{Late-time Dynamics and Dark Energy}

The $\Lambda$CDM model is up-to-date quite successful in fitting the
pre-2000 observational data, and specifically it predicted the
location and existence of the baryon acoustic oscillations. Since
this model is quite important in cosmology, in this chapter we shall
demonstrate that it can be realized in the context of modified
gravity. Also we shall discuss various phenomenological aspects of
the late-time evolution of the Universe, which is now known as the
dark energy era. Special emphasis shall be given on an especially
appealing feature of modified gravity, which enables one to provide
a unified theoretical description of early and late-time
acceleration. Moreover, we shall discuss in brief the possibility of
having a phantom dark energy era in the context of modified gravity,
and we shall present in brief the implications of a phantom dark
energy era in modified gravity. Finally, we shall present a vital
feature of the $F(R)$ gravity description of the late-time era,
namely, the dark energy oscillations era, by using various
matter-fluids, not necessarily of non-interacting matter. Also the
evolution of the growth index in $F(R)$ gravity is briefly addressed
too.

\subsection{$\Lambda$CDM Epoch from $F(R)$ Gravity}

In this section we shall demonstrate how the $\Lambda$CDM model can
be realized in the context of $F(R)$ gravity. This issue has been
addressed in detail in Refs.~\cite{Nojiri:2009kx,Dunsby:2010wg}. We
shall consider the approach of Ref.~\cite{Nojiri:2009kx}, so the
$\Lambda$CDM Hubble rate is equal to,
\begin{equation}
\label{lcdmhubblerate1}
\frac{3}{\kappa^2}H^2=\frac{3}{\kappa^2}H_0^2+\rho_0a^{-3}\, .
\end{equation}
We shall use the reconstruction technique we developed in section
IV-B, so it is vital to find the function $G(N)$ which is equal to
$G(N)=H^2(N)$, which in the case at hand is,
\begin{equation}
\label{gnlcdm1}
G(N)=H_0^2+\frac{\kappa^2}{3}\rho_0a_0^{-3}\e^{-3N}\, ,
\end{equation}
and by using the fact that the scalar curvature in terms of $G(N)$ is
given in Eq.~(\ref{riccinrelat}), we can express the $e$-foldings
number as a function of $R$, so we get,
\begin{equation}
\label{nrfiunctionlcdm}
N=-\frac{1}{3}
\left (
\frac{R-12H_0^2}{\kappa^2\rho_0a_0^{-3}}\right )\, ,
\end{equation}
and therefore, Eq.~(\ref{newfrw1modfrom}) becomes in our case,
\begin{equation}
\label{eqnsfrwthistime}
3(R-9H_0^2)(R-12H_0^2)F''(R)-(\frac{R}{2}
-9H_0^2)F'(R)-\frac{F(R)}{2}=0\, ,
\end{equation}
where the prime this time denotes differentiation with respect to
the scalar curvature $R$. The differential equation
(\ref{eqnsfrwthistime}) can solved analytically since it is the
hypergeometric differential equation, which has as solution the
Gauss hypergeometric function $F(\alpha,\beta,\gamma ; x)$, with the
parameters $\alpha$, $\beta$, $\gamma$ and $x$ being equal to,
\begin{equation}
\label{parameterslcdmthiscase}
\alpha+\beta=-\frac{1}{6},\quad \alpha \beta
=-\frac{1}{6},\quad \gamma=-\frac{1}{6}\, ,\quad
x=\frac{R}{3H_0^2}-3\, .
\end{equation}
The analysis performed in Ref.~\cite{Dunsby:2010wg}, has the same
result for the $F(R)$ gravity that realizes the $\Lambda$CDM model,
see \cite{Dunsby:2010wg} for details.

\subsection{$\Lambda$CDM Epoch from Modified Gauss-Bonnet Gravity}

The $\Lambda$CDM epoch can be realized by a modified Gauss-Bonnet
gravity, and we now present the essential features of this
realization. The action of Eq.~(\ref{GB1b}) can be rewritten by
introducing an auxiliary scalar field $\phi$, as follows:
\begin{equation}
\label{ma32}
S=\int d^4 x \sqrt{-g}\left[ \frac{R}{2\kappa^2}
 - V(\phi) - \xi(\phi) \mathcal{G}
+ \mathcal{L}_\mathrm{matter}\right]\, .
\end{equation}
In fact, upon variation
of the above action with respect to the auxiliary scalar field
$\phi$, yields the following algebraic equation,
\begin{equation}
\label{ma33}
0=V'(\phi) + \xi'(\phi) \mathcal{G} \, .
\end{equation}
The equation
(\ref{ma33}) can be solved with respect to the scalar field $\phi$,
to yield the function $\phi=\phi(\mathcal{G})$.

By substituting the obtained expression of $\phi(\mathcal{G})$ into
the action of Eq.~(\ref{ma32}), we reobtain the action of
Eq.~(\ref{GB1b}), where,
\begin{equation}
\label{33c}
f(\mathcal{G}) \equiv - V\left(\phi(\mathcal{G})\right)
+ \xi\left(\phi(\mathcal{G})\right)\mathcal{G}\, .
\end{equation}
We employ the
formalism of
Refs.~\cite{Nojiri:2005jg,Nojiri:2005am,Cognola:2006eg}, and we
review the reconstruction method of scalar-$f(\mathcal{G})$ gravity.
The equations of motion are obtained by the varying the action with
respect to the metric, and for the FRW Universe, these have the
following form,
\begin{align}
\label{ma34}
0=& - \frac{3}{\kappa^2}H^2 + V(\phi)
+ 24 H^3 \frac{d \xi (\phi(t))}{dt} + \rho_\mathrm{matter} \, ,\nn
0=& \frac{1}{\kappa^2}\left(2\dot H + 3 H^2 \right) - V(\phi) - 8H^2
\frac{d^2 \xi (\phi(t))}{dt^2} - 16H \dot H \frac{d\xi
(\phi(t))}{dt} - 16 H^3 \frac{d \xi (\phi(t))}{dt}
+ p_\mathrm{matter}\, ,
\end{align}
which can be easily integrated in the following way,
\begin{align}
\label{ma35}
\xi (\phi(t)) =& \frac{1}{8}\int^t dt_1 \frac{a(t_1)}{H(t_1)^2} W(t_1)\, ,\nn
V(\phi(t)) =& \frac{3}{\kappa^2}H(t)^2 - 3a(t) H(t) W(t)
+ \rho_\mathrm{matter} (t) \, ,\nn
W(t) \equiv& \int^{t} \frac{dt_1}{a(t_1)} \left( \frac{2}{\kappa^2} \dot H (t_1)
+ \rho_\mathrm{matter} (t_1) + p_\mathrm{matter} (t_1) \right)
\, .
\end{align}
We should note here, that there is no kinetic term for the auxiliary
scalar field $\phi$ in the action (\ref{ma32}), and therefore we can
redefine the scalar field $\phi$ properly, and identify $\phi$ with
the cosmological time, that is, $\phi=t$. In effect, if we consider
the following potential $V(\phi)$ and $\xi$, given in terms of a
single function $g$ as follows,
\begin{align}
\label{ma36}
V(\phi) =& \frac{3}{\kappa^2}g'\left(\phi
\right)^2 - 3g'\left(\phi\right)
\e^{g\left(\phi\right)} U(\phi)
+ \rho_\mathrm{matter} (t) \, , \nn
\xi (\phi) =& \frac{1}{8}\int^\phi d\phi_1
\frac{\e^{g\left(\phi_1\right)}}{g'(\phi_1)^2} U(\phi_1)\, ,\nn
U(\phi) \equiv& \int^\phi d\phi_1 \left( \frac{2}{\kappa^2}
\e^{-g\left(\phi_1\right)} g''\left(\phi_1\right)
+ \rho_\mathrm{matter} (t_1) + p_\mathrm{matter} (t_1) \right)
\, ,
\end{align}
we can find the following explicit solution,
\begin{equation}
\label{ma37}
a=a_0\e^{g(t)}\ \left(H= g'(t)\right)\, .
\end{equation}

The energy density $\rho_\mathrm{matter}$ and the pressure density
$p_\mathrm{matter}$ of the matter fluids present, can be often
expressed as the sum of the contributions from various different
matter fluids, with different but constant EoS parameters $w_i$.
Since the energy density $\rho_{\mathrm{matter}\, i}$ with a
constant EoS parameter $w_i$ behaves as, $\rho_{\mathrm{matter}\, i}
= \rho_{0i} a^{- 3 \left( 1 + w_i \right)}$ and also since the
pressure $p_{\mathrm{matter}\, i}$ with a constant EoS parameter
$w_i$ is given by $p_{\mathrm{matter}\, i}=w_i
\rho_{\mathrm{matter}\, i}$, we may rewrite Eq.~(\ref{ma36}) as
follows,
\begin{align}
\label{ma36b}
V(\phi) =& \frac{3}{\kappa^2}g'\left(\phi
\right)^2 - 3g'\left(\phi\right)
\e^{g\left(\phi\right)} U(\phi)
+ \sum_i \rho_{0i} a_0^{- 3 \left( 1 + w_i \right)}
\e^{- 3 \left( 1 + w_i \right) g(t)} \, , \nn
\xi (\phi) =& \frac{1}{8}\int^\phi d\phi_1
\frac{\e^{g\left(\phi_1\right)}}{g'(\phi_1)^2} U(\phi_1)\, ,\nn
U(\phi) \equiv& \int^\phi d\phi_1 \left( \frac{2}{\kappa^2}
\e^{-g\left(\phi_1\right)} g''\left(\phi_1\right)
+ \sum_i \left(1 + w_i \right) \rho_{0i} a_0^{- 3 \left( 1 + w_i \right)}
\e^{- 3 \left( 1 + w_i \right) g(t_1)} \right)\, ,
\end{align}
where we have used the expression in Eq.~(\ref{ma37}).

Instead of the cosmic time $t$, we can use the e-folding number $N$,
and we may rewrite Eq.~(\ref{ma36}) as follows,
\begin{align}
\label{ma38}
\xi (\phi(N)) =& \frac{1}{8}\int^N dN_1 \frac{\e^{N_1}}{H(N_1)^3}
\tilde W(N_1) \, ,\nn
V(\phi(N)) =& \frac{3}{\kappa^2}H(N)^2 - 3\e^N H(N) \tilde W(N)
+ \sum_i \rho_{0i} a_0^{- 3 \left( 1 + w_i \right)}
\e^{- 3 \left( 1 + w_i \right) N)} \, , \nn
\tilde W(N) \equiv& \int^{N} \frac{dN_1}{\e^{N_1}} \left( \frac{2}{\kappa^2}
\dot H (N_1)
+ \sum_i \left(1 + w_i \right) a_0^{- 3 \left( 1 + w_i \right)} \rho_{0i}
\e^{- 3 \left( 1 + w_i \right) N_1} \right) \, .
\end{align}
We can now identify the auxiliary scalar $\phi$ with the e-foldings
number $N$, that is, $\phi=N$, instead of the cosmic time $\phi=t$,
and therefore if $V(\phi)$ and $\xi$ are chosen as follows,
\begin{align}
\label{ma39}
V(\phi) =& \frac{3}{\kappa^2}h\left(\phi\right)^2 - 3h\left(\phi\right)
\e^{\phi} \tilde U(\phi)
+ \sum_i \rho_{0i} a_0^{- 3 \left( 1 + w_i \right)}
\e^{- 3 \left( 1 + w_i \right) \phi)}
\, , \nn
\xi (\phi) =& \frac{1}{8}\int^\phi d\phi_1
\frac{\e^{\phi_1} }{h(\phi_1)^3} \tilde U(\phi_1)\, ,\nn
U(\phi) \equiv& \int^\phi d\phi_1 \e^{-\phi_1} \left( \frac{2}{\kappa^2}
h'\left(\phi_1\right)
+ \sum_i \left(1 + w_i \right) \rho_{0i} a_0^{- 3 \left( 1 + w_i \right)}
\e^{- 3 \left( 1 + w_i \right) \phi_1} \right) \, ,
\end{align}
we obtain the solution $H=h(N)$.

We now consider the evolution in $\Lambda$CDM cosmology, without CDM
(Cold Dark Matter). In case of the $\Lambda$CDM model, we have
\begin{equation}
\label{LambdaCDM}
\frac{3}{\kappa^2} H^2 = \Lambda + \rho_0 \e^{-3N}\, .
\end{equation}
Here $\Lambda$ is a constant corresponding to the
cosmological constant and $\rho_0$ is simply a constant. The second
term in Eq.~(\ref{LambdaCDM}) expresses the contribution from the
CDM. Therefore, we find,
\begin{equation}
\label{LambdaCDMh}
h (\phi) = \sqrt{
\frac{\kappa^2}{3} \left( \Lambda + \rho_0 \e^{-3\phi} \right) }\, .
\end{equation}
In effect, by substituting the expression from
Eq.~(\ref{LambdaCDMh}) into Eq.~(\ref{ma39}), we obtain the model which
reproduces the $\Lambda$CDM universe.

\subsection{Unification of Inflation with Dark Energy Era in $F(R)$ Gravity}

As we already discussed in the introduction, in the context of
modified gravity it is possible to construct cosmological models
which describe in a unified way the early-time acceleration with the
late-time acceleration of the Universe. In the literature there
exist various proposals on this, see for example
\cite{Nojiri:2007as,Nojiri:2007cq,Cognola:2007zu}. In this section
we shall be interested in the unified description of the
inflationary era with the dark energy era, in the context of $F(R)$
gravity. Before we start, let us summarize the conditions that need
to hold true in order for the unified $F(R)$ models, to be viable:

\begin{enumerate}
\item The first condition is of course that the model can generate the
inflationary era. This condition could be given by Eq.~(\ref{Uf1}).
Here we suppose that $F(R) = R + f(R)$. The viability condition
indicates that $f(R)$ goes to a constant $\Lambda_i$, which plays
the role of an effective cosmological constant in the early
Universe. In effect, the following requirement must be satisfied
$\Lambda_i \gg \left(10^{-33}\mathrm{eV}\right)^2$ and we may
naturally assume that $\Lambda_i\sim 10^{20 \sim 38}\left(
\mathrm{eV}\right)^2$.

Since the inflationary era should not be eternal, but it should stop
at some point, the de Sitter space-time described by Eq.~(\ref{Uf1})
should not be the stable de-Sitter point, but a quasi-stable de Sitter
point. We assume that the de Sitter space-time is realized in the
limit of large curvature, that is $R\sim R_I \sim \left( 10^{16 \sim
19}\, \mathrm{GeV}\right)^2$, and in effect, the quasi-stability
condition requires Eq.~(\ref{E3}). Note that, $f_{0I}$ and $f_{1I}$
are positive constants and also $m$ is a positive integer. If the
following condition is satisfied,
\begin{equation}
\label{R0I}
 - R_I + \frac{F'(R_I)}{F''(R_I)} < 0\, ,
\end{equation}
the de Sitter space-time solution is rendered strongly unstable.

In order to generate the inflationary era, there are other
possibilities, for example, we may assume that,
\begin{equation}
\label{UU2}
\lim_{R\to\infty} f (R) = \alpha R^m \, .
\end{equation}
Here the parameter $m$
is a positive integer greater than unity, $m>1$ and also $\alpha$ is
a constant. In order to avoid an antigravity era, which would imply
that $f'(R)>-1$, we need to require that $\alpha>0$, which indicates
that $f(R)$ should be positive during the inflationary era.

\item The second condition is the condition that the $F(R)$ gravity model should generate
the present-time cosmic acceleration. This condition could be
realized by simply requiring that the $f(R)$ function in the current
Universe, behaves as a small constant, as in Eq.~({Uf3}). We denote
the curvature of the present Universe by $R_0$, $R_0\sim
\left(10^{-33}\mathrm{eV}\right)^2$. We should note that $R_0>
\tilde R_0$ due to a small extra contribution to the curvature by
the matter fluids present.

Eq.~(\ref{Uf3}) indicates that $f(R_0)$ behaves as an effective
cosmological constant and by using the Einstein equations, we
effectively obtain $R_0=\tilde R_0 - \kappa^2 T_\mathrm{matter}$.
Note that $T_\mathrm{matter}$ expresses the trace of the matter
energy-momentum tensor.

Since we are now considering the time scale to be a few billion
years, this implies that $f'(R_0)$ does not need to vanish
completely but instead it should behave as, $\left| f'(R_0) \right|
\ll \left(10^{-33}\,\mathrm{eV}\right)^4$.

We should note that Eq.~(\ref{Uf3}) indicates that $f(R)$ could be
negative at the present Universe. Then, if we consider the model of
Eq.~(\ref{UU2}) in the early Universe, $f(R)$ should cross zero in
the past.

The de Sitter space-time given by Eq.~(\ref{Uf3}) should be stable.
This requires that when $R\sim R_0 \sim \left( 10^{-33}\,
\mathrm{eV}\right)^2$, the function $F(R)$ behaves as in
Eq.~(\ref{E2}). In some cases this condition might not be strictly
necessary, but we can use this condition in order to avoid a future
finite-time singularity. In the case $n=0$, the condition of
Eq.~(\ref{E2}) can be expressed as follows,
\begin{equation}
\label{R00}
 - R_0 + \frac{F'(R_0)}{F''(R_0)} > 0\, .
\end{equation}

\item The third condition is given by Eq.~(\ref{Uf4}),
which indicates that there is a flat space-time solution. Furthermore
if we require that in the small curvature limit, that is when $R\to
0$, the Einstein gravity solution should be recovered, we require
Eq.~(\ref{E1}). If the condition of Eq.~(\ref{E1}) is satisfied,
then the condition Eq.~(\ref{Uf4}) is automatically satisfied.

\item In some models of $F(R)$ gravity, a curvature singularity is easily
generated, when the density of the matter present is large. In order
to avoid this issue, we may require that in the large curvature
limit, the $F(R)$ gravity behaves as in Eq.~(\ref{E4}). Instead of
Eq.~(\ref{E4}), $F(R)$ may behaves as in Eq.~(\ref{E5}).

\item Since the expression $1/F'(R)$ plays the role of the effective gravitational
constant, that is $1/F'(R)\sim \kappa^2$, in order to avoid the
antigravity era, we should require (\ref{E6}) or (\ref{E7}). If the
antigravity era occurs, the graviton becomes a ghost and in effect
it generates negative norm states in the corresponding quantum
theory.

\item\label{req6B}
By combining conditions (\ref{E1}) and (\ref{E6}), we find the
following condition, that is, (\ref{E8}), and we need to note that
this is not an independent one.
\item As pointed in Ref.~\cite{Dolgov:2003px}, an instability could
occur due to matter itself, that is a matter instability. When one
considers matter perturbations of the scalar curvature, in the Einstein
gravity, the perturbations often increase rapidly. In order to avoid
the matter instabilities, we need to require the following,
\begin{equation}
\label{E8B}
U(R_b) \equiv \frac{R_b}{3} - \frac{F^{(1)}(R_b)
F^{(3)}(R_b) R_b} {3 F^{(2)}(R_b)^2} - \frac{F^{(1)}(R_b)}{3F^{(2)}(R_b)}
+ \frac{2 F(R_b) F^{(3)}(R_b)}{3
F^{(2)}(R_b)^2} - \frac{F^{(3)}(R_b) R_b}{3 F^{(2)}(R_b)^2}< 0\, .
\end{equation}
If the function $U(R_b)$ is positive, the time scale of the
instability is given by the expression $1/\sqrt{U(R_b)}$. Then, if
we consider the model $F(R) = R - \frac{\mu^4}{R}$ for example, the
time scale becomes of the order $10^{-26}$sec on the earth, and in
effect the gravitational field would be very strong.

\item If we rewrite the $F(R)$ gravity in a scalar-tensor form,
it can be seen that the corresponding scalar field
$\sigma$ has a direct coupling with matter. If the mass of the
scalar field $\sigma$ has the same order as the typical scale of the
accelerating expansion of the Universe at present time, the
propagation of the scalar field $\sigma$ generates long range forces
and hence it generates large corrections to the Newton law. In
effect, if the Chameleon mechanism is used, the mass of the scalar
field $\sigma$ becomes large enough, and in effect, the force
induced by the propagation of the scalar $\sigma$ becomes short
range. Hence, there would be no conflict with the present solar
system experiments or observations. Therefore, the mass of the
scalar field $\sigma$ in Eq.~(\ref{FRV3}), should become large
enough owing to the Chameleon mechanism.

It should be noted however, that a very large mass could be in
conflict with the matter abundance, see for example
Ref.~\cite{Katsuragawa:2016yir}. If we quantize the scalar field
$\sigma$, a scalar particle is produced, which may be created by the
reheating process. If this scalar field is very heavy, the particle
may decay to the standard model particles very excessively, which
could be in conflict with observations.

\end{enumerate}

If we require both the conditions of Eqs.~(\ref{E2}) and (\ref{E1}),
the extra, unstable de Sitter solution at $R=R_e$ $\left( 0< R_e <
R_0 \right)$ may occur. As it is known, the evolution of the
Universe will stop at $R=R_0$, since the de Sitter solution $R=R_0$
should be stable, and therefore the curvature never becomes smaller
than $R_0$. In effect, the extra de Sitter solution could not be
realized in the evolution of the Universe.

In Ref.~\cite{Nojiri:2010ny}, an $F(R)$ gravity model was proposed,
which satisfies all the above conditions, and it is the following,
\begin{align}
\label{EE1}
\frac{F(R)}{R^2} =&
\left\{ \left(X_m \left(R_I;R\right) - X_m\left(R_I;R_1\right) \right)
\left(X_m\left(R_I;R\right) - X_m\left(R_I;R_0\right) \right)^{2n+2}
\right.
\nn
& \left. + X_m\left(R_I;R_1\right) X_m\left(R_I;R_0\right)^{2n+2}
+ f_\infty^{2n+3} \right\}^{\frac{1}{2n+3}} \, ,
\end{align}
where,
\begin{equation}
\label{EE1b}
X_m\left(R_I;R\right) \equiv \frac{\left(2m
+1\right) R_I^{2m}}{\left( R - R_I \right)^{2m+1} + R_I^{2m+1}} \, .
\end{equation}
We choose the parameters $n$ and $m$ to be integers greater or
equal to unity, that is $n,m\geq 1$, and also the parameter $R_1$ is
related to the curvature $R_e$ as follows,
\begin{equation}
\label{EE2}
X\left(R_I;R_e\right) = \frac{\left(2n+2\right)
X\left(R_I;R_1\right)X\left(R_I;R_1\right)
+ X\left(R_I;R_0\right)}{2n+3}\, .
\end{equation}
In addition, it is compelling
to assume that,
\begin{equation}
\label{EEE1}
0<R_1<R_0 \ll R_I \, .
\end{equation}

Let us see how the inflationary era can be realized by the model
(\ref{EE1}). The behavior of (\ref{E3}) in the limiting case when
$R\sim R_I$, is given below,
\begin{align}
\label{EE6}
f_{0I} =& \left\{
\left(X_m \left(R_I;R_I\right) - X_m\left(R_I;R_1\right) \right)
\left(X_m\left(R_I;R_I\right) - X_m\left(R_I;R_0\right) \right)^{2n+2}
\right. \nn
& \left. + X_m\left(R_I;R_1\right) X_m\left(R_I;R_0\right) ^{2n+2}
+ f_\infty^{2n+3} \right\}^{\frac{1}{2n+3}} \nn
f_{1I} =& \frac{2m+1}{R_I^{2m+2}}
\left\{ \left(X_m \left(R_I;R_I\right) - X_m\left(R_I;R_1\right) \right)
\left(X_m\left(R_I;R_I\right) - X_m\left(R_I;R_0\right) \right)^{2n+2}
\right.
\nn
& \left. + X_m\left(R_I;R_1\right) X_m\left(R_I;R_0\right)^{2n+2}
+ f_\infty^{2n+3} \right\}^{\frac{2\left(n+1\right)}{2n+3}} \left\{
\left(X_m\left(R_I;R_I\right) - X_m\left(R_I;R_0\right) \right)^{2n+2}
\right. \nn
& \left. + \left(2n+2\right)\left(X_m \left(R_I;R_I\right) -
X_m\left(R_I;R_1\right) \right)
\left(X_m\left(R_I;R_I\right) - X_m\left(R_I;R_0\right) \right)^{2n+1}
\right\}\, .
\end{align}
Therefore, the condition (\ref{E3}) is satisfied. In effect, a de
Sitter space-time solution exists, and this solution describes the
inflationary era. Due to the fact that the condition (\ref{E3}) is
satisfied, the de Sitter solution is quasi-stable, and specifically
it is stable at large curvatures, but as the curvature decreases, it
gradually becomes unstable. Due to the quasi-stability, the exit
from inflation occurs in this theory, something guaranteed by the
gradual decrease of the curvature during the inflationary era.

Now we demonstrate how the model at hand can realize the
present-time acceleration of the Universe. In the limit $R\sim R_0$,
we have,
\begin{align}
\label{EE5}
f_{0L} =& \left\{ X_m\left(R_I;R_1\right) X_m\left(R_I;R_0\right)^{2n+2}
+ f_\infty^{2n+3} \right\}^{\frac{1}{2n+3}} \, , \nn
f_{1L} =& \frac{1}{2m+3} \left\{ X_m\left(R_I;R_1\right)
X_m\left(R_I;R_0\right)^{2n+2}
+ f_\infty^{2n+3} \right\}^{- \frac{2\left(n+1\right)}{2n+3}}
\left(X_m \left(R_I;R_1\right) - X_m\left(R_I;R_0\right) \right) \nn
& \times \frac{\left( 2m + 1 \right)^{4\left(m+1\right)}
\left\{ R_I \left(R_0 - R_I\right)\right\}^{4m\left(m+1\right)} }
{\left\{ \left( R_0 - R_I \right)^{2m+1} + R_I^{2m+1}
\right\}^{4\left(m+1\right)}} \, .
\end{align}
In effect, the condition of Eq.~(\ref{E2}) is satisfied and
therefore the stable de Sitter solution exists, which corresponds to
the current accelerating expansion of the Universe.

We now discuss the conditions of Eqs.~(\ref{Uf4}) or (\ref{E1}). The
function $X\left(R_I;R\right)$ should be a monotonically decreasing
function of the curvature $R$ and we should also require that, in
the limit $R\to 0$, the function $X\left(R_I;R\right)$ behaves as
follows,
\begin{equation}
\label{EE3}
X\left(R_I;R\right) \to \frac{1}{R}\, .
\end{equation}
Therefore, in the limit $R\to 0$, the $F(R)$ gravity in (\ref{EE1})
actually reproduces the behavior of Eq.~(\ref{E1}) and in addition
it satisfies the condition (\ref{E1}).

An important issue must be discussed at this point, having to do
with curvature singularities, generated when the matter density is
large. In the large curvature limit, that is when $R\to \infty$, we
find the following,
\begin{equation}
\label{EE4}
X\left(R_I;R\right) \to
\frac{\left(2m+1\right) R_I^{2m}}{R^{2m+1}} \to 0\, ,
\end{equation}
which
tells that $F(R)$ behaves as (\ref{E4}) and therefore it is not
possible for a curvature singularity to occur.

Also we can easily check the conditions for the absence of an
antigravity era, namely (\ref{E6}) or (\ref{E7}), since the $F(R)$
function in (\ref{EE1}) satisfies Eqs.~(\ref{E1}) and (\ref{E4}) in
the limiting cases $R\to 0$ or $R\to \infty$. In the interval
$R_e<R<R_0$, Eq.~(\ref{E6}) or (\ref{E7}) are trivially satisfied
because the function $\frac{F(R)}{R^2}$ is a monotonically
increasing function of the curvature $R$. Even in the interval
$R_1<R_0 \ll R \ll R_I$, Eq.~(\ref{E6}) or (\ref{E7}) are again
satisfied, since the function $\frac{F(R)}{R^2}$ behaves as
(\ref{EE8}). Therefore, an antigravity does not appear for all
curvature values that are considered.

Now let us investigate whether the model is free from matter
instabilities as in Ref.~\cite{Dolgov:2003px}, an issue completely
determined by the condition (\ref{E8B}). Since $R_1<R_0 \ll R \ll
R_I$ on the earth or the sun, we find that the behavior of the
$F(R)$ function is given as in Eq.~(\ref{EE8}) and hence we find the
following expression for $U(R_b)$ in Eq.~(\ref{E8B}),
\begin{equation}
\label{URb}
U(R_b) \sim - \frac{f_n}{6f_\infty} < 0\, ,
\end{equation}
which
indicates that the matter instability does not occur for the $F(R)$
model at hand, due to the fact that the function $U(R_b)$ is
negative.

At this point it would be interesting to check whether the Chameleon
mechanism \cite{Khoury:2003rn,Brax:2004qh} applies in the case at
hand, or not. In order to see this, we should study the behavior of
the mass of the scalar field $\sigma$ in Eq.~(\ref{FRV3}). Since we
are interested for the mass of the scalar field $\sigma$ in the
present-time Universe, we may focus our investigation in the
interval $R_1<R_0 \ll R \ll R_I$, where we can approximate the
function $X_m\left(R_I;R\right)$ as follows,
\begin{equation}
\label{EE7}
X_m\left(R_I;R\right) \sim \frac{1}{R}\, ,\quad
X_m\left(R_I;R_1\right) \sim \frac{1}{R_1}\, ,\quad
X_m\left(R_I;R_0\right) \sim \frac{1}{R_0}\, .
\end{equation}
By using Eq.~(\ref{EE7}), we can approximate the functional form
of the $F(R)$ gravity as follows,
\begin{equation}
\label{EE8}
\frac{F(R)}{R^2} \sim f_\infty + \frac{f_n}{R}\, ,\quad
f_n \equiv \frac{R_1 + \left(2n+2\right) R_0
f_\infty^{-2n -2}} {\left(2n+3\right) R_1 R_0^{2n+2}} \sim
\frac{f_\infty^{-2n -2}}{R_0^{2n+2}}\, .
\end{equation}
Note that we have
assumed $R_1\sim R_0$ in the last equation of (\ref{EE8}). In the
limit $f_{\infty} R \ll f_n$, that is,
\begin{equation}
\label{EE9}
\frac{1}{f_\infty^{2n+3}} \gg R R_0^{2n+2}\, ,
\end{equation}
by using Eq.~(\ref{FRV3}), we obtain,
\begin{equation}
\label{EE10}
m_\sigma^2 \sim \frac{3}{4f_\infty}\, .
\end{equation}
In the region inside the earth, since
$1\,{g}\sim 6\times 10^{32}\,\mathrm{eV}$ and also,
$1\,\mathrm{cm}\sim \left(2\times 10^{-5}\,\mathrm{eV}\right)^{-1}$,
the density of the matter is approximately equal to
$\rho_\mathrm{matter}\sim 1 \mathrm{g/cm^3} \sim 5\times 10^{18}\,
\mathrm{eV}^4$ and in effect, the curvature is approximated by $R
\sim \kappa^2 \rho_\mathrm{matter} \sim
\left(10^{-19}\,\mathrm{eV}\right)^2$. On the other hand, in the
atmosphere of the earth, we find, $\rho_\mathrm{matter}\sim 10^{-6}
\mathrm{g/cm^3} \sim 10^{12}\, \mathrm{eV}^4$ and therefore $R \sim
\kappa^2 \rho_\mathrm{matter} \sim
\left(10^{-25}\,\mathrm{eV}\right)^2$. In the solar system, where
the interstellar gas exists, since one proton (or a hydrogen atom)
is contained in a volume of the order $1\,\mathrm{cm}^3$ in the
interstellar gas, we find $\rho_\mathrm{matter}\sim
10^{-5}\,\mathrm{eV}^4$, $R_0\sim 10^{-61}\, \mathrm{eV}^2$. Then,
we find that the condition of Eq.~(\ref{EE9}) is satisfied. As an
example, we may choose $\frac{1}{f_\infty}\sim \mathrm{MeV}^2$,
which yields a very small Compton length of the scalar field and in
effect the resulting corrections to the Newton law are practically
negligible.

Now we discuss in brief some general cosmological features of the
model (\ref{EE1}). Particularly, after the exit from inflation at
$R=R_I$, radiation and matter could be produced, during the
reheating process. It is possible that matter and radiation may
dominate over the contribution from the $f(R)$ gravity (recall that
the $f(R)$ gravity is the modified gravity part of $F(R)=R+f(R)$)
and therefore the cosmological evolution thereafter is identical
with that of the Einstein gravity. In effect, the radiation
domination and the matter domination era of the Universe may be
realized. Eventually, when the curvature of the Universe becomes of
the order $R\sim R_0$, the late-time acceleration of the Universe is
realized.

An interesting possibility of having a double-inflationary era could
be realized, if the condition (\ref{E5}) holds true. Recall that the
condition (\ref{E5}) expresses the fact that large curvatures cannot
be generated, and hence an $R^2$ term may be generated in the very
large curvature limit of the theory. It is well known that such
$R^2$ terms are responsible for inflation. This new inflationary era
might be added to the inflationary era which occurs when $R=R_I$, if
the Universe started its evolution with a very high curvature, that
is $R\gg R_I$.

In some models in the literature, the phenomenological implications
are quite interesting, although not all the viability conditions are
satisfied. For example, the following realistic model was proposed
in Ref.~\cite{Nojiri:2007as}
\begin{equation}
\label{UU2d}
f(R)= \frac{\alpha R^{2n} - \beta R^n}{1 + \gamma R^n}\, ,
\end{equation}
where, $\alpha$,
$\beta$, and $\gamma$ are positive constants and the parameter $n$
is a positive integer. The model of Eq.~(\ref{UU2d}) has the de Sitter
space-time solution, with its curvature being, \cite{Nojiri:2007as}
\begin{equation}
\label{UUU7}
R_0=\left\{ \left(\frac{1}{\gamma}\right) \left(1
+ \sqrt{ 1 + \frac{\beta\gamma}{\alpha} }\right)\right\}^{1/n}\, ,
\end{equation}
and, therefore, the quantity $\tilde {R_0}$ in Eq.~(\ref{Uf3}) takes
the following form,
\begin{equation}
\label{UU6}
f(R_0) \sim -2 \tilde R_0 = \frac{\alpha}{\gamma^2} \left( 1
+ \frac{\left(1 - \frac{\beta\gamma}{\alpha} \right) \sqrt{ 1
+ \frac{\beta\gamma}{\alpha}}}{2 + \sqrt{ 1
+ \frac{\beta\gamma}{\alpha}}} \right) \, .
\end{equation}
Then we easily find that,
\begin{equation}
\label{UU9}
\alpha \sim 2 \tilde R_0 R_0^{-2n}\, ,\quad
\beta \sim 4 {\tilde R_0}^2 R_0^{-2n} R_I^{n-1}\, ,\quad \gamma \sim
2 \tilde R_0 R_0^{-2n} R_I^{n-1}\, .
\end{equation}
As it was shown in
\cite{Nojiri:2007as}, the correction to the Newton law is small for
the model (\ref{UU2d}). This is due to the fact that the mass
$m_{\sigma}$ appearing in Eq.~(\ref{JGRG24}), is given by
$m_\sigma^2 \sim 10^{-160 + 109 n}\,\mathrm{eV}^2$ in the solar
system and $m_\sigma^2 \sim 10^{-144 + 98 n}\,\mathrm{eV}^2$ in the
atmosphere of the earth, and therefore it is large enough
\cite{Nojiri:2007as}.


Now we discuss another model which has some interesting cosmological
features, which was proposed in Ref.~ \cite{Elizalde:2010ts} (see
also \cite{Cognola:2007zu} and \cite{Bamba:2012qi}) which has the
following form:
\begin{equation}
\label{total}
F(R)=R-2\Lambda\left(1-\mathrm{e}^{-\frac{R}{R_{0}}}\right)
 -\Lambda_{i}\left(1-\mathrm{e}^{-\left(\frac{R}{R_i}\right)^n}\right)
+\gamma R^\alpha\, . \end{equation}For later convenience, we may define,
\begin{equation}
\label{fi}
f_{i} = -\Lambda_{i}\left(1-\e^{-\left(\frac{R}{R_i}\right)^n}\right)\, ,
\end{equation}
We choose the constants $R_{i}$ and $\Lambda_{i}$ to have the
typical values of the scalar curvature and of the expected
cosmological constant during the inflationary era, respectively,
that is, $R_{i}$, $\Lambda_i$ $\simeq 10^{20-38} \text{eV}^2$. We
also assume the parameter $n$ to be $n>1$. In the last term of
(\ref{total}), namely $\sim \gamma R^\alpha$, the parameter $\gamma$
is assumed to be a positive constant and $\alpha$ is a real number.
We need to note that the latter term is important in order to
achieve the exit from inflation.

Let us first discuss how inflation and the exit from the inflation
can be realized in the context of the model (\ref{total}). Since
the curvature $R$ should be large during the inflationary era, we
may assume that $R \gg R_i$. Then, $f_i$ behaves as a cosmological
constant $f_i \sim -\Lambda_i$, which effectively generates the
early-time acceleration of the Universe. In order to satisfy the
condition of Eq.~(\ref{E6}), and to avoid antigravity, we must have,
\begin{equation}
\label{uno}
R_i>\Lambda_i n
\left(\frac{n-1}{n}\right)^{\frac{n-1}{n}}\mathrm{e}^{-\frac{n-1}{n}}\, .
\end{equation}
Let us illustrate the previous arguments by using a simple
example and by choosing $n=4$, Eq.~(\ref{uno}) is satisfied for
$R_i>1.522\, \Lambda_i$. Although $f_i$ behaves as an effective
cosmological constant, we shall investigate if the de Sitter space-time
solution actually exists. We denote the de Sitter space-time
curvature as $R=R_{\mathrm{dS}}$ and when the curvature
$R_{\mathrm{dS}}$ is very large, that is, $R_{\mathrm{dS}} \gg R_i$,
since $\Lambda_i\sim R_i$, the condition for the existence of the de
Sitter solution, namely, $0=2 F\left( R_{\mathrm{dS}} \right)
 - R_{\mathrm{dS}} F' \left( R_{\mathrm{dS}} \right)$ has a non-trivial solution if
$\Lambda_i\sim R_i$, and in turn this holds true when,
\begin{equation}
\label{fi5}
2<\alpha<3\, .
\end{equation}
By using Eq.~(\ref{R0I}), we find that
the resulting solution describing the de Sitter space-time, is
strongly unstable and due to this fact, the graceful exit from
inflation comes as an outcome. Indeed, the instability of the de
Sitter space-time is the following,
\begin{equation}
\label{fi7}
5/2\leq\alpha<3\, ,
\end{equation}
and we can estimate the characteristic time of the instability
as follows,
\begin{equation}
\label{fi6}
t_i\sim\frac{1}{\sqrt{R_{\mathrm{dS}}}}\sim 10^{-10}-10^{-19}\,
\mathrm{sec} \, .
\end{equation}
Hence, the graceful exit from inflation surely occurs.

The inflationary era study of the model (\ref{total}) has also been
investigated in detail in Ref.~\cite{Bamba:2012qi} and the following
estimates of the $e$-foldings number have been found,
\begin{equation}
\label{stime}
N \simeq 107 \quad (\mathrm{for} \quad \alpha=5/2)\, ,\quad
N \simeq 64 \quad (\mathrm{for} \quad \alpha=8/3)\, ,\quad
N \simeq 51 \quad (\mathrm{for} \quad \alpha=11/4)\, .
\end{equation}
More concretely, if we specify the parameters in
Eq.~(\ref{total}) as follows,
\begin{equation}
\label{fi8}
n=4\, ,\quad
\alpha=\frac{5}{2}\, ,
\end{equation}
and also if we choose the value of the
curvature $R_i$ as,
\begin{equation}
\label{fi9}
R_i=2\Lambda_i\, ,
\end{equation}
then we
find that $R_i$ satisfies the condition (\ref{uno}) and in effect,
the antigravity era is avoided. The de Sitter space-time solution can
be found by solving the equation, $0=2 F\left( R_{\mathrm{dS}}
\right)
 - R_{\mathrm{dS}} F' \left( R_{\mathrm{dS}} \right)$ and this has as solution,
\begin{equation}
\label{mitiko}
R_{\mathrm{dS}}=4\Lambda_i\, .
\end{equation}
Since we choose
$n$ to be large enough, the curvature $R_{\mathrm{dS}}$ becomes
sufficiently large, when it is compared with $R_i$, and we find that
$f_i(R_{\mathrm{dS}}) \simeq -\Lambda_i$. In addition, it can be
seen that the only non-trivial solution of the equation $0=2 F\left(
R_{\mathrm{dS}} \right)
 - R_{\mathrm{dS}} F' \left( R_{\mathrm{dS}} \right)$, is given in
Eq.~(\ref{mitiko}), and it can be seen that this solutions
satisfies the de Sitter instability condition of Eq.~(\ref{R0I}).
In turn, this indicates that the Universe ultimately exits from
inflation, and tends towards a minimal attractor at $R=0$.
Alternatively, it is possible that the Universe might develop a
singularity when $R\rightarrow \infty$. In order to avoid the
curvature singularity, we need to require that the condition (\ref{E3}) for the
quasi-stability is satisfied. In effect, the de Sitter solution is
stable at large curvatures, but as the curvature gradually drops,
the de Sitter solution becomes unstable.

Let us now discuss how the late-time acceleration era can be
realized in the context of the $F(R)$ gravity model at hand. First
let us point out that prior to the late-time acceleration era, the
matter domination era occurs, and the parameter $n$ in
Eq.~(\ref{fi}) was introduced in order to avoid the effects of inflation
during the matter domination era. In fact, since $R \ll R_{i}$, if
$n>1$, we find
\begin{equation}
\label{fi2}
R \gg |f_i(R)|\simeq \frac{R^{n}}{R_i^{n-1}} \, ,
\end{equation}
and therefore $f_i$ is suppressed
during the matter domination era. The last term $\gamma R^\alpha$ in
Eq.~(\ref{total}) was introduced in order to generate the graceful
exit from inflation, as we already discussed. If we choose
$\gamma\sim 1/R_i^{\alpha-1}$ and $\alpha>1$, the contribution from
this term becomes practically negligible, when the small curvature
regime is considered, and hence during the matter domination era,
when $R \ll R_i$. In fact, we find that,
\begin{equation}
\label{fi3}
R \gg \frac{R^\alpha}{R_i^{\alpha-1}}\, .
\end{equation}
It is easy to check that
Eq.~(\ref{Uf4}) or (\ref{E1}) are satisfied.

In order for the late-time acceleration to occur, it is compelling
to have the stable asymptotic de Sitter solution in the theory. This
is guaranteed if the condition in Eq.~(\ref{R00}) is satisfied, and
this happens when,
\begin{equation}
\label{fi4}
n>\alpha\, .
\end{equation}
Also the de Sitter
vacuum exists at late time, with curvature $R=R_{\mathrm{dS}}$
\cite{Bamba:2012qi}. For the model (\ref{fi8}), $R_i$ satisfies the
condition in Eq.~(\ref{uno}), and hence the antigravity era is
avoided.

A variant model of that developed in \cite{Elizalde:2010ts}, was
constructed in Ref.~\cite{Cognola:2007zu}, in which case the $F(R)$
gravity contains only the first two terms of (\ref{total}), namely,
$R-2\Lambda\left(1-\mathrm{e}^{-\frac{R}{R_{0}}}\right)$. During the
late-time era, the other terms in (\ref{total}) can be neglected,
and the model of Ref.~\cite{Cognola:2007zu} coincides with that in
Eq.~(\ref{total}). In the limit $R \gg R_{0}$, we have $F(R)\simeq
R-2\Lambda$, and hence the model mimics the $\Lambda$CDM model. For
simplicity, consider the model of Eq.~(\ref{fi8}). In the late-time
era, we have,
\begin{equation}
\label{l1}
F(R)= R -2\Lambda(1-\mathrm{e}^{-R/R_0})\, ,
\end{equation}
and therefore,
\begin{align}
\label{l2}
F'(R)= & 1 - \frac{2\Lambda}{R_{0}}\mathrm{e}^{-R/R_0}\, , \\
\label{l3}
F''(R)= & \frac{2\Lambda}{R_{0}^2}\mathrm{e}^{-R/R_0}\, .
\end{align}
In the limit $R \gg R_0$, the second term in $F'(R)$ appearing in
Eq.~(\ref{l2}), can be safely neglected. In effect, we find $F'(R)
\sim 1$, and therefore the antigravity is avoided in the present
model at late times. Since $F(R) \simeq R$ in the limit $R \gg R_0$,
we expect that the matter dominated era can be realized as in
the Einstein-Hilbert action. At late times, a non-trivial solution of
the equation $0=2 F\left( R_{\mathrm{dS}} \right)
 - R_{\mathrm{dS}} F' \left( R_{\mathrm{dS}} \right)$, is
$R=4\Lambda$, as in the case of (\ref{mitiko}). In contrast to the
inflationary era de Sitter solution, the late-time de Sitter vacuum
satisfies the condition (\ref{R00}) for small curvatures, and hence
this solution is stable. Hence, this late-time de Sitter vacuum is
the final and stable attractor of the model at hand.

By using the expression in Eq.~(\ref{FRV3}), the mass of the scalar
field during the late-time era is estimated to be equal to,
\begin{equation}
\label{MM1}
m_\sigma^2 \sim \frac{R_0^2
\e^{\frac{R}{R_0}}}{4\Lambda}\, ,
\end{equation}
which is positive, and hence
the late-time asymptotic de Sitter space-time is stable. Since
$\Lambda$ in Eq.~(\ref{total}) plays the role of an effective
cosmological constant in the late-time Universe, we have,
\begin{equation}
\label{AA2}
\Lambda \sim R_0 \sim \left(10^{-33}\,\mathrm{eV}\right)^2\, .
\end{equation}
In the solar system, we find $R\sim 10^{-61}\,\mathrm{eV}^2$, and
in effect we find that $m_\sigma^2 \sim 10^{1,000}\,\mathrm{eV}^2$,
which is ultimately heavy. In effect, the corrections to the Newton
law are negligible. Indeed, in the atmosphere of the earth, we have
$R \sim 10^{-50}\,\mathrm{eV}^2$, and even in the case we choose $1/b
\sim R_0 \sim \left(10^{-33}\,\mathrm{eV}\right)^2$, we find that
$m_\sigma^2 \sim 10^{10,000,000,000}\,\mathrm{eV}^2$. Therefore the
corrections to the Newton law are negligible. There is an issue
however having to do with the effects of having such a large mass,
due to the excess of the particles produced by the scalar during the
reheating, and this could be in conflict with current observations.
In addition, if the Compton length of the scalar particle becomes
smaller that the interatomic distance, the matter fluids cannot be
approximated in the general relativistic way, that is as perfect
fluids, and hence, the Chameleon mechanism cannot work in this case.
This however could set an upper limit constraint on the mass of the
scalar particle $\sigma$.

Let us work out some essential features of the dark energy era
corresponding to the model of Eq.~(\ref{total}). The focus will be
on the dark energy density evolution, which is equal to
$\rho_{\mathrm{DE}}=\rho_{\mathrm{eff}}-\rho/F'(R)$. From the FRW
equation we can identify the dark energy density as follows,
\begin{equation}
\label{EeffFRW}
\frac{3}{\kappa^2}H^2 = \rho_{\mathrm{eff}} \equiv
\frac{1}{F'(R)}\left\{\rho+\frac{1}{2\kappa^2}
\left[(F'(R)R-F(R))-6H\dot{F}'(R)\right]\right\}\, .
\end{equation}
We also use
the new variable $y_\mathrm{H}$ defined by
\begin{equation}
\label{y}
y_\mathrm{H}\equiv\frac{\rho_{\mathrm{DE}}}{\rho_m^{(0)}}
=\frac{H^2}{\tilde{m}^2}-a^{-3}-\chi a^{-4}\, ,
\end{equation}
where $\rho_m^{(0)}$ stands for the energy density of matter at
present time, and $\tilde{m}^2$ is the mass scale
\begin{equation}
\label{tildem}
\tilde{m}^2\equiv\frac{\kappa^2\rho_m^{(0)}}{3}\simeq 1.5 \times
10^{-67}\text{eV}^2\, .
\end{equation}
Also we define $\chi$ as follows,
\begin{equation}
\label{chi}
\chi\equiv\frac{\rho_r^{(0)}}{\rho_m^{(0)}}\simeq 3.1
\times 10^{-4}\, ,
\end{equation}
where $\rho_r^{(0)}$ is the energy density of
radiation at present time.

By using the conservation law,
\begin{equation}
\label{acons}
\frac{d \rho_\mathrm{DE}}{d \left( \ln a \right)} + 3 \left(
\rho_\mathrm{DE} + p_\mathrm{DE} \right)=0 \, ,
\end{equation}
we obtain the
following expression for the EoS dark energy EoS parameter
$w_{\mathrm{DE}}$,
\begin{equation}
\label{eins}
w_{\mathrm{DE}} = \frac{p_\mathrm{DE}}{\rho_\mathrm{DE}}
= -1-\frac{1}{3}\frac{1}{y_\mathrm{H}} \frac{d y_\mathrm{H}}{d
\left(\ln a\right)}\, .
\end{equation}
By combining Eq.~(\ref{EeffFRW}) and the
expression of the scalar curvature for the FRW Universe, which is,
$R=12 H^2 + 6 \dot H$, we find the following differential equation,
\begin{equation}
\label{superEq}
\frac{d^2 y_\mathrm{H}}{d (\ln a)^2}+J_1\frac{d y_\mathrm{H}}{d (\ln a)}
+J_2 y_\mathrm{H}+J_3=0\, ,
\end{equation}
which completely determines the evolution. The functions $J_1$,
$J_2$, and $J_3$ are defined as follows,
\begin{align}
\label{Js}
& J_1=4+\frac{1}{y_\mathrm{H}+a^{-3}
+\chi a^{-4}}\frac{1-F'(R)}{6\tilde{m}^2 F''(R)}\, , \quad
J_2=\frac{1}{y_\mathrm{H}+a^{-3}
+\chi a^{-4}}\frac{2-F'(R)}{3\tilde{m}^2 F''(R)}\, , \nn
& J_3=-3 a^{-3}-\frac{(1-F'(R))(a^{-3}+2\chi a^{-4})
+(R-F(R))/(3\tilde{m}^2)}{y_\mathrm{H}+a^{-3}
+\chi a^{-4}}\frac{1}{6\tilde{m}^2 F''(R)}\, .
\end{align}
In Eq.~(\ref{Js}), the scalar curvature $R$ is expressed in terms of
the quantity $y_\mathrm{H}$ as follows,
\begin{equation}
\label{Ricciscalar}
R=3\tilde{m}^2 \left(\frac{d y_\mathrm{H}}{d \ln a}
+4y_\mathrm{H}+a^{-3}\right)\, .
\end{equation}
Also we choose the free
parameters appearing in Eq.~(\ref{total}) as follows,
\begin{align}
\label{paratotal}
& \Lambda=(7.93)\tilde{m}^2\, , \quad
\Lambda_i=10^{100}\Lambda\, , \quad
R_i=2\Lambda_i\, ,\quad \phantom{sp}n=4\, , \nn
& \alpha=\frac{5}{2}\, ,\quad
\phantom{sp}\gamma=\frac{1}{(4\Lambda_i)^{\alpha-1}}\, , \quad
R_0=0.6\Lambda\, ,\quad 0.8\Lambda\, ,\quad \Lambda\, .
\end{align}
In Ref.~\cite{Elizalde:2010ts}, Eq.~(\ref{superEq}) was solved
numerically, in the range of $R_0\ll R\ll R_i$, which includes the
matter domination era and the current acceleration epoch. Then
$y_\mathrm{H}$ can be expressed as a function of the redshift, which
is equal to $z=\frac{1}{a}-1$. For the numerical calculations, the
initial conditions at $z=z_i$ were chosen in
Ref.~\cite{Elizalde:2010ts}, as follows,
\begin{equation}
\label{iniyH}
\left. \frac{d y_\mathrm{H}}{d (z)} \right|_{z_i}=0\, , \quad
\left. y_\mathrm{H}\right|_{z_i}=\frac{\Lambda}{3\tilde{m}^2}\, ,
\end{equation}
which
are practically the initial conditions of the $\Lambda$CDM model. By
using these initial conditions, the cosmological evolution of the
model at hand, is very close to that of the $\Lambda$CDM, when the
large redshift regime is considered. We can choose the values of
$z_i$, in such a way so that, $RF''(z = z_i) \sim 10^{-5}$, by
simply assuming that $R= 3\tilde{m}^2 (z+1)^3$. In effect, we have,
$z_i=1.5$, $2.2$, $2.5$ for $R_0=0.6\Lambda$, $0.8\Lambda$,
$\Lambda$, respectively. Note that the choice of the parameters were
made in view of the latest WMAP results, in combination with Baryon
Acoustic Oscillations and Supernovae surveys \cite{Komatsu:2008hk}.
The results of the numerical analysis indicate that the dark energy
EoS parameter is very close to the phantom divide line $-1$. At the
present epoch, where $z=0$, we find that $w_\mathrm{DE}=-0.994$,
$-0.975$, $-0.950$ for $R_0=0.6\Lambda$, $0.8\Lambda$, $\Lambda$
respectively. When $R_0$ decreases, the behavior of the model at
hand (\ref{total}), becomes indistinguishable from the $\Lambda$CDM
model, in which case $w_\mathrm{DE}=-1$. Also the numerical analysis
shows that the phantom divide crossing does not cause any serious
phenomenological issues to the model at hand (\ref{total}).
Particularly, in the limit $z\rightarrow -1^+$, the scalar curvature
$R$, becomes approximately equal to
$12\tilde{m}^2y_\mathrm{H}(z\rightarrow -1^+)$, and it acts like an
effective cosmological constant. Practically, this effective
cosmological constant is the stable late-time de Sitter attractor of
the theory.

It is known that for some viable models, the dark energy density
oscillates during the matter domination epoch, and the corresponding
frequencies might diverge \cite{Bamba:2012qi}. These dark energy
oscillations were also pointed out in Ref.~\cite{Matsumoto:2013sba}.
Due to the fact that the $F(R)$ gravity equations of motion include
fourth order derivatives, the scalar modes oscillations occur, in
addition to the graviton oscillations. In the scalar-tensor form of
the $F(R)$ gravity, the corresponding oscillations are quantified in
terms of the scalar field $\sigma$, and hence the frequency of the
oscillations are affected by the mass of the scalar field $\sigma$.
Practically, the scalar field $\sigma$ is the scalar curvature $R$,
and therefore any high-frequency oscillations occur, the background
will also oscillate rapidly. This in effect could make the
perturbative analysis inapplicable \cite{Bamba:2012qi}, due to the
occurrence of non-linearities. In a later section we shall discuss
the dark energy oscillations issue in a more concrete way.

\subsection{Unification of Inflation with Dark Energy Era in Modified
Gauss-Bonnet Gravity}

We may use the formulation of Eq.~(\ref{ma39}) for the model that
unifies the inflationary era with the accelerating expansion in the
present Universe. Then, we may consider a model, for which the
Hubble rate $H$ becomes a large constant during early times, that
is, for small $a=\e^N$, and at late times, the Hubble rate becomes a
small constant. For example, consider the following model,
\begin{equation}
\label{uH}
H(N) = \frac{H_e \e^{-N} + H_l \e^{N}}{\e^{-N} + \e^{N}}\, .
\end{equation}
Then at early times, where $N\to - \infty$, we have
$H\to H_e$. On the other hand, in at late times, where $N\to
+\infty$, we have, $H \to H_l$. Then if we choose $H_e\gg H_l$, the
inflationary era and the late-time acceleration can be realized.
Then, by choosing $h(\phi)$ in Eq.~(\ref{ma39}) as follows,
\begin{equation}
\label{uh}
h(\phi) = \frac{H_e \e^{-\phi} + H_l
\e^{\phi}}{\e^{-\phi} + \e^{\phi}}\, ,
\end{equation}
the unification of early
and late-time acceleration can be achieved easily for the
Gauss-Bonnet gravity, by using the formalism we presented in
sections II and III. We do not proceed in the details of this for
brevity, since details of this can be found in the previous
sections.

\subsection{Phantom Dark Energy Era}

An intriguing possibility for the late-time Universe is that the
effective equation of state (EoS) of the Universe, namely $w_\mathrm{eff}$,
ultimately crosses the phantom divide line which is $w_\mathrm{eff}=-1$.
This possible evolution era of our Universe is known as phantom dark
energy era, and in this section we shall present how this era can be
realized in the context of $F(R)$ gravity. The motivation for
studying a phantom dark energy era comes from the latest Planck data
\cite{Ade:2015xua}, but also from the earlier WMAP data
\cite{Spergel:2006hy} and from Type Ia supernovae gold dataset,
where the possibility of crossing the phantom divide line is
stressed. Particularly, with regard to the Planck data, the EoS
parameter is constrained to be $w_\mathrm{eff}=-1.54^{+0.62}_{-0.50}$,
which is approximately $2\sigma$ in the phantom domain, so this
clearly is a significant motivation to study a phantom dark energy
era.

The possibility of having a phantom dark energy era in the context
of modified gravity was stressed in Ref.~\cite{Bamba:2008hq}, and in
this section we shall present the most significant outcomes of this
work. Apart from the modified gravity approach, there exist in the
literature various different approaches with regard to the phantom
dark energy era, for example in Ref.~\cite{Nojiri:2003vn} a
superacceleration era is realized, in
Refs.~\cite{Perivolaropoulos:2005yv,Li:2005fm} a phantom dark energy
era is realized in the context of scalar-tensor theories with
non-minimal gravitational coupling between the scalar field and the
scalar curvature, and in Refs.~\cite{Nojiri:2005vv,Sami:2005zc} a
non-minimal coupling between the scalar field and the Gauss-Bonnet
term. Also in Ref.~\cite{Vikman:2004dc} the problem was addressed in
the context of a scalar field with non-linear kinetic terms, and in
Ref.~\cite{Nojiri:2005sx} phantom dark energy was coupled with dark
matter. Also in Ref.~\cite{Nojiri:2005sr} a thermodynamical
inhomogeneous dark energy model was introduced, and in
Ref.~\cite{Chimento:2006xu} a multiple $k$-essence model was
studied. Models with multiple scalar fields were studied in
Refs.~\cite{Elizalde:2008yf,Caldwell:2005ai}, and in
Refs.~\cite{Zhang:2005eg,Feng:2004ad,Elizalde:2004mq}, quintom
models were studied, which consisted of canonical and phantom scalar
fields. Finally, string inspired models were studied in
Refs.~\cite{Cai:2007gs,delaCruzDombriz:2006fj}.

Obviously the literature is rich on this issue, but in this section
we shall briefly discuss how the phantom dark energy era can be
realized by $F(R)$ gravity.

We start off our presentation by briefly mentioning the
reconstruction method we shall use, which was developed in
\cite{Nojiri:2006gh}. Consider an $F(R)$ gravity with action,
\begin{equation}
\label{action1dse}
\mathcal{S}=\frac{1}{2\kappa^2}\int d^4x
\sqrt{-g}F(R)+S_m(g_{\mu \nu},\Psi_m)\, ,
\end{equation}
where $S_m$ denotes
the action of all the matter fluids present. For the FRW metric, the
first FRW can be obtained by varying the action with respect to the
metric, so it reads,
\begin{equation}
\label{frwf1}
 -18\left ( 4H(t)^2\dot{H}(t)+H(t)\ddot{H}(t)\right )F''(R)
+3\left(H^2(t)+\dot{H}(t) \right )F'(R)-\frac{F(R)}{2}+\kappa^2\rho
=0\, ,
\end{equation}
where $F'(R)=\frac{d F(R)}{d R}$. By using an
auxiliary scalar field $\phi $, the $F(R)$ gravity action
(\ref{action}) can be written as follows,
\begin{equation}
\label{neweqn123}
S=\int d ^4x\sqrt{-g}\left ( P(\phi )R+Q(\phi )
+\mathcal{L}_\mathrm{matter} \right )\, ,
\end{equation}
where the term $\mathcal{L}_m$ stands for the matter Lagrangian. The
auxiliary functions $P(\phi )$ and $Q(\phi )$ will eventually
enable us to determine the $F(R)$ gravity which realizes a specific
cosmological evolution. By varying the action (\ref{neweqn123}) with
respect to the auxiliary scalar field $\phi $, we obtain,
\begin{equation}
\label{auxiliaryeqns}
P'(\phi )R+Q'(\phi )=0\, ,
\end{equation}
which will yield the function $\phi (R)$. Then, the resulting $F(R)$
gravity is,
\begin{equation}
\label{r1}
F(\phi( R))= P (\phi (R))R+Q (\phi (R))\, .
\end{equation}
The function $P(\phi )$ satisfies the following differential
equation,
\begin{align}
\label{r2}
& -6H^2P(\phi (t))-Q(\phi (t) )-6H\frac{d P\left (\phi
(t)\right )}{d t}+\rho_i=0 \, . \nn
& \left ( 4\dot{H}+6H^2
\right ) P(\phi (t))+Q(\phi (t) )+2\frac{d ^2P(\phi
(t))}{\mathrm {d}t^2}+\frac{d P(\phi
(t))}{d t}+p_i=0 \, ,
\end{align}
where $H(t)$ is the Hubble rate of a given evolution. By eliminating
the function $Q(\phi (t))$, we obtain,
\begin{equation}
\label{r3}
2\frac{d ^2P(\phi (t))}{\mathrm {d}t^2}-2H(t) P(\phi
(t))+4\dot{H}\frac{d P(\phi (t))}{d t}+\rho_i+p_i
=0 \, ,
\end{equation}
where $\rho_i,p_i$ stand for the total energy density and pressure
of the matter fluids which are present, respectively. As it is
proved in the Appendix of Ref.~\cite{Nojiri:2006gh}, due to the
mathematical equivalence of the actions (\ref{action}) and
(\ref{neweqn123}), the auxiliary scalar field is identified with the
cosmic time, that is $\phi =t$. Then, by assuming that the given
cosmological evolution has the following scale factor,
\begin{equation}
\label{r4}
a=a_0\e^{g(t)}\, ,
\end{equation}
we can rewrite the differential equation (\ref{r3}) as follows,
\begin{equation}
\label{r5}
2\frac{d ^2P(\phi (t))}{\mathrm {d}t^2}-2g'(\phi
)\frac{d P(\phi (t))}{d t}+4g''(\phi ) P(\phi
(t))+\sum_i(1+w_i)\rho_{i0}a_0^{-3(1+w_i)}\e^{-3(1+w_i)g(\phi )}=0
\, ,
\end{equation}
where $w_i$ are the effective equation of state parameters for the
perfect matter fluids that are present. Eventually, by finding the
function $P(\phi )$, we may find easily $Q(\phi )$, by using
Eq.~(\ref{r2}), and its final form is,
\begin{equation}
\label{r5a}
Q(\phi )=-6g'(\phi )^2P(\phi )-6g'(\phi )\frac{d P(\phi
)}{d \phi
}+\sum_i(1+w_i)\rho_{i0}a_0^{-3(1+w_i)}\e^{-3(1+w_i)g(\phi )}
\end{equation}
Let us apply this method in order to find the $F(R)$ gravity that
realizes a cosmology which crosses the phantom divide line. Consider
the vacuum $F(R)$ gravity case, in which case a solution to the
differential equation (\ref{r3}) is,
\begin{align}
\label{friendswacall}
& P(\phi)=\e^{g(\phi)/2}p(\phi)\, , \quad
g(\phi)=-10\ln \left[ \left (\frac{\phi}{t_0} \right)^{-\gamma}
 -C \left( \frac{\phi}{t_0}\right)^{\gamma+1}\right] \, , \nn
& p(\phi)=p_+\phi^{\beta_+}+p_-\phi^{\beta_{-}}\, , \quad
\beta_{\pm}=\frac{1+\pm \sqrt{1+100\gamma
(\gamma+1)}}{2}\, ,
\end{align}
with $\gamma$ and $C$ being arbitrary positive constants, $t_0$ is
the present time and $p_{\pm}$ being arbitrary constants. The
function $g(\phi)$ diverges at the time instance $t_s$ which is
equal to,
\begin{equation}
\label{tstimeinstance}
t_s=t_0C^{-1/(2\gamma+1)}\, ,
\end{equation}
so a Big Rip singularity occurs at $t_s$. Therefore we consider only
the era $0<t<t_s$, before the Big Rip singularity occurs. When $t\ll
t_s$, the Hubble rate is approximately,
\begin{equation}
\label{hubblesmallt1}
H(t)\sim \frac{10\gamma}{t}\, ,
\end{equation}
and by using the reconstruction method we just presented, the
resulting $F(R)$ gravity for $t\ll t_s$ is \cite{Bamba:2008hq},
\begin{equation}
\label{frphantomres1}
F(R)\sim \left[
\frac{(\frac{1}{t_0}\sqrt{60\gamma(20\gamma-1)
R^{-1/2}})^{\gamma}}{1-[\frac{1}{t_0}\sqrt{60\gamma
(20\gamma-1)R^{-1/2}}]^{2\gamma+1}}\right]^5R 
\sum_{j=\pm}\frac{5\gamma-1-\beta_j}{20\gamma-1}p_j
[60\gamma(20\gamma-1)]^{\beta_j/2}R^{-\beta_j/2}\, .
\end{equation}
Accordingly, when $t\to t_s$, the Hubble rate reads,
\begin{equation}
\label{hubbleratenerarip}
H(t)\sim \frac{10}{t_s-t}\, ,
\end{equation}
and therefore the resulting $F(R)$ gravity is
\cite{Bamba:2008hq},
\begin{align}
\label{frresulting}
F(R)\sim &\left(
\frac{\frac{1}{t_0^{\gamma}}\left[t_s-3\sqrt{140}R^{-1/2}
\right]^{\gamma}}{1-\left[1-\frac{3\sqrt{140}R^{-1/2}}{t_s}
\right]}\right)^5R\sum_{j=\pm}p_j\left[t_s-3\sqrt{140}
R^{-1/2}\right]^{\beta_j} \nn
&\times \left[
1-\sqrt{\frac{20}{7}}\left[\frac{15}{84}t_s+(\beta_j-15)
R^{-1/2}\right]\frac{1}{t_s-3\sqrt{140}R^{-1/2}}\right]\, .
\end{align}
It can be easily checked that the phantom divide is crossed for the
above cosmology, since the EoS parameter
$w_\mathrm{eff}=-1-\frac{2\dot{H}}{H^2}$ for $t\to 0$ is equal to,
\begin{equation}
\label{weffttotzero}
w_\mathrm{eff}=-1+\frac{1}{15\gamma}\, ,
\end{equation}
which satisfies $w_\mathrm{eff}>-1$. Accordingly, for $t>t_s$, the EoS is
equal to $w_\mathrm{eff}=-16/15$, which satisfies
$w_\mathrm{eff}<-1$, so the
phantom divide line is crossed. It can be found that the phantom
divide line is crossed at the time instance $t_c$, which is equal
to,
\begin{equation}
\label{tscrit}
t_c=t_s\left(-2\gamma+\sqrt{4\gamma^2}+\frac{\gamma}{\gamma+1}
\right)^{1/(2\gamma+1)}\, .
\end{equation}
The phenomenological implications of a phantom scalar-tensor theory,
when this is considered in the Jordan frame, are quite interesting
and these were studied in \cite{Bamba:2008hq}, and as it was shown,
a phantom scalar-tensor theory corresponds to a complex $F(R)$
gravity, when this theory is considered in the Jordan frame.

\subsection{Dark Energy Oscillations in $F(R)$ Gravity
Theories and Growth Index}

In this section we shall address two important issues relevant to
late-time dynamics, namely the dark energy oscillations issue
\cite{Linder:2005dw,Nojiri:2006ww,Kurek:2007bu,Liu:2009nv,
Jain:2007fa,Gorbunova:2009zz,Lazkoz:2010gz,Elizalde:2011ds,
Bamba:2012qi,Oikonomou:2014gsa} and the evolution of the growth
index \cite{Bamba:2012qi,Motohashi:2011wy}, in the context of $F(R)$
gravity. Both issues are related to higher derivatives of the Hubble
rate, so these are higher order effects in the late-time dynamics,
and therefore these are non-trivial effects.

\subsubsection{Dark Energy Oscillations}

In order to reveal the dark energy oscillations issue, and to study
it more formally, we rewrite the $F(R)$ gravity gravitational
equations as follows,
\begin{align}
\label{eq:flrw}
& 3F'H^2=k^2\rho_\mathrm{matter}+\frac{1}{2}(F'R-F)-3H\dot{F'} \nn
& -2F'\dot{H}=k^2\left (\rho_\mathrm{matter}+P_\mathrm{matter} \right)
+\ddot{F}-H\dot{F} \, ,
\end{align}
where $\rho_\mathrm{matter}$ is the total mass-energy density, which
contains all the matter fluids present. In order to provide a
generalized picture, we shall not restrict ourselves to perfect
matter fluids, but we shall assume that the matter fluid consists of
collisional matter \cite{Oikonomou:2014lua,Kleidis:2011ga} and
radiation, and therefore the total matter energy density is in the
case at hand,
\begin{equation}
\label{totalmattenergdensmf}
\rho_\mathrm{matter}=\varepsilon_m+\rho_{r}^{(0)}a^{-4}\, ,
\end{equation}
where $\varepsilon_m$ is,
\begin{equation}
\label{totenergydens}
\varepsilon_m=\rho_m\left[1+\Pi_0+w \ln \left(
\frac{\rho_m}{\rho_m^{(0)}} \right) \right]\, ,
\end{equation}
and also
$\rho_m$ is,
\begin{equation}
\label{evlotuniomatt}
\rho_m=\rho_\mathrm{matter}^{(0)}\left( \frac{a_0}{a} \right)^3\, .
\end{equation}
It is conceivable that the
non-collisional matter case is described by solely $\rho_m$. So by
combining Eqs.~(\ref{totenergydens}) and (\ref{evlotuniomatt}), we
eventually have,
\begin{equation}
\label{totalmattenergdensmf1}
\rho_\mathrm{matter}=\rho_m^{(0)}a^{-3}\left[ 1+\Pi_0+3w \ln (a)
\right] +\rho_{r}^{(0)}a^{-4}\, .
\end{equation}
Moreover, $P_\mathrm{matter}$
stands for the total pressure of the matter fluids present. We can
rewrite the first equation in Eq.~(\ref{eq:flrw}), as follows,
\begin{equation}
\label{eq:modifiedeinsteineqns2}
H^2-(F'-1)\left (H\frac{d H}{d \ln{a}} +H^2 \right )
+\frac{1}{6}(F-R)+H^2F''\frac{d R}{d \ln{a}}
=\frac{\rho_\mathrm{matter}}{3}\, ,
\end{equation}
with $R$ denoting the scalar
curvature as usual, which can be rewrite as follows,
\begin{equation}
\label{eq:ricciscal2}
R=12H^2+6H\frac{d H}{d \ln{a}}\, .
\end{equation}
In
order to provide a more generalized picture regarding the various
types of collisional matter, we assume that $\rho_m$ has the
following form,
\begin{equation}
\rho_\mathrm{matter}=\rho_m^{(0)}g(a)+\rho_{r}^{(0)}a^{-4}
=\rho_m^{(0)}\left( g(a)+\chi a^{-4}\right )\, ,
\end{equation}
where
$\rho_m^{(0)}$ and $\rho_r^{(0)}=\chi\rho_m^{(0)}$ denote the values
of the mass energy density and of radiation, at present time. Also,
$\chi\simeq3.1\times10^{-4}$, and it is defined to be the ratio
$\rho_r^{(0)}/\rho_m^{(0)}$. Effectively, the non-collisional nature
of matter is quantified in terms of $g(a)$, and in the case of
collisional non-relativistic matter, it is,
\begin{equation}
\label{fgd}
g(a)=a^{-3}\left[ 1+\Pi_0-3w \ln (a) \right] \, .
\end{equation}
The dark
energy oscillations are higher order effects, and in order to
perfectly quantify these effects, we need to introduce new variables
in Eqs.~(\ref{eq:modifiedeinsteineqns2}), with the characteristic
property that these new variables will vanish at high redshift
regimes. During these regimes the $F(R)$ gravity modifications are
expected to be negligible \cite{Elizalde:2011ds}. The new variables
are,
\begin{subequations}
\begin{align}
\label{eq:yH}
y_H&\equiv\frac{\rho_\mathrm{DE}}{\rho_m^{(0)}}=\frac{H^2}{\tilde{m}^2}
 -g(a)-\chi a^{-4} \, , \\
\label{eq:yR}
y_R&\equiv\frac{R}{\tilde{m}^2}-\frac{d g(a)}{d \ln{a}}\, ,
\end{align}
\end{subequations}
where $\rho_\mathrm{DE}$ the total dark energy, energy density. Hence, the
variable that quantifies the evolution is $y_H(z)$, which is the
scaled dark energy density with the scaling factor being
$\rho_m^{(0)}$. By using,
\begin{equation}
\frac{H^2}{\tilde{m}^2}-\frac{\rho_M}{2\tilde{m}^2}
=\frac{H^2}{\tilde{m}^2}-g(a)-\chi a^{-4}=y_H \, ,
\end{equation}
from (\ref{eq:yH}), and by dividing
Eqs.~(\ref{eq:modifiedeinsteineqns2}) by $\tilde{m}^2$, we get,
\begin{equation}
\label{eq:dRdlna}
\frac{1}{\tilde{m}^2}\frac{d R}{d \ln{a}}
=\left[-y_H+(F'-1)(\frac{H}{\tilde{m}^2}\frac{d H}{d \ln{a}}
+\frac{H^2}{\tilde{m}^2})-\frac{1}{6\tilde{m}^2}(F-R)\right]
\frac{1}{H^2F''}\, .
\end{equation}
Differentiating Eq.~(\ref{eq:yR}), with respect to the variable
$\ln{a}$ we obtain,
\begin{equation}
\frac{d y_R}{d \ln{a}}=\frac{1}{\tilde{m}^2}\frac{d R}{d \ln{a}}
 -\frac{d ^2g(a)}{d \ln{a}^2}\, ,
\end{equation}
and by using (\ref{eq:dRdlna}), we have,
\begin{equation}
\label{eq:dyR}
\frac{d y_R}{d \ln{a}}=-\frac{d ^2g(a)}{d \ln{a}^2}
+\left[-y_H+(F'-1)(\frac{H}{\tilde{m}^2}\frac{d H}{d \ln{a}}
+\frac{H^2}{\tilde{m}^2})-\frac{1}{6\tilde{m}^2}(F-R)\right]
\frac{1}{H^2F''} \, .
\end{equation}
Differentiating Eq.~(\ref{eq:yH}) again with respect to the variable
$\ln{a}$, we have,
\begin{equation}
\label{eq:dHlna}
\frac{H}{\tilde{m}^2}\frac{d H}{d \ln{a}}+\frac{H^2}{\tilde{m}^2}
= \frac{1}{2}\frac{d y_H}{d \ln{a}}
+\frac{1}{2}\frac{d g(a)}{d \ln{a}}+y_H+g(a)-\chi a^{-4}\, ,
\end{equation}
where we also used of Eq.~(\ref{eq:yH}). By using
Eqs.~(\ref{eq:dHlna}) and (\ref{eq:dyR}), we obtain,
\begin{align}
\label{eq:dyR2}
\frac{d y_R}{d \ln{a}}=&-\frac{d ^2g(a)}{d \ln{a}^2}
+ \left[-y_H+(F'-1)\left(\frac{1}{2}\frac{d y_H}{d \ln{a}}
+\frac{1}{2}\frac{d g(a)}{d \ln{a}}
+y_H+g(a)-\chi a^{-4}\right)\right. \nn
&\left.-\frac{1}{6\tilde{m}^2}(F-R)\right]
\frac{1}{\tilde{m}^2F''(y_H+g(a)+\chi a^{-4})}\, .
\end{align}
So by differentiating Eq.~(\ref{eq:dRdlna}) with respect to the
variable $\ln{a}$, we obtain,
\begin{equation}
\label{eq:dyH}
\frac{d y_H}{d \ln{a}}=\frac{2H}{\tilde{m}^2}\frac{d H}{d \ln{a}}
 -\frac{d g(a)}{d \ln{a}}+4\chi a^{-4}\, .
\end{equation}
Upon combining Eqs.~(\ref{eq:ricciscal2}) and (\ref{eq:yH}) we
obtain,
\begin{equation}
\frac{2H}{\tilde{m}^2}\frac{d H}{d \ln{a}}
=\frac{R}{3\tilde{m}^2}-\frac{4H^2}{\tilde{m}^2}
=\frac{R}{3\tilde{m}^2}-4y_H-4g(a)-4\chi
a^{-4}\, ,
\end{equation}
and therefore Eq.~(\ref{eq:dyH}) becomes,
\begin{align}
\label{eq:dyH2}
\frac{d y_H}{d \ln{a}}v=&\frac{R}{3\tilde{m}^2}
 -\frac{d g(a)}{d \ln{a}}-4y_H-4g(a) \nn
=&\frac{y_R}{3}-4y_H-\frac{2d g(a)}{3d \ln{a}}-4g(a) \, .
\end{align}
Moreover we can rewrite the scalar curvature as follows,
\begin{equation}
R=3\tilde{m}^2\left(4y_H+4g(a)+\frac{d y_H}{d \ln{a}}
+\frac{d g(a)}{d \ln{a}}\right)\, .
\end{equation}
By differentiating Eq.~(\ref{eq:dyH2}) with respect
to the variable $\ln{a}$, we get,
\begin{equation*}
\frac{d ^2y_H}{d \ln{a}^2}=\frac{d y_R}{3d \ln{a}}-\frac{4d y_H}{d \ln{a}}
 -\frac{2}{3}\frac{d ^2g(a)}{d \ln{a}^2}-4\frac{d g(a)}{d \ln{a}} \, ,
\end{equation*}
so by using Eq.~(\ref{eq:dyR2}), we get,
\begin{align}
\label{eq:FRform}
&\frac{d ^2y_H}{d \ln{a}^2}+\left(4+\frac{1-F'}{6\tilde{m}^2F''(y_H+g(a)
+ \chi a^{-4})}\right)\frac{d y_H}{d \ln{a}}
+ \left(\frac{2-F'}{3\tilde{m}^2F''(y_H+g(a)+\chi a^{-4})}\right)y_H \nn
& +\left(\frac{d ^2g(a)}{d \ln{a}^2}+4\frac{d g(a)}{d \ln{a}}
+\frac{(1-F')\left(3\frac{d g(a)}{d \ln{a}}+6g(a)
 -6\chi a^{-4}\right)+\frac{F-R}{\tilde{m}^2}}
{18\tilde{m}^2F'' (y_H+g(a)+\chi a^{-4})}\right)=0\, .
\end{align}
By using the following relations,
\begin{subequations}
\begin{align}
\frac{d }{d \ln{a}}&=-(z+1)\frac{d }{d z}\, ,\\
\frac{d ^2}{d \ln{a}^2}&=(z+1)\frac{d }{d z}+(z+1)^2\frac{d ^2}{d z^2}\, .
\end{align}
\end{subequations}
we can express all the previous results in terms of the redshift
$z$, so by applying the above relations in Eq.~(\ref{eq:FRform}), we
obtain the following master differential equations, which dictates
the way that the Universe evolves, which is,
\begin{align}
&\frac{d ^2y_H}{d z^2}
+ \frac{1}{(z+1)}\left(-3-\frac{F'(R)}{6\tilde{m}^2F''(R)(y_H
+g(z)+\chi(z+1)^4)}\right)\frac{d y_H}{d z} \nn
&+\frac{1}{(z+1)^2}\frac{1-F'(R)}{3\tilde{m}^2F''(R)(y_H+g(Z)
+\chi(z+1)^4)}y_H+\left(\frac{
d ^2g(z)}{d z^2}-\frac{3}{(z+1)}\frac{d g(z)}{d z}\right. \nn
&\left.-\frac{1}{(z+1)}\frac{F'(R)\left(-(z+1)\frac{d g(z)}{
d z}+2g(z)-2\chi(z+1)^4\right)
+\frac{F}{3\tilde{m}^2}}{6\tilde{m}^2F''(R)(y_H
 -g(z)+\chi(z+1)^4)}\right)=0 \, .
\end{align}
It is conceivable that the above differential equation is not easy
to solve analytically, so we shall solve this numerically and we
study the late-time evolution in terms of the resulting numerical
solution. The matter profile shall be assumed to be that of
Eq.~(\ref{fgd}), and we shall compare the cases that the matter is
non-collisional to the collisional matter case. We assume that the
$F(R)$ gravity is a modified exponential gravity of the form
\cite{Elizalde:2011ds,Bamba:2012qi},
\begin{equation}
\label{expmodnocurvcorr}
F(R)=R-2\Lambda \left ( 1-\e^{\frac{R}{b\Lambda}}
\right)-\tilde{\gamma}\Lambda
\left(\frac{R}{3\tilde{m}^2} \right )^{1/3} \, ,
\end{equation}
where $\Lambda=7.93\tilde{m}^2$ and $\tilde{\gamma}=1/1000$
\cite{Elizalde:2011ds,Bamba:2012qi,Oikonomou:2014gsa}. Also we
assume that $\Omega_M=0.279$, where $\Omega_M$ is the present value
of the total energy density, and also that $w=0.2$ and $w=0.8$, so
we study both these cases for collisional matter. Also the initial
conditions are assumed to be
\cite{Elizalde:2011ds,Bamba:2012qi,Oikonomou:2014gsa},
\begin{equation}
\label{initialcond}
y_H(z)\mid_{z=z_{f}}=\frac{\Lambda
}{3\tilde{m}^2}\left(1+\frac{z_{f}+1}{1000}\right)\, ,\quad
y_\mathrm{H}'(z)\mid_{z=z_{f}}=\frac{\Lambda}{3\tilde{m}^2}
\frac{1}{1000} \, ,
\end{equation}
where $z_{f}=10$ and also $\Lambda $ is $\Lambda \simeq 11.89
$eV$^2$ .

\begin{figure}[h]
\centering
\includegraphics[width=15pc]{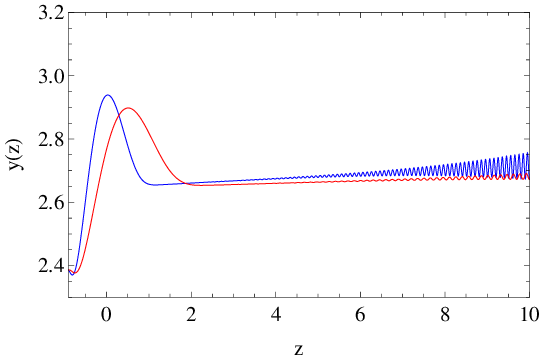}
\includegraphics[width=15pc]{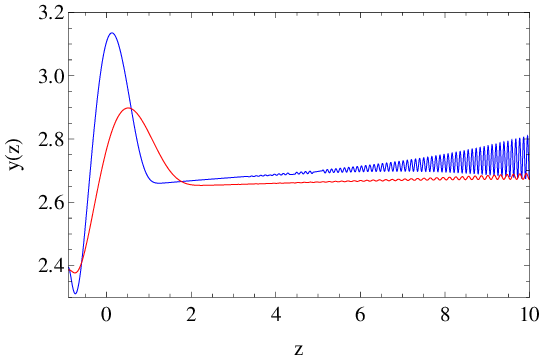}
\caption{Comparison of the scaled dark energy density
$y_H(z)=\frac{\rho_\mathrm{DE}}{\rho_m^{(0)}}$ over z, for $w=0.2$
(left) and $w=0.8$ (right). The red line corresponds to
non-collisional matter and the blue one to collisional
matter.}\label{ycomp}
\end{figure}

We start our analysis with the numerical solution for $y_H(z)$, both
for non-collisional and collisional matter, and we use the two
values for the parameter $w$, namely $w=0.2$ and $w=0.8$. In
Fig.~\ref{ycomp} we compare the resulting $y_H(z)$ collisional and
non-collisional matter cases. The blue curves correspond to
collisional matter with $g(a)$ being given in Eq.~(\ref{fgd}), and
the red curves correspond to non-collisional matter with
$g(a)=a^{-3}$. The left plot corresponds to $w=0.2$ and the right
plot to $w=0.8$.
As it can be seen, the scaled dark energy
$y_H(z)=\frac{\rho_\mathrm{DE}}{\rho_m^{(0)}}$ for the collisional matter,
has a strong oscillatory behavior in comparison to the
non-collisional matter. This behavior becomes more intense as the
parameter $w$ increases, and note that the two cases coincide when
$w=0$.

\begin{figure}[h]
\centering
\includegraphics[width=15pc]{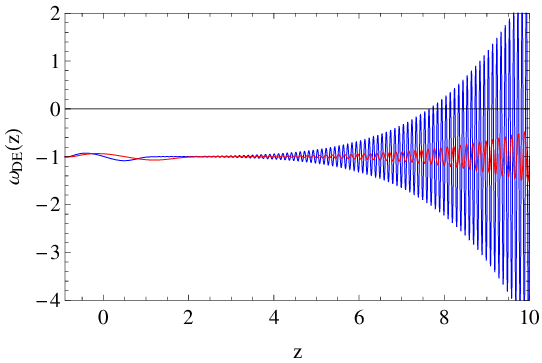}
\includegraphics[width=15pc]{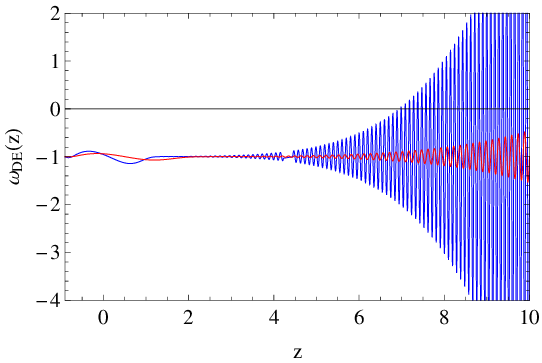}
\caption{Behavior of the dark energy equation of state parameter
$\omega_\mathrm{DE}(z)$ as a function of the redshift $z$, for
$w=0.2$ (left plot) and $w=0.8$ (right plot). The red curves
correspond to non-collisional matter while the blue curves
correspond to collisional matter.}\label{omegadecomp}
\end{figure}

Let us now investigate the behavior of the dark energy oscillations,
so we quantify these in terms of the dark energy equation of state
parameter, which is defined as
$\omega_\mathrm{DE}=P_\mathrm{DE}/\rho_\mathrm{DE}$. The
dark energy density in terms of the redshift and also in terms of
$y_H(z)$, is equal to,
\begin{equation}
\label{deeqnstateprm}
\omega_\mathrm{DE}(z)=-1+\frac{1}{3}(z+1)\frac{1}{y_H(z)}\frac{d y_H(z)}{d z}
\, .
\end{equation}
In Fig.~\ref{omegadecomp}, we have plotted the behavior of the dark
energy equation of state parameter, for collisional matter (blue
curves) and for non-collisional matter (red curves), for $w=0.2$
(left plot) and $w=0.8$ (right plot).
In this case, the same behavior as in the case of $y_H(z)$ occurs,
so as the value of the parameter $w$ increases, the oscillations
become more pronounced.

\begin{figure}[h]
\centering
\includegraphics[width=15pc]{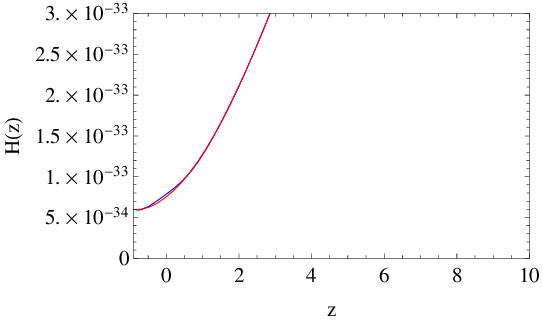}
\caption{Comparison of the Hubble parameter $H(z)$ over z for
$w=0.8$. The red line corresponds to non-collisional matter while
the blue corresponds to collisional matter.}\label{hubcurv}
\end{figure}

\begin{figure}[h]
\centering
\includegraphics[width=15pc]{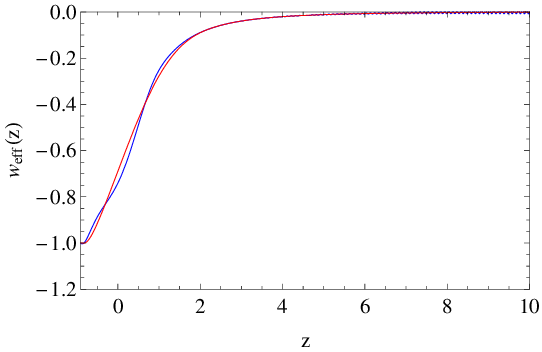}
\includegraphics[width=15pc]{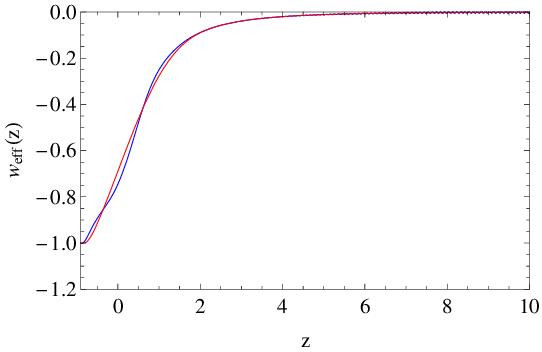}
\caption{Comparison of the effective equation of state parameter
$\omega_\mathrm{eff}(z)$ over z, for $w=0.2$ (left) and $w=0.8$
(right). The red line corresponds to non-collisional matter while
the blue corresponds to collisional matter.}\label{weffcomp}
\end{figure}

Finally, let us see how the scalar curvature $R(z)$, the Hubble rate
$H(z)$, and the total equation of state $\omega_\mathrm{eff}(z)$ behave as
a function of $z$. Starting with the $H(z)$, this can be expressed
in terms of $y_H(z)$ as follows,
\begin{equation}
\label{hubblepar}
H(z)=\sqrt{\tilde{m}^2y_H(z)+g(a(z))+\chi (z+1)^{4}}\, ,
\end{equation}
so by using the numerical solution we found for $y_H(z)$, in
Fig.~\ref{hubcurv} (left plot), we plotted the evolution of the Hubble
rate as a function of $z$, for $w=0.8$, both for collisional (blue)
and non-collisional matter (red). As it can be seen, the behavior of
$H(z)$ in both the collisional and non-collisional matter, is the
same.
The same applies for the scalar curvature, which can be written as,
\begin{equation}
R=3\tilde{m}^2\left(4y_H+4g(z)-(z+1)\frac{d y_H}{d z}
 -(z+1)\frac{d g(z)}{d z}\right)\, .
\end{equation}
Finally, let us discuss the behavior of the total equation of state
parameter $\omega_\mathrm{eff}(z)$, and in Fig.~(\ref{weffcomp}), we
compare the collisional (blue curves) and non-collisional (red
curves) matter cases, for $w=0.2$ (left plot) and $w=0.8$ (right
plots).
Note that we used the formula,
\begin{equation}
\label{effeqnofstateform}
\omega_\mathrm{eff}(z)
=-1+\frac{2(z+1)}{3H(z)}\frac{d H(z)}{d z}\, .
\end{equation}
By looking at the behavior of the dark energy equation of state
parameter and the total equation of state parameter, we can see that
the total equation of state parameter never crosses the phantom
divide line, however the dark energy equation of state parameter,
has strong oscillations around the phantom divide line $w=-1$.

\subsubsection{Growth Index Evolution}

\begin{figure}[h]
\centering
\includegraphics[width=15pc]{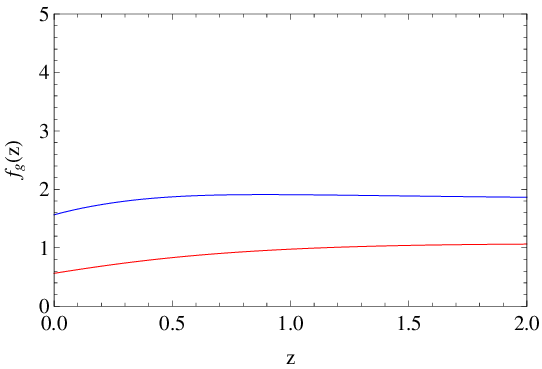}
\caption{Plots of the growth factor $f_g(z)=\frac{d \ln
\delta}{d \ln a}$, as a function of the redshift $z$, for
$k=0.1$Mpc$^{-1}$ and $w=0.8$. The blue curve corresponds to
collisional matter and the red curve to non-collisional
matter.}\label{fig:oikonomouplots1}
\end{figure}

As a final issue, we shall study the evolution of the growth index
in the context of $F(R)$ gravity, assuming a general profile for the
matter fluids. It is known that the cosmological evolutions
corresponding to various $F(R)$ gravities can be distinguished by
using the cosmological perturbation theory. Particularly, it might
happen that two $F(R)$ gravity models might generate similar
cosmological evolutions in terms of the Hubble rate or in terms of
the evolution of the scalar curvature in the two cases. In the
previous section we came across with one situation like this, since
the exponential $F(R)$ gravity model for collisional matter was
indistinguishable from the non-collisional evolution, when the
Hubble rate was considered. However, the cosmological perturbations
may distinguish the various modified gravity models, since the
cosmological perturbations differentiate the evolution caused by
each model, from the background evolution
\cite{Bamba:2012qi,Matsumoto:2011ne,Fu:2010zza}. In this section we
shall focus on the matter density perturbations, in the subhorizon
approximation, where the theory is consistent with the Newtonian
limit \cite{Bamba:2012qi,Matsumoto:2011ne}. We shall address the
matter perturbations issue in the context of $F(R)$ gravity with a
general collisional mass profile. In the context of the subhorizon
approximation, the wavelengths that are considered comoving with the
spacelike hypersurface which describes the evolution, are much
shorter in comparison to the Hubble radius $R_H=1/(aH)$ of this
hypersurface, that is, \cite{Bamba:2012qi,Matsumoto:2011ne},
\begin{equation}
\label{subhorapprx}
\frac{k^2}{a^2}\gg H^2 \, ,
\end{equation}
with $k$ and $a$
being the wavenumber an the scale factor respectively. The
subhorizon approximation breaks down before the matter domination
era, so we focus on the matter domination era and onwards. In the
case of a generalized collisional matter profile, the matter density
perturbations are quantified in terms of $\delta =\frac{\delta
\varepsilon_m}{\varepsilon_m}$, where $\varepsilon_m$ is the
collisional matter-energy density given in
Eq.~(\ref{totenergydens}). The non-relativistic non-interacting
matter case, can be easily obtained from the collisional case, if
the self interactions are zero in (\ref{totenergydens}), which means
if $\Pi_0$ and $w$ are zero in Eq.~(\ref{totenergydens}). The
parameter $\delta$ satisfies the differential equation
\cite{Bamba:2012qi,Motohashi:2011wy,delaCruzDombriz:2008cp},
\begin{equation}
\label{matterperturb}
\ddot{\delta}+2H\dot{\delta}
 -4\pi G_\mathrm{eff}(a,k)\varepsilon_m\delta =0 \, ,
\end{equation}
where $G_\mathrm{eff}(a,k)$ is the $F(R)$ gravity theory effective
gravitational constant, which is,
\begin{equation}
\label{geff}
G_\mathrm{eff}(a,k)=\frac{G}{F'(R)}\left[ 1
+\frac{\frac{k^2}{a^2}\frac{F''(R)}{F'(R)}}{1
+3\frac{k^2}{a^2}\frac{F''(R)}{F'(R)}}
\right] \, ,
\end{equation}
where $G$ is the Newtonian gravity gravitational constant. By using
the quantity $f_g(z)=\frac{d \ln \delta}{d \ln a}$,
and also the relations,
\begin{align}
\label{basicrelations}
& \dot{H}=\frac{d H}{d z}(z+1)H(z)\, , \nn
& \dot{\delta}=H\dot{f_g}\delta \, , \nn
& \ddot{\delta}=\dot{H}f_g\delta+H\dot{f_g}\delta
+Hf_g\dot{\delta}\, ,
\end{align}
we can express the differential equation (\ref{matterperturb}) in
terms of the redshift $z$,
\begin{equation}
\label{presenceofcoll}
\frac{d f_g(z)}{d z}
+\left(\frac{1+z}{H(z)}\frac{d H(z)}{d z}-2
 -f_g(z)\right)\frac{f_g(z)}{1+z}
+\frac{4\pi}{G}
\frac{G_\mathrm{eff}(a(z),k)}{(z+1)H^2(z)}\varepsilon_m=0 \, .
\end{equation}
The term $G_\mathrm{eff}(a(z),k)$ determines the effects of $F(R)$ gravity
on the matter perturbations, and note that it depends explicitly on
the wavenumber, something that does not happen in the general
relativistic case. Now we numerically solve (\ref{presenceofcoll})
and we use the same initial conditions we used in the previous
section, with $z_\mathrm{fin}=10$. In order to respect the subhorizon
approximation, we need to be cautious by choosing the values of the
wavenumber, so by using the present time values of the Hubble rate
and of the scale factor, we easily find that the wavenumber has to
satisfy $k>0.000156$, for non-collisional matter, while in the
collisional case, we have $k>0.0001174$. So by disregarding
radiation, in which case,
\begin{equation}
\label{ga}
g(a)=a^{-3} \left[ 1+\Pi_0+3w \ln (a) \right] \, ,
\end{equation}
in Fig.~\ref{fig:oikonomouplots1} we plotted the growth factor
$f_g(z)$ as a function of the redshift, for collisional (blue curve)
and non-collisional matter (red curve), by using $k=0.1$Mpc$^{-1}$
and $w=0.8$
As we can see in Fig.~(\ref{fig:oikonomouplots1}), unlike in the
Hubble rate comparison for collisional and non-collisional matter,
where the two cases are indistinguishable, the matter perturbations
differ significantly in the collisional and non-collisional cases.
This justifies our statement that the matter perturbations are
valuable since these can be used in order to distinguish the effects
of different modified gravity theories.

Before closing we need to mention that in the case of mimetic $F(R)$
gravity, if the potential is appropriately chosen, the dark energy
oscillations do not occur, see
Refs.~\cite{Odintsov:2016oyz,Oikonomou:2016pkp} for more details on this
issue.

Thus in this section we discussed the dark energy issue and how to
describe inflation and dark energy in a unified way in the context
of modified gravity.

We need to note that our presentation intended to describe the
procedure of model building in the context of modified gravity, and
how modified gravity models can harbor the two acceleration eras. So
we did not attempt to solve any of the phenomenological models that
the $\Lambda$CDM model suffers from, like the coincidence problem.
This issue would require more advanced techniques, unknown so far,
so this issue is out of the scope of this review. We believe that a
potential way to solve such intriguing problems such as the
coincidence problem\footnote{The coincidence problem refers to the
question why dark energy and dark matter energy densities are
similar in magnitude at present time.} is to use cosmographic
principles\cite{Bamba:2012cp,Dunsby:2015ers,Capozziello:2008qc,Capozziello:2011hj},
or similar approaches, that may provide insightful hints. However,
we mainly focused on describing certain aspects of late-time
evolution and also to theorize how the unification of early and
late-time acceleration can be achieved. It is conceivable that a the
presentation of a complete cosmological model that describes all the
evolution eras, is what most theoretical cosmologists seek for the
moment, so it is not possible to present it in a review article,
since this issue is the focus of the research of many cosmologists
today (2017). Some insights may be provided by modified gravity, see
for example Ref.~\cite{Odintsov:2016plw}, in which the unified
description of all cosmological eras of our Universe was achieved in
the context of $F(R)$ gravity, but in the model of Ref.
\cite{Odintsov:2016plw}, partial solutions were given due to the
lack of analyticity. But the challenge in every cosmological theory
is to find the ultimate theory which will combine all the features
we just discussed. The path towards this theory is long and light
will be shed by the observational data. As theoretical physics
history dictates us, the cosmologists working today are particularly
lucky since the existence of petabytes of cosmological data show the
way on how to construct viable models, unlike what happens in string
theories, due to the lack of experimental data.

\section{Astrophysical Applications}

A complete review on modified gravity could not be complete without
even a small chapter devoted on astrophysical applications of
modified gravity. Hence, although we deviate conceptually from the
title of this review, in this chapter we shall briefly present the
latest developments of modified gravity on astrophysical objects and
solutions. We shall be interested in neutron and quark stars, the
anti-evaporation phenomena in black holes and finally wormholes, in
the context of $F(R)$ gravity.

\subsection{Neutron and quark stars from $F(R)$ gravity}

Modified gravity offers the possibility of having new types of
neutron stars allowed, that cannot be described by the standard
Einstein gravity. Recently, in
Refs.~\cite{Barziv:2001ad,Rawls:2011jw,Nice:2005fi,Demorest:2010bx},
massive compact neutron stars with mass $M_\mathrm{NS} \sim
2M_{\odot}$ have been found, where $M_{\odot}$ is the solar mass. In
the contexts of the standard Einstein gravity and modern hadron
physics, it is rather impossible to realize such super-massive
neutron stars. The conceptual approaches towards understanding the
neutron stars, are mainly realized by two streams of research,
namely, the particle physics approach, and the gravitational theory
approach. From the viewpoint of particle physics approach, the EoS
of the matter which consists the neutron star, practically
determines the physics of the neutron star. On the other hand, the
gravitational approach may have new insights to offer since this is
a much more fundamental approach, affecting the inner structure of
the neutron star in a geometric way. Since the size of the compact
stars is determined by the balance between two competing forces, the
degeneracy force and the gravitation force, there are three
approaches that may explain the mass of the neutron star, which we
list here:
\begin{itemize}
\item[(I)] The repulsive force is considered to be stronger in
comparison to the force corresponding to Quantum-Chromo-Dynamics
originating equations of state. Such examples of EoS, can be due
to new types of interactions, as in
Refs.~\cite{Miyatsu:2013yta,Tsubakihara:2012ic}.
\item[(II)] The attractive force is weaker in comparison to the
one predicted from the standard Einstein gravity.
\item[(III)] Both cases (I) and (II) apply.
\end{itemize}

Here we shall consider the viewpoint of the approach (II), that is,
we shall try to explain the massive neutron stars solutions, by
using $F(R)$ gravity, as in
Refs.~\cite{Capozziello:2011nr,Astashenok:2014pua,
Astashenok:2014nua,Capozziello:2015yza,
Yazadjiev:2015zia,Staykov:2014mwa,Fiziev:2014rla,Henttunen:2007bz,Resco:2016upv,Multamaki:2007jk}
or by using the dRGT massive gravity coupled with matter, which is
described by using standard equations of state
\cite{Katsuragawa:2015lbl}.

We shall base the $F(R)$ gravity approach on neutron stars on
Ref.~\cite{Astashenok:2013vza}, and for similar approaches on
relativistic stars in the context of modified gravity, see
Refs.~\cite{Alavirad:2013paa,DeLaurentis:2013zv,delaCruzDombriz:2012xy,
Capozziello:2011gm,Gungor:2011vq,Bamba:2011sm,Arapoglu:2010rz,
Babichev:2009fi,Cooney:2009rr,Babichev:2009td,Kobayashi:2008tq}.

We consider the following form for the $F(R)$ gravity functions,
\begin{equation}
\label{fR}
F(R)=R+\alpha h(R)\, .
\end{equation}
At a later point we shall
assume that $\alpha$ is small enough, but for the moment $\alpha$ is
arbitrary. We shall consider a spherically symmetric and static
metric of the form,
\begin{equation}
\label{metric}
ds^2= - \e^{2\phi}c^2 dt^2 + \e^{2\lambda}dr^2
+r^2 (d\theta^2 +\sin^2\theta d\phi^2)\, .
\end{equation}
Also the energy-momentum tensor corresponding to the perfect fluid
which consists the neutron star, is
$T_{\mu\nu}=\mbox{diag}(\e^{2\phi}\rho c^{2}, \e^{2\lambda}p, r^2p,
r^{2}\sin^{2}\theta p)$, were $\rho$ and $p$ are the energy density
and the pressure of the perfect fluid respectively. In effect, we
find the following components of the field equations,
\begin{align}
\label{f-tt}
\frac{ -8\pi G}{c^2} \rho =& -r^{-2} + \e^{-2\lambda}(1-2r\lambda')r^{-2}
+\alpha h_R(-r^{-2} + \e^{-2\lambda}(1-2r\lambda')r^{-2}) \nn
& -\frac12\alpha(h-h_{R}R) + \e^{-2\lambda}\alpha
\left[ h_R'r^{-1}(2-r\lambda')+h_R'' \right] \, , \\
\frac{8\pi G}{c^4} P =& -r^{-2} + \e^{-2\lambda}(1+2r\phi')r^{-2}
+\alpha h_R(-r^{-2} + \e^{-2\lambda}(1+2r\phi')r^{-2}) \nn &
-\frac12\alpha(h-h_{R}R) + \e^{-2\lambda}\alpha
h_R'r^{-1}(2+r\phi')\, ,
\label{f-rr}
\end{align}
where the prime denotes the derivative with respect to the radial
coordinate $r$. We also introduce the function $M(r)$, which
corresponds to the mass variable for the Schwarzschild space-time,
which is equal to,
\begin{equation}
\label{mass}
\e^{-2\lambda}=1-\frac{2G M(r)}{c^2 r}\, .
\end{equation}
We shall identify the value of $M(r)$ on the
surface of neutron star, to be the gravitational mass of the neutron
star. By using the conservation of the energy momentum tensor
$\nabla^\mu T_{\mu\nu}=0$, we get in component form the continuity
equation,
\begin{equation}
\label{hydro}
\frac{dp}{dr}=-(\rho +P/c^2)\frac{d\phi}{dr}\, .
\end{equation}
In the same way as in the standard
Einstein gravity, we can obtain the second
Tolman-Oppenheimer-Volkov (TOV) equation by substituting $d\phi/dr$
from (\ref{hydro}) in Eq.~(\ref{f-rr}). In order to write down the
TOV equation, for convenience reasons, we shall introduce the
dimensionless variables in the following way,
\begin{equation}
\label{NS1}
M=m M_{\odot},\quad r\rightarrow r_{g}r\, , \quad \rho\rightarrow\rho
M_{\odot}/r_{g}^{3}\, ,\quad p\rightarrow p
M_{\odot}c^{2}/r_{g}^{3}\, , \quad R\rightarrow {R}/r_{g}^{2}\, .
\end{equation}
Recall that $M_{\odot}$ denotes the mass of the Sun and $r_{g}$
is equal to $r_{g}=GM_{\odot}/c^{2}=1.47473$ km. Then, by using
these new variables, Eqs.~(\ref{f-tt}) and (\ref{f-rr}) can be
rewritten as in the following way,
\begin{align}
\label{TOV-1}
&\left(1+ \alpha r_{g}^{2} h_{{R}}
+\frac{1}{2}\alpha r_{g}^{2} h'_{{R}} r\right)\frac{dm}{dr}
=4\pi{\rho}r^{2}-\frac{1}{4}\alpha r^2 r_{g}^{2}\left(h-h_{{R}}{R}
 -2\left(1-\frac{2m}{r}\right)\left(\frac{2h'_{{R}}}{r}+h''_{{R}}\right)
\right)\, , \\
\label{TOV-2}
& 8\pi p=-2\left(1+\alpha r_{g}^{2}h_{{R}}\right)\frac{m}{r^{3}}
 -\left(1-\frac{2m}{r}\right)\left(\frac{2}{r}(1+\alpha r_{g}^{2} h_{{R}})
+\alpha r_{g}^{2} h'_{{R}}\right)({\rho}+p)^{-1}\frac{dp}{dr} \nn
& -\frac{1}{2}\alpha r_{g}^{2}\left(h-h_{{R}}{R}-4\left(1-\frac{2m}{r}\right)
\frac{h'_{{R}}}{r}\right)\, .
\end{align}
For a non-vanishing $\alpha$, it is compelling to include the
equation corresponding to the trace of the field equations, which
has the following form,
\begin{align}
\label{TOV-3}
& 3\alpha
r_{g}^{2}\left(\left(\frac{2}{r}-\frac{3m}{r^{2}}-\frac{dm}{rdr}
-\left(1-\frac{2m}{r}\right)\frac{dp}{(\rho+p)dr}\right)\frac{d}{dr}
+\left(1-\frac{2m}{r}\right)\frac{d^{2}}{dr^{2}}\right)h_{{R}}
+\alpha r_{g}^{2} h_{{R}}{R}-2\alpha r_{g}^{2}
h-{R} \nn
& =-8\pi({\rho}-3p)\, .
\end{align}
At this point it is necessary to
specify the EoS inside the neutron star, in order to find a solution
to the above equations. The simplest choice for an EoS is to use a
polytropic EoS of the form $p\sim \rho^{\gamma}$, but we can use
more realistic EoS for the matter of the neutron star.

We now perform a perturbation of the pressure, the mass density and
the curvature in terms of the parameter $\alpha$ in (\ref{fR}) as
follows,
\begin{equation}
\label{expnsn}
p=p^{(0)}+\alpha p^{(1)}+ \cdots \, ,\quad
\rho=\rho^{(0)}+\alpha \rho^{(1)}+ \cdots \, ,\quad m=m^{(0)}+\alpha
m^{(1)}+ \cdots \, , \quad R=R^{(0)}+\alpha R^{(1)}+ \cdots \, .
\end{equation}
Then by
keeping terms of the order of $\mathcal{O}(\alpha)$, we obtain the
following equation,
\begin{align}
\label{dmdr}
\frac{dm}{dr}=&4\pi\rho r^2-\alpha r^{2}\left(4\pi \rho^{(0)}h_{R}
+\frac{1}{4}\left(h-h_{R}R\right)\right) \nn
& +\frac{1}{2}\alpha\left(\left(2r-3m^{(0)}
 -4\pi\rho^{(0)}r^{3}\right)\frac{d}{dr}
+r(r-2m^{(0)})\frac{d^{2}}{dr^{2}}\right) h_{R} \, , \\
\label{dpdr}
\frac{r-2m}{\rho+p}\frac{dp}{dr}=&4\pi r^2 p+\frac{m}{r}
 -\alpha r^2\left(4\pi p^{(0)}h_{R}
+\frac{1}{4}\left(h-h_{R}R\right)\right)
 - \alpha \left(r-3m^{(0)}
+2\pi p^{(0)}r^{3}\right)\frac{dh_{R}}{dr}\, .
\end{align}
Note that we disregard terms of the order ${\mathcal
O}\left(\alpha^2 \right)$ as $m=m^{(0)}+\alpha m^{(1)}$ and
$p=p^{(0)}+\alpha p^{(1)}$. Since the scalar curvature appears only
via the terms $h_{R}$ and $h$, at order ${\mathcal O}(\alpha)$, we
only need to keep terms of the order ${\mathcal O}(1)$, for $R$, in
the following way,
\begin{equation}
\label{prtR}
R \thickapprox R^{(0)}=8\pi(\rho^{(0)}-3p^{(0)})\, .
\end{equation}
If the EoS for the nuclear
matter for large densities is specified, the equations (\ref{dmdr})
and (\ref{dpdr}) can easily be solved. For example, the cases of SLy
\cite{Chabanat:1997un,Douchin:2000kx,Douchin:2001sv} and FPS
\cite{FPS} equations of state can be expressed in a unified manner,
\begin{align}
\label{FPS}
\zeta= & \frac{a_{1}+a_{2}\xi+a_{3}\xi^3}{1+a_{4}\xi}f(a_{5}(\xi-a_{6}))
+(a_{7}+a_{8}\xi)f(a_{9}(a_{10}-\xi)) \nn &
+(a_{11}+a_{12}\xi)f(a_{13}(a_{14}-\xi))+(a_{15}
+a_{16}\xi)f(a_{17}(a_{18}-\xi))\, ,
\end{align}
where $\xi$, $\zeta$ and $f(x)$ in the above equation are,
\begin{equation}
\label{zetaxif}
\zeta=\log(p/\mbox{dyn} \mbox{cm}^{-2})\, , \qquad
\xi=\log(\rho/\mbox{g}\mbox{cm}^{-3})\, , \qquad
f(x)=\frac{1}{\exp(x)+1}\, .
\end{equation}
The detailed form of the
coefficients $a_{i}$ for the SLy and FPS equations of state, can be
found in Ref.~\cite{Camenzind}. We can also consider a neutron star
model, with a quark core, in which case the EoS called Tis, is as
follows,
\begin{equation}
\label{EoSQM}
p_{Q}=a(\rho-4B)\, ,
\end{equation}
where $a$ is a constant and the parameter
$B$ takes values from $\sim 60$ to $90$ Mev/fm$^{3}$. The value of
the parameter $a$ in the case of quark matter, with a massless
strange quark, is $a=1/3$, but here we shall assume that $a=0.28$,
which corresponds to $m_{s}=250$ MeV. For the numerical calculations
performed in Ref.~\cite{Astashenok:2013vza}, Eq.~(\ref{EoSQM}) was
used under the assumption that $\rho \geq \rho_\mathrm{tr}$, with
$\rho_\mathrm{tr}$ being the transition density for which the
pressure of the quark matter is equal to the pressure of ordinary
dense matter. For the case of an FPS EoS, the transition density is
equal to $\rho_\mathrm{tr}=1.069\times 10^{15}$ g/cm$^{3}$ ($B=80$
Mev/fm$^{3}$), while for a SLy EoS, it is
$\rho_\mathrm{tr}=1.029\times 10^{15}$ g/cm$^{3}$ ($B=60$
Mev/fm$^{3}$).

In Ref.~\cite{Astashenok:2013vza}, various $F(R)$ models for neutron
stars have been studied, for example the exponential model in which
case the $F(R)$ gravity is,
\begin{equation}
\label{EXP}
F(R)=R+\beta R(\exp(-R/R_{0})-1)\, .
\end{equation}
In this case, for the neutron star
models of (\ref{EoSQM}), with a quark core assumed, there is no
significant difference in comparison with the Einstein
gravity. For the case of a simplified EoS as in Eq.~(\ref{FPS}),
interesting results are found, although the model with a FPS EoS is
ruled out by the latest observational data
\cite{Antoniadis:2013pzd,Demorest:2010bx}. In the case $\beta<0$,
and for large central densities, stable stellar configurations occur
when $dM/d\rho_{c}>0$. Studying the model with a SLy EoS, could
potentially be more interesting, since the upper limit of neutron
star mass is of the order $\sim 2M_{\odot}$. In addition, there
exists a second branch of stable stellar configurations, in the
large central density regime, which describes the observational data
better, in comparison to the model with a SLy EoS, in the context of
the Einstein gravity. Another interesting model studied in
Ref.~\cite{Astashenok:2013vza}, has the following $F(R)$ gravity
function,
\begin{equation}
\label{CUB}
F(R)=R+\alpha R^{2}(1+\gamma R)\, .
\end{equation}
For
this model, the neutron star configuration is stable, and neutron
star solutions can be generated, which result to smaller radii in
comparison to the Einstein gravity solutions. If a SLy EoS
is assumed, the model at hand can describe neutron stars with mass
$\sim 2M_{\odot}$.

Let us now discuss in brief the massive gravity description of
neutron stars, in which case the corresponding Einstein equations
are $G_{\mu \nu} + m^{2}_{0}I_{\mu \nu} = \kappa^{2}T_{\mu \nu}$. By
using the Bianchi identity $\nabla^\mu G_{\mu\nu}=0$ and in addition
the conservation of the energy-momentum tensor $\nabla^\mu
T_{\mu\nu}$, we obtain the constraint $\nabla^\mu I_{\mu\nu}=0$.
This constraint plays an important role in massive gravity and it
drastically changes the qualitative behavior of static and
spherically symmetric solutions.

We assume that the dynamical metric $g_{\mu \nu}$ and the reference
metric $f_{\mu \nu}$ are both static and spherically symmetric, with
the corresponding line elements having the following form,
\begin{align}
\label{metricg}
g_{\mu \nu}dx^{\mu}dx^{\nu} =&
 - \e^{2\phi(\rho)}dt^{2} + \e^{2\lambda(\rho)} d\rho^{2} + D^{2}(\rho) \left(
d\theta^{2} + \sin^{2}\theta d \varphi^{2} \right) \, , \\
\label{metricf}
f_{\mu \nu}dx^{\mu}dx^{\nu} =&
 - dt^{2} + d\rho^{2} + \rho^{2} \left( d\theta^{2}
+ \sin^{2}\theta d \varphi^{2} \right) \, ,
\end{align}
where the radial coordinate is represented by using the variable
$\rho$. By introducing a new variable $r$, defined in such a way so
that $D(\rho) = r^{2}$, the line elements of Eqs.~(\ref{metricg})
and (\ref{metricf}), can be rewritten as follows,
\begin{align}
\label{metricg2}
g_{\mu \nu}dx^{\mu}dx^{\nu} =&
 - \e^{2\phi}dt^{2} + \e^{2\lambda} dr^{2} + r^{2}\left( d\theta^{2} +
\sin^{2}\theta d \varphi^{2} \right) \, , \\
\label{metricf2}
f_{\mu \nu}dx^{\mu}dx^{\nu} =&
 - dt^{2} + \left( \chi^{\prime}(r) \right)^{2} dr^{2} +
\chi^{2}(r) \left( d\theta^{2} + \sin^{2}\theta d \varphi^{2} \right) \, .
\end{align}
Note that the scalar function $\chi(r)$ quantifies the Stuckelberg
field degree of freedom. In the case of minimal dRGT massive
gravity, the parameters $\beta_{n}$ are chosen in the following way,
\begin{equation}
\beta_{0} = 3\, , \quad \beta_{1} = -1\, , \quad
\beta_{2} = 0\, , \quad \beta_{3} = 0 \, .
\label{minimal}
\end{equation}
Then the following equations are obtained,
\begin{align}
\label{tov1-3}
m^{\prime}(r) =& 4 \pi \tilde{\rho} (r) r^{2}
+ \frac{1}{2} \alpha^{2} \left(r_{g}M_{\odot} \right)^{2} r^{2}
\left[ 3 - \frac{2 \chi (r)}{r} - \chi^{\prime} (r)
\left( 1 - \frac{2m(r)}{r}\right)^{1/2} \right] \, , \\
\label{tov2-3}
8\pi p(r) =& - \frac{1}{r^2}
+ \frac{1}{r^2} \left[1 - 2(p+\tilde{\rho})^{-1}p^{\prime}r \right]
\left(1 - \frac{2m(r)}{r} \right)
+ \alpha^{2} \left(r_{g}M_{\odot} \right)^{2}
\left( 3 - \frac{2 \chi (r)}{r} - \e^{\int \left( p
+ \tilde{\rho} \right)^{-1} p^{\prime} dr} \right) \, , \\
\label{tov3-3}
8 \pi p(r) =&
\left[ - \left( \left( p + \tilde{\rho} \right)^{-1} p^{\prime} \right)^{\prime}
+ (\left( p + \tilde{\rho} \right)^{-1} p^{\prime})^{2}
 - \frac{1}{r}\left( p + \tilde{\rho} \right)^{-1} p^{\prime} \right]
\left( 1 - \frac{2m(r)}{r} \right)
\nonumber \\
& - \frac{1}{2} \left[ \left( p + \tilde{\rho} \right)^{-1} p^{\prime} -
\frac{1}{r} \right] \left( 1 - \frac{2m(r)}{r} \right)^{\prime}
\nonumber \\
& + \alpha^{2} \left(r_{g}M_{\odot} \right)^{2}
\left( 3 - \frac{\chi (r)}{r}
 - \chi^{\prime} (r) \left( 1 - \frac{2m(r)}{r} \right)^{1/2}
 - \e^{\int \left( p + \tilde{\rho} \right)^{-1} p^{\prime} dr} \right) \, .
\end{align}
In addition, the constraint is given as follows,
\begin{equation}
0= \left( - \left( p + \tilde{\rho} \right)^{-1} p^{\prime}
+ \frac{2}{r} \right) \left( 1 - \frac{2m(r)}{r} \right)^{1/2} - \frac{2}{r} \, .
\label{bianchi2}
\end{equation}
A detailed numerical study can show that the
mass-radius relation in the context of dRGT massive gravity, is more
constrained in comparison to the standard Einstein gravity,
and we showed that the maximal mass becomes smaller for quark and
neutron stars. In effect, the compact massive neutron stars cannot
be explained by using this specific version of dRGT massive gravity,
a result which is in contrast with the $F(R)$ gravity case
\cite{Capozziello:2015yza,Astashenok:2013vza}. We need to note that
the results obtained in Ref.~\cite{Katsuragawa:2015lbl}, do not
always exclude the dRGT massive gravity description from being a
viable one, due to the observational data. Indeed, for the
calculations we took into account the standard EoS in order to find
the maximal mass, and in addition the study was specialized by using
a particular minimal model of massive gravity. It may be the case
that a massive neutron star can be realized in the context of
massive gravity, if alternative EoS are used, or more complicated
versions of massive gravity are used, at the expense of having more
free parameters in the theory.

\subsection{Black holes in $F(R)$ Gravity and anti-evaporation}

In this section we shall present some recent studies on black holes
solutions, in the context of $F(R)$ gravity, based on
Ref.~\cite{Nojiri:2013su}. For other astrophysical solutions and various
metric solutions from $F(R)$ gravity and modified gravity in
general, see for example
Refs.~\cite{Capozziello:2009jg,Reijonen:2009hi,Babichev:2009fi,
Sharif:2009xa,Lobo:2009ip,Faraoni:2009xb,Nzioki:2009av,
delaCruzDombriz:2009et,Arbuzova:2010iu,Addazi:2016prb,
Henttunen:2007bz,Kainulainen:2007bt,Addazi:2016hip,
Oikonomou:2016fxb}.

In Ref.~\cite{Nojiri:2013su}, it was demonstrated that
anti-evaporation of the Nariai space-time may possibly occur in the
context of $F(R)$ gravity. In the usual descriptions of black holes
in the Einstein gravity, the horizon radius of the black
hole, in vacuum, decreases in size, since the black hole looses
energy via the Hawking radiation. However, as was demonstrated by
Hawking and Bousso, if someone includes quantum corrections, then it
is possible that the Nariai black hole \cite{Nariai1,Nariai2} may
increase its size, with this phenomenon being known ever since as
anti-evaporation of black holes \cite{Bousso:1997wi}. The Nariai
black hole is a limiting case of a Schwarzschild- de-Sitter
space-time, and it can be obtained if the cosmological horizon of the
Schwarzschild- de-Sitter black hole coincides with that of the black
hole horizon. The metric of the Nariai space-time has the following
line element,
\begin{equation}
\label{Nr1}
ds^2 = \frac{1}{\Lambda^2 \cosh^2 x}
\left( - dt^2 + dx^2 \right) + \frac{1}{\Lambda^2}d\Omega^2\, ,\quad
d\Omega^2 \equiv d\theta^2 + \sin^2 \theta d\phi^2\, .
\end{equation}
The
parameter $\Lambda$ stands for a mass scale and also $d\Omega^2$
denotes the metric of the two dimensional unit sphere. Since the
scalar curvature is equal to, $R=R_0\equiv 4\Lambda^2$, by using
Eq.~(\ref{JGRG16}), we obtain,
\begin{equation}
\label{Nr2}
0 = F\left(
4\Lambda^2 \right) - 2 \Lambda^2 F' \left( 4\Lambda^2 \right)\, .
\end{equation}
We shall consider the perturbations of Eq.~(\ref{Nr2}), so we
assume,
\begin{equation}
\label{Nr3}
ds^2 = \e^{2\rho\left(x,t\right)} \left( - dt^2 + dx^2 \right)
+ \e^{-2 \varphi\left(x,t\right)} d\Omega^2\, .
\end{equation}
By considering that the perturbations have the following form,
$\rho = - \ln \left( \Lambda \cosh x \right) + \delta\rho$ and
$\varphi = \ln \Lambda + \delta \varphi$, in effect we find the
following perturbation equations,
\begin{align}
\label{Nr8}
0 = & \frac{- F'\left( R_0 \right) + 2 \Lambda^2 F''
\left( R_0 \right)}{2 \Lambda^2 \cosh^2 x} \delta R
 - \frac{F \left( R_0 \right)}{\Lambda^2 \cosh^2 x}\delta\rho
 - F' \left( R_0 \right) \left( - \delta \ddot\rho + 2 \delta \ddot\varphi
+ \delta \rho'' + 2 \tanh x \delta\varphi' \right) \nn
& + \tanh x
F'' \left( R_0 \right) \delta R' + F'' \left( R_0 \right) \delta
R''\, ,\nn
0 = & - \frac{- F'\left( R_0 \right) + 2 \Lambda^2 F''
\left( R_0 \right)}{2 \Lambda^2 \cosh^2 x} \delta R + \frac{F \left(
R_0 \right)}{\Lambda^2 \cosh^2 x}\delta\rho
 - F' \left( R_0 \right) \left( \delta \ddot\rho + 2 \delta \varphi''
 - \delta \rho''
+ 2 \tanh x \delta\varphi' \right) \nn
& + F'' \left( R_0 \right) \delta \ddot R
+ \tanh x F'' \left( R_0 \right) \delta R' \, ,\nn
0 =& - 2 \left( \delta{\dot\varphi}' + \tanh x \delta \dot\varphi
\right) + \frac{F'' \left( R_0 \right) }{F' \left( R_0 \right)
}\left( \delta {\dot R}' + \tanh x \delta \dot R\right)\, ,\nn
0 = & - \frac{- F'\left( R_0 \right) + 2 \Lambda^2 F'' \left( R_0
\right)}{2 \Lambda^2 } \delta R - \frac{F \left( R_0
\right)}{\Lambda^2} \delta\varphi
 - \cosh^2 x F' \left( R_0 \right) \left( - \delta \ddot\varphi + \delta
\varphi'' \right) \nn
& - \cosh^2 x F'' \left( R_0 \right) \left( - \delta \ddot R + \delta R''
\right) \, .
\end{align}
By setting $\delta\varphi = \varphi_0 \cosh \omega t \cosh^\beta x$
where $\varphi_0$, and $\omega$ are arbitrary constants, we find,
\begin{equation}
\label{NrR1}
\omega=\pm \beta\, ,\quad \beta = \frac{1}{2}\left(
1 \pm \sqrt{ \frac{19 \alpha - 8}{3\alpha} } \right)\, , \quad
\alpha \equiv \frac{2\Lambda^2 F'' \left(R_0\right)}{F' \left(R_0
\right)} = \frac{F \left(R_0\right) F'' \left(R_0\right)}{F'
\left(R_0 \right)^2}\, .
\end{equation}
The above relations indicate that if
$\alpha<0$ or $\alpha> \frac{8}{19}$, $\beta$ and in effect $\omega$
are real. Since the real part of the parameter $\omega$ is always
positive, the solution corresponding to the Nariai space-time is
rendered always unstable.

The horizon is determined by the following equation,
\begin{equation}
\label{Nr18}
g^{\mu\nu}\nabla_\mu \varphi \nabla_\nu\varphi = 0\, .
\end{equation}
Then we find
\begin{equation}
\label{Nr20}
\delta\varphi =
\delta\varphi_\mathrm{h} \equiv \varphi_0 \cosh^2 \beta t \, .
\end{equation}
Since Eq.~(\ref{Nr3}) indicates that $\e^{-\varphi}$ can be viewed
as a radius coordinate, by using Eq.~(\ref{Nr1}), the radius of the
horizon may be defined by $r_\mathrm{h} =
\e^{-\delta\varphi_\mathrm{h}}/\Lambda$ and then we obtain,
\begin{equation}
\label{Nr21}
r_\mathrm{h} = \frac{\e^{-\varphi_0 \cosh^2 \beta
t}}{\Lambda}\, .
\end{equation}
In the case that $\varphi_0 < 0$, the radius
$r_\mathrm{h}$ increases, and hence anti-evaporation occurs. If
$\beta$ and $\omega$ are complex, instead of $\delta\varphi =
\varphi_0 \cosh \omega t \cosh^\beta x$, we obtain the following
solution for $\varphi$
\begin{equation}
\label{Nr26}
\delta\varphi = \Re \left\{
\left( C_+ \e^{\beta t} + C_+ \e^{- \beta t} \right) \e^{\beta
x}\right\} \, ,
\end{equation}
where the parameters $C_\pm$ are complex
numbers, and we expressed the real part of numbers as $\Re$. Since
the real part of the parameter $\beta$ is always positive, if $C_+
\neq 0$, when $t$ increases, the parameter $\delta\varphi$ also
increases and in effect the perturbations increase. As a result, the
solution corresponding to the Nariai space-time is rendered unstable.

As a particular case, we may consider the following solution of
Eq.~(\ref{Nr26}), which is,
\begin{equation}
\label{NrRRR1}
\delta\varphi =
\delta{\tilde \varphi}_\mathrm{h} \equiv \delta\varphi_0 \left\{
\e^{\frac{t+x}{2}}\left( \cos \frac{\gamma\left(t+x\right)}{2}
 - \frac{1}{\gamma} \sin \frac{\gamma\left(t+x\right)}{2} \right)
+ \e^{\frac{- t+x}{2}}\left( \cos \frac{\gamma\left(t-x\right)}{2}
+ \frac{1}{\gamma} \sin \frac{\gamma\left(t-x\right)}{2} \right)
\right\}\, .
\end{equation}
It should be noted that the solution (\ref{NrRRR1})
is chosen to satisfy the initial condition $\delta \dot \varphi = 0$
at $t=0$. Also we expressed the parameter $\beta$ as,
\begin{equation}
\label{NrRRR2}
\beta = \frac{1}{2}\left( 1 \pm i \sqrt{ \frac{8 - 19
\alpha}{3\alpha} } \right) = \frac{1}{2} \left( 1 + i \gamma
\right)\, .
\end{equation}
Then Eq.~(\ref{Nr18}) takes the following form,
\begin{equation}
\label{NrRRR2B}
0 = \frac{\delta \varphi_0^2}{2} \gamma^2 A^2 \e^x
\sin\frac{\gamma\left(t+x\right)}{2}
\sin\frac{\gamma\left(t-x\right)}{2} \, ,
\end{equation}
and hence we find that
an infinite numbers of horizons occur,
\begin{equation}
\label{NrRRR3}
\mbox{(A)}\ x = - t + \frac{2n \pi}{\gamma}\ \mbox{or}\ \mbox{(B)}
\ x = t + \frac{2n \pi}{\gamma}\, .
\end{equation}
On the horizon, we find that,
\begin{align}
\label{NrRRR4}
\mbox{(A)}&\ \varphi = \delta\varphi_0 (-1)^n \left\{
\e^{\frac{n\pi}{\gamma}} + \e^{-t + \frac{n\pi}{\gamma} } \left(
\cos \left( \gamma t \right)
+ \frac{1}{\gamma} \sin \left( \gamma t \right) \right)\right\}\, , \\
\label{NrRRR5}
\mbox{(B)}&\ \varphi = \delta\varphi_0 (-1)^n \left\{
\e^{\frac{n\pi}{\gamma}} + \e^{t + \frac{n\pi}{\gamma} } \left( \cos
\left( \gamma t \right)
 - \frac{1}{\gamma} \sin \left( \gamma t \right) \right)\right\}\, .
\end{align}
Hence, the radius of horizon, which is defined by the condition
$r_\mathrm{h} = \e^{-\delta{\tilde\varphi}_\mathrm{h}}/\Lambda$,
oscillates, so we have a constant interplay of evaporation and
anti-evaporation. Particularly, for the case (B) of
Eq.~(\ref{NrRRR5}), the amplitude of the oscillations gradually becomes
larger.

\subsection{Wormholes in $F(R)$ Gravity}

As in most applications in modified gravity, the physical
description of a phenomenon or quantity that modified gravity offers
can be entirely different when compared to the Einstein-Hilbert
gravitational description. In this section we shall consider an
exotic and intriguing physical solution of the Einstein
gravity, namely that of a wormhole. The wormhole solution
\cite{Misner:1960zz,Morris:1988cz,Morris:1988tu,
Klebanov:1988eh,Hawking:1988ae} is a geometric space-time tunnel
through which an observer can theoretically travel through it. The
focus in this section will not be on the theoretical implications of
a wormhole geometry per se, but we shall focus on the fact that
modified gravity allows these solutions to exist without the need of
having some extra exotic matter present. Indeed, in the ordinary
Einstein-Hilbert description of wormholes, an exotic matter
component is needed in order the solution is consistently defined,
and this exotic matter is called exotic because it violates the null
energy condition $T_{\mu \nu}k^{\mu} k^{\nu} \geq 0$, where
$k^{\mu}$ is an arbitrary null vector. As it was shown in the
literature \cite{Lobo:2009ip}, in the context of $F(R)$ gravity,
wormhole geometries can consistently be realized. We shall base our
analysis and notation on Ref.~\cite{Lobo:2009ip} where the $F(R)$
gravity realization of wormholes geometry was performed in detail.
In the literature, several modified gravity and alternative
approaches to the wormhole geometry existence can be found. For
example in Refs.~\cite{Visser:1989kh,Barcelo:2000zf,Lobo:2005us,
Visser:2003yf,Hochberg:1997wp,Nojiri:1999pc} the possibility of
having traversable wormholes was studied in various contexts, see
also \cite{Ford:1995wg} for a quantum field theoretic approach on
traversable wormholes. Also in
Refs.~\cite{ArmendarizPicon:2002km,Lobo:2005yv,
Poisson:1995sv,Sushkov:2007me} the stability of wormholes was
studied in various theoretical contexts, and in
Refs.~\cite{Hochberg:1998ii,Hochberg:1998ha} the energy conditions
of wormholes were examined. The possibility of having a wormhole
supporting phantom energy was discussed in \cite{Sushkov:2005kj}, or
a phantom scalar \cite{Bolokhov:2012kn}, while in
Ref.~\cite{Bronnikov:2002rn}, various wormhole solutions were
studied in the context of brane worlds. Finally in
Refs.~\cite{Bronnikov:2016osp,Bronnikov:2016xyp} wormhole solutions
were studied in the context of extra dimensions.

In the rest of this section we follow the presentation developed in
Ref.~\cite{Lobo:2009ip}. Consider the spherically symmetric and
static metric with the following line element,
\begin{equation}
\label{metrworm}
d s^2=\e^{-2\Phi (r)}d t^2+\frac{1}{1-b(r)/r}d r^2+r^2\left (d
\theta^2+\sin^2\theta d \phi^2 \right)\, ,
\end{equation}
with $\Phi (r)$ and
$b(r)$ being arbitrary functions of the radius $r$, which are called
the redshift and the shape function respectively. The radius value
$r=r_0$ denotes the location of the throat, where $b(r_0)=r_0$. Two
compelling conditions that need to be satisfied, in order for a
wormhole solution to exist, are the following,
\begin{equation}
\label{cond1}
\frac{b(r)-b'(r)r}{b(r)^2}>0\, ,\quad 1-\frac{b(r)}{r}>0 \, ,
\end{equation}
where
the prime denotes differentiation with respect to $r$, and in
addition, at the throat, the shape function must satisfy
$b'(r_0)<1$. It is exactly due to the imposition of the above
conditions that in the ordinary Einstein-Hilbert description, an
exotic form of matter, violating the null energy condition, is
needed. As we shortly see, this violation of the null energy
condition is not needed in the case of $F(R)$ gravity. For
simplicity we set $\Phi (r)=\mathrm{const}$, so that $\Phi'(r)=0$.
Consider an $F(R)$ gravity with matter fluids present, in which case
the action is given as usual in (\ref{JGRG6}), and the corresponding
equations of motion can be found by varying the action with respect
to the metric, and the result can be written as follows,
\begin{equation}
\label{resultingeqns}
R_{\mu \nu}-\frac{1}{2}Rg_{\mu \nu}=T_{\mu\nu}^\mathrm{eff}\, ,
\end{equation}
with $T_{\mu\nu}=\frac{T_{\mathrm{matter}\, \mu \nu}}{F_R'}
+T_{\mu \nu}^c$, and
$F_R=\frac{\partial F}{\partial R}$. The energy momentum tensor
$T_{\mathrm{matter}\, \mu \nu}$ corresponds to the ordinary matter
fluids present, while $T_{\mu \nu}^c$ is equal to,
\begin{equation}
\label{tmnc}
T_{\mu \nu}^c=\frac{1}{F_R}\left (\nabla_{\mu}
\nabla_{\nu}F_R-\frac{1}{4}\left(RF_R'+\square F_R+T\right)
\right)\, ,
\end{equation}
where $T$ is the trace of $T_{\mathrm{matter}\, \mu\nu}$.
In the case of modified gravity, the total energy momentum
tensor of the matter and of the modified gravity that threads the
wormhole is supposed to satisfy all the energy conditions, and this
condition will eventually impose certain restrictions on the
functional form of the $F(R)$ gravity. Particularly, the energy
momentum tensor is assumed to have the following decomposition,
\begin{equation}
\label{energymomdecomp}
T_{\mu \nu}=(\rho+p_t)U_{\mu} U_{\nu}
+p_tg_{\mu\nu}+(p_r-p_t)\chi_{\mu}\chi_{\nu}\, ,
\end{equation}
with $U_{\mu}$
being the comoving four-velocity, and $\chi_{\mu}$ being the unit
spacelike vector along the radial direction, that is,
$\chi_{\mu}=\sqrt{1-\frac{b(r)}{r}}\delta_{\mu}$. Also $\rho (r)$ is
the energy density, $p_r(r)$ is the radial pressure along
$\chi_{\mu}$ and finally $p_t(r)$ is the transverse pressure along
the orthogonal direction to $\chi_{\mu}$. In effect, the energy
momentum is $T^{\mu}_{\nu}=\mathrm{diag}[-\rho
(r),p_r(r),p_t(r),p_t(r)]$, and the equations of motion become,
\begin{align}
\label{eqnsofmotionwormhole}
\frac{b'}{r^2}=& \frac{\rho+H}{F_R}\, ,\nn
 -\frac{b}{r^3}=& \frac{p_r}{F_R}+\frac{1}{F_R}\left(
\left(1-\frac{b}{r}\right)\left[ F_R''-F_R'\frac{b'r-b}{2r^2
\left(1-\frac{b}{r} \right)}\right]-H\right)\, , \nn
 -\frac{b'r-b}{2r^3}=&\frac{p_t}{F_R}+\frac{1}{F_R}\left[
\left(1-\frac{b}{r}\frac{F_R'}{r}-H \right) \right]\, ,
\end{align}
where $H$ is,
\begin{equation}
\label{hdef}
H(r)=\frac{1}{4}(F_R'R+\square
F_R'+T)\, ,
\end{equation}
and the prime this time denotes differentiation with
respect to the radius $r$. By using the line element
(\ref{metrworm}) for $\Phi=\mathrm{const}$, the scalar curvature
reads $R=\frac{2b'}{r^2}$, and also the term $\square F_R'$ reads,
\begin{equation}
\label{squeref}
\square F_R=\left(1-\frac{b}{r} \right)\left[
F_R''-F_R'\frac{b'r-b}{2r^2(1-\frac{b}{r})}+\frac{2F_R'}{r}\right]\, .
\end{equation}
In the case at hand, by taking into account the gravitational
equations, the null energy condition violation $T_{\mu
\nu}k^{\mu}k^{\nu}<0$ takes the following form,
\begin{equation}
\label{condgravworm1}
\frac{b'r-b}{r^3}<0 \, ,
\end{equation}
which holds true
due to relations (\ref{cond1}). Also at the throat of the wormhole
we have,
\begin{equation}
\label{cartmanfut}
\left. \frac{\rho+p_r}{F_R} \right|_{r=r_0}
+\left. \frac{1-b}{2r}\frac{F_R'}{F_R}\right|_{r=r_0}<0\, .
\end{equation}
Therefore,
in order for the conditions (\ref{cond1}) (actually the second
condition in Eq.~(\ref{cond1})), to be satisfied, the functional
form of the $F(R)$ gravity must be restricted at the location of the
throat $r=r_0$ as follows,
\begin{equation}
\label{restrictionsonfr}
\left. F_R' \right|_{r=r_0}< \left. -\frac{2r(\rho+\rho_r)}{1-b'} \right|_{r=r_0}\, ,
\end{equation}
if $F_R'>0$, while in the case $F_R'<0$, the functional form
of the $F(R)$ gravity is restricted as follows,
\begin{equation}
\label{restricfr2}
\left. F_R' \right|_{r=r_0}> \left. -\frac{2r(\rho+\rho_r)}{1-b'}
\right|_{r=r_0}\, .
\end{equation}
By assuming
that the matter fluids present, which are quantified by the energy
momentum tensor $T_{\mu \nu}$, satisfy the null energy conditions,
then, the functional form of the $F(R)$ gravity, must satisfy the
following inequalities,
\begin{equation}
\label{ineqssatisfied}
\frac{F_Rb'}{r^2}\geq 0\, ,\quad
\frac{(2F_R+rF_R')(b'r-b)}{2r^2}-F_R'' \left(1-\frac{b}{r} \right)\geq 0\, .
\end{equation}
Then if the shape function $b(r)$ is specified in a way that the
above are respected, we may easily find the $F(R)$ gravity that
satisfies the above conditions and also that it can generate the
wormhole geometry (\ref{metrworm}). Indeed, by specifying the shape
function $b(r)$, then the scalar curvature can be found as a
function of $r$, that is $r(R)$, by using the relation
$R=\frac{2b'}{r^2}$, and by also specifying the equation of state
for $\rho_r$ and $p_r$, we can obtain the functional form of
$F_R(r)$ and by using the function $r(R)$ and integrating once with
respect to the scalar curvature, we may obtain $F(R)$. In
Ref.~\cite{Lobo:2009ip} several examples where worked out, and we
shall briefly present one of them, for which the energy and pressure
density functions satisfy $-\rho+p_r+2p_t=0$, so the resulting
equations of motion become,
\begin{equation}
\label{eqnsofmotionspecific}
F_R'' \left(1-\frac{b}{r} \right)-\frac{b'r+b-2r}{2r^2}F_R'
 -\frac{b'r-b}{2r^3}F_R=0\, .
\end{equation}
Then, by assuming that the shape function is $b(r)=r_0^2/r$,
the resulting $F_R(r)$ solution is,
\begin{equation}
\label{exactsolqfwormh}
F_R(r)=C_1\sinh \left[ \sqrt{2}\arctan \left(
\frac{r_0}{\sqrt{r^2-r_0^2}}\right)\right]+C_2\cosh \left[
\sqrt{2}\arctan \left( \frac{r_0}{\sqrt{r^2-r_0^2}}\right) \right]\, ,
\end{equation}
where $C_i$, $i=1,2$, are integration constants, and the
energy and pressure density functions $\rho$, $p_r$ and $p_t$, can
be found by combing Eqs.~(\ref{exactsolqfwormh}) and
(\ref{eqnsofmotionwormhole}). Then, by using $R=-2r_0^2/r^4$, we can
find the resulting $F(R)$ gravity, which is \cite{Lobo:2009ip},
\begin{equation}
\label{finalfrforwormhole}
F(R)= -C_1R\sinh \left[ \sqrt{2}\arctan
\left( \frac{R_0}{\sqrt{R^2-R_0^2}}\right)\right]+C_2\cosh \left[
\sqrt{2}\arctan \left( \frac{R_0}{\sqrt{R^2-R_0^2}}\right) \right]\, ,
\end{equation}
where $R_0=-2/r_0^2$, so it is a negative value. In
Ref.~\cite{Lobo:2009ip} many other examples were worked out, but the
method is the same, so we refrain from going into further details.
In the literature there exist various studies on wormholes in the
context of modified gravity, and for important stream of papers we
refer the reader to
Refs.~\cite{Harko:2013yb,Bambi:2015zch,Lobo:2014zla,
Capozziello:2012hr,Bronnikov:2010tt,Saiedi:2012qk,
Mazharimousavi:2012xv,DeBenedictis:2012qz,
Darabi:2011hy,Saeidi:2011zz,Bejarano:2016gyv}, where the $F(R)$
gravity case was studied, both in metric and Palatini formalisms,
and in
Refs.~\cite{Azizi:2012yv,Garcia:2010xb,Bohmer:2011si,
Sushkov:2011jh,Battarra:2014naa,
Vacaru:2014cwa,Lobo:2010sb,Sushkov:2011zh} for other modified
gravity and Brans-Dicke approaches.

\section{Bouncing Cosmologies from Modified Gravity}

\subsection{General Features of Bouncing Cosmologies}

The cosmological bounces
\cite{Brandenberger:2012zb,Brandenberger:2016vhg,Battefeld:2014uga,Novello:2008ra,Cai:2014bea,deHaro:2015wda,Lehners:2011kr,Lehners:2008vx,
Cheung:2016wik,Cai:2016hea,Li:2014era,Cai:2013kja,Cai:2014zga,
deHaro:2014kxa,Qiu:2010vk,Qiu:2010ch,Cai:2012va,Cai:2007qw}, are
quite appealing alternative scenarios to the standard inflationary
paradigm, and the most appealing feature of these cosmologies is the
absence of the initial singularity. In principle, a successful
bouncing cosmology must solve all the problems that the inflationary
scenario solved, and also it is required that a nearly scale
invariant power spectrum is produced. There are many bouncing
cosmologies in the literature, and in this chapter we shall discuss
how these cosmologies can be realized by vacuum modified gravities.
In particular we shall investigate how these cosmologies can be
realized by $F(R)$, $f(\mathcal{G})$ and $F(T)$ gravities. For
studies in the literature on this issue, see
\cite{Odintsov:2015uca}. Before we proceed to this task, let us
recall some essential features of bouncing cosmology.

One of the most important feature that a viable cosmology should
have, is, the correct evolution of the Hubble horizon at early and
late times. Particularly, at early times the Hubble horizon should
contract and at intermediate and late times, the Hubble horizon
should expand again. This is due to the fact that at early times,
the Hubble horizon should contract in order for the primordial
quantum fluctuations to exit the horizon. Then the Hubble horizon in
the future should contract, so the initial modes will re-enter the
horizon and will become physically relevant for present times
observations.

\begin{figure}[h] \centering
\includegraphics[width=15pc]{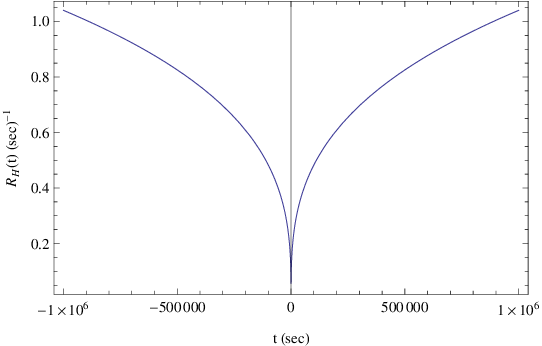}
\caption{The Hubble radius $R_H(t)$ as a function of the
cosmological time $t$, for the matter bounce scenario
$a(t)=\left(\frac{3}{2} \rho_c t^2+1\right)^{\frac{1}{3}}$.}
\label{matterbhubradius}
\end{figure}

The Universe's evolution in a general bouncing cosmology, consists
of two eras, an era of contraction and an era of expansion. In
between these two eras, the Universe reaches a bouncing point, at
which the scale factor reaches a minimum, non-zero size, and
eventually the Universe bounces off and starts to expand again. It
is due to the fact the Universe reaches a non-zero minimum size,
that the initial singularity problem is solved in the context of
bouncing cosmologies. During the contracting period, the scale
factor of the Universe decreases and it satisfies $\dot{a}<0$. As
the Universe reaches the bouncing point, the scale factor satisfies
the condition $\dot{a}=0$, and after that, and during the expanding
period, the scale factor satisfies $\dot{a}>0$. In terms of the
Hubble rate $H(t)$, the conditions for a bounce to occur is that
during the contraction, the Hubble rate satisfies $H(t)<0$, at the
bouncing point the Hubble rate is $H(t)=0$, and as the Universe
expands again, the Hubble rate is $H(t)>0$. Suppose that the
bouncing point occurs at $t=t_s$, then the conditions for a bounce
to occur are summarized below,
\begin{eqnarray}
\label{bounceconditions}
& \mbox{Before the bouncing point}\
t<t_s: &\dot{a}(t)<0,\quad H(t)<0 \, ,\nn
& \mbox{At the bouncing
point}\quad t=t_s: &\dot{a}(t)=0,\quad H(t)=0 \, , \nn
& \mbox{After the bouncing point\ }\ t>t_s:
&\dot{a}(t)>0,\quad H(t)>0 \, .
\end{eqnarray}
In the literature there exist various cosmological bounce scenarios,
which we now briefly describe. First we consider the matter bounce
scenario \cite{Brandenberger:2016vhg,Quintin:2014oea,Cai:2011ci,
Cai:2011zx,Bamba:2012ka,deHaro:2012xj,deHaro:2014jva,
deHaro:2015wda,Bamba:2013fha,Barragan:2009sq}, which occurs quite
often in the context of LQC \cite{Ashtekar:2011ni,Ashtekar:2007tv,
Corichi:2009pp,Singh:2009mz,Bojowald:2008ik}. The matter bounce
scale factor and the Hubble rate are the following,
\begin{equation}
\label{matterbouncescale}
a(t)=\left(\frac{3}{2} \rho_c
t^2+1\right)^{\frac{1}{3}}\, , \quad
H(t)=\frac{2 t \rho_c}{2+3 t^2\rho_c}\, ,
\end{equation}
where the parameter $\rho_c$ is critical energy
density related to the underlying LQC theory.
\begin{figure}[h] \centering
\includegraphics[width=15pc]{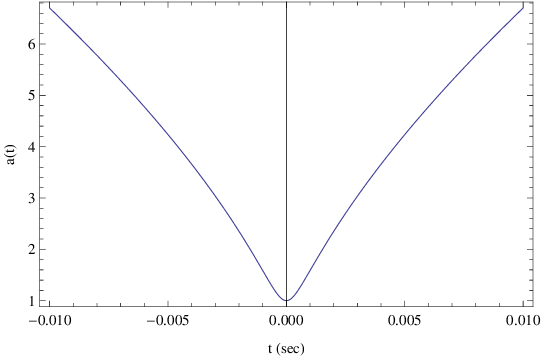}
\includegraphics[width=15pc]{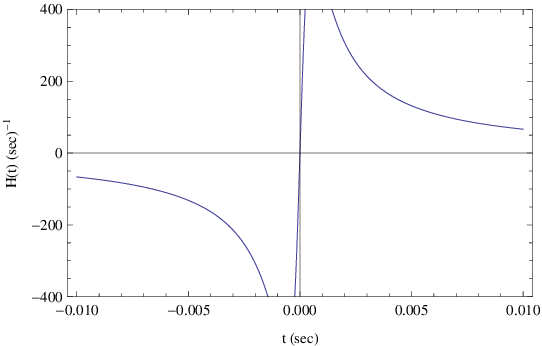}
\caption{The scale factor $a(t)$ (left plot) and the Hubble rate
(right plot) as functions of the cosmic time $t$, for the matter
bounce scenario $a(t)=\left(\frac{3}{2} \rho_c
t^2+1\right)^{\frac{1}{3}}$.} \label{mattbounce1}
\end{figure}
The cosmic time dependence of the scale factor and of the Hubble
rate can be found in Fig.~\ref{mattbounce1}. As it can be seen, the
bouncing cosmology conditions are satisfied, since $H>0$ for $t>0$
and $H<0$ for $t<0$, and the bouncing point occurs at $t=0$. In the
matter bounce scenario, the primordial perturbations are generated
early at the beginning of the contraction era. A complete study of
the Hubble horizon can describe this perfectly, so in
Fig.~\ref{matterbhubradius} we can see the time dependence of the
Hubble radius $R_H=1/(aH)$, for $\rho_c=2\times 10^6\,
\mathrm{sec}^{-2}$. As we can see, for $t<0$ the Hubble radius
decreases from an infinite size at $t\rightarrow -\infty$, until the
bouncing point is reached. At the bouncing point the horizon reaches
a minimal size and eventually it starts expanding again, for $t>0$.
Thus the primordial modes relevant for present day observations are
generated at $t\to -\infty$, that is, at the beginning of the
contracting phase.
Another scenario with interesting phenomenology is the superbounce
scenario \cite{Koehn:2013upa,Odintsov:2015uca,Oikonomou:2014yua},
which firstly appeared in the context of ekpyrotic cosmological
scenarios \cite{Koehn:2013upa}. For the superbounce scenario, the
scale factor and the corresponding Hubble rate can be found below,
\begin{equation}
\label{basicsol1}
a(t)=(-t+t_s)^{\frac{2}{c^2}}\, , \quad
H(t)=-\frac{2}{c^2 (-t+t_s)}\, ,
\end{equation}
where the parameter $c$ is an
arbitrary parameter. Note that the bouncing point is at $t=t_s$.
\begin{figure}[h] \centering
\includegraphics[width=15pc]{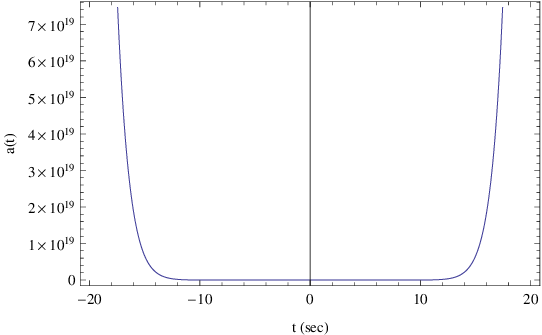}
\includegraphics[width=15pc]{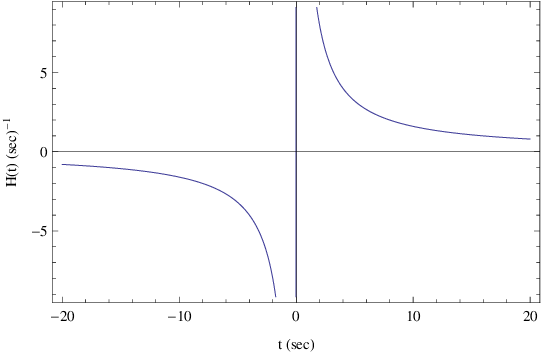}
\caption{The scale factor $a(t)$ (left plot) and the Hubble rate
(right plot) as a function of the cosmological time $t$, for the
superbounce scenario $a(t)=(-t+t_s)^{\frac{2}{c^2}}$.}
\label{superbounce1}
\end{figure}
The cosmic time dependence of the superbounce scenario can be found
in Fig.~\ref{superbounce1}, for $c=\sqrt{8}$ and $t_s=10^{-36}$sec,
and as it can be seen, the cosmological bounce conditions are
satisfied in this case too, with the contraction occurring when
$t<0$ and the expansion occurring when $t>0$. In addition, the
Hubble horizon behaves in the same way as in the matter bounce case,
since during the contraction phase it decreases, and during the
expansion phase it increases.

Another interesting scenario which we now briefly mention is the
singular bounce scenario, which was developed and thoroughly studied
in
\cite{Odintsov:2015ynk,Oikonomou:2015qfh,
Oikonomou:2015qha,Odintsov:2015zza}.
The new novel feature of this bouncing cosmology is that although
the initial singularity is avoided, a Type IV singularity
\cite{Odintsov:2015ynk} occurs at the bouncing point. In the
singular bounce scenario, the scale factor and the corresponding
Hubble rate are,
\begin{equation}
\label{singularbscale}
a(t)=\e^{\frac{f_0}{\alpha +1} (t-t_s)^{\alpha +1 }} \, ,\quad
H(t)=f_0\left( t-t_s\right)^{\alpha }\, ,
\end{equation}
where the parameter
satisfies $f_0>0$. The condition in order to have a Type IV
singularity at $t=t_s$, is that $\alpha>1$, and also for physical
consistency, $\alpha$ should be chosen as,
\begin{equation}
\label{alpahhacond}
\alpha=\frac{2n+1}{2m+1}\, ,
\end{equation}
with the integers $n$ and $m$ being
chosen in such a way, so that $\alpha>1$. The singular bounce has a
peculiar evolution of the Hubble radius, since for $t\to -\infty$,
it has an infinite size and starts to decrease until near the
bouncing point, where it blows up. After the bouncing point it
starts to decrease again. As it was demonstrated in
Refs.~\cite{Odintsov:2015ynk,Oikonomou:2015qfh,
Oikonomou:2015qha,Odintsov:2015zza}, the power spectrum of the
singular bounce for the modes near the bounce is not scale
invariant, and therefore it has to be combined with alternative
scenarios, like the ones studied in
Refs.~\cite{Cai:2014jla,Odintsov:2015zua} in order to be a viable
cosmological bounce scenario.

In the following sections we shall present how the superbounce
cosmology can be realized with $F(R)$, $f(\mathcal{G})$ and
$F(T)$ gravity.

\subsection{Bounce Cosmology from $F(R)$ Gravity}

In order to realize the superbounce of Eq.~(\ref{basicsol1}), we
shall use a reconstruction method introduced in \cite{Nojiri:2009kx}
(for alternative methods see \cite{Carloni:2010ph}), according to
which the $e$-foldings number is used instead of the cosmic time,
with the $e$-foldings number being related to the scale factor as
follows,
\begin{equation}
\label{efoldpoar}
\e^{-N}=\frac{a_0}{a}\, .
\end{equation}
In terms of the variable $N$, we can re-express the $F(R)$ gravity
equations of motion for the FRW Universe, as follows,
\begin{align}
\label{newfrw1}
& -18\left [
4H^3(N)H'(N)+H^2(N)(H')^2+H^3(N)H''(N) \right ]F''(R) \\
& +3\left [H^2(N)+H(N)H'(N)
\right]F'(R)-\frac{F(R)}{2}+\kappa^2\rho=0\, .
\end{align}
Introducing the function $G(N)=H^2(N)$, for the flat FRW Universe we
eventually have that,
\begin{equation}
\label{riccinrelat}
R=3G'(N)+12G(N)\, .
\end{equation}
The above relation will enable us to determine the function $N(R)$.
By writing the Hubble rate as a function of the scale factor we
have,
\begin{equation}
\label{hpscf}
H=\frac{2}{c^2}a^{-\frac{c^2}{2}}\, ,
\end{equation}
and therefore by expressing $a$ in terms of $N$ by making use of
Eq.~(\ref{efoldpoar}) we get,
\begin{equation}
\label{gnfunction}
G(N)=A \e^{-c^2N} \, ,
\end{equation}
with the parameter $A$ being $A=\frac{4}{c^4}a_0^{-c^2}$. Combining
Eqs.~(\ref{riccinrelat}) and (\ref{gnfunction}), we can express $N$
as a function of $R$, as follows,
\begin{equation}
\label{efoldr}
N=-\frac{1}{c^2}\ln \left[\frac{R}{3A(4-c^2)}\right]\, ,
\end{equation}
and by substituting in (\ref{newfrw1}), we can rewrite it as
follows,
\begin{equation}
\label{newfrw1modfrom}
 -9G(N(R))\left[ 4G'(N(R))+G''(N(R)) \right]F''(R) 
+\left[3G(N)+\frac{3}{2}G'(N(R))
\right]F'(R)-\frac{F(R)}{2}+\kappa^2\rho_\mathrm{tot}=0\, ,
\end{equation}
where $G'(N)=d G(N)/d N$ and
$G''(N)=d ^2G(N)/d N^2$. The total matter energy
density $\rho_\mathrm{tot}$ should be expressed in terms of $N(R)$. By
assuming that the perfect matter fluids have constant equation of
state parameters, so these satisfy $\dot{\rho}_i+3H(1+w_i)\rho_i=0$,
eventually we have,
\begin{equation}
\label{mattenrgydens}
\rho_\mathrm{tot} =\sum_i\rho_{i0}
a_0^{-3(1+w_i)}\e^{-3N(R)(1+w_i)}\, .
\end{equation}
By using the above relations, the FRW equation
(\ref{newfrw1modfrom}) becomes,
\begin{align}
\label{bigdiffgeneral1}
&a_1 R^2\frac{d ^2F(R)}{d R^2} +a_2R\frac{d F(R)}{d R}-\frac{F(R)}{2}
+\sum_iS_{i}R^{ \frac{3(1+w_i) }{c^2}}=0\, ,
\end{align}
where the parameters $a_1$ and $a_2$ are equal to,
\begin{equation}
\label{apara1a2}
a_1=\frac{c^2}{4-c^2} \, , \quad
a_2=\frac{2-c^2}{2(4-c^2)}\, ,
\end{equation}
and also,
\begin{equation}
\label{gfdgfdgf}
S_i =\frac{\kappa^2
\rho_{i0}a_0^{-3(1+w_i)}}{[3A(4-c^2)]^{\frac{3(1+w_i)}{c^2}}} \, .
\end{equation}
By solving the differential equation (\ref{bigdiffgeneral1}), we can
obtain the exact $F(R)$ which can generate the superbounce cosmology
(\ref{basicsol1}).

We are interested in the vacuum $F(R)$ cosmology, so the
differential equation (\ref{bigdiffgeneral1}) becomes a second order
homogeneous Euler equation, which has the following solution,
\begin{equation}
\label{frgenerlargetssss}
F(R)=c_1R^{\rho_1}+c_2R^{\rho_2}\, ,
\end{equation}
with $c_1$, $c_2$ being arbitrary integration constants, and also
$\rho_1$ and $\rho_2$ are,
\begin{equation}
\label{rho12}
\rho_1=\frac{-(a_2-a_1)+\sqrt{(a_2-a_1)^2+2a_1}}{2a_1}\, , \quad
\rho_2=\frac{ -(a_2-a_1)-\sqrt{(a_ 2-a_1)^2+2a_1}}{2a_1}\, .
\end{equation}
In the case that $c\gg 1$, the $F(R)$ gravity becomes,
\begin{equation}
\label{frgenerlargetsssslargec}
F(R)\simeq c_1R+c_2R^{-1/2}\, ,
\end{equation}
so the resulting $F(R)$ gravity is a modification of the Einstein
gravity. Let us now consider the case that matter fluids are taken
into account too. In this case, by solving (\ref{bigdiffgeneral1}),
we obtain,
\begin{align}
\label{newsolutionsnoneulerssss}
F(R) = & \left[\frac{c_2\rho_1}{\rho_2}
 -\frac{c_1\rho_1}{\rho_2(\rho_2-\rho_1+1)}\right]R^{\rho_2+1}
+\sum_i \left[\frac{c_1S_i}{\rho_2(\delta_i+2+\rho_2-\rho_1)}\right]
R^{\delta_i+2+\rho_2} \nn
& -\sum_iB_ic_2R^{\delta_i+\rho_2}+c_1R^{\rho_1}
+c_2R^{\rho_2} \, ,
\end{align}
with $c_1$, $c_2$ being arbitrary integration constants, and also
$\delta_i$ and $B_i$ are defined as follows,
\begin{equation}
\label{paramefgdd}
\delta_i=\frac{3(1+w_i)-2c^2}{c^2}-\rho_2+2 \, , \quad
B_i=\frac{S_i}{\rho_2\delta_i}\, .
\end{equation}
In the limit $c\gg 1$, the $F(R)$ gravity above becomes,
\begin{align}
\label{newsolutionsnoneulersssslargec}
F(R)\simeq R+\alpha R^{2}+c_1R^{-1/2}+\Lambda\, ,
\end{align}
where we assumed that $c_2=1$,
$\alpha=\frac{c_1}{3}-2+2c_1\sum_i\frac{S_i}{3}$ and
$\Lambda=-\sum_iA_ i$. Note that this resulting $F(R)$ gravity,
resembles a higher order $R^2$ gravity. At early times it is
approximated by the $R+\alpha R^2$ (Starobinsky inflation) gravity,
while at late times it is $R-R^{-1/2} +\Lambda$, so it resembles the
$\Lambda$CDM model.

\subsubsection{Stability in the $F(R)$ reconstructions}

Having at hand the $F(R)$ gravity solutions that realize the
superbounce, now we shall examine the stability of the solutions.
Particularly we are interested in the linear stability of the
dynamical system that the FRW constitute. More details on this issue
can be found in Ref.~\cite{Odintsov:2015uca}. So we consider linear
perturbations of the function $G(N)$, of the following form,
\begin{equation}
\label{pert1}
G(N)=g(N)+\delta g(N)\, ,
\end{equation}
and so by substituting this in the FRW equations
(\ref{newfrw1modfrom}) we obtain,
\begin{align}
0= & g(N) \left. \frac{d ^2F(R)}{d R^2}\right|_{R=R_1}\delta ''g(N) \nn
& +\left\{3g(N)\left[4g'(N)+g''(N)\right]
\left. \frac{d ^3F(R)}{d R^3}\right|_{R=R_1}
+\left [3g(N)-\frac{1}{2}g'(N)\right ]
\left. \frac{d ^2F(R)}{d R^2}\right|_{R=R_1}\right\}\delta 'g(N) \nn
& +\left\{
12g(N)\left[4g'(N)+g''(N)\right]
\left. \frac{d ^3F(R)}{d R^3}\right|_{R=R_1}
\right. \nn
& \left. +\left[-4g(N)+2g'(N)+g''(N)\right]
\left. \frac{d ^2F(R)}{d R^2} \right|_{R=R_1 }
+\frac{1}{3} \left. \frac{d F(R)}{d R}\right.|_{R=R_1}\right\}
\delta g(N)\, ,
\label{stabpert1}
\end{align}
where $R_1=3g'(N)+12g(N)$. The stability conditions for the
perturbations of $G(N)$, are the following,
\begin{equation}
\label{st0}
J_1=\frac{6[4 g'(N)+g''(N)]F'''(R)}{F''(R)}+6-\frac{g'(N)}{g(N)}>0
\, ,
\end{equation}
and also,
\begin{equation}
\label{st01}
J_2=\frac{36 [4 g'(N)+g''(N)] F'''(R)}{F''(R)}-12+\frac{6
g'(N)}{g(N)}+\frac{3 g''(N)}{g(N)}+\frac{ F'(R)}{g(N) F''(R)}>0\, .
\end{equation}
For the superbounce realizing vacuum $F(R)$ gravity solution of
Eq.~(\ref{frgenerlargetssss}), the stability conditions of
Eq.~(\ref{st0}) and (\ref{st01}) read,
\begin{align}
J_1= & 6+c^2-2 c^2
\left[3^{\rho_1} c_1 Q_1 ^{\rho_1} (\rho_1-2) (\rho_1-1) \rho_1
+ 3^{\rho_2} c_2 Q_1^{\rho_2}
(\rho_2-2) (\rho_2-1) \rho_2\right] \nn
& \times \left[3^{\rho_1} c_1 Q_1^{\rho_1}
(\rho_1-1) \rho_ 1+3^{\rho_2} c_2Q_1^{\rho_2} (\rho_2-1)
\rho_2\right]>0\, ,
\label{stab1}
\end{align}
with $Q_1=-A (c^2-4) \e^{-c^2 N}$, and
\begin{align}
J_2 = & -12-6 c^2+3 c^4+\e^{c^2 N}
\left( c_1Q_2^{\rho_1-1} \rho_1 +c_2 Q_2^{\rho_2-1}
\rho_2\right) A^{-1} Q_3^{-1} \nn
& -12c^2 Q_2 \left[c_1Q_2^{\rho_1-3} (\rho_1-2) (\rho_1-1)
\rho_1 +c_2 Q_2^{\rho_2-3} (\rho_2-2)
(-1+\rho_ 2) \rho_2\right] Q_3^{-1}>0 \, ,
\label{st2}
\end{align}
with $Q_2=3 A \e^{-c^2 N}(4 - c^2 )$ and $Q_3=c_1 Q_2^{\rho_1-2}
(\rho_1-1) \rho_1+c_2 Q_2^{\rho_2-2} (\rho_2-1) \rho_2$. In the
large-$c$ limit, the stability conditions read,
\begin{equation}
\label{st41}
J_1= 4 c^2+6>0 \, , \quad
J_2= 3 c^4+18 c^2 -12>0\, ,
\end{equation}
and therefore the $F(R)$ gravity which realizes the superbounce is a
stable solution.

Now we study the stability of the case that perfect matter fluids
are present, so the $F(R)$ gravity has the form
(\ref{newsolutionsnoneulerssss}). In this case, the stability
conditions of Eqs.~(\ref{st0}) and (\ref{st01}), read,
\begin{align}
\label{st5}
J_1=&6+c^2-6c^2 Q_1 \left[3^{\rho_1-3} c_1
 (\rho_1-2) (\rho_1-1) \rho_1Q_1^{\rho_1-3}+3^{\rho_2-3} c_2
(\rho_2-2) (\rho_2-1) \rho_2Q_1^{\rho_2-3} \right. \nn
&+3^{\rho_2-2} q_1 (\rho_2-1) \rho_2 (\rho_2+1)Q_1^{\rho_2-2}
-3^{\delta_i+\rho_2-3} Q_1^{\delta_i+\rho_2-3} q_2
(\delta_i+\rho_2-2) (\delta_i+\rho_2-1) (\delta_i+\rho_2) \nn
& \left. +3^{\delta_i+\rho_2-1} Q_1^{\delta_i+\rho_2-1} q_2
(\delta_i+\rho_2) (\delta_i+\rho_2+1) (\delta_i+\rho_2+2)\right]
\nn
& \times \left[3^{\rho_1-2} c_1 (\rho_1-1) \rho_1Q_1^{\rho_1-2}
+3^{\rho_2-2} c_2 (\rho_2-1) \rho_2 Q_1^{\rho_2-2}
 +3^{\rho_2-1} q_1 \rho_2 (\rho_2+1)Q_1^{\rho_2-1} \right. \nn
& \left. -3^{\delta_i+\rho_2-2} q_2 (\delta_i+\rho_2-1)
(\delta_i+\rho_2)Q_1^{\delta_i+\rho_2-2} +3^{\delta_i+\rho_2} q_2
(\delta_i+\rho_2+1) (\delta_i+\rho_2+2)Q_1^{\delta_i+\rho_2}
\right]^{-1} \, ,
\end{align}
where $Q_1=-A (c^2-4) \e^{-c^2 N}$, and also,
\begin{align}
J_2=&-12-6 c^2+3 c^4+ A \left(Q_3 +Q_4\right)
\left\{ \e^{c^2 N} \left[c_1 \rho_1 Q_2^{ \rho_1-1}
 +c_2\rho_2 Q_2^{\rho_2-1} \right. \right. \nn
& \left. \left. +Q_2^{\rho_2} q_1 (1+\rho_2)-Q_2^{\delta_i+\rho_2-1}
q_2 (\delta_i+\rho_2) +Q_2^{\delta_i+\rho_2+1} q_2
(2+\delta_i+\rho_2) \right] \right\} \nn
& + 36 A c^2 \e^{- c^2 N} (c^2-4)\left(Q_3 +Q_4\right)^{-1}
\nn
& \times\left[ c_1 Q_2^{\rho_1-3} (\rho_1-2) (\rho_1-1) \rho_1 +c_2
Q_2^{\rho_2-3} (\rho_2-2) (\rho_2-1) \rho_2 \right. \nn
& +Q_2^{\rho_2-2} q_1 (\rho_2-1) \rho_2 (\rho_2+1)
 -Q_2^{\delta_i+\rho_2-3} q_2
(\delta_ i+\rho_2-2) (\delta_i+\rho_2-1) (\delta_i+\rho_2) \nn
& \left. +Q_2^{\delta_i+\rho_2-1} q_2 (\delta_i+\rho_2)
(\delta_i+\rho_2+1) (\delta_i+\rho_2+2) \right]\, ,
\label{st6}
\end{align}
with $Q_2=3 A \e^{-c^2 N}(4 - c^2 )$, $Q_3=c_1 Q_2^{\rho_1-2}
(\rho_1-1) \rho_1+c_2 Q_2^{\rho_2-2} (\rho_2-1) \rho_2$ and
$Q_4= Q_2^{\rho_2-1} q_1 \rho_2 (1+\rho_2)
 -Q_2^{\delta_i+\rho_2-2} q_2 (\delta_
i+\rho_2-1) (\delta_i+\rho_2)
 +Q_2^{\delta_i+\rho_2}
 q_2 (1+\delta_i+\rho_2) (2+\delta_i+\rho_2)$.
By taking the large-$c$ limit, we obtain,
\begin{equation}
\label{st7}
J_1\approx 6c^2\, , \quad
J_2\approx c^4\, ,
\end{equation}
so in this case too, the $F(R)$ gravity solution is stable.

\subsection{Bounce Cosmology from $f(\mathcal{G})$ Gravity}

In this section we shall investigate how the superbounce cosmology
of Eq.~(\ref{basicsol1}) can be realized by using an
$f(\mathcal{G})$ gravity. Using the reconstruction method we
developed in the previous chapter, we introduce two functions $P(t)$
and $Q(t)$, which are chosen in such a way so these satisfy
Eq.~(\ref{ebasc}) and therefore the $f(\mathcal{G})$ the
gravitational action (\ref{actionfggeneral}) is given in
Eq.~(\ref{actionfrg}). By varying with respect to $t$, we get the
differential equation in Eq.~(\ref{auxeqnsvoithitiko}), which when
solved with respect to the function $t=t(\mathcal{G})$ it enables us
to find the $f(\mathcal{G})$ function, by simply substituting
$t=t(\mathcal{G})$ in Eq.~(\ref{ebasc}). By combining
Eqs.~(\ref{ebasc}) and (\ref{eqnsfggrav}), we obtain (\ref{ak}) so
by further combining Eqs.~(\ref{ebasc}) and (\ref{ak}) we get
(\ref{diffept}) and hence we have the function $P(t)$. For the case
at hand, the Hubble rate is, hence, the differential equation
(\ref{diffept}) reads,
\begin{equation}
\label{diffept1}
(t-t_*) \frac{d ^2P}{dt^2}
 - (2+a)\frac{d P}{d t}-\frac{c^4}{8}(t-t_*) =0\, ,
\end{equation}
which can be analytically solved, and the solution is,
\begin{equation}
\label{partsol}
P(t)=\frac{c^4\,t (2t_* -t) }{8 (c^2+2)}\, .
\end{equation}
By substituting the above in Eq.~(\ref{ak}) we get the function
$Q(t)$,
\begin{equation}
\label{finawxpqsup}
Q(t)= \frac{32(c^2-1)}{c^4(c^2+2)(t- t_*)^2}\, .
\end{equation}
Combining Eqs.~(\ref{auxeqnsvoithitiko}), (\ref{partsol}), and
(\ref{finawxpqsup}) we get the following,
\begin{equation}
\label{finalsolutionsbef}
t_1(\mathcal{G})= t_*-\frac{\sqrt{2}
\left[8(2-11c^2)
G\right]^{1/4}}{c^2\sqrt\mathcal{G}} \, , \quad
t_2(\mathcal{G})= t_*+\frac{\sqrt{2} \left[8(2-11c^2)
\mathcal{G}\right]^{1/4}}{c^2\sqrt\mathcal{G}}\, .
 \end{equation}
Therefore, we substitute the functions $t_{1}$ and $t_2$ from
Eq.~(\ref{finalsolutionsbef}) in Eq.~ (\ref{ebasc}),
so finally we have the two $f(\mathcal{G})$ gravities that
realize the superbounce cosmology.
\begin{equation}
\label{finalsuperbouncefgs}
F_1(\mathcal{G})= \frac{c^4\,t_*^2 \,\mathcal{G}
 -8 \sqrt{2(2-11c^2) \mathcal{G}}}{8(c^2+2)} \, , \quad
F_2(\mathcal{G})= \frac{c^4\,t_*^2 \,\mathcal{G}
 -4 \sqrt{2(2-11c^2) \mathcal{G}}
(1+\mathcal{G})}{8(2c^2+2)\mathcal{G}}\, .
\end{equation}

\subsubsection{Stability in the $f(\mathcal{G})$ reconstructions}

Now we examine the linear stability of the gravitational equations
for the solutions (\ref{finalsuperbouncefgs}), so we adopt the
method used in Ref.~\cite{Odintsov:2015uca}. Consider the
perturbation (\ref{pert1}), so by inserting this in the
gravitational equation (\ref{eqnsfggrav}), we obtain the following
stability conditions,
\begin{equation}
\label{stcondgg}
\frac{J_2}{J_1}>0\, , \quad \frac{J_3}{J_1}>0\, ,
\end{equation}
with $J_1$ and $J_2$ being equal to,
\begin{align}
\label{st11}
J_1=&288 g(N)^3 F''(\mathcal{G}) \, , \\
\label{st12}
J_2=&432 g(N)^{2 }\left\{(2 g(N)+g'(N)) F''(\mathcal{G})\right. \nn
& \left. +8 g(N) \left[g'(N)^2+g(N) (4
g'(N)+g''(N))\right] F''(\mathcal{G})\right\} \, , \\
\label{st13}
J_3=&6 \left\{1+24
g(N)\left\{-8 g(N)^2+3 g'(N)^2+6 g(N) [3 g'(N)+g''(N)]\right\}
F''(\mathcal{G}) \right. \nn
& \left. +24 g(N) [4 g(N)+g'(N)] \left\{g'(N)^2+g(N) [4
g(N)+g''(N)]\right\}F''(\mathcal{G}) \right\} \, .
\end{align}
For the superbounce (\ref{basicsol1}), we showed in the previous
section that there are two $f(\mathcal{G})$ gravity solutions which realize
this cosmology, namely $F_i(\mathcal{G})$, $i=1,2$, so for the function
$F_1(\mathcal{G})$, we have,
\begin{align}
\label{st17}
\frac{J_2}{J_1}=\frac{3}{2} (c^2-2) \left(16 A^2 c^2 \e^{-2 c^2
N}-1\right)>0 \, ,
\end{align}
and also,
\begin{align}
\frac{J_3}{J_1}=& \frac{\left(-\frac{99}{2}c^6+ 108c^4+ 26c^2 -8
\right)}{2-11c^2} +8A \e^{-c^2 N}+\frac{Ac^2 \e^{-c^2 N}
\left( 11c^6 -46c^4 + 30c^2-4 \right)}{2-11c^2} \nn
& - \sqrt{6}(4-c^4) \sqrt{ \frac{c^2-2}{11c^2-2} } >0\, ,
\label{st18}
\end{align}
where $A=\frac{4}{c^4}a_0^{-c^2}$. Hence, the solution $F_1(\mathcal{G})$ is
conditionally stable. In the same way, for the function $F_2(\mathcal{G})$ we
have,
\begin{equation}
\label{st23}
\frac{J_2}{J_1}=\frac{3}{2} \left(c^2-2\right)\left(16
A^2 c^2 \e^{-2 c^2 N}-1\right)>0 \, ,
\end{equation}
and also,
\begin{align}
\frac{J_3}{J_1}=& \frac{
4A^2\left(16-60c^2-190 c^4+207 c^6 \right)}{Q_a} \nn
& +\frac{4A^3 \e^{-c^2 N}\left(-32+ 200 c^2
 -152 c^4+122 c^6 -68 c^8+11 c^{10}\right)}{Q_a} \nn
&
+\frac{ A \e^{c^2 N}\left(16-92c^2+30 c^4-46 c^6+11c^8 \right)}{Q_a}
+\frac{ \e^{2 c^2 N}\left(-8 +26c^2 +117 c^4-99c^6 \right)}{Q_a}
\nn
& +\frac{16 \sqrt{6} A^2 (2+c^2) (4-4c^2+c^4) \sqrt{(2-11c^2)
(c^2-2) }} {c^2Q_a}>0\, ,
\label{tor1}
\end{align}
with $Q_a=(2-11c^2)\left[4 A^2 (c^2-2) +\e^{2 c^2 N}\right]$.
Therefore, in this case too, conditional stability is ensured.

\subsection{Bounce Cosmology from $F(T)$ Gravity}

In this section we shall briefly investigate how the superbounce
cosmology can be realized by an $F(T)$ gravity. We use the formalism
we developed in chapter II, for $F(T)$, so for the flat FRW space-time,
the field equations read,
\begin{align}
\label{FTFR1}
&\frac{TF_T}{3}-\frac{F}{6} +\frac{\kappa^2}{3}\rho_m=0\, , \\
\label{FTFR2}
&\dot{H}(F_T+2T F_{TT})=-\frac{\kappa^2}{2}(\rho_m+p_m)\, ,
\end{align}
with $\rho_m$ and $p_m$ being the matter energy density and the
pressure, and also for the flat FRW Universe we have,
\begin{equation}
\label{TH2}
T=-6H^2\, .
\end{equation}
The Hubble rate as a function of the scale factor is
$H=2 a^{-c^2/2}/2$. So we can express $N$ as a function of $T$,
\begin{equation}
\label{efoldNFT}
N=-\frac{1}{c^2}\ln \left(\frac{T}{\tilde{A}}\right)\, ,
\end{equation}
where,
\begin{equation}
\label{tildeA}
\tilde{A}=-\frac{24}{c^4} a_0^{-c^2}\, .
\end{equation}
In addition, by using
the continuity equation for the matter fluids, we can write,
\begin{equation}
\label{mattenrgydensBB}
\rho_m
=\sum_i\rho_{i0}a_0^{-3(1+w_i)}\e^{-3N(T)(1+w_i)}\, .
\end{equation}
By
substituting the above in Eq.~(\ref{FTFR1}) we have,
\begin{equation}
\label{bigdiffgeneral1FT}
T\frac{d F(T)}{d
T}-\frac{F(T)}{2}+\sum_iS_{i}T^{ \frac{3(1+w_i) }{c^2}}=0\, ,
\end{equation}
where $S_i$ is defined by (\ref{gfdgfdgf}). Hence the solution of
Eq.~(\ref{bigdiffgeneral1FT}) is actually the exact $F(T)$ gravity
that realizes the superbounce. Particularly, the vacuum $F(T)$
solution reads,
\begin{equation}
\label{FTsol1}
F(T)=c_1
\left(-T\right)^{\frac{1}{2}}\, ,
\end{equation}
with $c_1$ being an arbitrary
integration constant. In the case that the matter fluids are
present, the solution reads,
\begin{equation}
\label{newsolutionsnoneulerssssFT}
F(T)=c_1 \left(-T\right)^{\frac{1}{2}}- \sum_i Q_i
\left(-T\right)^{\frac{3(1+w_i)}{c^2}}\, ,
\end{equation}
with
\begin{equation}
Q_i=\frac{2\kappa^2 c^2 \rho_{i0} a_0^{-3(1+w_i)}}{6(1+w_i)-c^2}\, .
\end{equation}
As it can be seen, for $c=3(1+w_i)$ ($c=3$ for dust matter), the
superbounce cosmology can be realized by the teleparallel equivalent
of General Relativity.

Hence in this chapter we demonstrated how bouncing cosmology
scenarios can be realized in the context of $F(R)$, $f(\mathcal{G})$
and $F(T)$ gravity. We focused on the superbounce case and we also
provided a stability analysis of the resulting modified gravities,
with the stability referring to the dynamical system corresponding
to the cosmological equations. For further details on these issues
we refer the reader to Ref.~\cite{Odintsov:2015uca}.

Before closing this chapter we need to briefly discuss an important
feature of the bouncing cosmology evolution which we did not
address. Particularly, the reheating issue should be carefully
addressed for bouncing cosmologies in the context of modified
gravity. As we mentioned earlier the reheating era is extremely
important for a cosmological evolution, since even in the case of
bouncing cosmologies, the Universe should be reheated and in effect
the particle content of a cosmological theory should be excited.
However, in the literature this issue is not addressed appropriately
in the context of the modified gravities we discussed in this
review. To our knowledge, the only serious approach to the reheating
issue in the context of bouncing cosmology was performed in Ref.
\cite{Quintin:2014oea} (see also section 6 of Ref.
\cite{deHaro:2015wda} for an interesting approach on the matter
bounce scenario), where the authors used a two-field scalar model in
order to realize the bounce. Particularly, the study was performed
in the context of the ekpyrotic scenario, and some fields were
assumed to have $K$-essence terms, see \cite{Quintin:2014oea}. The
reheating issue should be appropriately addressed in the context of
the modified gravities we presented in this review, but this study
is beyond the scopes of this work and should be studied in a focused
research article.

\section{Conclusion}

With this review we aimed to provide the latest developments in
modified gravity, and also to provide a virtual toolbox for studying
inflation, dark energy and bouncing cosmologies in the context of
modified gravity. We tried to make the article self-contained, so we
discussed most of the existing forms of modified gravity that appear
in the literature and we presented in detail the theoretical
framework of each modified gravity model. We also presented the
essential information related to inflationary dynamics with various
modified gravity theories, and also we demonstrated how various
bouncing cosmologies can be realized by using various popular
modified gravity theories. Now the question is, what can be the
future in modified gravity theories, did we solved every existing
problem, so that modified gravity can be considered a theory very
well studied? The answer is categorically no, since even after many
years of extensive research, we believe that we are still in the
start of a long journey, and the theoretical challenges that need
firstly to be understood, and then correctly modeled, are quite
many. Practically, where the $\Lambda$CDM model fails to provide a
consistent description of the observable Universe, modified gravity
could fill in the gap. The $\Lambda$CDM model has many successes in
being compatible with the observational data, at least in the
pre-2000 observational data, since firstly, it predicted
successfully the location and the existence of the baryon acoustic
oscillations. Secondly, the $\Lambda$CDM model fits very well the
statistics of weak gravitational lensing. Thirdly, the polarization
of the Cosmic Microwave Background predicted by the $\Lambda$CDM
model, fits very well the 2015 Planck data \cite{Ade:2015lrj}.
Particularly, the temperature power spectrum (TT) peak, the
temperature polarization cross spectrum peaks (TE), and the
polarization spectrum (EE) peaks, are very conveniently accommodated
in the $\Lambda$CDM model predictions.

However, there are many challenging open problems and observational
data, that the $\Lambda$CDM model does not suffice to offer a
successful description for these
\cite{Perivolaropoulos:2008ud,Mendoza:2014qza}. Mainly, the dark
matter and dark energy problems haunt the $\Lambda$CDM model, since
no dark matter particle has yet been observed in a particle
accelerators or indirectly from direct dark matter searches. Also,
the dark energy seems to be undetectable by particle accelerators or
from any other laboratory apparatus, at least for the moment.
Moreover, the $\Lambda$CDM model predictions on large scales are
very successful, however, on sub-galactic scales, many issues occur
with regards to the predictions of many dwarf galaxies and an excess
of dark matter present in the innermost regions of the galaxies.
However, it is still debatable if these issues are actually problems
of the model itself or a problem of statistical analysis of the
current research approach on these issues. In addition, there are
several issues that are still in question with regard to the
$\Lambda$CDM model, since these are measured by taking into account
several imposed conditions. It is therefore possible that a modified
gravity extension of the $\Lambda$CDM model might offer consistent
answers to these problems. For example the redshift magnitude of
type Ia supernovae, depends strongly on the properties of stars that
exploded during the early-time era of the Universe, at least the
first stages of the evolution. Also the gravitational dynamics of
galaxies, gas and stars, and especially their relative motion,
strongly depends on the assumption that the light coming from stars
is the source of mass in the Universe. The same assumption applies
when one considers gravitational lensing. Hence in some sense, the
current data that fit the $\Lambda$CDM model offer an accurate
cosmological description, but not a precision cosmology. It is thus
possible that future data might not fit so well the $\Lambda$CDM
model. Also a future challenging task for modified gravity is to
explain why the cosmological constant is very small, which
conveniently allows for galaxies to form and not to tear apart due
to the gravitational acceleration.

There are two highly non-trivial problems that eventually modified
gravity might provide consistent answers. The first has to do with
the physics before the inflationary era, and in some studies, this
issue has already been addressed, since it is speculated that a
bounce preexisted the inflationary era. This issue is however highly
non-trivial and it requires much effort in order to find a
consistent answer on this. The second issue is related to the
question if the cosmos comes to a crushing end eventually. In
modified gravity it is possible that the Universe might abruptly end
its evolution on a crushing type finite-time singularity, like a Big
Rip or a Big Crush. The effects of these singularities on galactic
scales might be evident even some million years before the
occurrence of the singularity \cite{Caldwell:2003vq}, so the
question is if such a scenario is possible.

In addition, modified gravity provides an alternative view of
classical particle physics problems, like the baryogenesis issue.
Particularly, it is possible to generate non-zero baryon to entropy
ratio in the Universe by using the gravitational baryogenesis
mechanism \cite{Davoudiasl:2004gf}. Then, in the context of modified
gravity it is possible to generalize the gravitational baryogenesis
mechanism, and various proposals towards this issue have appeared in
the literature
\cite{Lambiase:2006dq,Odintsov:2016hgc,Oikonomou:2015qfh,
Saaidi:2010ey,Sadjadi:2007dx,Lambiase:2006ft,Bento:2005xk,
Feng:2004mq,Oikonomou:2016jjh,Oikonomou:2016pnq,
Odintsov:2016apy,Arbuzova:2016cem,Lima:2016cbh}.
It is conceivable that the phenomenology provided by modified
gravity is quite rich and research on this topic is still ongoing.

Moreover, modified gravity is one of the few theoretical frameworks
for which inflation and dark energy can be described in an unified
way, and we believe that the most successful model will be the one
which is most compatible with observations and theoretical
predictions.

Also a mysterious era to current cosmological research is the
pre-inflationary era. In the existing literature there exist various
proposal describing this pre-inflationary era, for example there are
considerations for a superinflationary era \cite{Cai:2015nya} which
occurs before the slow-roll inflationary era. Also a bouncing phase
before the inflationary era is also an appealing alternative to the
superinflationary scenario \cite{Cai:2015nya}. In both these
scenarios, the initial singularity is absent, so these are quite
appealing alternative scenarios to the standard Big Bang cosmology.
The motivation to look for a pre-inflationary era is supported by
certain features of the cosmic microwave background, and
particularly from the large scale power deficit of the TT-mode of
the cosmic microwave background. This issue is verified by the
latest Planck data \cite{Ade:2015lrj}, as was pointed out in
Ref.~\cite{Cai:2015nya}. Actually the large scale anomalies can occur due
to an existing contracting or expanding phase before the slow-roll
era, or some sort of an superinflationary phase. The corresponding
power spectrum of primordial curvature perturbations admit a large
scale cutoff, which can potentially explain the power deficit of the
TT-mode. The bounces preceding the inflationary phase, were proposed
in
Refs.~\cite{Cai:2015nya,Piao:2003zm,Piao:2005ag,Saidov:2010wx,
Cai:2008qb,Bamba:2016gbu},
and these are known as the ``bounce inflation scenario''. An
interesting proposal that appeared in the literature was the
singular bounce \cite{Odintsov:2015ynk} in which case a soft Type IV
singularity occurs at the bounce point. The power spectrum in the
context of $F(R)$ gravity is not scale invariant, so this scenario
could be combined with a slow-roll inflationary scenario which
occurs after the bounce. In these cases modified gravity offers a
conceptually simple description, so it is interesting to look for
combined scenarios in the context of modified gravity. It is
interesting to note that in the context of $f(\mathcal{G})$ gravity, the
singular bounce yields a nearly scale invariant power spectrum
\cite{Oikonomou:2015qha}, so this is intriguing, since a different
modified gravity description of the same cosmological evolution
yields entirely different results.

Moreover, with regard to the mysterious dark matter component of our
Universe, it needs to be verified whether it is a particle, in which
case the lightest supersymmetric particle is a viable candidate
\cite{Jungman:1995df} or it has a modified gravity geometric origin
\cite{Capozziello:2006uv,Nojiri:2008nt,Boehmer:2007kx}. Among many
possible experimental ways that may determine the nature of dark
matter, one promising to our opinion is the direct dark matter
determination \cite{Bai:2010hh} and especially the experiments that
use collisions of dark matter onto nuclei
\cite{Prezeau:2003sv,Oikonomou:2006mh,Cheung:2014pea}. The latter
method has the appealing feature of being costless in comparison to
other dark matter experiments. In some cases, it is possible that
direct dark matter searches may constraint the Big Bounce cosmology
\cite{Cheung:2014pea,Li:2015egy}, therefore the scientific future of
dark matter is full of surprises for cosmologists.

Another interesting perspective comes from the future observational
data on gravitational waves, which might bring new physics along,
related to modified gravity to some extend. As it was noted in
Refs.~\cite{Capozziello:2010iy,Basini:2016hkr} (see also
\cite{Smith:2005mm,Corda:2010zza,Kamionkowski:1997av,
Capozziello:2008fn,Bellucci:2008jt,Ananda:2007xh,Kamionkowski:1996zd}),
any newly discovered degrees of freedom of gravitational waves, may
in some sense reveal the modified gravity theory which may govern
such new degrees of freedom. It is certainly an arena for future
gravity tests, and for alternative gravity tests at a cosmological
scale, see \cite{Jain:2010ka} and also Ref.~\cite{Corda:2010zza} for
a study on massive relics of $F(R)$ gravity gravitational waves.
Along with primordial gravity waves and the $F(R)$ gravity degrees
of freedom, the issue of primordial magnetic fields can be viewed as
a consequence of $F(R)$ gravity, since magnetic fields are known to
become amplified in $F(R)$ gravity \cite{Lambiase:2008zz}. It is
important to note that a cosmological bounce and also the Big-Bang
inflationary scenario are just two out of many proposed cosmological
scenarios, like for example the emergent Universe scenario
\cite{Ellis:2003qz,Cai:2012yf}. The future cosmological observations
will indicate in a consistent way which cosmological scenario fits
the observational data in a successful way and also will determine
how inflation and dark energy are constrained
\cite{Feng:2004ad,Xia:2013dea}.

We need to note that one of the ultimate aims and challenges for
theoretical physicists at least the last 100 years is to find the
theory that unifies gravity with all other interactions. This could
be equivalent to finding a consistent quantum gravity theory, which
for some physicists is a rather intangible dream or a wishful
thought. To our opinion, nature represents its simplicity in various
physical systems, so it could be that microphysics systems, like
condensed matter physics, might eventually show the way on how
gravity manifests its quantum self. It can be that the
micro-geometry of space-time mimics to some extend microsystems and
modified gravity might play an important role in this interplay of
disciplines, see for example
Refs.~\cite{Lobo:2014nwa,Klinkhamer:2009fu} for an insightful study
on this. In principle we could go deeper from the level of curvature
fluctuations, and we could speak for and understand principles like
the metric fluctuations. There is a long way for completely
understanding these principles, so the future might bring surprises.
Also, the existence of the mysterious cold spots
\cite{Cruz:2008sb,Rudnick:2007kw,Larson:2004vm} in the cosmic
microwave background spectrum is an intriguing feature of the
late-time observations of our Universe, and the possible
explanations \cite{Cruz:2008sb} offer a fruitful background for
theoretical cosmology research. Eventually, modified gravity might
play an important role towards the understanding of the underlying
physics.

We believe that modified gravity offers aspects of possible
realities for our Universe, but in general it is an established
research principle in cosmology that the Universe will eventually
surprise us. Hence what we can do is to humbly try to find pieces of
the reality of our Universe. The time is right for new studies in
the cosmology of modified gravity, since the theoretical
cosmologists confront the same situation as the theoretical
physicists in the beginning of the twentieth century did, that is,
too many data need to be correctly interpreted. Also at the time
being, modified gravity is mainly a phenomenological theory aiming
for describing the evolution of the Universe. However, as we saw in
chapter II (super)string theory quite frequently induces
modifications to General Relativity. Furthermore, after the
discovery of gravitational waves, we also tend to believe in a
quantum gravity theory yet to be found. However, any quantum gravity
differs from General Relativity and in effect, it may be considered
as a sort of modified gravity. Also, by reconsidering standard
approaches in cosmology, like for example the slow-roll era, and by
replacing these with other conceptual approaches, like for example
an era of constant-roll during inflation, may bring new insights in
modified gravity cosmology \cite{Nojiri:2017qvx}. In effect,
modified gravity cosmology is a timely theoretical arena for the
young scientists minds.

Finally, in this work we reviewed some aspects of modified gravity
which are mainly related with our research interests. Clearly, quite
a number of topics and models were not discussed or addressed
properly. This is not strange, as it is impossible to cover all the
aspects of a vast and still quickly growing subject such as modified
gravity, in a single review. We aimed to address some important
features of inflation, dark energy and bouncing cosmology, and we
hope that we contributed positively in the field.

\section*{Acknowledgments}

This work is supported by MINECO (Spain), project
 FIS2013-44881, FIS2016-76363-P and by CSIC I-LINK1019 Project (S.D.O), also by Russian Ministry of Education and Science, Project No.
3.1386.2017 (S.D.O and V.K.O), and also by MEXT KAKENHI Grant-in-Aid
for Scientific Research on Innovative Areas ``Cosmic Acceleration''
(No. 15H05890) (S.N).

\end{document}